\documentclass[12pt]{oxarticle}
\pdfoutput=1

\usepackage{hyperref}
\usepackage[usenames, dvipsnames]{xcolor}
\usepackage{graphicx, float, array, xspace, amscd, amsthm, amssymb, latexsym, longtable, bbm, fancyhdr}
\usepackage[letterpaper, body={6.5in, 9.25in}, top=1.0in, left=1.5in]{geometry}
\usepackage[bbgreekl]{mathbbol}
\usepackage[utf8]{inputenc}
\usepackage[sort&compress, comma, square, numbers]{natbib}

\DeclareSymbolFontAlphabet{\mathbb}{AMSb}
\DeclareSymbolFontAlphabet{\mathbbl}{bbold}

\usepackage{amsmath}

\usepackage{multirow}
\usepackage{simplewick}
\usepackage{slashed}

\usepackage{pifont}
\newcommand{\cmark}{\ding{51}}%
\newcommand{\xmark}{\ding{55}}%

\usepackage[T1]{fontenc}
\usepackage{lmodern}

\usepackage{mmacells}
\usepackage{color}
\usepackage{float}

\usepackage{tikz}

\usepackage[many]{tcolorbox}

\usepackage{listings}
\DeclareFontFamily{U}{bbold}{}
\DeclareFontShape{U}{bbold}{m}{n}
{  <-5.5> s*[1.05] bbold5
	<5.5-6.5> s*[1.05] bbold6
	<6.5-7.5> s*[1.05] bbold7
	<7.5-8.5> s*[1.05] bbold8
	<8.5-9.5> s*[1.05] bbold9
	<9.5-11.5> s*[1.05] bbold10
	<11.5-16> s*[1.05] bbold12
	<16-> s*[1.05] bbold17
}{}

\newcommand{\Ii}{\mathbbm{i}}

\mmaDefineMathReplacement[≤]{<=}{\leq}
\mmaDefineMathReplacement[≥]{>=}{\geq}
\mmaDefineMathReplacement[≠]{!=}{\neq}
\mmaDefineMathReplacement[→]{->}{\to}[2]
\mmaDefineMathReplacement[⧴]{:>}{:\hspace{-.2em}\to}[2]
\mmaDefineMathReplacement{∉}{\notin}
\mmaDefineMathReplacement{∞}{\infty}
\mmaDefineMathReplacement{𝕕}{\mathbbm{d}}
\mmaDefineMathReplacement{α}{\alpha}
\mmaDefineMathReplacement{β}{\beta}
\mmaDefineMathReplacement{γ}{\gamma}

\mmaSet{
  morefv={gobble=2},
  linklocaluri=mma/symbol/definition:#1,
  morecellgraphics={yoffset=1.9ex}
}

\newtcolorbox{defn}{colback=red!5!white,colframe=red!75!black}
\newtcolorbox{funcDefn}{colback=red!5!white,colframe=red!75!black}
\newtcolorbox{objDefn}{colback=blue!5!white,colframe=blue!75!black}
\newtcolorbox{optDefn}{colback=green!5!white,colframe=green!75!black}

\newtcolorbox[auto counter,number within=section]{funcDefnN}[2][]{%
	colback=red!5!white,colframe=red!75!black,
	fonttitle=\bfseries,
	segmentation style={line width=0.5mm,solid, red!75!black},
	title=Function ~\thetcbcounter: #2,#1}
\newtcolorbox[auto counter,number within=section]{optDefnN}[2][]{%
	colback=green!5!white,colframe=green!75!black,fonttitle=\bfseries,title=Option ~\thetcbcounter: #2,#1}
\newtcolorbox[auto counter,number within=section]{objDefnN}[2][]{%
	colback=blue!5!white,colframe=blue!75!black,fonttitle=\bfseries,title=Object ~\thetcbcounter: #2,#1}

\newcommand{\m}[1]{\mmaUnd{\(#1\)}}
\newcommand{\mtu}[1]{\texttt{\textcolor{blue}{#1}}}
\newcommand{\mtd}[1]{\texttt{#1}}

\font\eightrm=cmr8 at 8pt
\font\csc=cmcsc10

\begin{document}

\pagestyle{empty}      
\begin{center}
	\null\vskip0.1in
	{\Huge "Gamma MaP" - A Mathematica Package for Clifford Algebras, Gamma Matrices and Spinors\\[0.2in]}
	\vskip15pt
	{\csc Pyry Kuusela\\[0.4in]}
	{\centering
		\it Mathematical Institute\\
		University of Oxford\\
		Andrew Wiles Building\\
		Radcliffe Observatory Quarter\\
		Oxford, OX2 6GG, UK\\}
	\vspace{1cm}
	{\centering	
	\href{mailto:pyry.r.kuusela@gmail.com}{\mtd{pyry.r.kuusela@gmail.com}}\\
	}

	\vskip30pt
	\begin{abstract}
	\vskip-5pt
	\noindent We present a Mathematica package for doing computations with gamma matrices, spinors, tensors and other objects, in any dimension and signature. The approach we use is based on defining the commutation relations of the relevant matrices, and is thus general and flexible. As examples, we reproduce torsion conditions for AdS$_3$ compactification of type IIB supergravity from literature, and check the vanishing of supersymmetry variations of 10-dimensional supersymmetric Yang-Mills theory.
	\end{abstract}
\end{center}

\newpage
{\renewcommand{\baselinestretch}{1.3}\tableofcontents}
\newpage

\pagestyle{fancy}
\setlength{\headheight}{14.5pt}
\renewcommand{\headrulewidth}{0pt}
\rhead{\thepage}
\lhead{\ifproofmode\eightrm draft: \today, \hourandminute \else {}\fi}
\cfoot{}

\setcounter{page}{1}

\section{Introduction}
\vspace{-5pt}
We present a Mathematica package, "Gamma MaP" (\underline{Gamma} \underline{Ma}trix \underline{P}ackage), for performing calculations involving gamma matrices, spinors and tensors in any dimension and signature. The main utility of the package is making these computations without needing to specify an explicit representations, but the package can also be used to perform computations with explicit representations. Our approach is substantially different from earlier packages, such as \cite{Poghosyan:2005xs,Gran:2001yh,Saulina:1995dq}, in various ways: the approach we use is based on defining basic objects, such as spinors, gamma matrices and tensors. Then we define functions implementing various operations involving these objects, including index contraction, non-commutative multiplication, and more. The leading idea is that any function can be used on any expression that can be constructed using the basic objects. This allows for a relatively simple and intuitive implementation of even complex operations, and ensures that the program can be easily extended. Furthermore, this approach allows for user-defined preferences for example concerning, where various intertwiner matrices should be commuted when simplifying expressions containing these. This makes the package ideal for many different types of calculations, since it does not make many assumptions about the form in which the user gets the simplified output. Instead, the user can modify most elements of this behaviour.

The approach which we have used to create the basic functionality of the package is based on essentially defining the commutation properties that the gamma matrices and different intertwiners satisfy. Then, based on the preferences of the user, matrices are automatically (anti-)commuted with each other, resulting in greatly simplified expressions in most cases. This approach could also be easily used in other cases that require commuting operators with given (anti-)commutation rules to a pre-defined order, such as when calculating CFT correlation functions.

In addition to the basic operations on gamma matrices, the package contains various operations familiar from many tensor packages (for example \cite{Wang:2013mea}) such as upper and lower indices compatible with Einstein summation convention, tensors with different symmetries, and partial derivatives. In addition to the new functionality, there is also some functionality that already exists in Mathematica, but that has been modified to better suit the needs of gamma matrix calculations. For example, we have defined new functions for making and using assumptions, and dealing with real and imaginary parts of expressions, which have been implemented with gamma matrix calculations in mind. 

Below we will present a brief review of relevant properties of gamma matrices, spinors and tensors following mainly \cite{VanProeyen:1999ni}, and explain how various properties and operations are implemented in this package. As examples, we reproduce a few results from literature \cite{Couzens:2017nnr,Brink:1976bc}. The structure of this paper is as follows. We begin in section \ref{sect_Review} with a review of the most important properties of spinors and gamma matrices in various dimensions. In section \ref{sect_Setup} we give instructions for installing and running the package, and defining the Clifford algebra, whose representation the gamma matrices then form. In the following section \ref{sect_Objects} we introduce basic objects such as antisymmetrised products of gamma matrices, and show how they are implemented and used in this package. In section \ref{sect_Functions} we list basic functions, such as multiplication of matrices, and show how to use these. Then, in section \ref{sect_Explicit_reps}, we explain how to use explicit representation for the gamma matrices, and other objects instead of using just the abstract definitions. In section \ref{sect_Subalgebras} we discuss using features related to subalgebras that are useful, for instance, when considering dimensional reduction. Finally, in section \ref{sect_Examples}, a few examples are provided by reproducing some lengthy calculations from literature. 

In appendices \ref{app_Objects}, \ref{app_Functions}, \ref{app_Options} and \ref{App_Reserved_Expressions}, we provide a reference list of objects, functions, options and other reserved expressions. In appendix \ref{app_Rep_Properties}, we give a list of properties satisfied by gamma matrices and intertwiners, and summarise the most important formulas from the review of section \ref{sect_Review}.
\newpage
\section{Review of Spinors and Gamma Matrices in Different Dimensions} \label{sect_Review}
\vspace{-5pt}
In this section, we review the basic properties of spinors and gamma matrices in different dimensions, and show how these are related, mainly following \cite{VanProeyen:1999ni}. This section also serves to fix the notation and conventions that we will be largely following in the following sections, where we show how these properties are implemented in the package.
\vspace{-5pt}
\subsection{Gamma Matrices} \label{sect_Review_gamma_matrices}
\vspace{-5pt}
Gamma matrices are matrices that satisfy the following anticommutation relations.
\begin{align}
\{\gamma^\mu, \gamma^\nu\} = 2 \eta^{\mu \nu} \label{Clifford_algebra_anticommutation}
\end{align}
for some diagonal matrix $\eta$ that has components $\eta^{\mu \nu}$, which take values $\pm 1$. We call $\eta^{\mu \nu}$ the metric, and the above anticommutation relation the Clifford algebra. Thus gamma matrices form representations of Clifford algebra. We call the indices $\alpha$ with $\eta^{\alpha \alpha} = -1$ timelike and denote their number by $t$. The indices $a$ with $\eta^{a a} = 1$ are called spacelike, and their number is denoted by $s$. We call the total number of diagonal entries $\eta^{\mu \nu}$ the dimension $d=s+t$. This is often called the dimension of the representation even though it is not the dimension of the representation of the Clifford algebra in the usual sense. For ease of notation we will assume throughout the review (but not the in the following sections) that the signature is of the form $(-\dotsc-+\dotsc+)$. The generalisation for any permutations of this form is straightforward.\\

In two (and three) dimensions with signature $(++)$ ($(+++)$), we note that one representation is given by the sigma matrices $\sigma_1$ and $\sigma_2$ (and $\sigma_3$) that are defined as
\begin{align}
\sigma_1 = \begin{pmatrix}
0 & 1 \\
1 & 0 
\end{pmatrix},~~~
\sigma_2 = \begin{pmatrix}
0 & -i \\
i & 0 
\end{pmatrix},~~~
\sigma_3 = \begin{pmatrix}
1 & 0 \\
0 & -1 
\end{pmatrix}.
\end{align}
Since these anticommute with each other and square to one, these satisfy the Clifford algebra for Euclidean signature.
\begin{align}
\{\sigma_i,\sigma_j\} = 2 \delta_{ij}.
\end{align}
For different signatures involving timelike directions, we can also use sigma matrices, since by multiplying the matrices corresponding to the timelike directions by $i$. For example, to form a three dimensional representation with signature $(-++)$, we can use the representation given by $\gamma^1 = i \sigma_1$, $\gamma^2 = \sigma_2$ and $\gamma^3 = \sigma_3$. 

As is well known, we can use sigma matrices to build higher dimensional representations as well by taking tensor product of these. In particular, there is a useful representation that we will in the following call the "special representation". For the case with no timelike directions, $(+\dotsc+)$, it is given by
\begin{equation}
\begin{split}
\gamma^1 &= \sigma_1 \otimes I \otimes I \otimes \dotsc,\\
\gamma^2 &= \sigma_2 \otimes I \otimes I \otimes \dotsc,\\
\gamma^3 &= \sigma_3 \otimes \sigma_1 \otimes I \otimes \dotsc,\\
\gamma^4 &= \sigma_3 \otimes \sigma_2 \otimes I \otimes \dotsc,\\
\gamma^5 &= \sigma_3 \otimes \sigma_3 \otimes \sigma_1 \otimes \dotsc,\\
&\dotsc\\
\gamma^d &= \sigma_3 \otimes \sigma_3 \otimes \dotsc \otimes \sigma_2 \text{ in even dimensions,}\\
\gamma^d &= \sigma_3 \otimes \sigma_3 \otimes \dotsc \otimes \sigma_3 \text{ in odd dimensions.} \label{gamma_matrices_special_rep}
\end{split}
\end{equation}
In case with timelike directions, the matrices corresponding to the timelike directions can be just multiplied by $i$ to get a representation with the correct signature. When speaking of the special representation, we assume that the signature is of the form $(-\dotsc-+\dotsc+)$ so that all the timelike directions are before the spacelike directions.

This representation is actually somewhat more general than it may appear at first. Namely, we easily see that the matrices $\gamma'_\mu$ that are related to the original matrices $\gamma_\mu$ by an unitary transformation $U$ still satisfy the same algebra, and thus give still a representation of the Clifford algebra. Therefore any matrices that are given by the expression of the form
\begin{align}
\gamma'_\mu = U \gamma_\mu U^{-1} \label{gamma_matrices_unitary_transformation}
\end{align}
also form a valid representation. In fact, it can be shown \cite{Freedman:2012zz} that up to unitary transformations there is one unique representation in even dimensions, and two inequivalent representations in odd dimensions. Therefore actually \textit{any} representation in an even dimension can be written in the form \eqref{gamma_matrices_unitary_transformation}. In odd dimensions the first of the two representations is exactly the special representation \eqref{gamma_matrices_special_rep}, and the second can be gotten by reversing the sign of the last matrix, $\gamma^d \to -\gamma^d$. We note that for these representations, we have exactly that $\gamma^d = \zeta \gamma_*$, where $\zeta=\pm 1$, and $\gamma_*$ is given by the product of all other gamma matrices, $\gamma_* = \gamma_0...\gamma_{d-1}$.
\vspace{-5pt}
\subsection{A-intertwiner and Hermiticity}
\vspace{-5pt}
Armed with the knowledge of the general form of the gamma matrices, we can immediately deduce the following Hermiticity properties:
\begin{equation}
\begin{split}
\gamma_\mu^\dagger &= - \gamma_\mu \text{ for timelike directions,}\\
\gamma_\mu^\dagger &= \gamma_\mu \text{ for spacelike directions.}
\end{split}
\end{equation}
Using the fact that the timelike gamma matrices square to minus identity, and that gamma matrices with different indices anticommute, we can easily verify, that these two relations are equivalent to
\begin{align}
\gamma_\mu^\dagger = (-1)^t ~(\gamma_0 \gamma_1 \dotsc \gamma_{t-1}) \, \gamma_\mu \, (\gamma_0 \gamma_1 \dotsc \gamma_{t-1})^{-1}.
\end{align}
where $t$ denotes the number of timelike directions $0,\dotsc,t-1$. We call the product of timelike gamma matrices appearing here the $A$-intertwiner.
\begin{align}
A = \gamma_0\dotsc\gamma_{t-1} = \frac{1}{t!} ~\gamma_{[0}\dotsc\gamma_{t-1]}. \label{A}
\end{align}
With the help of this matrix, we can write the Hermiticity relations of gamma matrices compactly as
\begin{align}
\gamma_\mu^\dagger = (-1)^t~A \gamma_\mu A^{-1}. \label{App_A_A_commutation}
\end{align}
It is easy to verify the following properties of the $A$-intertwiner that are true in any representation.
\begin{equation}
\begin{split}
A^\dagger &= A^{-1} = (-1)^{t(t+1)/2} A,\\
A^{*} &= (-1)^{t(t+1)/2} A^T.
\end{split} \label{A_properties_general}
\end{equation}
In addition, by using the expressions \eqref{gamma_matrices_special_rep}, we can verify the following properties for special representations.
\begin{equation}
\begin{split}
A^*&= (-1)^{\left\lfloor{t/2}\right\rfloor+t} A,\\
A^T &= (-1)^{\left\lfloor{t/2}\right\rfloor + t(t-1)/2 } A.
\end{split}
\label{A_properties_special}
\end{equation}
\vspace{-5pt}
\subsection{C-intertwiner and Symmetry}
\vspace{-5pt}
It is always possible to define a matrix $C$, called charge conjugation intertwiner (or $C$-intertwiner), that satisfies the following relations.
\begin{equation}
\begin{split}
\gamma_\mu^T &= - \eta \,C \gamma_\mu C^{-1},\\
C^T &= - \epsilon \,C, \label{C_relations}
\end{split}
\end{equation}
for some signs $\epsilon, \eta = \pm 1$.

For example, for the special representation mentioned before, these matrices are given by the following expressions
\begin{equation}
\begin{split}
C &= \sigma_2 \otimes \sigma_1 \otimes \sigma_2 \otimes \dotsc \text{ for $\eta=1$},\\
C &= \sigma_1 \otimes \sigma_2 \otimes \sigma_1 \otimes \dotsc \text{ for $\eta=-1$}. \label{C_special_rep}
\end{split}
\end{equation}
In even dimensions either of these can be used, but in odd dimensions only one of these satisfies the required properties \eqref{C_relations}. In other representations that are related to the special representation by some unitary transformations, a suitable $C$-intertwiner is given by
\begin{align}
C' = U^T C U.
\end{align}
Therefore, the $C$-intertwiner has in general the following properties.
\begin{equation}
\begin{split}
C^T &= -\epsilon \,C,\\
C^\dagger &= C^{-1} =-\epsilon \,C^*. \label{C_properties general}
\end{split}
\end{equation}
In addition for the special representation, we have.
\begin{equation}
\begin{split}
C&= C^\dagger = C^{-1},\\
C^* &= C^T = - \epsilon \, C. \label{C_properties_special}
\end{split}
\end{equation}
The signs $\epsilon$ and $\eta$ are related to the symmetry properties of the gamma matrices. Namely, notice that from the first relation \eqref{C_relations} follows that
\begin{align}
(C\gamma^{(n)})^T = - \epsilon  (-1)^{n(n-1)/2} (-\eta)^n \, C \gamma^{(n)}. \label{Cgamma_symmetry}
\end{align}
We can use this observation, together with a simple counting argument \cite{VanProeyen:1999ni} to determine, which signs $\epsilon$ and $\eta$ are allowed in different dimensions. We begin by noting that it is easy to show that in even dimensions the antisymmetrised products of gamma matrices $\{\gamma_{(n)} \equiv \gamma_{[\mu_1}\dotsc\gamma_{\mu_n]}|n=0,\dotsc,d\}$ form a basis in the space of all $2^{\left\lfloor{d/2}\right\rfloor} \times 2^{\left\lfloor{d/2}\right\rfloor}$ matrices (see section \ref{sect_Review_Basis}). Since The $C$-intertwiner is invertible, the set $\{C\gamma_{(n)}|n=0,\dotsc,d\}$ is also a basis. In odd dimensions a basis is formed by $\{C\gamma_{(n)}|n=0,\dotsc,(d-1)/2\}$. Therefore the number of (anti-)symmetric matrices in these bases must be the same as the number of (anti-)symmetric $2^{\left\lfloor{d/2}\right\rfloor} \times 2^{\left\lfloor{d/2}\right\rfloor}$ matrices. The number of antisymmetric $2^{\left\lfloor{d/2}\right\rfloor} \times 2^{\left\lfloor{d/2}\right\rfloor}$ matrices is
\begin{align}
2^{\left\lfloor{d/2}\right\rfloor-1}(2^{\left\lfloor{d/2}\right\rfloor}-1). \label{number_antisymmetric}
\end{align}
And the number of symmetric matrices is
\begin{align}
2^{\left\lfloor{d/2}\right\rfloor-1}(2^{\left\lfloor{d/2}\right\rfloor}+1). \label{number_symmetric}
\end{align}
On the other hand, the symmetry or antisymmetry of the matrices $C\gamma^{(n)}$ is determined by $\epsilon$ and $\eta$ via \eqref{Cgamma_symmetry}. The sign $- \epsilon  (-1)^{n(n-1)/2} (-\eta)^n$ is invariant under shifts of $n$ by 4. Therefore we have four cases $C\gamma^{(4j+0)},C\gamma^{(4j+1)},C\gamma^{(4j+2)}$, and $C\gamma^{(4j+3)}$ ($j=0,\dotsc$) which can be a priori symmetric or antisymmetric independently of one another. We can in principle choose $\epsilon$ and $\eta$ so that we can choose the sign for every one of these independently. However, the consistency requires that there must be \eqref{number_antisymmetric} antisymmetric and \eqref{number_symmetric} matrices. The number of matrices in the set $\{C\gamma^{(4j+0)}|j=0,\dotsc,\left\lfloor{d/4}\right\rfloor\}$ is given by
\begin{align}
{d \choose 0} + {d \choose 4} + \dotsc  &= 2^{d-2} + 2^{d/2-1}\cos\left(\frac{d \pi}{4}\right), \nonumber
\end{align}
and the number of matrices in the other sets is similarly. 
\begin{align}
\{C\gamma^{(4j+1)}\}: {d \choose 1} + {d \choose 5} + \dotsc  &= 2^{d-2} + 2^{d/2-1}\sin\left(\frac{d \pi}{4}\right), \nonumber\\
\{C\gamma^{(4j+1)}\}: {d \choose 2} + {d \choose 6} + \dotsc  &= 2^{d-2} - 2^{d/2-1}\cos\left(\frac{d \pi}{4}\right),\nonumber\\
\{C\gamma^{(4j+1)}\}: {d \choose 3} + {d \choose 7} + \dotsc  &= 2^{d-2} - 2^{d/2-1}\sin\left(\frac{d \pi}{4}\right).\nonumber
\end{align}
We must therefore find a way of choosing which of the sets $\{C\gamma^{(4j+0)}\}$, $\{C\gamma^{(4j+1)}\}$, $\{C\gamma^{(4j+2)}\}$ and $\{C\gamma^{(4j+3)}\}$ contain antisymmetric and which symmetric matrices. The consistency requires that the number of antisymmetric and symmetric matrices is given by the correct formulae \eqref{number_antisymmetric}, \eqref{number_symmetric}. The task is made significantly easier by noting that the problem is basically invariant under $d \to d+8$, so we need to only consider dimension modulo 8. We tabulate the possible solutions in table \ref{Table_epsilon_eta}.

\begin{table}[]
	\setlength{\doublerulesep}{3pt}
	\begin{center}
		\begin{tabular}{|c|c|c|r|r|}
			\hline
			d (mod 8)          & A & S & $\epsilon$ & $\eta$  \\ \hline \hline
			\multirow{2}{*}{0} & 1,2  &  0,3 & -1 & 1 \\ \cline{2-5} 
			& 2,3  &  0,1 & -1 & -1 \\ \hline
			1                  & 2,3 & 0,1  & -1 & -1 \\ \hline
			\multirow{2}{*}{2} &  2,3 & 0,1  & -1 & -1 \\ \cline{2-5} 
			& 0,3 & 1,2  & 1 & 1 \\ \hline
			3                  &  0,3 & 1,2  & 1 & 1 \\ \hline
			\multirow{2}{*}{4} & 0,3  & 1,2  & 1 & 1 \\ \cline{2-5} 
			& 0,1  & 2,3  & 1 & -1 \\ \hline
			5                  & 0,1  & 2,3  & 1 & -1 \\ \hline
			\multirow{2}{*}{6} &  0,1 & 2,3  & 1 & -1 \\ \cline{2-5} 
			& 1,2  & 0,3  & -1 & 1 \\ \hline
			7                  & 1,2  & 0,3  & -1 & 1 \\ \hline
		\end{tabular}
		\caption{The allowed values of $\epsilon$ and $\eta$ \cite{VanProeyen:1999ni}. The columns A ans S indicate, which of the matrices $\gamma_{(n)}$ ($n$ is taken modulo 4) are antisymmetric and which are symmetric. $\epsilon$ and $\eta$ are the possible values for the signs appearing in conditions \eqref{C_properties general}.} \label{Table_epsilon_eta}
	\end{center}
\end{table}
\vspace{-5pt}
\subsection{B-intertwiner and Complex Conjugation}
\vspace{-5pt}
Having discussed hermitian conjugation and transposition of the gamma matrices, the only basic operation that we are missing\footnote{The inversion properties of gamma matrices are determined by the Clifford algebra \eqref{Clifford_algebra_anticommutation} itself.} is the complex conjugation. It is natural to inquire whether there is an intertwiner that implements the complex conjugation on gamma matrices. The answer is obviously yes, since we can just combine the effects of the $A$-intertwiner and $C$-intertwiner to get this missing matrix, called the $B$-intertwiner. We therefore define
\begin{align}
B &= (A^{-1} C)^T. \label{B}
\end{align}
It is straightforward to verify that the $B$-intertwiner defined this way satisfies the following relation.
\begin{align}
\gamma_\mu^* &= - \eta (-1)^t \, B \gamma_\mu B^{-1}. \label{B_commutation}
\end{align}
Using the relevant properties of the $A$- and $C$-intertwiners, we can easily establish the following identities for $B$.
\begin{equation}
\begin{split}
B^\dagger &= B^{-1} = - \epsilon \eta^t (-1)^{t(t+1)/2}B^*,\\
B &=  - \epsilon \eta^t (-1)^{t(t+1)/2} B^T.
\end{split} \label{B_properties_general} 
\end{equation}
For the special representation the $B$-intertwiner can be expressed in terms of $A$ and $C$ as $B=(-1)^{t^2+1+\left\lfloor{t/2}\right\rfloor} \epsilon \, A C$, giving the following additional relations.
\begin{equation}
\begin{split}
B^{-1} &= \eta^t (-1)^{\left\lfloor{t/2}\right\rfloor + t(3t+1)/2} B,\\
B^T &= -(-\eta)^t(-1)^{3/2t(t-1)} \epsilon \, B,\\
B^* &= -(-1)^{(\left\lfloor{t/2}\right\rfloor +t)}\epsilon \, B,\\
B^\dagger &= (-\eta)^t(-1)^{(\left\lfloor{t/2}\right\rfloor +3t(t-1)/2+t)}B.
\end{split} \label{B_properties_special}
\end{equation}
\vspace{-5pt}
\subsection{Highest Rank Gamma Matrix $\gamma_*$}
\vspace{-5pt}
Finally, in even dimensions the product of all gamma matrices is linearly independent of all the gamma matrices as will be seen in section \ref{sect_Review_Basis}. To get as nice properties as possible, we define the matrix, called $\gamma_*$ (or sometimes $\gamma_{d+1}$, although we will not use this notation) by
\begin{align}
\gamma_* = (-i)^{d/2+t} \, \gamma^0\dotsc\gamma^{d-1}.
\end{align}
This matrix is Hermitian, anticommutes with all the gamma matrices, and squares to identity, giving us the following list of properties.
\begin{equation}
\begin{split}
\gamma_*\gamma_n \gamma_*^{-1} &= - \gamma_n,\\
\gamma_*^\dagger &= \gamma_*^{-1} = \gamma_*,\\
\gamma_*^T &= \gamma_*^*.
\end{split}
\end{equation}
In addition, in the special representation also the transpose and complex conjugation are equal to $\gamma_*$ itself.
\begin{align}
\gamma_*^T = \gamma_*^*  = \gamma_*.
\end{align}
\vspace{-5pt}
\subsection{Clifford Algebra and Lorentz Group}
\vspace{-5pt}
The main application of Clifford algebras in physics comes about from the fact that we can use representations of Clifford algebras to generate representations of Lorentz (Special Orthogonal) algebras. Namely, given a representation (a set of gamma matrices) of Clifford algebra $\text{Cliff}(t,d-t)$, the rank 2 antisymmetric products of the gamma matrices form a representation of $\text{SO}(t,d-t)$. The generators of the representation are given by
\begin{align}
\Sigma_{\mu \nu} = \frac{1}{4} \, \gamma_{[\mu \nu]}.
\end{align}
With a little work, it is easy to check that $\Sigma_{\mu \nu}$ defined so satisfy the correct commutation relations of $\text{SO}(t,d-t)$.
\begin{align}
[\Sigma_{\mu \nu}, \Sigma_{\alpha \beta}] = \eta_{\nu \alpha} \, \Sigma_{\mu \beta} - \eta_{\mu \alpha} \, \Sigma_{\nu \beta} - \eta_{\nu \beta} \, \Sigma_{\mu \alpha} + \eta_{\mu \beta} \, \Sigma_{\nu \alpha}.
\end{align}
The objects that transform under the transformations generated by $\Sigma_{\mu \nu}$ are called ($\text{Spin}(t,d-t)$) spinors. Explicitly the transformation rules of a spinor $\xi$ under Lorentz transformation parametrised by $\omega^{\mu \nu}$ is given by
\begin{align}
\xi \to e^{\frac{1}{2}\omega^{\mu \nu}\Sigma_{\mu \nu}} \xi.
\end{align}
The components of the spinors can be ordinary, commuting numbers, but often they are taken to be anticommuting Grassmann numbers, since this takes correctly the fermionic statistics of the spinors into account in many cases. For the most part, both types of spinors are handled similarly. The only difference is that when exchanging the order of spinors, for example when transposing a bilinear (see section \ref{sect_Review_BL_relations}), we have an additional minus sign coming from exchanging the order of the spinors, if they are anticommuting. In the remainder of this section, we will assume that the spinors are anticommuting, but the generalisation to anticommuting spinors should be obvious.\\

Under the Lorentz transformations, the antisymmetrised products of gamma matrices can be considered invariant tensors \cite{Georgi:1999wka} with the Lorentz indices transforming as a Lorentz vector, and the two spinor indices transforming under the spinor representation. To see this, let us start by noting that using the anticommutation relation \eqref{Clifford_algebra_anticommutation}, we can derive the following relation.
\begin{align}
-[\Sigma^{\alpha \beta},\gamma^\mu] = \eta^{\alpha \mu} \, \gamma^{\beta} + \eta^{\beta \mu} \, \gamma^\alpha.
\end{align}
The right-hand side can be written in a form from which we can recognise the components of the generators of the vector representation of the Lorentz group.
\begin{align}
-[\Sigma^{\alpha \beta},\gamma^\mu] = (\eta^{\alpha \mu} \, \delta^\beta_\nu + \eta^{\beta \mu} \, \delta^\alpha_\nu) \, \gamma^\nu = (M_{\alpha \beta})^\mu_{\phantom{\mu}\nu} \, \gamma^\nu,
\end{align}
where $M_{\alpha \beta}$ are the generators of the vector representation of the Lorentz group. Then this relation is nothing but the infinitesimal version of the following relation.
\begin{align}
e^{-\frac{1}{2}\omega^{\alpha \beta}\Sigma_{\alpha \beta}} \, \gamma^\mu \, e^{\frac{1}{2}\omega^{\alpha \beta}\Sigma_{\alpha \beta}} - \Lambda^\mu_{\phantom{\mu}\nu} \, \gamma^\nu = 0. 
\end{align}
So gamma matrix can indeed be interpreted as an invariant tensor with two indices transforming under spinor representation and one index transforming under the vector representation. By iterating this relation, we immediately get the transformation rules for all antisymmetric products of gamma matrices.
\begin{align}
e^{-\frac{1}{2} \omega^{\alpha \beta} \Sigma_{\alpha \beta}} \, \gamma^{\mu_1 \dotsc \mu_n} \, e^{\frac{1}{2} \omega^{\alpha \beta}\Sigma_{\alpha \beta}} - \Lambda^{\mu_1}_{\phantom{\mu_1}\nu_1}\dotsc \, \Lambda^{\mu_n}_{\phantom{\mu_n}\nu_n} \gamma^{\nu_1\dotsc\nu_n} = 0. \label{App_Lorentz_group_gamma_tensors}
\end{align}
This relation essentially allows us to commute the factor representing the Lorentz transformation on a spinor with an antisymmetrised product of gamma matrices, at a cost of introducing a factor representing a Lorentz transformation on the vector indices.
\vspace{-5pt}
\subsubsection{Spinor Bilinears and Tensors} \label{sect_Review_Spinor_BLs_and_Tensors}
\vspace{-5pt}
We can utilise the relations \eqref{App_Lorentz_group_gamma_tensors} to use spinors and gamma matrices to construct objects that transform as tensors under the Lorentz group. We begin by noting that from the Hermiticity relations \eqref{App_A_A_commutation} follows that
\begin{align}
(e^{\frac{1}{2}\omega^{\mu \nu}\Sigma_{\mu \nu}})^\dag = A \, e^{-\frac{1}{2}\omega^{\mu \nu}\Sigma_{\mu \nu}} \, A^{-1}.
\end{align}
Therefore the following combination is clearly a Lorentz scalar.
\begin{align}
\xi_i^\dag \, A \, \xi_j.
\end{align}
We call the combination $\xi_i^\dag A \equiv \bar{\xi}_i$ the Dirac conjugate of spinor $\xi_i$. Then, by using the relation \eqref{App_Lorentz_group_gamma_tensors}, we note that if we sandwich gamma matrices between a Dirac conjugated spinor and another spinor, we get objects that transform as tensors. Specifically, we have that under Lorentz transformations.
\begin{align}
\bar{\xi}_i \, \gamma^{\mu_1\dotsc\mu_n} \, \xi_j \to \Lambda^{\mu_1}_{\phantom{\mu_1}\nu_1}\dotsc \, \Lambda^{\mu_n}_{\phantom{\mu_n}\nu_n} \bar{\xi}_i \, \gamma^{\nu_1\dotsc\nu_n} \, \xi_j.
\end{align}
We call tensors built out of spinors in this fashion bilinears. In even dimensions, we also have $\gamma_*$ that commutes with $\Sigma_{\mu \nu}$, so that we can define spinor bilinears by including also $\gamma_*$ between the spinors. The objects defined this way transform under a Lorentz transformation in exactly the same way as the bilinears that do not include $\gamma_*$
\begin{align}
\bar{\xi}_i \, \gamma^{\mu_1\dotsc\mu_n} \gamma_* \, \xi_j \to \Lambda^{\mu_1}_{\phantom{\mu_1}\nu_1}\dotsc \, \Lambda^{\mu_n}_{\phantom{\mu_n}\nu_n} \bar{\xi}_i \, \gamma^{\nu_1\dotsc\nu_n} \gamma_* \, \xi_j.
\end{align}
The Lorentz scalar and vector formed thus are often called pseudoscalar and -vector, respectively due to the sign they pick up under parity transformations \cite{Peskin:1995ev}.

In addition to the Dirac conjugate, we can define so-called ("right") charge conjugate, denoted $\xi^{c}$ by
\begin{align}
\xi^{c} = B^{-1} \xi^*.
\end{align}
Using the defining relations \eqref{B_commutation} of the $B$-intertwiner, it is easy to see that the charge conjugate transforms under Lorentz transformations as
\begin{align}
\xi^c \to e^{\frac{1}{2}\omega^{\mu \nu}\Sigma_{\mu \nu}} \, \xi^c .
\end{align}
Therefore we can use the charge conjugate to define bilinears of the form
\begin{align}
\overline{\xi}_1 \, \gamma^{(n)} \, \xi_2^{c},
\end{align}
which transform as antisymmetric tensors under Lorentz transformations. Since $\xi^c$ transforms like $\xi$ under Lorentz transformations, we can immediately deduce that $\bar{\xi}^c = \xi^T (B^{-1})^\dagger A$ transforms like $\overline{\xi}$ under Lorentz transformations. We call $\overline{\xi}^c$ the left charge conjugate of $\xi$, and can naturally use it to build bilinears. By using the relations in appendix \ref{app_Rep_Properties}, we find the following proportionality.
\begin{align}
\xi^T (B^{-1})^\dagger A = -\epsilon (- \eta)^t (-1)^{t(t-1)/2} \xi^T C.
\end{align}
The latter expression is also often used as the definition of the "left" charge conjugate, and it differs from our definition at most by a sign. 
\vspace{-5pt}
\subsection{Irreducible Spinors} \label{sect_Irreducible_Spinors}
\vspace{-5pt}
By previous discussion, we know that gamma matrices can be used to build generators of a Lorentz algebra. The modules on which these generators act are called spinors. It is an interesting question, then, to find the irreducible modules i.e. minimal spinors for every such representation of the Lorentz algebra.
\vspace{-5pt}
\subsubsection{Weyl Spinors}
\vspace{-5pt}
We can formulate every irreducible spinor with some condition on the spinor. In even dimensions, we have an obvious candidate for such condition. Recalling that the last gamma matrix $\gamma_*$ anticommutes with every gamma matrix, it is then obvious that it commutes with the generators $M_{\mu \nu}$ of the Lorentz group. Therefore the spinors corresponding to some eigenvalue of $\gamma_*$ form an irreducible spinor. Since $\gamma_*^2 = I$, the possible eigenvalues are $\pm 1$, and since $\text{Tr}(\gamma_*) = 0$, these values must occur in pairs. Therefore there are two eigenspaces, corresponding to positive and negative eigenvalues of $\gamma_*$. We call these spinors left and right chiral (or positive and negative chiral) spinors, respectively. These both are called Weyl spinors. We can define projectors to the left and right chiral parts by
\begin{equation}
\begin{split}
\xi_L &= P_L \, \xi = \frac{1}{2}(1 + \gamma_*) \, \xi,\\
\xi_R &= P_R \, \xi = \frac{1}{2}(1 - \gamma_*) \, \xi.
\end{split}
\end{equation}
\vspace{-5pt}
\subsubsection{Majorana Spinors}
\vspace{-5pt}
The second type of condition we can try to impose on a spinor is a reality condition. To be completely general, we will assume that the complex conjugated spinor is related to some other transformation on the spinor, i.e. that
\begin{align}
\xi^* = D \, \xi
\end{align}
for some matrix $D$. In order to define an irreducible spinor, this condition must be compatible with Lorentz transformations (i.e. a "real" spinor should be mapped to a "real" spinor by Lorentz transformations). This means that we must have
\begin{align}
\left(\frac{1}{4} \gamma_{ab}\, \xi \right)^* = D \, \frac{1}{4} \, \gamma_{ab} \, \xi.
\end{align}
Using the reality condition for the second time, we can express this as
\begin{align}
\gamma_{ab}^* \, D \, \xi = D \, \gamma_{ab} \, \xi.
\end{align}
This is clearly equivalent to
\begin{align}
D^{-1}\gamma_{ab}^* \,D = \gamma_{ab}.
\end{align}
This condition is exactly satisfied by the matrix $B$ in \eqref{B_commutation}. Therefore we can, without loss of generality, take the reality condition, also called the \textit{Majorana condition}, to be
\begin{align}
\xi^* = B \, \xi.
\end{align}
To see if this condition can be imposed on a spinor without making the spinor vanish, we must consider the consistency condition $(\xi^*)^* = \xi$. This immediately leads to a condition on $B$, $B B^* = 1$. By comparing this to the earlier equation \eqref{B_properties_general}, we get a condition on the signs $\epsilon$ and $\eta$ and the number of timelike directions, $t$.
\begin{align}
- \epsilon \eta^t (-1)^{t(t+1)/2} = 1. \label{App_Irrep_Spinors_Majorana_consistency}
\end{align}
We recall from earlier that $\epsilon$ and $\eta$ can take values that depend on the dimension modulo 8, and $(-1)^{t(t+1)/2}$ clearly depends on $t$ modulo 4. Therefore, we can consider the cases one by one. It turns out that the condition is satisfied in the following cases:
\begin{equation}
\begin{split}
d-2t = 0,1,7 \text{ mod } 8,\\
d-2t = 2 \text{ mod } 8 \text{ if } (-1)^{d/2} \eta = 1,\\
d-2t = 6 \text{ mod } 8 \text{ if } (-1)^{d/2} \eta = -1.
\end{split}
\end{equation}
If one of these conditions is satisfied, then we can consistently define a spinor satisfying the Majorana condition. These spinors are called Majorana spinors.
\vspace{-5pt}
\subsubsection{Symplectic Majorana Spinors}
\vspace{-5pt}
If the consistency condition \eqref{App_Irrep_Spinors_Majorana_consistency} is not satisfied so that the left hand side is equal to -1, we can still impose a condition similar to Majorana condition on spinors if we have more than one spinor in our disposal. Namely, if we denote our spinors as $\xi_i$ we can use matrix $\Omega$ that operates on the index $i$ labelling the number of the spinor and impose a condition.
\begin{align}
\xi_i^* = \Omega_i^{\phantom{i}j} B \, \xi_j.
\end{align}  
The consistency condition $(\xi_i^*)^* = \xi_i$ is satisfied if $\Omega$ is an antisymmetric matrix that satisfies $\Omega^2 = - I$. Spinors satisfying this condition are called \textit{Symplectic Majorana spinors}. 
\vspace{-5pt}
\subsubsection{Majorana-Weyl Spinors}
\vspace{-5pt}
In some cases Majorana or Weyl spinors are not irreducible. This happens when it is possible to impose both Majorana and Weyl conditions simultaneously. Requiring that the Majorana condition preserves chirality gives us an additional consistency condition, 
\begin{align}
B P_{L/R} \, \xi_i &= (P_{L/R} \, \xi_i)^* = P_{L/R}^* \, \xi_i^* = \frac{1}{2}(1 \pm \gamma_*^*) \, \xi_i^*,\nonumber \\
&= (1 \pm (-1)^{d/2+t} \gamma_*^* ) \, B \, \xi_i =  (1 \pm (-\eta)^d (-1)^{d/2+t}  \gamma_*^* ) \, \xi_i^*.
\end{align}
Recalling that the Weyl spinors can only be defined in even dimensions, $d=2m$, we see that the Majorana condition respects chirality if and only if
\begin{align}
(-1)^{d/2+t}  =1.
\end{align}
This is satisfied exactly if
\begin{align}
d-2t = 0 \text{ mod } 4.
\end{align}
\vspace{-5pt}
\subsubsection{Symplectic Majorana-Weyl Spinors}
\vspace{-5pt}
We can similarly impose the symplectic Majorana condition and a Weyl condition simultaneously if the symplectic Majorana condition preserves chirality. Since the matrix $\Omega$ does not act on the spinor indices, the consistency condition in this case is exactly the same as for Majorana-Weyl spinors i.e.
\begin{align}
d-2t = 0 \text{ mod } 4.
\end{align}
We summarise the possible conditions on spinors in table \ref{Table_possible_spinors}.
\begin{table}[h!]
	\setlength{\doublerulesep}{3pt}
	\begin{tabular}{|l|l|l|l|l|l|}
		\hline
		\multicolumn{6}{|c|}{\textbf{t=0 (mod 4)}}                                                                              \\ \hline \hline
		d   (mod 8)      & M & W                 & MW               & SM               & SMW             \\ \hline 
		0                  & \cmark  &  \cmark  &    \cmark    &     \xmark &   \xmark   \\ \hline
		1                  &  \cmark &   \xmark   &  \xmark    &  \xmark & \xmark    \\ \hline
		\multirow{2}{*}{2} ($\eta = \phantom{+}1$)& \xmark   & \multirow{2}{*}{\cmark} & \multirow{2}{*}{\xmark} & \cmark & \multirow{2}{*}{\xmark} \\ \cline{2-2} \cline{5-5}
		~~\,($\eta = -1$)& \cmark   &                   &                   &   \xmark  &                   \\ \hline
		3                  & \xmark  &  \xmark   &  \xmark   &  \cmark  &  \xmark      \\ \hline
		4                  & \xmark &  \cmark  &   \xmark  &  \cmark &  \cmark    \\ \hline
		5                  & \xmark  &  \xmark     & \xmark  &  \cmark  &  \xmark     \\ \hline
		\multirow{2}{*}{6} ($\eta=\phantom{+}1$)&  \cmark  & \multirow{2}{*}{\cmark} & \multirow{2}{*}{\xmark} & \xmark & \multirow{2}{*}{\xmark} \\ \cline{2-2} \cline{5-5}
		~~\,($\eta=-1$)&  \xmark  &                   &                   &        \cmark           &                   \\ \hline
		7                  &  \cmark &   \xmark  &  \xmark  &   \xmark &  \xmark   \\ \hline
	\end{tabular}
	\begin{tabular}{|l|l|l|l|l|l|}
		\hline
		\multicolumn{6}{|c|}{\textbf{t=1 (mod 4)}}                                                       \\ \hline \hline
		d    (mod 8)  & M & W                 & MW                & SM & SMW               \\ \hline
		\multirow{2}{*}{0} ($\eta=\phantom{+}1$)& \xmark  &  \multirow{2}{*}{\cmark} & \multirow{2}{*}{\xmark} &  \cmark  & \multirow{2}{*}{\xmark} \\ \cline{2-2} \cline{5-5}
		~~\,($\eta=-1$)& \cmark  &                   &       &   \xmark &           \\ \hline
		1                  & \cmark  &  \xmark       & \xmark    &  \xmark  &  \xmark      \\ \hline
		2                  & \cmark  &  \cmark  &  \cmark   &  \xmark &   \xmark     \\ \hline
		3                  & \cmark  & \xmark    &  \xmark &  \xmark &   \xmark    \\ \hline
		\multirow{2}{*}{4} ($\eta=\phantom{+}1$)&  \cmark  & \multirow{2}{*}{\cmark} & \multirow{2}{*}{\xmark} &  \xmark  & \multirow{2}{*}{\cmark} \\ \cline{2-2} \cline{5-5}
		~~\,($\eta=-1$)& \xmark  &                   &                   &  \cmark &  \\ \hline
		5                  & \xmark  & \xmark     &   \xmark   &  \cmark &   \xmark   \\ \hline
		6                  &  \xmark &  \cmark    &  \xmark &  \cmark  &   \cmark  \\ \hline
		7                  & \xmark  &  \xmark  &  \xmark  &  \cmark  &  \xmark  \\ \hline
	\end{tabular}	\\
\\
\\
	\begin{tabular}{|l|l|l|l|l|l|}
		\hline
		\multicolumn{6}{|c|}{\textbf{t=2 (mod 4)}}                                                                              \\ \hline \hline
		d    (mod 8)   & M & W                 & MW               & SM               & SMW             \\ \hline 
		0                  & \xmark  &  \cmark  &    \xmark    &     \cmark &   \cmark   \\ \hline
		1                  &  \xmark &   \xmark   &  \xmark    &  \cmark & \xmark    \\ \hline
		\multirow{2}{*}{2} ($\eta =\phantom{+}1$)& \cmark   & \multirow{2}{*}{\cmark} & \multirow{2}{*}{\xmark} & \xmark & \multirow{2}{*}{\xmark} \\ \cline{2-2} \cline{5-5}
		~~\,($\eta = -1$)& \xmark   &      &    &   \cmark  &        \\ \hline
		3                  & \cmark  &  \xmark   &  \xmark   &  \xmark  &  \xmark      \\ \hline
		4                  & \cmark &  \cmark  &   \cmark  &  \xmark &  \xmark    \\ \hline
		5                  & \cmark  &  \xmark     & \xmark  &  \xmark  &  \xmark     \\ \hline
		\multirow{2}{*}{6} ($\eta=\phantom{+}1$)&  \xmark  & \multirow{2}{*}{\cmark} & \multirow{2}{*}{\xmark} & \cmark & \multirow{2}{*}{\xmark} \\ \cline{2-2} \cline{5-5}
		~~\,($\eta=-1$)&  \cmark  &  &   &   \xmark   	&      \\ \hline
		7                  &  \xmark &   \xmark  &  \xmark  &   \cmark &  \xmark   \\ \hline
	\end{tabular}
	\begin{tabular}{|l|l|l|l|l|l|}
		\hline
		\multicolumn{6}{|c|}{\textbf{t=3 (mod 4)}}                                                       \\ \hline \hline
		d   (mod 8)    & M & W                 & MW                & SM & SMW               \\ \hline
		\multirow{2}{*}{0} ($\eta=\phantom{+}1$)& \cmark  &  \multirow{2}{*}{\cmark} & \multirow{2}{*}{\xmark} &  \xmark  & \multirow{2}{*}{\xmark} \\ \cline{2-2} \cline{5-5}
		~~\,($\eta=-1$)& \xmark  &                   &       &   \cmark &           \\ \hline
		1                  & \xmark  &  \xmark       & \xmark    &  \cmark  &  \xmark      \\ \hline
		2                  & \xmark  &  \cmark  &  \xmark   &  \cmark &   \cmark     \\ \hline
		3                  & \xmark  & \xmark    &  \xmark &  \cmark &   \xmark    \\ \hline
		\multirow{2}{*}{4} ($\eta=\phantom{+}1$)&  \xmark  & \multirow{2}{*}{\cmark} & \multirow{2}{*}{\xmark} &  \cmark  & \multirow{2}{*}{\xmark} \\ \cline{2-2} \cline{5-5}
		~~\,($\eta=-1$)& \cmark  &                   &                   &  \xmark &  \\ \hline
		5                  & \cmark  & \xmark     &   \xmark   &  \xmark &   \xmark   \\ \hline
		6                  &  \cmark &  \cmark    &  \cmark &  \xmark  &   \xmark  \\ \hline
		7                  & \cmark  &  \xmark  &  \xmark  &  \xmark  &  \xmark  \\ \hline
	\end{tabular}\\
	\caption{Summary of the allowed conditions on spinors in different dimensions $d$ (mod 8) with different number, $t$, of timelike directions (mod 4). M stands for the Majorana condition, W for the Weyl conditions and MW for the case when both of these can be imposed simultaneously. SM stands for the symplectic Majorana condition in case there are multiple spinors, and SMW for the case when both the symplectic Majorana condition and the Weyl conditions can be imposed simultaneously.}
	\label{Table_possible_spinors}
\end{table}
\vspace{-5pt}
\subsection{Relations for Bilinears} \label{sect_Review_BL_relations}
\vspace{-5pt}
We can derive various relations for bilinears by using the fact that being scalars in spinor indices, transpose acts as identity on bilinears. Similarly, the Hermitian conjugation acts only as complex conjugation on bilinears. In this section, we assume that the spinors are commuting. The relations for anticommuting spinors have a different sign in each case.

For example, consider the following:
\begin{align}
(\overline{\xi}_1 \, \gamma^{(n)} \, \xi_2)^* = (\overline{\xi}_1  \, \gamma^{(n)} \, \xi_2)^\dagger = \xi_2^\dagger \, (\gamma^{(n)})^\dagger A^\dagger \, \xi_1 = (-1)^{nt}(-1)^{n(n-1)/2}(-1)^{t(t+1)/2} \, \overline{\xi}_2 \, \gamma^{(n)} \, \xi_1. \label{Eq_Complex_conjugate_Bilinear}
\end{align}
This relations basically allows us to change the order of two spinors in a bilinear at the cost of introducing complex conjugation, and possibly a sign. In addition, we can use this relation to determine whether a bilinear of the form $\overline{\xi} \, \gamma^{(n)} \, \xi$ is real or imaginary. 
\begin{align}
\overline{\xi} \, \gamma^{(n)} \, \xi \text{ is } 
\begin{cases}
\text{ real if } (-1)^{nt}(-1)^{n(n-1)/2}(-1)^{t(t+1)/2}=1,\\[3pt]
\text{ imaginary if } (-1)^{nt}(-1)^{n(n-1)/2}(-1)^{t(t+1)/2}=-1.
\end{cases}
\end{align}
It is also easy to derive a similar relation for bilinears that are of the form $\overline{\xi}_1 \, \gamma_* \gamma^{(n)} \, \xi_2$. We summarise these relations in appendix \ref{app_bilinear_relations}.

There are a few relations that simplify the expressions for bilinears in case they are formed out of irreducible spinors. For Weyl spinors, we note that from the definition of $\gamma_*$ it follows that spinors with different chiralities are orthogonal to each other, and that multiplying a spinor by a gamma matrix changes the chirality of the spinor. Using this idea, we easily derive the following relation for a bilinear constructed out of two spinors $\xi_{1}$ and $\xi_{2}$, whose chiralities we denote by $(\pm)_1$ and $(\pm)_2$
\begin{equation}
\begin{split}
\overline{\xi}_2 \, \gamma^{(n)} \, \xi_1 &= (\pm)_1 \xi_2^\dag \, A \gamma^{(n)} \gamma_* \, \xi_1 = (-1)^{n+t} (\pm)_1 (\gamma_* \, \xi_2)^\dag A \gamma^{(n)} \, \xi_1 \\
&= (-1)^{n+t} (\pm)_1 (\pm)_2 \overline{\xi}_2 \, \gamma^{(n)} \, \xi_1,
\end{split}
\end{equation}
where in the second step we used the fact that we have defined $\gamma_*$ so that $\gamma_*^\dag = \gamma_*$. From this we can infer when a bilinear formed out of two Weyl spinors vanishes.
\begin{align}
\overline{\xi}_2 \, \gamma^{(n)} \, \xi_1 = 0 \text{ if }
\begin{cases}
\xi_1 \text{ and } \xi_2 \text{ have the same chirality, and } (-1)^{n+t}=-1,\\[3pt]
\xi_1 \text{ and } \xi_2 \text{ have the oposite chiralities, and } (-1)^{n+t}=1.
\end{cases}
\end{align}
We can derive a similar relation for simplifying expressions with Majorana spinors. This time we will make use of the relations
\begin{equation}
\begin{split}
\overline{\xi}^T &= -\epsilon C \, \xi,\\
\xi^T &= \phantom{+}\overline{\xi} \, C^{-1},
\end{split}
\end{equation}
which follow directly from the Majorana condition and the definition of the $B$-intertwiner. Using this, we have the following.
\begin{equation}
\begin{split}
\overline{\xi}_1 \, \gamma^{(n)} \, \xi_2 &= (\overline{\xi}_1 \, \gamma^{(n)} \, \xi_2)^T = \xi_2^T \, (\gamma^{(n)})^T \, \overline{\xi}_1^T = -\epsilon \overline{\xi}_2 \, C^{-1} \gamma^{(n)} C \, \xi_1\\
&= -\epsilon (-\eta)^n (-1)^{n(n-1)/2} \overline{\xi_2} \, \gamma^{(n)} \, \xi_1. \label{Eq_Majorana_spinor_exchange}
\end{split}
\end{equation}
In general, this relation allows us to change the order of the spinors in a bilinear, provided that both are Majorana, but in a special case, where the spinors on both sides are equal, this allows us to determine when the bilinear vanishes automatically.
\begin{equation}
\overline{\xi} \, \gamma^{(n)} \, \xi = 0 \text{ if $\xi$ is Majorana and } \epsilon (-\eta)^n (-1)^{n(n-1)/2}=1
\end{equation}
There are also similar relations for bilinears containing $\gamma_*$, and other conjugates. We summarise these in appendix \ref{app_bilinear_relations}.
Especially nice relations are those for charge conjugate bilinears. For these, we can derive relations allowing us to basically change the order of the bilinears, at the cost of possibly introducing a sign. To derive these relations, we simply express the bilinear in a convenient form, and then transpose it.
\begin{equation}
\begin{split}
\overline{\xi}_1^c \, \gamma^{(n)} \, \xi_2 &= \pm \xi_1^T \, C \gamma^{(n)} \, \xi_2 = \pm \xi_2^T \, (\gamma^{(n)})^T C^T \, \xi_1 = \pm(-\epsilon)(-\eta)^n(-1)^{n(n-1)/2} \xi_2^T \, C \gamma^{(n)} \, \xi_1\\[3pt]
&= -\epsilon(-\eta)^n(-1)^{n(n-1)/2} \, \overline{\xi}_2^c \, \gamma^{(n)} \, \xi_1.
\end{split}
\end{equation}
In particular this allows us to deduce when certain charge conjugate bilinears vanish automatically.
\begin{align}
\overline{\xi}^c \, \gamma^{(n)} \, \xi = 0 \text{ if } \epsilon(-\eta)^n(-1)^{n(n-1)/2} = 1.
\end{align}
Furthermore, the bilinears that have charge conjugate spinors on both sides, i.e. those of the form $\overline{\xi}_1^c \, \gamma^{(n)} \, \xi_2^c$ are related to "normal" bilinears of the form $\overline{\xi}_2 \, \gamma^{(n)} \, \xi_1$ by the following relation.
\begin{align}
\overline{\xi}_1^c \, \gamma^{(n)} \, \xi_2^c = \eta^{n+t}(-1)^{3t(t+1)/2}(-1)^{n(n+1)/2} \, \overline{\xi}_2 \, \gamma^{(n)} \, \xi_1.
\end{align}
\vspace{-5pt}
\subsection{Basis for Clifford Algebra} \label{sect_Review_Basis}
\vspace{-5pt}
The full representation of the Clifford algebra $\text{Cliff}(t,d-t)$ consists not only from gamma matrices but also from their rank-$n$ antisymmetric products, which are defined for $n=1,\dotsc,d$ as $\gamma^{(n)} = \gamma_{[\mu_1}\dotsc\gamma_{\mu_n]}$ with $\mu_1,\dotsc,\mu_n$ taking values $0,\dotsc,d-1$ (or $1,\dotsc,d$) with $\mu_1 < \mu_2 < \dotsc < \mu_n$ and for $n=0$ as $\gamma_{(0)} = I$. In the case of even dimensions these are even all linearly independent. One easy way to see this is to introduce the following inner product.
\begin{align}
\langle \gamma_{\mu_1\dotsc\mu_n},\gamma^{\nu_1\dotsc\nu_m} \rangle = \frac{(-1)^{n(n-1)/2}}{2^{\left\lfloor{d/2}\right\rfloor}}\text{Tr}(\gamma_{\mu_1\dotsc\mu_n}\gamma^{\nu_1\dotsc\nu_m}) =\delta_{m n} \, \delta_{\mu_1}^{\nu_1}\dotsc \, \delta_{\mu_n}^{\nu_n},
\end{align}
The set of all rank-n antisymmetrised products of gamma matrices also forms a basis of the set of all $2^{\left\lfloor{d/2}\right\rfloor} \times 2^{\left\lfloor{d/2}\right\rfloor}$ matrices, which is easy to see by considering a counting argument: The set of all rank-$n$ matrices contains ${d\choose m}$ matrices, so the basis of the Clifford algebra consists of
\begin{align}
{d \choose 0} + {d \choose 1} + \dotsc + {d \choose d} = 2^d
\end{align}
matrices. This is clearly also the number of (complex) components of $2^{\left\lfloor{d/2}\right\rfloor} \times 2^{\left\lfloor{d/2}\right\rfloor}$ matrices, so the set must form a basis of all $2^{\left\lfloor{d/2}\right\rfloor} \times 2^{\left\lfloor{d/2}\right\rfloor}$ matrices.

In odd dimensions, the story is very similar, but with one crucial difference. In odd dimensions the matrix $\gamma^{d-1}$ (or $\gamma^{d}$) corresponds to $\gamma_*$ in a representation with dimension $d-1$. Thus not all the matrices are independent, since we have, for example, the following inner product that is non-zero.
\begin{align}
\langle \gamma_{d-1},\gamma_{0\dotsc d-2} \rangle = \langle (-i)^{(d-1)/2+t} \gamma^{0\dotsc d-2},\gamma_{0\dotsc d-2} \rangle = (-i)^{(d-1)/2+t} \neq 0.
\end{align}
Therefore, even though the gamma matrices are again complete in the space of $2^{\left\lfloor{d/2}\right\rfloor} \times 2^{\left\lfloor{d/2}\right\rfloor}$ matrices, not all the antisymmetrised products are independent, so that to form a basis, we must choose some subset of the matrices. 

It is relatively straightforward to show that we have the following relation between rank-$r$ and rank-$(d-r)$ matrices in odd dimensions.
\begin{align}
\gamma^{\mu_1\dotsc\mu_r} &= (-1)^t \zeta \frac{i^{(d-1)/2+t}}{(d-r)!} \epsilon^{\mu_1\dotsc\mu_d} \, \gamma_{\mu_d\dotsc\mu_{r+1}}. \label{Basis_dual_odd}
\end{align}
Therefore, to get a basis, it is enough to consider the set $\{\gamma_{(0)},\dotsc,\gamma_{(d-1)/2+1}\}$ in odd dimensions. 
As an aside, we note that there is a similar relation in even dimensions.
\begin{equation}
\gamma^{\mu_1\dotsc\mu_r} = (-1)^t\frac{i^{d/2+t}}{(d-r)!} \epsilon^{\mu_1\dotsc\mu_d} \gamma_* \gamma_{\mu_d\dotsc\mu_{r+1}} \label{Basis_dual_Even}
\end{equation}
\vspace{-5pt}
\subsubsection{Fierz Rearrangement}
\vspace{-5pt}
Using the completeness and orthogonality of the bases discussed above, we can easily derive a rule for expressing any matrix in these bases. Namely, we clearly have that
\begin{align}
M = \sum_\Gamma \langle M,\Gamma \rangle \Gamma,
\end{align}
where the sum runs over $\{\gamma_{(0)},\dotsc,\gamma_{(d)}\}$ in even dimensions and over  $\{\gamma_{(0)},\dotsc,\gamma_{(d-1)/2+1}\}$ in odd dimensions. One important special case of this formula can be derived by considering the situation where the matrix $M$ is given by an outer product of spinors, such as $M = \xi \chi^\dag$. Then, by plugging this in, and using the fact that $Tr(\xi \chi^\dag \gamma_{(n)}) = \chi^\dagger \gamma_{(n)} \xi$, we get the following formula in odd dimensions $d=2m+1$.
\begin{equation}
\begin{split}
\xi \, \chi^\dag &= \frac{1}{2^m} \big(\chi^\dagger \, \xi +\gamma_{\mu_1} \, \chi^\dagger \, \gamma^{\mu_1} \xi - \sum_{\mu_1 < \mu_2} \gamma_{\mu_1 \mu_2} \chi^\dagger \, \gamma^{\mu_1 \mu_2} \, \xi + \dotsc \\[3pt]
&\phantom{=}\hskip25pt \,+\sum_{\mu_1 < \mu_2 <\dotsc<\mu_m} (-1)^{m(m-1)/2}\, \gamma_{\mu_1\dotsc\mu_m} \, \chi^\dagger \, \gamma^{\mu_1\dotsc\mu_m} \, \xi \big). 
\end{split}
\end{equation}
A slightly more convenient form can be achieved by letting the sum run over all indices, and compensating this with a combinatorial factor.
\begin{equation}
\begin{split}
\xi \, \chi^\dag &= \frac{1}{2^m} \big(\chi^\dagger \, \xi +\gamma_{\mu_1} \, \chi^\dagger \, \gamma^{\mu_1} \, \xi -\frac{1}{2} \gamma_{\mu_1 \mu_2} \, \chi^\dagger \, \gamma^{\mu_1 \mu_2} \, \xi + \dotsc \\[3pt]&
\phantom{=}\hskip25pt \,+ \frac{(-1)^{m(m-1)/2}}{m!} \,\gamma_{\mu_1\dotsc\mu_m} \, \chi^\dagger \, \gamma^{\mu_1\dotsc\mu_m} \, \xi \big). 
\end{split}
\label{sect_Review_Basis_Fierz_basic_odd}
\end{equation}
This basic identity can be multiplied by gamma matrices and spinors to get a set of relations between different spinor bilinears. For example, we can derive
\begin{equation}
\begin{split}
\beta^\dagger \, \gamma^{a_1\dotsc,a_i} \, \xi \, \chi^\dag \, \gamma^{b_1\dotsc,b_j} \, \delta &= \frac{1}{2^m} \big(\beta^\dagger \, \gamma^{a_1\dotsc,a_i} \, \gamma^{b_1\dotsc,b_j} \, \delta \, \chi^\dagger \, \xi +\beta^\dagger \, \gamma^{a_1\dotsc,a_i} \, \gamma_{\mu_1} \, \gamma^{b_1\dotsc,b_j} \, \delta \, \chi^\dagger \, \gamma^{\mu_1} \, \xi + \dotsc\\[3pt]
&\phantom{=} \hskip25pt \, + \frac{(-1)^{m(m-1)/2}}{m!} \, \beta^\dagger \, \gamma^{a_1\dotsc,a_i} \, \gamma_{\mu_1\dotsc\mu_m} \, \gamma^{b_1\dotsc,b_j} \, \delta \, \chi^\dagger \, \gamma^{\mu_1\dotsc\mu_m} \, \xi \big). 
\end{split}\label{sect_Review_Basis_Fierz_odd}
\end{equation}
This is usually called the Fierz rearrangement formula.

In even dimensions we have a completely analogous relation, which can be derived in a completely similar fashion. The basic formula for expressing spinor outer products is
\begin{equation}
\begin{split}
\xi \, \chi^\dag =& \frac{1}{2^m} \big(\chi^\dagger \, \xi + \chi^\dagger \, \gamma_* \, \xi +\gamma_{\mu_1} \, \chi^\dagger \, \gamma^{\mu_1} \, \xi +\gamma_{\mu_1} \, \chi^\dagger \, \gamma^{\mu_1} \gamma_* \, \xi + \dotsc \\[3pt] 
& \hskip25pt  \, + \frac{(-1)^{m(m-1)/2}}{m!} \, \gamma_{\mu_1\dotsc\mu_m} \gamma_* \, \chi^\dagger \, \gamma^{\mu_1\dotsc\mu_m} \gamma_* \, \xi \big). \label{sect_Review_Basis_Fierz_basic_even}
\end{split}
\end{equation}
This can be again used to derive the Fierz rearrangement formula in even dimensions.
\begin{equation}
\begin{split}
\beta^\dagger \, \gamma^{a_1\dotsc,a_i} \, \xi \, \chi^\dag \, \gamma^{b_1\dotsc,b_j} \, \delta &= \frac{1}{2^m} \big(\beta^\dagger \, \gamma^{a_1\dotsc,a_i}\gamma^{b_1\dotsc,b_j} \, \delta \, \chi^\dagger \, \xi +\beta^\dagger \, \gamma^{a_1\dotsc,a_i}\gamma^{b_1\dotsc,b_j} \, \delta \, \chi^\dagger \gamma_{*} \xi + \dotsc\\[3pt]
&\phantom{=}\hskip25pt  + \frac{(-1)^{m(m-1)/2}}{m!}  \beta^\dagger \, \gamma^{a_1\dotsc,a_i} \gamma_{\mu_1\dotsc\mu_m} \gamma_* \gamma^{b_1\dotsc,b_j} \, \delta  \chi^\dagger \, \gamma^{\mu_1\dotsc\mu_m} \gamma_* \, \xi \big). \label{sect_Review_Basis_Fierz_even}
\end{split}
\end{equation}
\vspace{-5pt}
\subsubsection{Expressing Gamma Matrix Products in Terms of the Basis}
\vspace{-5pt}
Since we have above derived a basis for all $2^{\left\lfloor{d/2}\right\rfloor} \times 2^{\left\lfloor{d/2}\right\rfloor}$ matrices, and in particular those forming the representation of the Clifford algebra, it is useful to be able to express all the gamma matrix products in this basis. This can be done by basically iterating the relation \eqref{Clifford_algebra_anticommutation} defining the Clifford algebra. The most basic example would be to express the product of two gamma matrices $\gamma^{\mu}\gamma^{\nu}$ in terms of the basis for the Clifford algebra. By adding $\gamma^{\mu \nu}$ to both sides of \eqref{Clifford_algebra_anticommutation}, we arrive immediately at expression.
\begin{align}
\frac{1}{2}\{ \gamma^{\mu}, \gamma^\nu \} + \gamma^{\mu \nu} = \gamma^{\mu} \gamma^\nu = \eta^{\mu \nu} + \gamma^{\mu \nu}. \label{App_Basis_antisymmetrisation_relation}
\end{align} 
Then, with more complicated matrix products, we can just iterate this rule. For example, expressing the product of four gamma matrices $\gamma^{\mu} \gamma^\nu \gamma^\sigma \gamma^\lambda$ can be done as follows:
\begin{equation}
\begin{split}
\gamma^{\mu} \gamma^\nu \gamma^\sigma \gamma^\lambda 
&= \gamma^\mu \gamma^\nu (\gamma^{\sigma \lambda} + \eta^{\sigma \lambda})\\[3pt]
&= \gamma^\mu (\eta^{\nu \sigma} \gamma^{\lambda} - \eta^{\nu \lambda} \gamma^\sigma + \gamma^{\nu \sigma \lambda} + \gamma^\nu  \eta^{\sigma \lambda}) \\[3pt] 
&= \eta^{\nu \sigma} \gamma^\mu \gamma^{\lambda} - \eta^{\nu \lambda} \gamma^\mu  \gamma^\sigma + \gamma^\mu \gamma^{\nu \sigma \lambda} + \gamma^\mu \gamma^\nu  \eta^{\sigma \lambda}\\[3pt]
&= \eta^{\nu \sigma} (\eta^{\mu \lambda} + \gamma^{\mu \lambda}) - \eta^{\nu \lambda} (\eta^{\mu \sigma} + \gamma^{\mu \sigma}) + (\eta^{\mu \nu} \gamma^{\sigma \lambda} - \eta^{\mu \sigma} \gamma^{\nu \lambda} + \eta^{\mu \lambda} \gamma^{\nu \sigma} + \gamma^{\mu \nu \sigma \lambda}) \\[3pt] 
&\phantom{=} \, \, \, + (\eta^{\mu \nu} + \gamma^{\mu \nu})  \eta^{\sigma \lambda} \\[3pt]
&= \eta^{\mu \nu} \eta^{\sigma \lambda} - \eta^{\mu \sigma} \eta^{\nu \lambda} + \eta^{\mu \lambda} \eta^{\nu \sigma} + \eta^{\mu \nu} \gamma^{\sigma \lambda} - \eta^{\mu \sigma} \gamma^{\nu \lambda} + \eta^{\mu \lambda} \gamma^{\nu \sigma} + \eta^{\nu \sigma} \gamma^{\mu \lambda} \\[3pt] 
&\phantom{=} \, \, \,- \eta^{\nu \lambda} \gamma^{\mu \sigma} + \eta^{\sigma \lambda} \gamma^{\mu \nu} + \gamma^{\mu \nu \sigma \lambda}.
\end{split}\raisetag{40pt}
\end{equation}
This rather cumbersome and unwieldy computation can also be expressed using more compact diagrammatical rules, using notation similar to Wick contractions. We represent the relation \eqref{App_Basis_antisymmetrisation_relation} by the following.
\begin{align}
\gamma^\mu \gamma^\nu  = \contraction{}{\gamma^\mu}{}{\gamma^\nu} \gamma^\mu \gamma^\nu + \gamma^\mu \gamma^\nu = \eta^{\mu \nu} + \gamma^{\mu \nu}.
\end{align}
Note a possible source of confusion: on the right-hand side the term $\gamma^{\mu} \gamma^\nu$ denotes now the antisymmetrised product of gamma matrices. The exact rules are:
\begin{enumerate}
	\item Add together all possible contractions.
	\item The contractions $\contraction{}{\gamma^\mu}{}{\gamma^\nu} \gamma^\mu \gamma^\nu$ correspond to $\eta^{\mu \nu}$.
	\item Uncontracted gamma matrices correspond to gamma matrices with all appearing indices antisymmetrised.
	\item The sign of each term is determined by anticommuting matrices in every contraction next to each other before using rule 2.
\end{enumerate}
To illustrate the usage of these rules, let us redo the example above using, contractions this time.
\begin{equation}
\begin{split}
\gamma^\mu \gamma^\nu \gamma^\sigma \gamma^\lambda &= \contraction{}{\gamma^\mu}{}{\gamma^\nu} \contraction{\gamma^\mu \gamma^\nu}{\gamma^\sigma}{}{\gamma^\lambda} \gamma^\mu \gamma^\nu \gamma^\sigma \gamma^\lambda + \contraction{}{\gamma^\mu}{\gamma^\nu}{\gamma^\sigma} \contraction[2ex]{\gamma^\mu}{\gamma^\nu}{\gamma^\sigma}{\gamma^\lambda} \gamma^\mu \gamma^\nu \gamma^\sigma \gamma^\lambda + \contraction[2ex]{}{\gamma^\mu}{\gamma^\nu \gamma^\sigma}{ \gamma^\lambda} \contraction{\gamma^\mu}{\gamma^\nu}{}{\gamma^\sigma} \gamma^\mu \gamma^\nu \gamma^\sigma \gamma^\lambda + \contraction{}{\gamma^\mu}{}{ \gamma^\nu} \gamma^\mu \gamma^\nu \gamma^\sigma \gamma^\lambda + \contraction{}{\gamma^\mu}{\gamma^\nu}{\gamma^\sigma} \gamma^\mu \gamma^\nu \gamma^\sigma \gamma^\lambda + \contraction{}{\gamma^\mu}{\gamma^\nu \gamma^\sigma}{\gamma^\lambda} \gamma^\mu \gamma^\nu \gamma^\sigma \gamma^\lambda \\
&\phantom{=}+\contraction{\gamma^\mu}{\gamma^\nu}{}{\gamma^\sigma} \gamma^\mu \gamma^\nu \gamma^\sigma \gamma^\lambda +\contraction{\gamma^\mu}{\gamma^\nu}{\gamma^\sigma}{\gamma^\lambda} \gamma^\mu \gamma^\nu \gamma^\sigma \gamma^\lambda +\contraction{\gamma^\mu\gamma^\nu}{\gamma^\sigma}{}{\gamma^\lambda} \gamma^\mu \gamma^\nu \gamma^\sigma \gamma^\lambda + \gamma^\mu \gamma^\nu \gamma^\sigma \gamma^\lambda \\[5pt]
&= \eta^{\mu \nu} \eta^{\sigma \lambda} - \eta^{\mu \sigma} \eta^{\nu \lambda} + \eta^{\mu \lambda} \eta^{\nu \sigma} + \eta^{\mu \nu} \gamma^{\sigma \lambda} - \eta^{\mu \sigma} \gamma^{\nu \lambda} + \eta^{\mu \lambda} \gamma^{\nu \sigma} + \eta^{\nu \sigma} \gamma^{\mu \lambda}\nonumber \\ &\phantom{=}- \eta^{\nu \lambda} \gamma^{\mu \sigma} + \eta^{\sigma \lambda} \gamma^{\mu \nu} + \gamma^{\mu \nu \sigma \lambda}.
\end{split}
\end{equation}
These rules have to be modified only slightly to be able to accommodate cases in which we have products of antisymmetrised products of gamma matrices. We have to introduce one new rule to avoid contractions that vanish due to antisymmetrisation.
\begin{enumerate}
	\setcounter{enumi}{4}
	\item Do not contract indices that are antisymmetrised with each other.
\end{enumerate}
To illustrate this, let us consider the case of $\gamma^{\mu \nu} \gamma^{\sigma \lambda}$.
\begin{equation}
\begin{split}
\gamma^{\mu \nu} \gamma^{\sigma \lambda} &= \contraction{}{\gamma^{[\mu}}{\gamma^{\nu]}}{\gamma^{[\sigma}} \contraction[2ex]{\gamma^{[\mu} }{\gamma^{\nu]}}{\gamma^{[\sigma}}{\gamma^{\lambda]}} \gamma^{[\mu} \gamma^{\nu]} \gamma^{[\sigma} \gamma^{\lambda]} + \contraction[2ex]{}{\gamma^{[\mu}}{\gamma^{\nu]}\gamma^{[\sigma}}{\gamma^{\lambda]}} \contraction{\gamma^{[\mu} }{\gamma^{\nu]}}{}{\gamma^{[\sigma}} \gamma^{[\mu} \gamma^{\nu]} \gamma^{[\sigma} \gamma^{\lambda]} +\contraction{}{\gamma^{[\mu}}{\gamma^{\nu]}}{\gamma^{[\sigma}} \gamma^{[\mu} \gamma^{\nu]} \gamma^{[\sigma} \gamma^{\lambda]} +\contraction{}{\gamma^{[\mu}}{\gamma^{\nu]}\gamma^{[\sigma}}{\gamma^{\lambda]}} \gamma^{[\mu} \gamma^{\nu]} \gamma^{[\sigma} \gamma^{\lambda]} +\contraction{\gamma^{[\mu}}{\gamma^{\nu]}}{}{\gamma^{[\sigma}} \gamma^{[\mu} \gamma^{\nu]} \gamma^{[\sigma} \gamma^{\lambda]} \\ &\phantom{=}+\contraction{\gamma^{[\mu}}{\gamma^{\nu]}}{\gamma^{[\sigma}}{\gamma^{\lambda]}} \gamma^{[\mu} \gamma^{\nu]} \gamma^{[\sigma} \gamma^{\lambda]} + \gamma^{[\mu} \gamma^{\nu]} \gamma^{[\sigma} \gamma^{\lambda]} \\[5pt]
&=- \eta^{\mu \sigma} \eta^{\nu \lambda} + \eta^{\mu \lambda} \eta^{\nu \sigma} - \eta^{\mu \sigma} \gamma^{\nu \lambda} + \eta^{\mu \lambda} \gamma^{\nu \sigma} + \eta^{\nu \sigma} \gamma^{\mu \lambda} - \eta^{\nu \lambda} \gamma^{\mu \sigma} + \gamma^{\mu \nu \lambda \sigma}.
\end{split}
\end{equation}
There is also a completely general formula for computing expressions like this, given in \cite{vanHolten:1982mx}.
\begin{equation}
\begin{split}
&\gamma_{\mu_1 \dotsc \mu_n} \gamma^{\nu_1 \dotsc \nu_m} = \sum_{i=|n-m|}^{n+m} \frac{n! m!}{s_i! v_i! u_i!} \, \delta^{[\nu_1}_{[\mu_n} \, \dotsc \, \delta^{\nu_{s_i}}_{\mu_{v_i+1}} \gamma_{\mu_1\dotsc\mu_{v_i}]}^{\phantom{\mu_1\dotsc\mu_{v_i}]}\nu_{s_i+1}\dotsc\nu_m]},\\[3pt]
&s_i = \frac{1}{2}(n+m-i), v_i = \frac{1}{2}(n-m+i), u_i = \frac{1}{2}(m-n+i)
\end{split}
\end{equation}
We can even generalise these rules to cases where we have repeated indices that we sum over, although the most general rules will get quite involved. We need to introduce contractions of more than two indices. Namely, we can contract the repeated indices to arbitrarily many indices. Therefore we add a new rule:
\begin{enumerate}
	\setcounter{enumi}{5}
	\item When there are repeated indices, contract the index pair with one another, and then add all contractions, where arbitrarily many other indices are be contracted with other indices.
\end{enumerate}
To show how this rule works, we consider the product $\gamma^\mu \gamma^\nu \gamma^\sigma \gamma_\mu$. Then all the contractions in this case are given by
\begin{align}
\gamma^\mu \gamma^\nu \gamma^\sigma \gamma_\mu = \contraction{}{\gamma^\mu}{\gamma^\nu\gamma^\sigma}{\gamma_\mu} \gamma^\mu \gamma^\nu \gamma^\sigma \gamma_\mu + \contraction[2ex]{}{\gamma^\mu}{\gamma^\nu\gamma^\sigma}{\gamma_\mu} \contraction{\gamma^ \mu}{\gamma^\nu}{}{\gamma^\sigma} \gamma^\mu \gamma^\nu \gamma^\sigma \gamma_\mu + \contraction{}{\gamma^\mu}{\gamma^\nu\gamma^\sigma}{\gamma_\mu} \contraction{\gamma^ \mu}{\gamma^\nu}{\gamma^\sigma}{\gamma_\mu} \gamma^\mu \gamma^\nu \gamma^\sigma \gamma_\mu + \contraction{}{\gamma^\mu}{\gamma^\nu\gamma^\sigma}{\gamma_\mu} \contraction{\gamma^ \mu\gamma^\nu}{\gamma^\sigma}{}{\gamma_\mu} \gamma^\mu \gamma^\nu \gamma^\sigma \gamma_\mu + \contraction{}{\gamma^\mu}{\gamma^\nu \gamma^\sigma}{\gamma_\mu} \gamma^\mu \gamma^\nu \gamma^\sigma \gamma_\mu + \contraction{}{\gamma^\mu}{\gamma^\nu\gamma^\sigma}{\gamma_\mu} \contraction{\gamma^ \mu}{\gamma^\nu}{}{\gamma^\sigma} \gamma^\mu \gamma^\nu \gamma^\sigma \gamma_\mu
\end{align}
Then we also need an additional rules for determining what we should replace the new contractions with. 
\begin{enumerate}
	\setcounter{enumi}{6}
	\item Contractions that contain only repeated indices that are summed over, are replaced by a combinatorial factor given by
	\begin{equation}
	\begin{split}
		(d-m)(d-m-1)\dotsc(d-m-n),
	\end{split}
	\end{equation}
	where $m$ is the number of contractions in which we include "empty contractions" corresponding to gamma matrices that are not contracted with any other matrices, and $n$ is the number of contractions involving repeated indices. In case the first term is negative, we take the expression to be zero.
	\item In terms that involve contractions with both repeated sum indices and fixed indices that appear only once, the gamma matrices with sum indices are replaced by 1 after commuting them next to one another. Commuting a matrix past another matrix in the same contraction gives 1, and commuting matrix past index in another contraction (or uncontracted matrix) gives -1 as usual. Then in the remaining term, the contractions that involved gamma matrices with indices that were summed over, are removed. Then the remaining term is treated using the rules 2-4, and the following rule.
	\begin{align}
	\contraction{}{\gamma^{\mu_1}}{}{\gamma^{\mu_2}} \contraction{\gamma^{\mu_1}}{\gamma^{\mu_2}}{}{\gamma^{\mu_3}} \contraction{\gamma^{\mu_1}\gamma^{\mu_2}}{\gamma^{\mu_3}}{\dotsc}{\gamma^{\mu_n}} \gamma^{\mu_1} \gamma^{\mu_2} \gamma^{\mu_3} \dotsc\gamma^{\mu_n} =
	\begin{cases} 	\contraction{}{\gamma^{\mu_1}}{}{\gamma^{\mu_2}} \contraction{\gamma^{\mu_1}\gamma^{\mu_2}}{\gamma^{\mu_3}}{}{\gamma^{\mu_4}} \contraction{\gamma^{\mu_1}\gamma^{\mu_2}\gamma^{\mu_3}\gamma^{\mu_4}\dotsc}{\gamma^{\mu_{n-1}}}{}{\gamma^{\mu_n}} \gamma^{\mu_1} \gamma^{\mu_2} \gamma^{\mu_3} \gamma^{\mu_4} \dotsc\gamma^{\mu_{n-1}}\gamma^{\mu_n} = \eta^{\mu_1 \mu_2} \eta^{\mu_3 \mu_4} \dotsc \eta^{\mu_{n-1} \mu_n} \text{ if } n \text{ is even},\\[3pt]
	\contraction{}{\gamma^{\mu_1}}{}{\gamma^{\mu_2}} \contraction{\gamma^{\mu_1}\gamma^{\mu_2}}{\gamma^{\mu_3}}{}{\gamma^{\mu_4}} \gamma^{\mu_1} \gamma^{\mu_2} \gamma^{\mu_3} \gamma^{\mu_4} \dotsc\gamma^{\mu_n} = \eta^{\mu_1 \mu_2} \eta^{\mu_3 \mu_4} \dotsc \gamma^{\mu_n} \text{ if } n \text{ is odd}.
	\end{cases}\raisetag{25pt}
	\end{align}
	Finally, to every such term we add the condition that the indices that are contracted are equal, and those that are not contracted are not equal.
\end{enumerate}
To illustrate the rule 7, we compute below the combinatorial factors corresponding to a few different contractions.

\begin{equation}
\begin{tabular}{ l c l }
$\contraction{}{\gamma^\mu}{\gamma^\nu\gamma^\sigma}{\gamma_\mu} \gamma^\mu \gamma^\nu \gamma^\sigma \gamma_\mu:$ & \qquad & $d-2$, \\[6pt]
$\contraction{}{\gamma^\mu}{\gamma^\nu\gamma^\sigma}{\gamma_\mu} \gamma^\mu \gamma^\nu \gamma^\sigma \gamma_\mu \gamma^\lambda :$ & \qquad & $d-3,,$ \\[6pt]
$\contraction[2ex]{}{\gamma^\mu}{\gamma^\nu\gamma^\sigma}{\gamma_\mu} \contraction{\gamma^ \mu}{\gamma^\nu}{}{\gamma^\sigma} \gamma^\mu \gamma^\nu \gamma^\sigma \gamma_\mu:$ & \qquad & $d-1,$ \\[6pt]
$\contraction[2ex]{}{\gamma^\mu}{\gamma^\nu\gamma^\sigma}{\gamma_\mu} \contraction{\gamma^\mu}{\gamma^\nu}{}{\gamma^\sigma} \gamma^\mu \gamma^\nu \gamma^\sigma  \gamma_\mu \gamma^\lambda:$ & \qquad & $d-2,$\\[6pt]
$\contraction{}{\gamma^\mu}{\gamma^\nu\gamma^\sigma}{\gamma_\mu} \contraction{\gamma^ \mu}{\gamma^\nu}{\gamma^\sigma}{\gamma_\mu} \gamma^\mu \gamma^\nu \gamma^\sigma \gamma_\mu:$ & \qquad & $1,$\\[6pt]
$\contraction{\gamma^\nu}{\gamma^\mu}{\gamma^\sigma \gamma^\lambda}{\gamma_\mu} \contraction[2ex]{}{\gamma^\nu}{\gamma^\mu \gamma^\sigma \gamma^\lambda \gamma_\mu}{\gamma_\nu}  \gamma^\nu \gamma^\mu \gamma^\sigma \gamma^\lambda \gamma_\mu \gamma_\nu$ & \qquad & $(d-2)(d-3),$\\[6pt]
$\contraction{\gamma^\nu}{\gamma^\mu}{\gamma^\sigma \gamma^\lambda}{\gamma_\mu} \contraction{}{\gamma^\nu}{\gamma^\mu \gamma^\sigma \gamma^\lambda \gamma_\mu}{\gamma_\nu}  \gamma^\nu \gamma^\mu \gamma^\sigma \gamma^\lambda \gamma_\mu \gamma_\nu:$ & \qquad & $(d-2).$\\
\end{tabular}
\end{equation}
Especially the last point of rule 8 probably requires additional explanation, so we give an example of its usage, and calculate the product $\gamma^\mu \gamma^\alpha \gamma^\beta \gamma^\delta \gamma_\mu$. It is perhaps easiest to think of terms in this products falling into five different cases, depending on the contractions of free indices that are not summed over. These cases are exactly the conditions referred in the last point of rule 8.
\begin{equation}
\begin{split}
\gamma^\mu \gamma^\nu \gamma^\sigma \gamma^\delta \gamma_\mu &=
\begin{cases}
\contraction{}{\gamma^\mu}{\gamma^\alpha\gamma^\beta \gamma^\delta}{\gamma_\mu} \gamma^\mu \gamma^\alpha \gamma^\beta \gamma^\delta \gamma_\mu +\contraction{}{\gamma^\mu}{\gamma^\alpha\gamma^\beta \gamma^\delta}{\gamma_\mu} \contraction{}{\gamma^\mu}{}{\gamma_\alpha} \gamma^\mu \gamma^\nu \gamma^\beta \gamma^\delta \gamma_\mu +\contraction{}{\gamma^\mu}{\gamma^\nu\gamma^\beta \gamma^\delta}{\gamma_\mu} \contraction{}{\gamma^\mu}{\gamma_\alpha}{\gamma^\beta} \gamma^\mu \gamma^\alpha \gamma^\beta \gamma^\delta \gamma_\mu +\contraction{}{\gamma^\mu}{\gamma^\alpha\gamma^\beta \gamma^\delta}{\gamma_\mu} \contraction{}{\gamma^\mu}{\gamma_\alpha\gamma^\beta}{\gamma^\delta} \gamma^\mu \gamma^\alpha \gamma^\beta \gamma^\delta \gamma_\mu, \text{ if }\alpha \neq \beta \neq \delta, \alpha \neq \delta, \\[3pt]
\contraction[2ex]{}{\gamma^\mu}{\gamma^\alpha\gamma^\beta\gamma^\delta}{\gamma_\mu} \contraction{\gamma^\mu\gamma^\alpha}{\gamma^\beta}{}{\gamma^\delta}\gamma^\mu \gamma^\alpha \gamma^\beta \gamma^\delta \gamma_\mu + \contraction{}{\gamma^\mu}{\gamma^\alpha\gamma^\alpha\gamma^\alpha}{\gamma_\mu} \contraction{\gamma^\mu\gamma^\alpha}{\gamma^\alpha}{}{ \gamma^\alpha}\gamma^\mu \gamma^\alpha \gamma^\beta \gamma^\delta \gamma_\mu + \contraction{}{\gamma^\mu}{\gamma^\alpha\gamma^\alpha\gamma^\alpha}{\gamma_\mu} \contraction{}{\gamma^\mu}{}{\gamma^\alpha}\gamma^\mu \gamma^\alpha \gamma^\beta \gamma^\delta \gamma_\mu, \text{ if }\alpha \neq \beta = \delta,\\[3pt]
\contraction[2ex]{}{\gamma^\mu}{\gamma^\alpha\gamma^\beta\gamma^\delta}{\gamma_\mu} \contraction{\gamma^\mu}{\gamma^\alpha}{}{\gamma^\beta}\gamma^\mu \gamma^\alpha \gamma^\beta \gamma^\delta \gamma_\mu + \contraction{}{\gamma^\mu}{\gamma^\alpha\gamma^\alpha\gamma^\alpha}{\gamma_\mu} \contraction{\gamma^\mu}{\gamma^\alpha}{}{\gamma^\alpha} \gamma^\mu \gamma^\alpha \gamma^\beta \gamma^\delta \gamma_\mu + \contraction{}{\gamma^\mu}{\gamma^\alpha\gamma^\alpha\gamma^\alpha}{\gamma_\mu} \contraction{}{\gamma^\mu}{\gamma^\alpha\gamma^\alpha}{\gamma^\alpha}\gamma^\mu \gamma^\alpha \gamma^\beta \gamma^\delta \gamma_\mu, \text{ if }\alpha = \beta \neq \delta,\\[3pt]
\contraction[2ex]{}{\gamma^\mu}{\gamma^\alpha\gamma^\beta\gamma^\delta}{\gamma_\mu} \contraction{\gamma^\mu}{\gamma^\alpha}{\gamma^\beta}{\gamma^\delta}\gamma^\mu \gamma^\alpha \gamma^\beta \gamma^\delta \gamma_\mu + \contraction{}{\gamma^\mu}{\gamma^\alpha\gamma^\alpha\gamma^\alpha}{\gamma_\mu} \contraction{\gamma^\mu}{\gamma^\alpha}{\gamma^\alpha}{\gamma^\alpha} \gamma^\mu \gamma^\alpha \gamma^\beta \gamma^\delta \gamma_\mu + \contraction{}{\gamma^\mu}{\gamma^\alpha\gamma^\alpha\gamma^\alpha}{\gamma_\mu} \contraction{}{\gamma^\mu}{\gamma^\alpha}{\gamma^\alpha}\gamma^\mu \gamma^\alpha \gamma^\beta \gamma^\delta \gamma_\mu, \text{ if }\alpha = \delta \neq \beta,\\[3pt]
\contraction[2ex]{}{\gamma^\mu}{\gamma^\alpha\gamma^\beta\gamma^\delta}{\gamma_\mu} \contraction{\gamma^\mu}{\gamma^\alpha}{\gamma^\beta}{\gamma^\delta} \contraction{\gamma^\mu}{\gamma^\alpha}{}{\gamma^\beta}\gamma^\mu \gamma^\alpha \gamma^\beta \gamma^\delta \gamma_\mu + \contraction{}{\gamma^\mu}{\gamma^\alpha\gamma^\beta\gamma^\delta}{\gamma_\mu} \contraction{\gamma^\mu}{\gamma^\alpha}{\gamma^\alpha}{\gamma^\alpha} \contraction{\gamma^\mu}{\gamma^\alpha}{}{\gamma^\alpha} \gamma^\mu \gamma^\alpha \gamma^\beta \gamma^\delta \gamma_\mu, \text{ if }\alpha = \beta = \delta,
\end{cases}\\[5pt]
&=
\begin{cases}
-(d-3) \gamma^{\alpha \beta \delta} + \gamma^{\alpha \beta \delta} + \gamma^{\alpha \beta \delta} + \gamma^{\alpha \beta \delta}, \text{ if }\alpha \neq \beta \neq \delta, \alpha \neq \delta,  \\[3pt]
-(d-2) \gamma^\alpha \eta^{\beta \delta} - \gamma^\alpha \eta^{\beta \delta} + \gamma^\alpha \eta^{\beta \delta}, \text{ if }\alpha \neq \beta = \delta,\\[3pt]
-(d-2) \gamma^\delta \eta^{\alpha \beta} - \gamma^\delta \eta^{\alpha \beta} + \gamma^\delta \eta^{\alpha \beta}, \text{ if }\alpha = \beta \neq \delta,\\[3pt]
-(d-2) \gamma^\beta \eta^{\alpha \delta} - \gamma^\beta \eta^{\alpha \delta}  + \gamma^\beta \eta^{\alpha \delta}, \text{ if }\alpha = \delta \neq \beta,\\[3pt]
-(d-1)\contraction{}{\gamma^\alpha}{\gamma^\beta}{\gamma^\delta} \contraction{}{\gamma^\alpha}{}{\gamma^\beta} \gamma^\alpha \gamma^\beta \gamma^\delta + \contraction{}{\gamma^\alpha}{\gamma^\beta}{\gamma^\delta} \contraction{}{\gamma^\alpha}{}{\gamma^\beta} \gamma^\alpha \gamma^\beta \gamma^\delta, \text{ if }\alpha = \beta = \delta
\end{cases}\\[5pt]
&=
\begin{cases}
-(d-6) \gamma^{\alpha \beta \delta}, \text{ if }\alpha \neq \beta \neq \delta, \alpha \neq \delta,\\[3pt]
-(d-2)\gamma^\alpha \eta^{\beta \delta}, \text{ if }\alpha \neq \beta = \delta,\\[3pt]
-(d-2)\gamma^\delta \eta^{\alpha \beta}, \text{ if }\alpha = \beta \neq \delta,\\[3pt]
-(d-2)\gamma^\beta \eta^{\alpha \delta}, \text{ if }\alpha = \delta \neq \beta,\\[3pt]
-(d-2) \gamma^\alpha \eta^{\beta \delta}, \text{ if }\alpha = \beta = \delta.
\end{cases}
\end{split}\raisetag{175pt}
\end{equation}
We could alternatively express this for example as
\begin{align}
-(d-2) \gamma^\alpha \eta^{\beta \delta} \delta_{\alpha \beta} -(d-6) \gamma^{\alpha \beta \delta} - \begin{cases}
(d-2)\gamma^\alpha \eta^{\beta \delta} \text{ if } \alpha \neq \beta = \delta,\\[3pt]
(d-2)\gamma^\delta \eta^{\alpha \beta} \text{ if } \alpha = \beta \neq \delta,\\[3pt]
(d-2)\gamma^\beta \eta^{\alpha \delta} \text{ if } \alpha = \delta \neq \beta,
\end{cases}
\end{align}
where no sum over the repeated index is implied. There are again some general formulae for computations like this. For example \cite{vanHolten:1982mx}:
\begin{equation}
\gamma_{\mu_1 \dotsc \mu_n} \gamma_{\nu_1 \dotsc \nu_m} \gamma^{\mu_1 \dotsc \mu_n} =(-1)^{n(n-1)/2+mn} n! \sum_{i=0}^{\text{min}[m,n]} (-1)^i {m\choose i} {d-m \choose k-i} \gamma_{\nu_1 \dotsc \nu_m}.
\end{equation}
\vspace{-5pt}
\subsection{Subalgebras} \label{sect_Review_Subalgebras}
\vspace{-5pt}
Often, such as when considering dimensional reduction, it is useful to consider subalgebras $\text{Cliff}(n) \subset \text{Cliff}(d)$ of the full Clifford algebra $\text{Cliff}(d)$. 
\vspace{-5pt}
\subsubsection{Gamma Matrices}
\vspace{-5pt}
In practice, to treat subalgebras, we decompose the gamma matrices forming a representation of $\text{Cliff}(d)$ in terms of the gamma matrices generating the subalgebras $\text{Cliff}(n)$ and $\text{Cliff}(d-n)$. For example, in the case where we start with an even-dimensional Clifford algebra $\text{Cliff}(d)$ and consider two odd-dimensional subalgebras $\text{Cliff}(n)$ and $\text{Cliff}(d-n)$. The gamma matrices decompose as.
\begin{equation}
\begin{split}
\gamma^\alpha &= \gamma_{(1)}^\alpha \otimes I_{(2)} \otimes \sigma_1 \text{ for $\alpha = 1,\dotsc,n$},\\[3pt]
\gamma^\mu &= I_{(1)} \otimes \gamma_{(2)}^\mu \otimes \sigma_2 \text{ for $\mu = n+1,\dotsc,d$}.
\end{split}
\end{equation}
Here the lower-dimensional gamma matrices satisfy the anticommutation relations
\begin{equation}
\begin{split}
\{\gamma_{(1)}^\alpha,\gamma_{(1)}^\beta\} = \eta^{\alpha \beta},\\[3pt]
\{\gamma_{(2)}^\mu,\gamma_{(2)}^\nu\} = \eta^{\mu \nu},
\end{split}
\end{equation}
where $\eta$ is the same as that appearing in \eqref{Clifford_algebra_anticommutation}. Note that these decompositions are not unique. We could have, for example, chosen the following representation, which also decomposes the $d$-dimensional representation into $n$- and $d-n$-dimensional ones.
\begin{equation}
\begin{split}
\gamma^\alpha &= \gamma_{(1)}^\alpha \otimes I_{(2)} \otimes \sigma_2 \text{ for $\alpha = 1,\dotsc,n$},\\[3pt]
\gamma^\mu &= I_{(1)} \otimes \gamma_{(2)}^\mu \otimes \sigma_3 \text{ for $\mu = n+1,\dotsc,d$}.
\end{split}
\end{equation}
However, by previous discussion we know that there is only one representation in even dimensions, up to unitary transformations, so every decomposition is equivalent.

There are similar relations for the other two qualitatively different cases, one in which an even-dimensional representation is decomposed in terms of two even dimensional representations, and the other in which an odd-dimensional representation is decomposed into an even-dimensional an odd-dimensional representation. We summarise these in appendix \ref{App_Matrix_Properties_Decomposition}.
\vspace{-5pt}
\subsubsection{Intertwiners}
\vspace{-5pt}
The intertwiner matrices can also be decomposed in terms of lower dimensional matrices. For the $A$-intertwiner the decomposition is simple, since the intertwiner can still be given in terms of gamma matrices, and the decomposition of these induces the desired decomposition. If $\mu_0,\dotsc,\mu_{t_1-1}$ denote the timelike indices of the first subalgebra, $t_1$ being the number of timelike directions corresponding to the first subalgebra, and $\alpha_0,\dotsc,\alpha_{t_2-1}$ the timelike indices of the second subalgebra, we have that
\begin{equation}
\begin{split}
A &= \gamma^{\mu_1}\dotsc\gamma^{\mu_{t_1-1}}\gamma^{\alpha_0}\dotsc\gamma^{\alpha_{t_2-1}} = \gamma^{\mu_0}_{(1)}\dotsc\gamma^{\mu_{t_1-1}}_{(1)}\gamma_{*(1)}^{t_2} \otimes \gamma^{\alpha_0}_{(2)}\dotsc\gamma^{\alpha_{t_2-1}}_{(2)} \otimes \sigma_1^{t_1} \sigma_2^{t_2} \\& = A_{(1)} \gamma_{*(1)}^{t_2} \otimes A_{(2)} \otimes \sigma_1^{t_1} \sigma_2^{t_2}.
\end{split}
\end{equation}
The decomposition of $C$-intertwiner is not quite as straightforward as that of the $A$-intertwiner given that there is no general expression for it in terms of gamma matrices. Instead, we will use the definition of the $C$-intertwiner through the relation \eqref{C_properties general}, and express $C$ and gamma matrices in terms of tensor products. It is clear that the decomposition must be of the form
\begin{align}
C = C_{(1)} \otimes C_{(2)} \otimes \sigma_i,
\end{align}
for some $2\times 2$ matrix that we suggestively denote as $\sigma_i$. Now, let us take a look at, what the commutation relations of gamma matrices and $C$ written in this form look like, and impose that these give the right relation \eqref{C_properties general}. This gives us the following conditions.
\begin{equation}
\begin{split}
\eta_1 &= \pm \eta, ~~~~~~ \sigma_i^{-1} \sigma_1 \sigma_i = \pm \sigma_1,\\
\eta_2 &= \pm \eta, ~~~~~~ \sigma_i^{-1} \sigma_2 \sigma_i = \pm \sigma_2.
\end{split}
\end{equation}
Which conditions we must choose depends on the allowed values of $\eta_1$ and $\eta_2$, which are uniquely determined by the dimensions $n$ and $d-n$. The solutions to the conditions on $\sigma_i$ are trivial to find, giving us the following possible expressions for the $C$-intertwiner.

\begin{equation}
\begin{tabular}{ l c l }
$C = C_{(1)} \otimes C_{(2)} \otimes I;$ & \qquad & $\eta_1 = \eta_2 = \eta$, \\[3pt]
$C = C_{(2)} \otimes C_{(2)} \otimes \sigma_2;$ & \qquad & $\eta_1 = -\eta, \eta_2 = \eta,$ \\[3pt]
$C = C_{(2)} \otimes C_{(2)} \otimes \sigma_1;$ & \qquad & $\eta_1 = \eta, \eta_2 = -\eta,$ \\[3pt]
$C = C_{(2)} \otimes C_{(2)} \otimes \sigma_3;$ & \qquad & $\eta_1 = \eta_2 = -\eta.$\\
\end{tabular}
\end{equation}
The $B$-intertwiner can still be written as $B=(A^{-1}C)^T$, which is now easy to express using the decompositions of $A$ and $C$. We again have four different cases, depending on the signs $\eta_1$ and $\eta_2$.

\begin{equation}
\begin{tabular}{ l c l }
$B = (-1)^{t_2} B_{(1)} \otimes B_{(2)} \otimes \sigma_1^{t_1} \sigma_2^{t_2};$ & \qquad & $\eta_1 = \eta_2 = \eta$, \\[3pt]
$B = (-1)^{t+1} B_{(1)} \otimes B_{(2)} \otimes \sigma_1^{t_1} \sigma_2^{t_2+1};$ & \qquad & $\eta_1 = -\eta, \eta_2 = \eta,$ \\[3pt]
$B = (-1)^{t_2} B_{(1)} \otimes B_{(2)} \otimes \sigma_1^{t_1+1} \sigma_2^{t_2};$ & \qquad & $\eta_1 = \eta, \eta_2 = -\eta,$ \\[3pt]
$B = (-1)^{t_2} B_{(1)} \otimes B_{(2)} \otimes \sigma_3 \sigma_1^{t_1} \sigma_2^{t_2};$ & \qquad & $\eta_1 = \eta_2 = -\eta.$\\
\end{tabular}
\end{equation}
In principle we could define new $B$-intertwiners so that the constant prefactor would not need to be included. However, to be consistent in our notation, we choose to include the prefactor here.

Finally, we decompose the highest rank Clifford algebra element $\gamma_*$. Using the definition of $\gamma_*$, we easily decompose it as
\begin{equation}
\gamma_* = (-i)^{d/2+t} i \, \gamma^0\dotsc\gamma^{d_1-1} \otimes \gamma^{d_1}\dotsc \gamma^{d_1+d_2} \otimes \sigma_3.
\end{equation}
Then we can use the duality relation \eqref{Basis_dual_odd} to express the gamma matrix products appearing here in a simpler form. It turns out that both of the products are proportional to the identity matrix, giving us a simple decomposition.
\begin{equation}
\gamma_* = (-i)^{d/2+t} i (\zeta_1  i^{(d_1-1)/2+t_1} I_{(1)}) \otimes ( \zeta_2 i^{(d_2-1)/2+t_2} I_{(2)}) \otimes \sigma_3 = \zeta_1 \zeta_2 \, I_{(1)} \otimes I_{(2)} \otimes \sigma_3.
\end{equation}
In other cases, we can find decompositions for the intertwiners in a completely similar fashion. We summarise the results in appendix \ref{App_Matrix_Properties_Decomposition}.
\vspace{-5pt}
\subsubsection{Representations of Lorentz Algebras}
\vspace{-5pt}
The decomposition $\text{Cliff}(d) \to \text{Cliff}(n)\times \text{Cliff}(d-n)$ also induces a corresponding decomposition on the associated special orthogonal representation $\text{SO}(d) \to \text{SO}(n) \times \text{SO}(d-n)$. Specifically, the representations of $\text{SO}(n)$ and $\text{SO}(n-d)$ are generated by
\begin{equation}
\begin{split}
SO(n): \frac{i}{4} \,\gamma^{\mu \nu},\\[3pt]
SO(d-n): \frac{i}{4} \, \gamma^{\alpha \beta}.
\end{split}
\end{equation}
For example, in the case of even-dimensional full algebra and two odd-dimensional subalgebras, the $SO$-generators are then.
\begin{equation}
\begin{split}
SO(n): \frac{i}{4} \, \gamma^{\mu \nu} = \frac{i}{4} \, \gamma_{(1)}^{\mu \nu} \otimes I_{(2)} \otimes I_{(3)},\\[3pt]
SO(d-n): \frac{i}{4} \, \gamma^{\alpha \beta} =  I_{(2)} \otimes \frac{i}{4} \, \gamma_{(2)}^{\alpha \beta} \otimes I_{(3)}.
\end{split}
\end{equation}
Then also the different objects that transform under $\text{SO}(d)$, in particular spinors and tensors, decompose into different representations of $\text{SO}(n) \times \text{SO}(d-n)$. 

Spinors decompose as $\text{Spin}(d) \to \text{Spin}(n) \times \text{Spin}(d-n)$, which means that we can compose the $\text{Spin}(d)$ spinors $\epsilon$ in terms of $\text{Spin}(n)$ and $\text{Spin}(d-n)$ spinors, $\eta_{(1)}$ and $\xi_{(2)}$, as a tensor product. In our example case, we also need an auxiliary spinor $\theta_{(3)}$ that does not transform under $SO(n)$ or $SO(d-n)$.
\begin{align}
\epsilon = \eta_{(1)} \otimes \xi_{(2)} \otimes \theta_{(3)}.
\end{align}
Also tensors can be decomposed under $\text{SO}(d) \to \text{SO}(n) \times \text{SO}(d-n)$. Under this they decompose into different representations basically based on how many indices are those transforming under $\text{SO}(n)$ and how many transform under $\text{SO}(d-n)$ \cite{Georgi:1999wka}. Let us consider for example a antisymmetric two-index tensor $F_{MN}$ transforming as $\mathbf{45}$ under $\text{SO}(10)$. Then, if we reduce to $\text{SO}(3)\times \text{SO}(7)$, we have the $\text{SO}(3)$ representation decomposes as follows:
\begin{align}
\mathbf{45} \to (\mathbf{1},\mathbf{21}) + (\mathbf{3},\mathbf{7}) + (\mathbf{3},\mathbf{1})  \label{F_rep_decomposition}
\end{align}
Here the first term corresponds to an object that transforms trivially under $\text{SO}(3)$, and as antisymmetric tensor under $\text{SO}(7)$. Therefore we associate this with $F_{\alpha \beta}$, the part of the tensor $F_{MN}$, whose both indices $\alpha$ and $\beta$ are associated to the $\text{SO}(7)$ subgroup, and transform non-trivially under it. The other parts have similar interpretations, so we can write this decomposition in terms of components of $F_{MN}$ as.
\begin{align}
F_{MN} \to F_{\alpha \beta} + (F_{\alpha \nu} + F_{\mu \beta}) + F_{\mu \nu}, \label{review_F_decomposition}
\end{align}
where $\alpha$ and $\beta$ are the indices that are associated to $SO(7)$, and $\mu$ and $\nu$ are associated to $SO(3)$. 

In case the representation of $F$ is known, this can lead into simplifications. For example, if it is known that $F$ transforms trivially under $\text{SO}(3)$, then only the first term in the decomposition \eqref{F_rep_decomposition} can appear. In terms of tensor components, this means that
\begin{align}
F_{MN} \to F_{\alpha \beta}.
\end{align}
\newpage
\section{Installation and Setup} \label{sect_Setup}

\subsection{Downloading and Installing}

The package comes with two Mathematica files. The file \mtd{GammaMaP.m} contains the package itself and  the file \mtd{GammaMaP\_examples.nb} contains instructions and examples. The most up-to-date versions of these files can be found at \url{https://github.com/PyryKuusela/GammaMaP}.

To install the package in Mathematica front end, go to menu \mtd{File}$\rightarrow$\mtd{Install...}. Then in the resulting dialog choose \mtd{Package} as \mtd{Type of Item to Install}, and as \mtd{Source}, choose \mtd{From File...}, navigate to the directory containing \mtd{GammaMaP.m}, and open it. After this, choose either to install the package for current user or all users.

Alternatively, to install the package manually, place the file \mtd{GammaMaP.m} to the directory \mtd{\$UserBaseDirectory/Applications}.

\subsection{Setup} \label{gSetRep}

The package can be loaded by using

\begin{mmaCell}{Input}
		<<\mmaDef{GammaMaP.m}
\end{mmaCell}

After doing this, we must specify the relevant Clifford algebra, whose presentation the gamma matrices form. As discussed in section \ref{sect_Review}, this is specified by giving the signature $(\pm 1,\pm 1,\dotsc,\pm1)$, corresponding to the diagonal elements of $\eta^{\mu \nu}$ appearing in \eqref{Clifford_algebra_anticommutation}, whose values are $\pm 1$.

After this, we need to specify the parameters $\epsilon$ and $\eta$, appearing in \eqref{C_relations}, which determine the properties of the intertwiners. In addition, for odd-dimensional representations, there are two representations that are inequivalent under unitary transformations. The sign of the last gamma matrix, $\zeta$, determines which one is used, as discussed in section \ref{sect_Review_gamma_matrices}.

Therefore, the Clifford algebra, together with the relevant intertwiners, is essentially completely defined by giving the signature of $\eta^{\mu \nu}$ together with $\epsilon,\eta$ and in odd dimensions $\zeta$.

These parameters can be set using function \mtd{gSetRep}, which has the following structure.
\begin{funcDefnN}[label=gSetRep_infobox]{\mtd{gSetRep}}
	\mtd{gSetRep[signature,$\epsilon$,$\eta$,$\zeta$,startFrom0]}\\
	Defines the Clifford algebra, and the properties of the intertwiners.\\[3pt]
	\textbf{Arguments}\\
	\mtd{signature} is a list with arguments $\pm 1$, corresponding to diagonal elements of $\eta^{\mu \nu}$,\\
	$\epsilon$ is the parameter appearing in \eqref{C_relations}, with values $\pm 1$,\\
	$\eta$ is the parameter appearing in \eqref{C_relations}, with values $\pm 1$.\\[3pt]
	\textbf{Optional Arguments}\\
	$\zeta$ is the parameter with values $\pm 1$ that determines which one of the two inequivalent representations in odd dimensions is used. Note that this is relevant only in odd dimensions. If this value is not specified for a representation in an odd dimension, 1 is used as the default value.\\
	\mtd{startFrom0} is a parameter, which takes values \mtd{True} or \mtd{False}, specifying whether the numbering of gamma matrices begins from 0 or 1. If this is not specified, the default value is set by whether the first element of \mtd{signature} is -1 or 1. In these cases the numbering starts from 0 and 1, respectively.
\end{funcDefnN}
For example, the following command sets a 4-dimensional representation with signature $(-+++)$. The corresponding gamma matrices are $\gamma^0,\gamma^1,\gamma^2,$ and $\gamma^3$. The parameters $\eta$ and $\epsilon$ are both $1$. Since this Clifford algebra has even dimension, $\zeta$ does not need to be specified, and since \mtd{startFrom0} is not specified, it uses its default value of \mtd{True}.
\begin{mmaCell}{Input}
		\mmaDef{\mmaDef{gSetRep}}[\{-1,1,1,1\},1,1]
\end{mmaCell}
If we would like to use gamma matrices $\gamma^1,\gamma^2,\gamma^3$ and $\gamma^4$ instead (where $\gamma^1$ would still be timelike), we would do this with the following command.
\begin{mmaCell}{Input}
		\mmaDef{\mmaDef{gSetRep}}[\{-1,1,1,1\},1,1,False]
\end{mmaCell}
Given a signature, and the values of $\eta$ and $\epsilon$, the representation of Clifford algebra is not unique. In particular, the gamma matrices and intertwiners can have some special properties for some representation that are not true in general. As discussed previously, there is a particularly nice representation in which the gamma matrices have expressions \eqref{gamma_matrices_special_rep}, and the intertwiners are given by \eqref{A}, \eqref{C_special_rep} and \eqref{B}. We refer to this representation as the special representation.

In general we do not assume that the representation has any of properties specific to certain representations. However, the properties of the above representation can be used if the option \mtd{gUseSpecialRep} is set to \mtd{True}. A list of properties of a completely generic representation and the special representation can be found in appendix \ref{app_Rep_Properties}.
\begin{optDefnN}[label=gUseSpecialRep_infobox]{\mtd{gUseSpecialRep}}
	\mtd{gUseSpecialRep[n,opt]}\\
	Determines whether the representation \eqref{gamma_matrices_special_rep} is used.\\[3pt]
	\textbf{Arguments}\\
	\mtd{opt} is one of the options listed below.\\[3pt]
	\textbf{Optional Arguments}\\
	\mtd{n} is the number of the subrepresentation for which the setting applies (see section \ref{sect_Subalgebras}). If this value is not specified, then this changes the setting for every subrepresentation.\\[3pt]
	\textbf{Options}\\
	\mtd{True} the special representation is used. This is the default value.\\
	\mtd{False} the special representation is not used. \label{gUseSpecialRep_infobox}
\end{optDefnN}
In practice, this means that if \mtd{gUseSpecialRep} is set to \mtd{True}, then in addition to those relations in appendix \ref{app_Rep_Properties} that are true for any representation, also the relations that are true only for the special representation are used, giving additional commutation relations. In particular, this relates the transposes, conjugates and inverses of the intertwiners to the intertwiners themselves. 

In the example below \mtd{gTimes} and $\star$ denote multiplication of matrices (see section \ref{sect_Functions_gTimes}). The intertwiners are automatically commuted as right as possible. This behaviour can, however, be modified using options (see section \ref{sect_Functions_gTimes_Options}).
\begin{mmaCell}{Input}
		\mmaDef{gUseSpecialRep}[False];	
		Conjugate[\mmaDef{m}[A]]
		\mmaDef{gTimes}[\mmaDef{y}[\{u[\m{$\mu$},\m{$\nu$}]\}],Conjugate[\mmaDef{m}[A]]]
		\mmaDef{gTimes}[\mmaDef{m}[C],\mmaDef{m}[C]]
		\mmaDef{gUseSpecialRep}[True];
		Conjugate[\mmaDef{m}[A]]
		\mmaDef{gTimes}[\mmaDef{y}[\{u[\m{$\mu$},\m{$\nu$}]\}],Conjugate[\mmaDef{m}[A]]]
		\mmaDef{gTimes}[\mmaDef{m}[C],\mmaDef{m}[C]]
		\mmaDef{gUseSpecialRep}[False];
\end{mmaCell}
\begin{mmaCell}{Output}
		\mmaSup{$\mathcal{A}$}{*}
\end{mmaCell}
\begin{mmaCell}{Output}
		\mmaSup{$\gamma$}{$\mu$,$\nu$}$\star$\mmaSup{$\mathcal{A}$}{*}
\end{mmaCell}
\begin{mmaCell}{Output}
		$\mathcal{C}$$\star$$\mathcal{C}$
\end{mmaCell}
\begin{mmaCell}{Output}
		-$\mathcal{A}$
\end{mmaCell}
\begin{mmaCell}{Output}
		\mmaSup{$\mathcal{A}$}{*}$\star$\mmaSup{(\mmaSup{$\gamma$}{$\mu$,$\nu$})}{$\dagger$}
\end{mmaCell}
\begin{mmaCell}{Output}
		1
\end{mmaCell}
As mentioned briefly before, the components of the spinors can either be ordinary, commutative, numbers or anticommutative Grassmann numbers. In the latter case, spinors anticommute with each other, and this introduces signs for example when taking a transpose of a spinor bilinear. Whether spinors are commutative or anticommutative is determined by option \mtd{gSpinorType}.
\begin{optDefnN}[label=gSpinorType_infobox]{\mtd{gSpinorType}}
	\mtd{gSpinorType[opt,n]}\\
	Determines whether spinors are commuting or anticommuting.\\[3pt]
	\textbf{Arguments}\\
	\mtd{opt} is one of the options listed below.\\[3pt]
	\textbf{Optional Arguments}\\
	\mtd{n} is the number of the subrepresentation for which the setting applies (see section \ref{sect_Subalgebras}). If this value is not specified, then this changes the setting for every subrepresentation.\\[3pt]
	\textbf{Options}\\
	\mtd{"Anticommutative"} anticommuting spinors are used. This is the default value for the representation of the full algebra, and the first subrepresentation.\\
	\mtd{"Commutative"} commuting spinors are used. The is the default value for all subrepresentations apart from the first one.
\end{optDefnN}
As mentioned above, this option affects for example, whether or not we pick up a sign when transposing an expression containing two spinors. Here \mtd{gT} denotes transpose.
\begin{mmaCell}{Input}
		\mmaDef{gSpinorType}["Commutative"]
		\mmaDef{gTimes}[\mmaDef{gT}[\mmaDef{s}[\m{$\xi$},\{\}]],\mmaDef{s}[\m{$\eta$},\{\}]]//\mmaDef{gT}
		\mmaDef{gSpinorType}["Anticommutative"]
		\mmaDef{gTimes}[\mmaDef{gT}[\mmaDef{s}[\m{$\xi$},\{\}]],\mmaDef{s}[\m{$\eta$},\{\}]]]//\mmaDef{gT}
\end{mmaCell}
\begin{mmaCell}{Output}
		\mmaSup{$\widetilde{\eta}$}{T}$\star$$\widetilde{\xi}$
\end{mmaCell}
\begin{mmaCell}{Output}
		-(\mmaSup{$\widetilde{\eta}$}{T}$\star$$\widetilde{\xi}$)
\end{mmaCell}
Note, however, that the program cannot always recognise the situations in which the expression vanishes automatically due to the anticommutation property of spinors.
\newpage
\section{Basic Objects} \label{sect_Objects}
\vspace{-5pt}
\subsection{Indices} \label{sect_Objects_Indices}
\vspace{-5pt}
We comment first on how the indices appearing in different objects are generically denoted, since the same form appears in almost all objects that have Lorentz indices. Usually a list of indices appearing inside curly brackets denotes indices that are antisymmetrised (or symmetrised, depending on the object in question) with each other, whereas indices inside different brackets usually have no special relation with each other. Upper indices are put inside \mtd{u}, so upper indices $\mu,\nu$ that are antisymmetrised (or symmetrised) would be denoted by \mtd{\mtu{u}[\mtu{$\mu$},\mtu{$\nu$}]}. Similarly, lower indices $\mu,\nu$ are denoted by \mtd{\mtu{d}[\mtu{$\mu$},\mtu{$\nu$}]}.

For example, suppose that we wish to define an object (such as a product of gamma matrices) that has indices $a,b,c,\mu,\nu,\sigma$, of which $a,b$ and $c$ are antisymmetrised with each other, and $\mu,\nu$ and $\sigma$ are similarly antisymmetrised with each other, but a priori we do not impose any relation between indices denoted by Roman and Greek letters. In addition we impose that $a,c,\mu$ and $\nu$ are upper indices and that the rest are lower indices. Then these indices are denoted by the following list.\\

\mtd{\{\mtu{u}[\mtu{a}],\mtu{d}[\mtu{b}],\mtu{u}[\mtu{c}]\},\{\mtu{u}[\mtu{$\mu$},\mtu{$\nu$}],\mtu{d}[\mtu{$\sigma$}]\}}
\vspace{-5pt}
\subsection{Gamma Matrices}
\vspace{-5pt}
Antisymmetrised products of gamma matrices are denoted by \mtd{y}.
\begin{objDefnN}[label=y_infobox]{\mtd{y}}
	\mtd{y[\{u[a1],d[a2]\},...,\{u[a(n-1)],d[an]\},n]}\\
	Denotes a product of gamma matrices.\\[3pt]
	\textbf{Arguments}\\
	\mtd{a1,...,an} are lists of antisymmetrised indices, which use the conventions explained in the previous section \ref{sect_Objects_Indices}.\\[3pt]
	\textbf{Optional Arguments}\\
	\mtd{n} is the number of the subrepresentation part of which the gamma matrix forms (see section \ref{sect_Subalgebras} for more information on subalgebras and -representations). If this is left unspecified then the default value 0, corresponding to the full representation, is used. In practice, this parameter can almost always be left unspecified, and the program will take care of this automatically. The only exceptions are the situations, in which the matrix appears on the left-hand side of a function assignment, in rules, or in assumptions. \\[3pt]
	\textbf{Notation}\\
	\mtd{\mmaSub{\mmaSup{\mmaSub{$\gamma$}{(n)}}{a1}}{a2}\mmaSub{\mmaSup{\mmaSub{$\gamma$}{(n)}}{a3}}{a4}...\mmaSub{\mmaSup{\mmaSub{$\gamma$}{(n)}}{a(n-1)}}{an}} 
\end{objDefnN}
For example, we can write the product $\gamma^{[\mu \nu}_{\phantom{\mu \nu}\sigma]} \gamma^{[a b]} \equiv \gamma^{\mu \nu}_{\phantom{\mu \nu}\sigma} \gamma^{a b}$ as follows.
\begin{mmaCell}{Input}
		\mmaDef{y}[\{u[\m{$\mu$},\m{$\nu$}],d[\m{$\sigma$}]\},\{u[a,b]\}]
\end{mmaCell}
\begin{mmaCell}{Output}
		\mmaSub{\mmaSup{$\gamma$}{$\mu,\nu$}}{$\sigma$}\mmaSup{$\gamma$}{a,b}
\end{mmaCell}
Note that since here the gamma matrices are those corresponding to the full Clifford algebra (i.e. they have \mtd{n=0}), the subscript denoting the subalgebra is not explicitly shown to avoid cluttering the expressions. The antisymmetrised indices are automatically ordered in alphabetical order with upper indices first, and lower indices next.
\begin{mmaCell}{Input}
		\mmaDef{y}[\{u[a],d[b],u[x]\},\{u[c]\}]
		-\mmaDef{y}[\{u[a,x],d[b]\},\{u[c]\}]
\end{mmaCell}
\begin{mmaCell}{Output}
		-(\mmaSub{\mmaSup{$\gamma$}{a,x}}{b}\mmaSup{$\gamma$}{c})	
\end{mmaCell}
\begin{mmaCell}{Output}	
		-(\mmaSub{\mmaSup{$\gamma$}{a,x}}{b}\mmaSup{$\gamma$}{c})
\end{mmaCell}
\vspace{-5pt}
\subsection{Other Matrices}
\vspace{-5pt}
Matrices other than the gamma matrices are denoted by \mtd{m}.
\begin{objDefnN}[label=m_infobox]{\mtd{m}}
	\mtd{m[X,n]}\\
	Denotes a matrix \mtd{X}.\\[3pt]
	\textbf{Arguments}\\
	\mtd{X} is the name of the matrix.\\[3pt]
	\textbf{Optional Arguments}\\
	\mtd{n} is the number of the subalgebra to which the matrix is associated (see section \ref{sect_Subalgebras} for more information on subalgebras and -representations). If this is left unspecified then the default value 0, corresponding to the full representation, is used. In practice this parameter can be almost always left unspecified, and the program will take care of this automatically. The only exceptions are the situations, in which the matrix appears on the left-hand side of a function assignment, in rules, or in assumptions.\\[3pt]
	\textbf{Notation}\\
	$\mathcal{X}_{(n)}$ 
\end{objDefnN}
Special cases of this are intertwiners $A,B$ and $B$, $\gamma_*$, sigma matrices $\sigma_1, \sigma_2$, and $\sigma_3$, and the identity matrix $I$, which all have predefined behaviour taking into account the properties \eqref{sigma_commutation_anticommutation}-\eqref{app_intertwiners_last}. $A,C$ and $B$-intertwiners are denoted by \mtd{m[\mtu{A}]}, \mtd{m[C]} and \mtd{m[\mtu{B}]}, respectively, $\gamma_*$ is denoted by \mtd{m[\mtu{$\gamma$5}]}, sigma matrices are denoted by \mtd{m[\mtu{$\sigma$1}]}, \mtd{m[\mtu{$\sigma$2}]} and \mtd{m[\mtu{$\sigma$3}]}, and the identity matrix is denoted by \mtd{m[\mtu{Id}]}. For example, $\gamma_*$ is written as follows.
\begin{mmaCell}{Input}
		\mmaDef{m}[\m{$\gamma$5}]
\end{mmaCell}
\begin{mmaCell}{Output}
		\mmaSub{$\gamma$}{*}
\end{mmaCell}
For more information on the behaviour of these matrices, see section \ref{sect_Functions_gTimes}.
\vspace{-5pt}
\subsection{Spinors}
\vspace{-5pt}
Spinors are denoted by \mtd{s}.
\begin{objDefnN}[label=s_infobox]{\mtd{s}}
	\mtd{s[x,\{i\},n]}\\
	Denotes a spinor with name \mtd{x} and an additional index \mtd{i}, transforming under the \mtd{n}:th subalgebra.\\[3pt]
	\textbf{Arguments}\\
	\mtd{x} is the name of the spinor.\\[3pt]
	\textbf{Optional Arguments}\\
	\mtd{i} is an additional index.\\
	\mtd{n} is the number of the subalgebra under which the spinor transforms (see section \ref{sect_Subalgebras} for more information on subalgebras and -representations). If this is left unspecified then the default value 0, corresponding to the full representation. In practice this parameter can be almost always left unspecified, and the program will take care of this automatically. The only exceptions are the situations, in which the spinor appears on the left-hand side of a function assignment (for an example, see section \ref{sect_Examples_D=4_SYM}), in rules or in assumptions.\\[3pt]
	\textbf{Notation}\\
	\mtd{$\widetilde{\text{\mtd{x}}}_\text{\mtd{i}}$} 
\end{objDefnN}
The additional index can be an $R$-symmetry index, for example, or just an index used to number different spinors. There is no pre-programmed significance on this index. The \mtd{TraditionalForm} output notation for spinors has tilde above the name of the spinor to distinguish it from other similar-looking objects. Note also that if the spinor transforms under the full algebra, corresponding to \mtd{n=0}, then the subscript \mtd{(0)} is omitted.
\begin{mmaCell}{Input}
		\mmaDef{s}[\m{$\xi$},\{a\}]
		\mmaDef{s}[\m{$\eta$},\{\}]
\end{mmaCell}

\begin{mmaCell}{Output}
		\mmaSub{$\widetilde{\xi}$}{a}
\end{mmaCell}
\begin{mmaCell}{Output}
		$\widetilde{\eta}$
\end{mmaCell}
There are also various ways of conjugating spinors, and the conjugated spinors have each their own notation.
\vspace{-5pt}
\subsubsection{Dirac Conjugate}
\vspace{-5pt}
The Dirac conjugate of a spinor (see section \ref{sect_Review_Spinor_BLs_and_Tensors}), $\bar{\xi} \equiv \xi^\dagger A$, is implemented by \mtd{dc}.
\begin{objDefnN}[label=dc_infobox]{\mtd{dc}}
	\mtd{dc[x,\{i\},n]}\\
	Denotes the Dirac conjugate of a spinor with name \mtd{x} and an additional index \mtd{i}, transforming under the \mtd{n}:th subalgebra.\\[3pt]
	\textbf{Arguments}\\
	\mtd{x} is the name of the spinor,\\[3pt]
	\textbf{Optional Arguments}\\
	\mtd{i} is an additional index.\\
	\mtd{n} is the number of the subalgebra under which the spinor transforms (see section \ref{sect_Subalgebras} for more information on subalgebras and -representations). If this is left unspecified then the default value 0, corresponding to the full representation. In practice this parameter can be almost always left unspecified, and the program will take care of this automatically. The only exceptions are the situations, in which the spinor appears on the left-hand side of a function assignment (for an example, see section \ref{sect_Examples_D=4_SYM}), in rules or in assumptions.\\[3pt]
	\textbf{Notation}\\
	\mtd{\mmaSub{$\overline{x}$}{i(n)}} 
\end{objDefnN} 
The arguments work in exactly the same way as those in the definition of spinor \mtd{s}, and the Dirac conjugate of a spinor is denoted by an overline in output.
\begin{mmaCell}{Input}
		\mmaDef{gAConvention}["ToLeft"];
		\mmaDef{dc}[\m{$\xi$},\{a\}]
		\mmaDef{dc}[\m{$\eta$},\{\}]
\end{mmaCell}
\begin{mmaCell}{Output}
		\mmaSub{$\bar{\xi}$}{a}
\end{mmaCell}
\begin{mmaCell}{Output}
		$\bar{\xi}$
\end{mmaCell}
Here \mtd{gAConvention} is a setting that determines the behaviour of $A$-intertwiners. In this case it makes sure that the $A$-intertwiner stays next to the spinor so that we may identify this product with the Dirac conjugated spinor. For more information on this and other similar options, see section \ref{sect_Functions_gTimes_Options}.	
\vspace{-5pt}	
\subsubsection{Charge Conjugate}
\vspace{-5pt}
Charge conjugate of a spinor is defined as the complex conjugate of the spinor times the inverse of the $B$-intertwiner. Also a Dirac conjugate of a charge conjugate can be defined.
\begin{align}
\xi^c &\equiv B^{-1} \xi^*,\\
\bar{\xi^c} &\equiv (\xi^c)^\dagger A = \xi^T B A.
\end{align}
Charge conjugated spinor is denoted by \mtd{rcc}, and the Dirac conjugate of a charge conjugated spinor is denoted by \mtd{lcc}.
\begin{objDefnN}[label=rcc_infobox]{\mtd{rcc}}
	\mtd{rcc[x,\{i\},n]}\\
	Denotes the charge conjugate of a spinor with name \mtd{x} and an additional index \mtd{i}, transforming under the \mtd{n}:th subalgebra.\\[3pt]
	\textbf{Arguments}\\
	\mtd{x} is the name of the spinor,\\[3pt]
	\textbf{Optional Arguments}\\
	\mtd{i} is an additional index.\\
	\mtd{n} is the number of the subalgebra under which the spinor transforms (see section \ref{sect_Subalgebras}). If this is left unspecified then the default value 0, corresponding to the full representation, is used. In practice this parameter can be almost always left unspecified, and the program will take care of this automatically. The only exceptions are the situations, in which the spinor appears on the left-hand side of a function assignment (for an example, see section \ref{sect_Examples_D=4_SYM}), in rules or in assumptions.\\[3pt]
	\textbf{Notation}\\
	\mtd{\mmaSub{\mmaSup{x}{c}}{i}}
\end{objDefnN}
\begin{objDefnN}[label=lcc_infobox]{\mtd{lcc}}
	\mtd{lcc[x,\{i\},n]}\\
	Denotes the Dirac conjugate of a charge conjugated spinor with name \mtd{x} and an additional index \mtd{i}, transforming under the \mtd{n}:th subalgebra.\\[3pt]
	\textbf{Arguments}\\
	\mtd{x} is the name of the spinor,\\[3pt]
	\textbf{Optional Arguments}\\
	\mtd{i} is an additional index.\\
	\mtd{n} is the number of the subalgebra under which the spinor transforms (see section \ref{sect_Subalgebras}). If this is left unspecified then the default value 0, corresponding to the full representation, is used. In practice this parameter can be almost always left unspecified, and the program will take care of this automatically. The only exceptions are the situations, in which the spinor appears on the left-hand side of a function assignment (for an example, see section \ref{sect_Examples_D=4_SYM}), in rules or in assumptions.\\[3pt]
	\textbf{Notation}\\
	\mtd{\mmaSub{\mmaSup{$\overline{x}$}{c}}{i}} 
\end{objDefnN}
The arguments work again in the same way as those in the definition of spinor \mtd{s}. Here \mtd{gBConvention} determines the behaviour of the $B$-intertwiner, see section \ref{sect_Functions_gTimes_Options} and below.
\begin{mmaCell}{Input}
		\mmaDef{gBConvention}["ToRight"];
		\mmaDef{rcc}[\m{$\xi$},\{a\}]
		\mmaDef{gBConvention}["ToLeft"];
		\mmaDef{lcc}[\m{$\eta$},\{\}]
\end{mmaCell}
\begin{mmaCell}{Output}
		\mmaSub{\mmaSup{$\xi$}{c}}{a}
\end{mmaCell}
\begin{mmaCell}{Output}
		\mmaSup{$\bar{\eta}$}{c}
\end{mmaCell}		
Depending on the options, such as \mtd{gBConvention}, the program automatically notices the product of an appropriate matrix and a spinor as a conjugate whenever possible, using the definitions of different conjugates. As an example, we compare here different cases. In the first case we select settings such that the simplifications are not done automatically, so that we see only a product of the $B$-intertwiner and a conjugated spinor. In the second case we change the settings to do the simplification automatically, and see that the product is automatically interpreted as a charge conjugate. Here \mtd{gTimes} (and its output notation $\star$) denotes multiplication of non-commutative objects, see section \ref{sect_Functions_gTimes}.
\begin{mmaCell}{Input}
		\mmaDef{gBConvention}["DoNothing"];
		\mmaDef{gTimes}[Inverse[\mmaDef{m}[B]],Conjugate[\mmaDef{s}[\m{$\xi$},\{\}]]]
		\mmaDef{gBConvention}["ToRight"];
		\mmaDef{gTimes}[Inverse[\mmaDef{m}[B]],Conjugate[\mmaDef{s}[\m{$\xi$},\{\}]]]
\end{mmaCell}
\begin{mmaCell}{Output}
		\mmaSup{$\mathcal{B}$}{-1}$\star$\mmaSup{$\widetilde{\xi}$}{*}
\end{mmaCell}
\begin{mmaCell}{Output}
		\mmaSup{$\xi$}{c}
\end{mmaCell}
Also the opposite is true. With appropriate settings, the charge conjugate spinor will be expressed as a product of a spinor and an intertwiner matrix.
\begin{mmaCell}{Input}
		\mmaDef{gBConvention}["ToLeft"];
		\mmaDef{rcc}[\m{$\xi$},\{\}]
\end{mmaCell}
\begin{mmaCell}{Output}
		\mmaSup{$\mathcal{B}$}{-1}$\star$\mmaSup{$\widetilde{\xi}$}{*}
\end{mmaCell}
\vspace{-5pt}
\subsection{Forms}
\vspace{-5pt}
Components of forms, i.e. tensors with antisymmetrised indices are denoted by \mtd{gForm}.
\begin{objDefnN}[label=gForm_infobox]{\mtd{gForm}} 
	\mtd{gForm[F,\{u[$\mu$],d[$\nu$]\}]}\\	
	Denotes components of a form F.\\[3pt]
	\textbf{Arguments}\\
	\mtd{F} is the name of the tensor,\\
	\mtd{$\mu$,$\nu$} are lists of Lorentz indices, which are up and down, respectively. All the indices in lists $\mu$ and $\nu$ are antisymmetrised with each other.\\[3pt]
	\textbf{Notation}\\
	\mtd{\mmaSub{\mmaSup{F}{$\mu$}}{$\nu$}} 
\end{objDefnN}
The upper and lower indices are ordered automatically in alphabetical order, taking into account possible overall sign from the change of signature. The indices are by default down.
\begin{mmaCell}{Input}
		\mmaDef{gForm}[G,\{u[c,a],d[b]\}]
		\mmaDef{gForm}[T,\{\m{$\mu$},\m{$\nu$}\}]
\end{mmaCell}

\begin{mmaCell}{Output}
		-\mmaSub{\mmaSup{G}{a,c}}{b}
\end{mmaCell}
\begin{mmaCell}{Output}
		\mmaSub{T}{$\mu$,$\nu$}
\end{mmaCell}		
As with all antisymmetric objects, forms are automatically zero if the same index appears twice, or if the number of indices exceeds the dimension.
\begin{mmaCell}{Input}
		\mmaDef{gSetRep}[\{-1,1,1,1\},1,1];
		\mmaDef{gForm}[G,\{a,b,a\}]
		\mmaDef{gForm}[G,\{\m{$\mu$},\m{$\nu$},\m{$\sigma$},\m{$\alpha$},\m{$\beta$}\}]
\end{mmaCell}
\begin{mmaCell}{Output}
		0
\end{mmaCell}
\begin{mmaCell}{Output}
		0
\end{mmaCell}
\vspace{-5pt}
\subsection{Symmetric Tensors}
\vspace{-5pt}
Components of symmetric tensors are denoted by \mtd{gSymm}.
\begin{objDefnN}[label=gSymm_infobox]{\mtd{gSymm}}
	\mtd{gSymm[F,\{u[$\mu$],d[$\nu$]\}]}\\	
	Denotes components of a symmetric tensor \mtd{F}.\\[3pt]
	\textbf{Arguments}\\
	\mtd{F} is the name of the tensor,\\
	\mtd{$\mu$,$\nu$} are lists of Lorentz indices, which are up and down, respectively. All the indices in lists $\mu$ and $\nu$ are symmetrised with each other.\\[3pt]
	\textbf{Notation}\\
	\mtd{\mmaSub{\mmaSup{$\hat{F}$}{$\mu$}}{$\nu$}} 
\end{objDefnN}
The indices are automatically ordered alphabetically, which results in no sign change, as the indices are symmetrised. Indices are by default down. In the \mtd{TraditionalForm} output notation, a symmetric tensor is distinguished by a hat.
\begin{mmaCell}{Input}
		\mmaDef{gSymm}[G,\{u[c,a],d[b]\}]
		\mmaDef{gSymm}[T,\{\m{$\mu$},\m{$\nu$}\}]
\end{mmaCell}
\begin{mmaCell}{Output}
		\mmaSub{\mmaSup{$\hat{G}$}{a,c}}{b}
\end{mmaCell}
\begin{mmaCell}{Output}
		\mmaSub{$\hat{T}$}{$\mu$,$\nu$}
\end{mmaCell}
\vspace{-5pt}
\subsection{Other Tensors}
\vspace{-5pt}
Tensors with no special symmetry or antisymmetry properties are denoted using \mtd{gTensor}, which has the following structure.
\begin{objDefnN}[label=gTensor_infobox]{\mtd{gTensor}} 
	\mtd{gTensor[F,\{u[$\mu$],d[$\nu$]\}]}\\
	Denotes components of a generic tensor \mtd{F}.\\	[3pt]
	\textbf{Arguments}\\
	\mtd{F} is the name of the tensor,\\
	\mtd{$\mu$,$\nu$} are lists of Lorentz indices, which are up and down, respectively.\\[3pt]
	\textbf{Notation}\\
	\mtd{\mmaSub{\mmaSup{\"F}{$\mu$}}{$\nu$}}
\end{objDefnN}
Now the indices are kept in the order in which they are in the input, and by default the indices are down. In output, \mtd{gTensor} is denoted by umlaut (double dot). 
\begin{mmaCell}{Input}
		\mmaDef{gTensor}[G,\{u[b,a],d[c]\}]
		\mmaDef{gTensor}[T,\{\m{$\mu$},\m{$\nu$}\}]
\end{mmaCell}
\begin{mmaCell}{Output}
		\mmaSub{\mmaSup{$\ddot{G}$}{b,a}}{c}
\end{mmaCell}
\begin{mmaCell}{Output}
		\mmaSub{$\ddot{T}$}{$\mu$,$\nu$}
\end{mmaCell}
Note that it is still possible to symmetrise or antisymmetrise some of the indices by defining the required (anti-)symmetry properties by hand. For example, the following code symmetrises the first and third index of a tensor $F$, while not specifying any symmetry properties for the second index.
\begin{mmaCell}{Input}
		\mmaDef{gTensor}[F,\{d[\mmaPat{a\_},\mmaPat{k\_},\mmaPat{b\_}]\}]:=\mmaDef{gTensor}[F,\{d[\mmaPat{b},\mmaPat{k},\mmaPat{a}]\}]/;
		  !TrueQ[Sort[\{\mmaPat{a},\mmaPat{b}\}]==\{\mmaPat{a},\mmaPat{b}\}]
\end{mmaCell}
Now the first and third index of $F$ are automatically put into alphabetical order, causing $F$ to be symmetric in its first and third indices. Note that even though we have defined relation only for the case in which all the indices are down, the program still recognises the symmetry property even when the indices are up. Indeed, all the custom properties should be defined with all indices down so that the program can recognise them.
\begin{mmaCell}{Input}
		\mmaDef{gTensor}[F,\{b,c,a\}]
		\mmaDef{gTensor}[F,\{b,c,a\}]//\mmaDef{gSimplify}
\end{mmaCell} 
\begin{mmaCell}{Output}
		\mmaSub{F}{a,c,b}
\end{mmaCell}
\begin{mmaCell}{Output}
		\mmaSup{F}{a,c,b}	
\end{mmaCell}
The custom tensor properties can be deleted cleared with \mtd{gClearTensorProperties}.
\begin{funcDefnN}[label=gClearTensorProperties_infobox]{\mtd{gClearTensorProperties}}
	\mtd{gClearTensorProperties[]}\\
	Deletes all user-defined properties for tensors.\\[3pt]
	\textbf{Arguments}\\
	None.
\end{funcDefnN}
After using this, the tensor $F_{abc}$, that previously was symmetrised in the first and last index does no longer have any special symmetry properties.
\begin{mmaCell}{Input}
		\mmaDef{gClearTensorProperties}[];
		\mmaDef{gTensor}[F,\{b,c,a\}]
\end{mmaCell} 
\begin{mmaCell}{Output}
		\mmaSub{F}{b,c,a}
\end{mmaCell}
\vspace{-5pt}
\subsection{Derivatives}
\vspace{-5pt}
Derivatives of spinors, tensors, scalars, and other objects are denoted by \mtd{gD}.
\begin{objDefnN}[label=gD_infobox]{\mtd{gD}} 
	\mtd{gD[$\delta$,\{u[$\mu$],d[$\nu$]\},expr]}\\
	Denotes a derivative $\partial^\mu \partial_\nu$ of expression \mtd{expr}, or a generic variation $\delta^\mu \delta_\nu$, if the first parameter is specified. \\[3pt]
	\textbf{Arguments}\\
	\mtd{expr} is the expression whose derivative \mtd{gD} denotes.\\
	\mtd{$\mu$,$\nu$} are lists of Lorentz indices, which are up and down, respectively.\\[3pt]
	\textbf{Optional Arguments}\\
	\mtd{$\delta$} is an parameter that can be used to make \mtd{gD} denote variations other that the usual partial derivative. These obey the sum and Leibniz rules, but do not in general commute among themselves, or conjugations etc.\\[3pt]
	\textbf{Notation}\\
	\mtd{\mmaSub{\mmaSup{\mmaSub{d}{$\delta$}}{$\mu$}}{$\nu$}expr}
\end{objDefnN}
Partial derivatives are defined simply by leaving the optional argument $\delta$ empty. For example, we would denote the derivative $\partial_\mu F_{ab}$ as.
\begin{mmaCell}{Input}
		\mmaDef{gD}[\{\m{$\mu$}\},\mmaDef{gForm}[F,\{a,b\}]]	
\end{mmaCell} 
\begin{mmaCell}{Output}
		\mmaSub{d}{$\mu$}\mmaSub{F}{a,b}
\end{mmaCell}
Since partial derivatives commute with one another, the indices are automatically ordered in alphabetical order with up-indices first and down-indices then, like in the case of a symmetric tensor. The indices are by default down.
\begin{mmaCell}{Input}
		\mmaDef{gD}[\{u[c],d[a],u[b]\},x]
		\mmaDef{gD}[\{\m{$\mu$}\},x]		
\end{mmaCell} 
\begin{mmaCell}{Output}
		\mmaSub{\mmaSup{d}{b,c}}{a}x
\end{mmaCell}
\begin{mmaCell}{Output}
		\mmaSub{d}{$\mu$}x
\end{mmaCell}
Derivatives can be taken of any expression, including tensors, matrices, spinors, scalars, and products and sums thereof. Complex numbers, gamma matrices, and intertwiners are by default considered to be constant. 
\begin{mmaCell}{Input}
		\mmaDef{gD}[\{\m{$\mu$}\},\mmaDef{gForm}[F,\{a,b\}]]
		\mmaDef{gD}[\{\m{$\mu$}\},\mmaDef{m}[X]]
		\mmaDef{gD}[\{u[a]\},3*\mmaDef{gForm}[F,\{a\}]*\mmaDef{gForm}[H,\{b\}]+3/4*\mmaDef{gSymm}[S,\{a,b\}]]//Expand
		\mmaDef{gD}[\{\m{$\mu$}\},\mmaDef{gTimes}[\mmaDef{dc}[\m{$\xi$},\{\}],\mmaDef{y}[\{u[a,b]\}],s[\m{$\xi$},\{\}]]]//Expand
\end{mmaCell} 
\begin{mmaCell}{Output}
		\mmaSub{d}{$\mu$}\mmaSub{F}{a,b}
\end{mmaCell}
\begin{mmaCell}{Output}
		\mmaSub{d}{$\mu$}\mmaSub{X}{(0)}
\end{mmaCell}
\begin{mmaCell}{Output}
		\mmaFrac{3}{4}(\mmaSup{d}{a}\mmaSub{$\hat{S}$}{a,b})+3(\mmaSup{d}{a}\mmaSub{H}{b})\mmaSub{F}{a}+3(\mmaSup{d}{a}\mmaSub{F}{a})\mmaSub{H}{b}
\end{mmaCell}
\begin{mmaCell}{Output}
		$\bar{\xi}$$\star$\mmaSup{$\gamma$}{a,b}$\star$(\mmaSub{d}{$\mu$}$\tilde{\xi}$)+(\mmaSub{d}{$\mu$}\mmaSup{$\tilde{\xi}$}{$\dagger$})$\star$A$\star$\mmaSup{$\gamma$}{a,b}$\star$$\tilde{\xi}$
\end{mmaCell}
Other objects can also be considered to be constant by using assumptions (see section \ref{sect_Functions_Assumptions}). For example, assuming that scalar $A$ is constant, can be done by using \mtd{gAddAssumptions}.
\begin{mmaCell}{Input}
		\mmaDef{gD}[\{\m{$\mu$}\},A]
		\mmaDef{gAddAssumptions}[A$\in$Constants]
		\mmaDef{gD}[\{\m{$\mu$}\},A]
\end{mmaCell} 
\begin{mmaCell}{Output}
		\mmaSub{d}{$\mu$}A
\end{mmaCell}
\begin{mmaCell}{Output}
		0
\end{mmaCell}
This works similarly on other objects, such as tensors, spinors and matrices.
\begin{mmaCell}{Input}
		\mmaDef{gAddAssumptions}[\{F$\in$Constants,m[X]$\in$Constants,s[\m{$\xi$},\{i\}]$\in$Constants\}]
		\mmaDef{gD}[\{\m{$\mu$}\},\mmaDef{gForm}[F,\{a,b\}]]
		\mmaDef{gD}[\{\m{$\mu$}\},\mmaDef{m}[X]]
		\mmaDef{gD}[\{\m{$\mu$}\},\mmaDef{dc}[\m{$\xi$},\{i\}]]
		\mmaDef{gClearAssumptions}[];
\end{mmaCell} 
\begin{mmaCell}{Output}
		0
\end{mmaCell}
\begin{mmaCell}{Output}
		0
\end{mmaCell}
\begin{mmaCell}{Output}
		0
\end{mmaCell}
Expressions for derivatives of objects can be defined manually. For example, making the program automatically recognise the relation $\partial_a F_{bc} = G_{abc}$ can be done with the following code.
\begin{mmaCell}{Input}
		\mmaDef{gD}[\{u[],d[\mmaPat{a\_}]\},\mmaDef{gForm}[F,\{\mmaPat{b\_},\mmaPat{c\_}\}]]:=\mmaDef{gForm}[G,\{\mmaPat{a},\mmaPat{b},\mmaPat{c}\}]
\end{mmaCell}
Note that here both \mtu{u} and \mtu{d} must be included in the expression for the derivative, even though one of them is empty. Now the program automatically simplifies any derivative $\partial_\mu F_{ab}$ to $G_{\mu ab}$.
\begin{mmaCell}{Input}
		\mmaDef{gD}[\{\m{$\mu$}\},\mmaDef{gForm}[F,\{\m{$\nu$},\m{$\sigma$}\}]]
\end{mmaCell}
\begin{mmaCell}{Output}
		\mmaSub{G}{$\mu$,$\nu$,$\sigma$}
\end{mmaCell}
Custom derivatives can be cleared by using \mtd{gClearDerivatives}.
\begin{funcDefnN}[label=gClearDerivatives_infobox]{\mtd{gClearDerivatives}}
	\mtd{gClearDerivatives[]}\\
	Deletes all user-defined expressions for derivatives.\\[3pt]
	\textbf{Arguments}\\
	None.
\end{funcDefnN}
\begin{mmaCell}{Input}
		\mmaDef{gClearDerivatives}[];
		\mmaDef{gD}[\{\m{$\mu$}\},\mmaDef{gForm}[F,\{\m{$\nu$},\m{$\sigma$}\}]]
\end{mmaCell}
\begin{mmaCell}{Output}
		\mmaSub{d}{$\mu$}\mmaSub{F}{$\nu$,$\sigma$}
\end{mmaCell}
More generic derivatives or variations, such as gauge covariant derivatives or supersymmetry variations, for example, can be denoted by defining value for the parameter $\delta$. A derivative defined in this way acts like a completely generic variation, so it satisfies for example the Leibniz rule, but does not commute with itself. It also treats complex numbers by default as constants.
\begin{mmaCell}{Input}
		\mmaDef{gD}[\m{$\delta$},\{\m{$\mu$}\},3*x*\mmaDef{gForm}[F,\{a,b\}]]
		\mmaDef{gD}[\m{$\delta$},\{\m{$\nu$}\},\mmaDef{gD}[\m{$\delta$},\{\m{$\mu$}\},\mmaDef{gForm}[F,\{a,b\}]]]
		  -\mmaDef{gD}[\m{$\delta$},\{\m{$\mu$}\},\mmaDef{gD}[\m{$\delta$},\{\m{$\nu$}\},\mmaDef{gForm}[F,\{a,b\}]]]
		\mmaDef{gD}[\m{$\delta$},\{\m{$\mu$}\},Conjugate[x]]
\end{mmaCell}
\begin{mmaCell}{Output}
		3(x(\mmaSub{$\delta$}{d$\mu$}\mmaSub{F}{a,b})+(\mmaSub{$\delta$}{d$\mu$}x)\mmaSub{F}{a,b})
\end{mmaCell}
\begin{mmaCell}{Output}
		-(\mmaSub{$\delta$}{d$\mu$}\mmaSub{$\delta$}{d$\nu$}\mmaSub{F}{a,b})+\mmaSub{$\delta$}{d$\nu$}\mmaSub{$\delta$}{d$\mu$}\mmaSub{F}{a,b}
\end{mmaCell}
\begin{mmaCell}{Output}
		\mmaSub{$\delta$}{$\mu$}(\mmaSup{x}{*})
\end{mmaCell}
This is particularly useful when defining custom values for the derivative, as it can then be used for example to denote supersymmetry variations. For an example of this usage, see section \ref{sect_Examples_D=4_SYM}.
\newpage
\section{Functions} \label{sect_Functions}
\vspace{-5pt}
\subsection{Multiplication} \label{sect_Functions_gTimes}
\vspace{-5pt}
Non-commutative objects, such as gamma matrices, intertwiners and spinors can be multiplied together by using \mtd{gTimes}, which is used in the same way as the ordinary \mtd{Times}, i.e. it takes as arguments all objects that are to be multiplied together. The difference is that \mtd{gTimes} automatically recognises which objects are non-commutative (i.e. have spinor indices), and does not change the order of those. On commutative objects, such as tensor components and scalars, \mtd{gTimes} works like \mtd{Times}.
\begin{funcDefnN}[label=gTimes_infobox]{\mtd{gTimes}}
	\mtd{gTimes[x1,x2,...,xn]}\\
	Multiplies expressions \mtd{x1,x2,...,xn} together, taking into account that some of the expressions may contain non-commutative objects.\\[3pt]
	\textbf{Arguments}\\
	\mtd{x1,x2,...,xn} are the expressions to be multiplied together.\\[3pt]
	\textbf{Notation}\\
	\mtd{x1$\star$x2$\star$...$\star$xn} 
\end{funcDefnN}
Writing a product of intertwiners, gamma matrices and spinors works as follows. Here the first line is an option, which governs the behaviour of $B$-intertwiner (see \ref{gBConvention_infobox}).
\begin{mmaCell}{Input}
		\mmaDef{gBConvention}["DoNothing"];
		\mmaDef{gTimes}[\mmaDef{m}[B],\mmaDef{y}[\{u[a,b]\}],\mmaDef{rcc}[\m{$\xi$},\{i\}]]
\end{mmaCell} 

\begin{mmaCell}{Output}
		$\mathcal{B}$$\star$\mmaSup{$\gamma$}{a,b}$\star$\mmaSup{\mmaSub{$\xi$}{i}}{c}
\end{mmaCell}
The notation $\star$ for multiplication can also be used in input.
\begin{mmaCell}{Input}
		\mmaDef{m}[B]$\star$\mmaDef{y}[\{u[a,b]\}]$\star$\mmaDef{rcc}[\m{$\xi$},\{i\}] //InputForm
\end{mmaCell} 

\begin{mmaCell}{Output}
		\mmaDef{gTimes}[\mmaDef{m}[B,0],\mmaDef{y}[\{u[a,b],d[]\},0],\mmaDef{rcc}[$\xi$,\{i\},0]]
\end{mmaCell}
Multiplication recognises commutative objects, and commutes them with the non-commutative objects.
\begin{mmaCell}{Input}
		\mmaDef{y}[\{u[a,b]\}]$\star$a$\star$\mmaDef{gForm}[F,\{a,b\}]$\star$\mmaDef{rcc}[\m{$\xi$},\{i\}]
\end{mmaCell} 
\begin{mmaCell}{Output}
		a\mmaSub{F}{a,b}(\mmaSup{$\gamma$}{a,b}$\star$\mmaSup{\mmaSub{$\widetilde{\xi}$}{i}}{c})
\end{mmaCell}
Also, the multiplication of sums works like the usual commutative multiplication.
\begin{mmaCell}{Input}
		\mmaDef{gAConvention}["ToRight"];
		\mmaDef{gTimes}[\mmaDef{y}[\{u[a,b]\}],3*a+\mmaDef{m}[A],\mmaDef{s}[\m{$\xi$},\{i\}]]
\end{mmaCell} 

\begin{mmaCell}{Output}
		3a(\mmaSup{$\gamma$}{a,b}$\star$\mmaSub{$\widetilde{\xi}$}{i})+\mmaSup{$\gamma$}{a,b}$\star$$\mathcal{A}$$\star$\mmaSub{$\widetilde{\xi}$}{i}
\end{mmaCell}
Spinor bilinears are also recognised automatically, and they acts as commutative objects.
\begin{mmaCell}{Input}
		\mmaDef{gTimes}[\mmaDef{gH}[\mmaDef{s}[\m{$\xi$},\{\}]],\mmaDef{y}[\{u[\m{$\mu$}]\}],\mmaDef{s}[\m{$\xi$},\{\}],\mmaDef{gH}[\mmaDef{s}[\m{$\eta$},\{\}]],\mmaDef{y}[\{d[\m{$\mu$}]\}],
		  \mmaDef{s}[\m{$\eta$},\{\}]]
\end{mmaCell} 

\begin{mmaCell}{Output}
		(\mmaSup{$\widetilde{\eta}$}{$\dagger$}$\star$\mmaSub{$\gamma$}{$\mu$}$\star$$\widetilde{\eta}$)(\mmaSup{$\widetilde{\xi}$}{$\dagger$}$\star$\mmaSup{$\gamma$}{$\mu$}$\star$$\widetilde{\xi}$)
\end{mmaCell}
\vspace{-5pt}
\subsubsection{Associated Options} \label{sect_Functions_gTimes_Options}
\vspace{-5pt}
The function \mtd{gTimes} attempts to automatically simplify the expression by using the known (anti-)commutation and multiplication properties of the matrices, such as those in appendix \ref{app_Rep_Properties}. The process can be controlled by various options, which determine, for example, if the preferred position of the $A$-intertwiner in on the left or on the right, or if the gamma matrices that are Hermitian conjugated should be expressed in the form with $A$-intertwiners instead. 
\begin{optDefnN}[label=gAConvention_infobox]{\mtd{gAConvention}}
	\mtd{gAConvention[n,opt]}\\
	Determines the behaviour of the $A$-intertwiners.\\[3pt]
	\textbf{Arguments}\\
	\mtd{opt} is one of the options listed below.\\[3pt]
	\textbf{Optional Arguments}\\
	\mtd{n} is the number of the subrepresentation for which the setting applies (see section \ref{sect_Subalgebras}). If this value is not specified, then the setting is changed for every subrepresentation.\\[3pt]
	\textbf{Options}\\
	\mtd{"ToLeft"}: $A$ is commuted to left whenever possible,\\
	\mtd{"ToRight"}: $A$ is commuted to right whenever possible. This is the default value.\\
	\mtd{"Unconjugated"}: $A$ is used to express Hermitian conjugated gamma matrices in a form without conjugations,\\
	\mtd{"DoNothing"}: $A$-intertwiner is not moved, unless doing so is required by other settings.
\end{optDefnN}
For the following examples, let us set up a seven-dimensional representation with signature $(---++++)$, so that we have $t=3$, and the parameters $\epsilon=1$ and $\eta=1$. In addition, we specify that we do not wish to use the special representation \eqref{gamma_matrices_special_rep}.
\begin{mmaCell}{Input}
		\mmaDef{gSetRep}[\{-1,-1,-1,1,1,1,1\},-1,1];
		\mmaDef{gUseSpecialRep}[False];
\end{mmaCell}
If \mtd{"ToLeft"} option is used, this commutes the $A$-intertwiners with gamma matrices using the formula \eqref{A_properties_general}.
\begin{mmaCell}{Input}
		\mmaDef{gAConvention}["ToLeft"];
		\mmaDef{gTimes}[\mmaDef{y}[\{u[a,b]\}],\mmaDef{m}[A]]
		\mmaDef{gTimes}[\mmaDef{gH}[\mmaDef{y}[\{u[\m{$\mu$}]\}]],\mmaDef{m}[A]]		
\end{mmaCell}
\begin{mmaCell}{Output}
		-($\mathcal{A}$$\star$\mmaSup{(\mmaSup{$\gamma$}{a,b})}{$\dagger$})
\end{mmaCell}
\begin{mmaCell}{Output}
		-($\mathcal{A}$$\star$\mmaSup{$\gamma$}{$\mu$})
\end{mmaCell}
Note that since the special representation \eqref{gamma_matrices_special_rep} is not used, the $A$-intertwiner cannot be commuted, for example, with transposed gamma matrices. 
\begin{mmaCell}{Input}
		\mmaDef{gTimes}[\mmaDef{gT}[\mmaDef{y}[\{u[\m{$\mu$}]\}]],\mmaDef{m}[A]]	
\end{mmaCell}
\begin{mmaCell}{Output}
		\mmaSup{(\mmaSup{$\gamma$}{$\mu$})}{T}$\star$$\mathcal{A}$
\end{mmaCell}
However, $A^T$ can of course be commuted with transposed or complex conjugated gamma matrices.
\begin{mmaCell}{Input}
		\mmaDef{gTimes}[\mmaDef{gT}[\mmaDef{y}[\{u[\m{$\mu$}]\}]],\mmaDef{gT}[\mmaDef{m}[A]]]	
		\mmaDef{gTimes}[\mmaDef{Conjugate}[\mmaDef{y}[\{u[\m{$\mu$}]\}]],\mmaDef{gT}[\mmaDef{m}[A]]]		
\end{mmaCell}
\begin{mmaCell}{Output}
		-(\mmaSup{$\mathcal{A}$}{*}$\star$\mmaSup{(\mmaSup{$\gamma$}{$\mu$})}{*})
\end{mmaCell}
\begin{mmaCell}{Output}
		-(\mmaSup{$\mathcal{A}$}{*}$\star$\mmaSup{(\mmaSup{$\gamma$}{$\mu$})}{T})
\end{mmaCell}
Here the transposed $A$-intertwiner is expressed in terms of the complex conjugated $A^*$ because of the relation \eqref{AT=Ac} and because the default value of option \mtd{gOperationOrder} (see section \ref{sect_functions_conjugations}) is such that the conjugate is preferred over the transpose.

If \mtd{gUseSpecialRep} is set to \mtd{True}, then the representation \eqref{gamma_matrices_special_rep} is used, which allows for more commutation relations, owing to the relations \eqref{special_A_relations}.
\begin{mmaCell}{Input}
		\mmaDef{gUseSpecialRep}[True];
		\mmaDef{gTimes}[\mmaDef{gT}[\mmaDef{y}[\{u[\m{$\mu$}]\}]],\mmaDef{m}[A]]	
		\mmaDef{gUseSpecialRep}[False];
\end{mmaCell}
\begin{mmaCell}{Output}
		-($\mathcal{A}$$\star$\mmaSup{(\mmaSup{$\gamma$}{$\mu$})}{*})
\end{mmaCell}
Hermitian conjugated spinor times the $A$-intertwiner is automatically interpreted as Dirac conjugated spinor.
\begin{mmaCell}{Input}
		\mmaDef{gTimes}[\mmaDef{gH}[\mmaDef{s}[\m{$\xi$},\{\}]],\mmaDef{m}[A]]
		\mmaDef{gTimes}[\mmaDef{gH}[\mmaDef{s}[\m{$\xi$},\{\}]],\mmaDef{y}[\{u[a,b]\}],\mmaDef{m}[A]]			
\end{mmaCell}
\begin{mmaCell}{Output}
		$\overline{\xi}$
\end{mmaCell}
\begin{mmaCell}{Output}
		-($\overline{\xi}$$\star$\mmaSup{(\mmaSup{$\gamma$}{a,b})}{$\dagger$})
\end{mmaCell}
If \mtd{"ToRight"} option is used, the $A$-intertwiners are commuted to right whenever possible. 
\begin{mmaCell}{Input}
		\mmaDef{gAConvention}["ToRight"];
		\mmaDef{gTimes}[\mmaDef{gH}[\mmaDef{m}[A]],\mmaDef{y}[\{u[a,b]\}]]
		\mmaDef{gTimes}[\mmaDef{m}[A],\mmaDef{y}[\{u[\m{$\mu$}]\}]]
		\mmaDef{gTimes}[\mmaDef{gT}[\mmaDef{m}[A]],\mmaDef{y}[\{u[\m{$\mu$}]\}]]	
\end{mmaCell}
\begin{mmaCell}{Output}
		-(\mmaSup{(\mmaSup{$\gamma$}{a,b})}{$\dagger$}$\star$$\mathcal{A}$)
\end{mmaCell}
\begin{mmaCell}{Output}
		-(\mmaSup{(\mmaSup{$\gamma$}{$\mu$})}{$\dagger$}$\star$$\mathcal{A}$)
\end{mmaCell}
\begin{mmaCell}{Output}
		\mmaSup{$\mathcal{A}$}{*}$\star$\mmaSup{$\gamma$}{$\mu$}
\end{mmaCell}
Now the Dirac conjugate $\overline{\xi}$ is automatically expressed as $\xi^\dagger A$, and $A$ is then commuted as left as possible.
\begin{mmaCell}{Input}
		\mmaDef{gTimes}[\mmaDef{dc}[\m{$\xi$},\{i\}],\mmaDef{y}[\{u[a,b]\}]]
\end{mmaCell}
\begin{mmaCell}{Output}
		-(\mmaSup{\mmaSub{$\widetilde{\xi}$}{i}}{$\dagger$}$\star$\mmaSup{(\mmaSup{$\gamma$}{a,b})}{$\dagger$}$\star$$\mathcal{A}$)
\end{mmaCell}
The option \mtd{"Unconjugated"} expresses the gamma matrices in a form that does not contain Hermitian conjugations.
\begin{mmaCell}{Input}
		\mmaDef{gAConvention}["Unconjugated"];
		\mmaDef{gH}[\mmaDef{y}[\{u[a,b]\}]]
\end{mmaCell}
\begin{mmaCell}{Output}
		-($\mathcal{A}$$\star$\mmaSup{$\gamma$}{a,b}$\star$$\mathcal{A}$)
\end{mmaCell}
In a case of complex conjugated and transposed gamma matrices, which can also be related to each other with the help of A-intertwiner, the option \mtd{gOperationOrder} determines which one is preferred (see section \ref{sect_functions_conjugations}). 
\begin{mmaCell}{Input}
		\mmaDef{gOperationOrder}[\{\mmaDef{Conjugate},\mmaDef{gT},\mmaDef{gH},\mmaDef{Inverse}\}];
		\mmaDef{gT}[\mmaDef{y}[\{u[a,b]\}]]
		\mmaDef{Conjugate}[\mmaDef{y}[\{u[a,b]\}]]
		\mmaDef{gOperationOrder}[\{\mmaDef{gT},\mmaDef{Conjugate},\mmaDef{gH},\mmaDef{Inverse}\}];
		\mmaDef{Conjugate}[\mmaDef{y}[\{u[a,b]\}]];
		\mmaDef{gOperationOrder}[\{\mmaDef{Inverse},\mmaDef{gH},\mmaDef{Conjugate},\mmaDef{gT}\}];
\end{mmaCell}
\begin{mmaCell}{Output}
		-(\mmaSup{$\mathcal{A}$}{*}$\star$\mmaSup{(\mmaSup{$\gamma$}{a,b})}{*}$\star$\mmaSup{(\mmaSup{$\mathcal{A}$}{*})}{-1})
\end{mmaCell}
\begin{mmaCell}{Output}
		\mmaSup{(\mmaSup{$\gamma$}{a,b})}{*}
\end{mmaCell}
\begin{mmaCell}{Output}
		-(\mmaSup{$\mathcal{A}$}{T}$\star$\mmaSup{(\mmaSup{$\gamma$}{a,b})}{T}$\star$\mmaSup{(\mmaSup{$\mathcal{A}$}{T})}{-1})
\end{mmaCell}
Again, $\xi^\dagger A$ is automatically interpreted as the Dirac conjugate $\overline{\xi}$, and $\overline{\xi} A$ is simplified by expressing the Dirac conjugate as $\overline{\xi}=\xi^\dagger A$, and then using the property $A^{-1} = (-1)^{1/2t(t+1)} A$.
\begin{mmaCell}{Input}
		\mmaDef{gTimes}[\mmaDef{gH}[\mmaDef{s}[\m{$\xi$},\{i\}]],\mmaDef{m}[A]]
		\mmaDef{gTimes}[\mmaDef{dc}[\m{$\xi$},\{i\}],\mmaDef{m}[A]]
\end{mmaCell}
\begin{mmaCell}{Output}
		\mmaSub{$\overline{\xi}$}{i}
\end{mmaCell}
\begin{mmaCell}{Output}
		\mmaSup{\mmaSub{$\xi$}{i}}{$\dagger$}
\end{mmaCell}
The option \mtd{"DoNothing"} makes the $A$-intertwiners not to move, except if it is required by other rules, such as \mtd{gIntertwinerOrder} that determines the preferred order of intertwiners (see \ref{gIntertwinerOrder_infobox}).
\begin{mmaCell}{Input}
		\mmaDef{gAConvention}["DoNothing"];
		\mmaDef{gTimes}[\mmaDef{gH}[\mmaDef{s}[\m{$\xi$},\{i\}]],\mmaDef{m}[A]]
		\mmaDef{gTimes}[\mmaDef{dc}[\m{$\xi$},\{i\}],\mmaDef{m}[A]]
		\mmaDef{gIntertwinerOrder}[\{\mmaDef{m}[\m{$\gamma$5}],\mmaDef{m}[A],\mmaDef{m}[C],\mmaDef{m}[B]\}];
		\mmaDef{gTimes}[\mmaDef{m}[A],\mmaDef{m}[\m{$\gamma$5}]]
\end{mmaCell}
\begin{mmaCell}{Output}
		\mmaSup{\mmaSub{$\widetilde{\xi}$}{i}}{$\dagger$}$\star$$\mathcal{A}$	
\end{mmaCell}
\begin{mmaCell}{Output}	
		\mmaSub{$\overline{\xi}$}{i}$\star$$\mathcal{A}$
\end{mmaCell}
\begin{mmaCell}{Output}
		\mmaSup{$\gamma$}{*}$\star$$\mathcal{A}$
\end{mmaCell}
\begin{optDefnN}[label=gCConvention_infobox]{\mtd{gCConvention}}
	\mtd{gCConvention[n,opt]}\\
	Determines the behaviour of the $C$-intertwiners.\\[3pt]
	\textbf{Arguments}\\
	\mtd{opt} is one of the options listed below.\\[3pt]
	\textbf{Optional Arguments}\\
	\mtd{n} is the number of the subrepresentation for which the setting applies (see section \ref{sect_Subalgebras}). If this value is not specified, then the setting is changed for every subrepresentation.\\[3pt]
	\textbf{Options}\\
	\mtd{"ToLeft"}: $C$ is commuted to left whenever possible,\\
	\mtd{"ToRight"}: $C$ is commuted to right whenever possible. This is the default value.\\
	\mtd{"Unconjugated"}: $C$ is used to express transposed gamma matrices in a form without transpositions,\\
	\mtd{"DoNothing"}: $C$-intertwiner is not moved, unless doing so is required by other settings.
\end{optDefnN}
The options for the $C$-intertwiner work in a completely similar way to those of the $A$-intertwiner, except that the relations \eqref{C_properties_first}-\eqref{C_properties_last} relevant to $C$ are used.

The option \mtd{"ToLeft"} commutes $C$ left whenever possible, transposing the gamma matrices.
\begin{mmaCell}{Input}
		\mmaDef{gCConvention}["ToLeft"];
		\mmaDef{gTimes}[\mmaDef{gT}[\mmaDef{y}[\{u[a,b]\}]],\mmaDef{m}[C]]
		\mmaDef{gTimes}[\mmaDef{y}[\{u[\m{$\mu$}]\}],\mmaDef{m}[C]]
		\mmaDef{gTimes}[\mmaDef{gH}[\mmaDef{y}[\{u[\m{$\mu$}]\}]],\mmaDef{m}[C]]	
\end{mmaCell}
\begin{mmaCell}{Output}
		-($\mathcal{C}$$\star$\mmaSup{$\gamma$}{a,b})
\end{mmaCell}
\begin{mmaCell}{Output}
		\mmaSup{$\gamma$}{$\mu$}$\star$$\mathcal{C}$
\end{mmaCell}
\begin{mmaCell}{Output}
		\mmaSup{(\mmaSup{$\gamma$}{$\mu$})}{$\dagger$}$\star$$\mathcal{C}$
\end{mmaCell}
Using the special representation gives again more commutation relations that can be used.
\begin{mmaCell}{Input}
		\mmaDef{gUseSpecialRep}[True];
		\mmaDef{gTimes}[\mmaDef{y}[\{u[\m{$\mu$}]\}],\mmaDef{m}[C]]
		\mmaDef{gTimes}[\mmaDef{gH}[\mmaDef{y}[\{u[\m{$\mu$}]\}]],\mmaDef{m}[C]]
		\mmaDef{gUseSpecialRep}[False];	
\end{mmaCell}
\begin{mmaCell}{Output}
		-($\mathcal{C}$$\star$\mmaSup{(\mmaSup{$\gamma$}{$\mu$})}{T})
\end{mmaCell}
\begin{mmaCell}{Output}
		-($\mathcal{C}$$\star$\mmaSup{(\mmaSup{$\gamma$}{$\mu$})}{*})
\end{mmaCell}
The option \mtd{"ToRight"} commutes the $C$-matrices to right whenever possible.
\begin{mmaCell}{Input}
		\mmaDef{gCConvention}["ToRight"];
		\mmaDef{gTimes}[\mmaDef{m}[C],\mmaDef{y}[\{u[a,b]\}]]
		\mmaDef{gTimes}[\mmaDef{m}[C],\mmaDef{gT}[\mmaDef{y}[\{u[\m{$\mu$}]\}]]]
		\mmaDef{gTimes}[\mmaDef{m}[C],\mmaDef{gH}[\mmaDef{y}[\{u[\m{$\mu$}]\}]]]	
\end{mmaCell}
\begin{mmaCell}{Output}
		-(\mmaSup{(\mmaSup{$\gamma$}{a,b})}{T}$\star$$\mathcal{C}$)
\end{mmaCell}
\begin{mmaCell}{Output}
		$\mathcal{C}$$\star$\mmaSup{(\mmaSup{$\gamma$}{$\mu$})}{T}
\end{mmaCell}
\begin{mmaCell}{Output}
		$\mathcal{C}$$\star$\mmaSup{(\mmaSup{$\gamma$}{$\mu$})}{$\dagger$}
\end{mmaCell}
In case we are using a generic representation, this works for conjugates of $C$, such as the complex conjugated $C^*$, only if they are next to a suitably conjugated gamma matrix.
\begin{mmaCell}{Input}
		\mmaDef{gOperationOrder}[\{Conjugate,\mmaDef{gT},\mmaDef{gH},Inverse\}];
		\mmaDef{gTimes}[Conjugate[\mmaDef{m}[C]],\mmaDef{y}[\{u[a,b]\}]]
		\mmaDef{gTimes}[Conjugate[\mmaDef{m}[C]],Conjugate[\mmaDef{y}[\{u[a,b]\}]]]	
\end{mmaCell}
\begin{mmaCell}{Output}
		\mmaSup{$\mathcal{C}$}{*}$\star$\mmaSup{$\gamma$}{a,b}
\end{mmaCell}
\begin{mmaCell}{Output}
		-(\mmaSup{(\mmaSup{$\gamma$}{a,b})}{$\dagger$}$\star$\mmaSup{$\mathcal{C}$}{*})
\end{mmaCell}
Finally, the setting \mtd{"Unconjugated"} makes it so that the transposes of gamma matrices are expressed in terms of $C$-intertwiners instead. In case of Hermitian conjugates and transposes, the option \mtd{gOpertationOrder} determines, which form is preferred.
\begin{mmaCell}{Input}
		\mmaDef{gCConvention}["Unconjugated"];
		\mmaDef{gT}[\mmaDef{y}[\{u[a,b]\}]]
		\mmaDef{gOperationOrder}[\{\mmaDef{gH},\mmaDef{Conjugate},\mmaDef{Inverse},\mmaDef{gT}\}];
		\mmaDef{Conjugate}[\mmaDef{y}[\{u[a,b]\}]]
\end{mmaCell}
\begin{mmaCell}{Output}
		-($\mathcal{C}$$\star$\mmaSup{$\gamma$}{a,b}$\star$\mmaSup{$\mathcal{C}$}{*})
\end{mmaCell}
\begin{mmaCell}{Output}
		-($\mathcal{C}$$\star$\mmaSup{(\mmaSup{$\gamma$}{a,b})}{$\dagger$}$\star$\mmaSup{$\mathcal{C}$}{$\dagger$})
\end{mmaCell}
Note that here we get the conjugates in terms of the $C$-intertwiner instead of the $A$-intertwiner, since we have set the \mtd{gAConvention} to \mtd{"DoNothing"}, earlier, and instead \mtd{gCConvention} is set to \mtd{"Unconjugated"}.
\begin{optDefnN}[label=gBConvention_infobox]{\mtd{gBConvention}}
	\mtd{gBConvention[opt,n]}\\
	Determines the behaviour of the $B$-intertwiners.\\[3pt]
	\textbf{Arguments}\\
	\mtd{opt} is one of the options listed below.\\[3pt]
	\textbf{Optional Arguments}\\
	\mtd{n} is the number of the subrepresentation for which the setting applies (see section \ref{sect_Subalgebras}). If this value is not specified, then the setting is changed for every subrepresentation.\\[3pt]
	\textbf{Options}\\
	\mtd{"ToLeft"}: $B$ is commuted to left whenever possible,\\
	\mtd{"ToRight"}: $B$ is commuted to right whenever possible. This is the default value.\\
	\mtd{"Unconjugated"}: $B$ is used to express complex conjugated gamma matrices in a form without conjugations,\\
	\mtd{"DoNothing"}: $B$-intertwiner is not moved, unless doing so is required by other settings. 
\end{optDefnN}
The settings for the $B$-intertwiner work in a completely analogous fashion to \mtd{gAConvention} and \mtd{gCConvention}. \mtd{"ToRight"} and \mtd{"ToLeft"} commute the $B$-intertwiners to right and left, respectively, whenever possible, complex conjugating the gamma matrices in the process.\\ 

Here we set \mtd{gCConvention} to \mtd{"DoNothing"} so that we will not get $C$-intertwiners appearing in the following examples. (For example now the program does not try to express \mtd{Conjugate[y[\{\mtu{u}[\mtu{a},\mtu{b}]\}]]} in terms of \mtd{gT[y[\{\mtu{u}[\mtu{a},\mtu{b}]\}]]} and $C$-intertwiners.)
\begin{mmaCell}{Input}
		\mmaDef{gCConvention}["DoNothing"];
		\mmaDef{gBConvention}["ToRight"];
		\mmaDef{gTimes}[\mmaDef{m}[B],\mmaDef{y}[\{u[a,b]\}]]	
		\mmaDef{gBConvention}["ToLeft"];
		\mmaDef{gTimes}[\mmaDef{Conjugate}[\mmaDef{y}[\{u[a,b]\}]],\mmaDef{m}[B]]
\end{mmaCell}
\begin{mmaCell}{Output}
		\mmaSup{(\mmaSup{$\gamma$}{a,b})}{*}$\star$$\mathcal{B}$
\end{mmaCell}
\begin{mmaCell}{Output}
		$\mathcal{B}$$\star$\mmaSup{$\gamma$}{a,b}
\end{mmaCell}
There are also related spinor conjugates, the left and right charge conjugates. These are recognised automatically, and expressed in the more compact short-hand notations \mtd{lcc} and \mtd{rcc} when the settings are appropriate. Conversely, these short-hand notations are expressed in their explicit form, if the settings require that.
\begin{mmaCell}{Input}
		\mmaDef{gBConvention}["ToRight"];
		\mmaDef{gTimes}[Inverse[\mmaDef{m}[B]],Conjugate[\mmaDef{s}[\m{$\xi$},\{\}]]]	
		\mmaDef{lcc}[\m{$\xi$},\{\}]
		\mmaDef{gBConvention}["ToLeft"];
		\mmaDef{gTimes}[\mmaDef{gT}[\mmaDef{s}[\m{$\xi$},\{\}]],\mmaDef{gH}[Inverse[\mmaDef{m}[B]]],\mmaDef{m}[A]]
		\mmaDef{rcc}[\m{$\xi$},\{\}]
\end{mmaCell}
\begin{mmaCell}{Output}
		\mmaSup{$\xi$}{c}
\end{mmaCell}
\begin{mmaCell}{Output}
		\mmaSup{$\widetilde{\xi}$}{T}$\star$B$\star$A
\end{mmaCell}
\begin{mmaCell}{Output}
		\mmaSup{$\overline{\xi}$}{c}
\end{mmaCell}
\begin{mmaCell}{Output}
		\mmaSup{$\mathcal{B}$}{$\dagger$}$\star$\mmaSup{$\widetilde{\xi}$}{*}
\end{mmaCell}
The option \mtd{"Unconjugated"} expresses the complex conjugated gamma matrices with the help of $B$-intertwiners.
\begin{mmaCell}{Input}
		\mmaDef{gBConvention}["Unconjugated"];
		Conjugate[\mmaDef{y}[\{u[\m{$\mu$},\m{$\nu$}]\}]]
\end{mmaCell}
\begin{mmaCell}{Output}
		$\mathcal{B}$$\star$\mmaSup{$\gamma$}{$\mu$,$\nu$}$\star$\mmaSup{$\mathcal{B}$}{$\dagger$}
\end{mmaCell}
Finally, as with other intertwiners, the option \mtd{"DoNothing"} will make the program not to attempt to move the $B$-intertwiner, unless required to do so by other settings.
\begin{optDefnN}[label=ggamma5Convention_infobox]{\mtd{g$\gamma$5Convention}}
	\mtd{g$\gamma$5Convention[n,opt]}\\
	Determines the behaviour of $\gamma_*$. This does not affect $\gamma^d$ in odd dimensions.\\[3pt]
	\textbf{Arguments}\\
	\mtd{opt} is one of the options listed below.\\[3pt]
	\textbf{Optional Arguments}\\
	\mtd{n} is the number of the subrepresentation for which the setting applies (see section \ref{sect_Subalgebras}). If this value is not specified, the setting is changed for every subrepresentation.\\[3pt]
	\textbf{Options}\\
	\mtd{"ToLeft"} $\gamma_*$ is commuted to left whenever possible. This is the default value.\\
	\mtd{"ToRight"} $\gamma_*$ is commuted to right whenever possible. \\
	\mtd{"DoNothing"} $\gamma_*$ is not moved, unless doing so is required by other settings. 
\end{optDefnN}
As with the intertwiners, \mtd{"ToLeft"} and \mtd{"ToRight"} attempt to commute $\gamma_*$ to left or right, respectively.  Note that here we need to use an even-dimensional representation of the Clifford algebra in order to be able to use $\gamma_*$
\begin{mmaCell}{Input}
		\mmaDef{gSetRep}[\{-1,1,1,1\},1,1];
		\mmaDef{g}$\gamma$\mmaDef{5Convention}["ToRight"];
		\mmaDef{gTimes}[\mmaDef{m}[\m{$\gamma$5}],\mmaDef{y}[\{u[\m{$\mu$},\m{$\nu$},\m{$\sigma$}]\}]]	
		\mmaDef{g}$\gamma$\mmaDef{5Convention}["ToLeft"];
		\mmaDef{gTimes}[\mmaDef{y}[\{u[\m{$\mu$},\m{$\nu$},\m{$\sigma$}]\}],\mmaDef{m}[\m{$\gamma$5}]]	
\end{mmaCell}
\begin{mmaCell}{Output}
		-(\mmaSup{$\gamma$}{$\mu$,$\nu$,$\lambda$}$\star$\mmaSub{$\gamma$}{*})
\end{mmaCell}
\begin{mmaCell}{Output}
		-(\mmaSub{$\gamma$}{*}$\star$\mmaSup{$\gamma$}{$\mu$,$\nu$,$\lambda$})
\end{mmaCell}
Unlike in the other cases with intertwiners, $\gamma_*$ does not represent any conjugation, so that there is no \mtd{"Unconjugated"} option. Also, as is the case with the intertwiners, the option \mtd{"DoNothing"} makes the program not to commute $\gamma_*$ automatically, unless required to do so by other options. 

Finally, the intertwiners can also commute among themselves. The preferred order of intertwiners is determined by \mtd{gIntertwinerOrder}.
\begin{optDefnN}[label=gIntertwinerOrder_infobox]{\mtd{gIntertwinerOrder}}
	\mtd{gIntertwinerOrder[n,opt]}\\
	Determines the order to which the special matrices that are next to each other are commuted.\\[3pt]
	\textbf{Arguments}\\
	\mtd{opt} is a list containing \mtd{m[A], m[C], m[B]} and \mtd{m[$\gamma5$]} in some order. These matrices are automatically commuted to the same order if possible. The default order is \mtd{\{m[B],m[A],m[$\gamma$5],m[C]\}} so that the B-intertwiner is commuted to left of all other intertwiners, if possible, then $A$ is commuted to left of all other intertwiners except $B$, and so on.\\[3pt]
	\textbf{Optional Arguments}\\
	\mtd{n} is the number of the subrepresentation for which the setting applies (see section \ref{sect_Subalgebras}). If this value is not specified, then the setting is changed for every subrepresentation.
\end{optDefnN}
For example, if we have a product of $\gamma_*$ and $A$, which can be commuted for any representation, we can set their preferred order by modifying \mtd{gIntertwinerOrder}. 
\begin{mmaCell}{Input}
		\mmaDef{gIntertwinerOrder}[\{\mmaDef{m}[\m{$\gamma$5}],\mmaDef{m}[A],\mmaDef{m}[B],\mmaDef{m}[C]\}];
		\mmaDef{gTimes}[\mmaDef{m}[\m{$\gamma$5}],\mmaDef{m}[A]]	
		\mmaDef{gTimes}[\mmaDef{m}[A],\mmaDef{m}[\m{$\gamma$5}]]	
		\mmaDef{gIntertwinerOrder}[\{\mmaDef{m}[A],\mmaDef{m}[\m{$\gamma$5}],\mmaDef{m}[B],\mmaDef{m}[C]\}];
		\mmaDef{gTimes}[\mmaDef{m}[\m{$\gamma$5}],\mmaDef{m}[A]]	
		\mmaDef{gTimes}[\mmaDef{m}[A],\mmaDef{m}[\m{$\gamma$5}]]	
\end{mmaCell}
\begin{mmaCell}{Output}
		\mmaSub{$\gamma$}{*}$\star$$\mathcal{A}$
\end{mmaCell}
\begin{mmaCell}{Output}
		-(\mmaSub{$\gamma$}{*}$\star$$\mathcal{A}$)
\end{mmaCell}
\begin{mmaCell}{Output}
		-($\mathcal{A}$$\star$\mmaSub{$\gamma$}{*})
\end{mmaCell}
\begin{mmaCell}{Output}
		$\mathcal{A}$$\star$\mmaSub{$\gamma$}{*}
\end{mmaCell}
As is the case with the gamma matrices, some intertwiners cannot be commuted for generic representations, but only for the special representation, for which the product of two intertwiners can actually be expressed in the term of the third.
\begin{mmaCell}{Input}
		\mmaDef{gUseSpecialRep}[False];
		\mmaDef{gTimes}[\mmaDef{m}[C],\mmaDef{m}[B]]	
		\mmaDef{gTimes}[\mmaDef{m}[B],\mmaDef{m}[C]]	
		\mmaDef{gUseSpecialRep}[True];
		\mmaDef{gTimes}[\mmaDef{m}[C],\mmaDef{m}[B]]	
\end{mmaCell}
\begin{mmaCell}{Output}
		$\mathcal{C}$$\star$$\mathcal{B}$
\end{mmaCell}
\begin{mmaCell}{Output}
		$\mathcal{B}$$\star$$\mathcal{C}$
\end{mmaCell}
\begin{mmaCell}{Output}
		-A
\end{mmaCell}
\vspace{-5pt}
\subsection{Products of Gamma Matrices}
\vspace{-5pt}
As explained in section \ref{sect_Review_Basis}, antisymmetrised products of gamma matrices $\gamma^{(1)},\dotsc,\gamma^{(d)}$ form a basis of all $2^{m} \times 2^{m}$ matrices, where $d=2m+1$ or $d=2m$. In particular, this implies that given any product of antisymmetrised products gamma matrices, such as $\gamma^{\mu\nu} \gamma^{ab}$, we can express that as a linear combination of antisymmetrised products of gamma matrices $\gamma^{(1)},\dotsc,\gamma^{(n)}$.

This operation is implemented by function \mtd{gOrd}.
\begin{funcDefnN}[label=gOrd_infobox]{\mtd{gOrd}}
	\mtd{gOrd[expr]}\\
	Expresses \mtd{expr} in a form in which gamma matrices appear only in antisymmetrised products.\\[3pt]
	\textbf{Arguments}\\
	\mtd{expr} is any expression
\end{funcDefnN}
For example, the product $\gamma^{\mu \nu}\gamma^{\sigma}$ can be expressed as a linear combination of antisymmetrised products of 3 and 1 gamma matrices. Here $\eta$ denotes the metric used to define the Clifford algebra \eqref{Clifford_algebra_anticommutation}. The symbol \mmaSup{$\delta$}{*} denotes different cases i.e. which indices are equal, and which are not (see the next section \ref{sect_Functions_products_gCase}).
\begin{mmaCell}{Input}
		\mmaDef{gOrd}[\mmaDef{y}[\{u[\m{$\mu$},\m{$\nu$}]\},\{u[\m{$\sigma$}]\}]]	
\end{mmaCell} 
\begin{mmaCell}{Output}
		\mmaSub{\mmaSup{$\delta$}{*}}{\{$\mu$\},\{$\sigma$\}}\mmaSup{$\hat{\eta}$}{$\nu$,$\sigma$}\mmaSup{$\gamma$}{$\mu$}-\mmaSub{\mmaSup{$\delta$}{*}}{\{$\nu$\},\{$\sigma$\}}\mmaSup{$\hat{\eta}$}{$\mu$,$\sigma$}\mmaSup{$\gamma$}{$\nu$}+\mmaSup{$\gamma$}{$\mu$,$\nu$,$\sigma$}
\end{mmaCell} 
\vspace{-5pt}
\subsubsection{Cases} \label{sect_Functions_products_gCase}
\vspace{-5pt}
The result of expressing products of gamma matrices in terms of antisymmetrised products depends on whether the indices appearing in various gamma matrices are equal or not. Therefore, to represent the most general result, we denote different cases by \mtd{gCase}, which appears in the previous output as $\delta^*$.
\begin{objDefnN}[label=gCase_infobox]{\mtd{gCase}}
	\mtd{gCase[\{X1\},...,\{Xn\}]}\\
	Denotes a case where indices in lists \mtd{X1,..,Xn} are all equal to the indices in the same list, and not equal to the indices in different lists. This can be though of as a kind of generalised Kronecker delta, which is equal to one when this condition is satisfied, and is otherwise zero.\\[3pt]
	\textbf{Arguments}\\
	\mtd{X1,..,Xn} are lists of indices that are assumed to be equal to other indices in the same list, and not equal to indices in different lists.\\[3pt]
	\textbf{Notation}\\
	\mtd{\mmaSub{\mmaSup{$\delta$}{*}}{\{X1\},...,\{Xn\}}}
\end{objDefnN}
To illustrate how \mtd{gCase} is used, let us consider as an example \mtd{\mmaSub{\mmaSup{$\delta$}{*}}{\{$\mu$\},\{$\nu$,$\sigma$\}}}. This has two lists, \mtd{\{$\mu$\}} and \mtd{\{$\nu$,$\sigma$\}}. The indices $\nu$ and $\sigma$ are therefore equal in the case where this $\delta^*$ is not zero. Similarly, $\mu$ and $\nu$ are in the different lists, so they must be unequal. Therefore, we see that we can read this $\delta^*$ as
\begin{align}
\text{\mtd{\mmaSub{\mmaSup{$\delta$}{*}}{\{$\mu$\},\{$\nu$,$\sigma$\}}}} = \begin{cases}
1 \text{ in case } \nu = \sigma \neq \mu,\\[3pt]
0 \text{ otherwise.}
\end{cases}\nonumber
\end{align}
As another example, we can read the whole expression in the previous example as
\begin{align}
\text{\mtd{\mmaSub{\mmaSup{$\delta$}{*}}{\{$\mu$\},\{$\sigma$\}}\mmaSup{$\eta$}{$\nu$,$\sigma$}\mmaSup{$\gamma$}{$\mu$}-\mmaSub{\mmaSup{$\delta$}{*}}{\{$\nu$\},\{$\sigma$\}}\mmaSup{$\eta$}{$\mu$,$\sigma$}\mmaSup{$\gamma$}{$\nu$} \mmaSup{$\gamma$}{$\mu$,$\nu$,$\sigma$}}} = \begin{cases}
\eta^{\nu \sigma} \gamma^\mu + \gamma^{\mu \nu \sigma}\text{ in case } \sigma \neq \mu,\\[3pt]
\eta^{\mu \sigma} \gamma^\nu + \gamma^{\mu \nu \sigma}\text{ in case } \sigma \neq \nu,\\[3pt]
\gamma^{\mu \nu \sigma} \text{ otherwise.}
\end{cases} \nonumber
\end{align}
As with other functions, \mtd{gOrd} can be used on expressions that contain many terms with gamma matrices appearing in different places. For example
\begin{mmaCell}{Input}
		2*\mmaDef{gTimes}[\mmaDef{dc}[\m{$\xi$},\{\}],\mmaDef{y}[\{u[a]\},\{u[b]\}],\mmaDef{s}[\m{$\xi$},\{\}]]+
		  \mmaDef{gTimes}[\mmaDef{y}[\{u[\m{$\mu$}]\},\{d[\m{$\nu$}]\}],\mmaDef{s}[\m{$\xi$},\{\}]]//\mmaDef{gOrd}
\end{mmaCell} 
\begin{mmaCell}{Output}
		2\mmaSup{$\hat{\eta}$}{a,b}($\overline{\xi}$$\star$$\widetilde{\xi}$)+\mmaSub{\mmaSup{$\gamma$}{$\mu$}}{$\nu$}$\star$$\widetilde{\xi}$+2($\overline{\xi}$$\star$\mmaSup{$\gamma$}{a,b}$\star$$\widetilde{\xi}$)+\mmaSub{\mmaSup{$\hat{\eta}$}{$\mu$}}{$\nu$}$\widetilde{\xi}$
\end{mmaCell}
If there are other objects, such as forms that are contracted with some indices appearing in the gamma matrices, the symmetry or antisymmetry properties of these objects is taken into account when expressing the gamma matrix products in a new form. For example, in the product $F_{\mu \nu}\gamma^\mu \gamma^\nu$, only the antisymmetric part $F_{\mu \nu}\gamma^{\mu \nu}$ can appear if $F_{\mu \nu}$ is antisymmetric. Similarly, only the symmetric part $F_{\mu \nu}\eta^{\mu \nu}$ can appear if $F_{\mu \nu}$ is symmetric.
\begin{mmaCell}{Input}
		\mmaDef{y}[\{u[\m{$\mu$}]\},\{u[\m{$\nu$}]\}] //\mmaDef{gOrd}
		\mmaDef{gForm}[F,\{\m{$\mu$},\m{$\nu$}\}]*\mmaDef{y}[\{u[\m{$\mu$}]\},\{u[\m{$\nu$}]\}] //\mmaDef{gOrd}
		\mmaDef{gSymm}[F,\{\m{$\mu$},\m{$\nu$}\}]*\mmaDef{y}[\{u[\m{$\mu$}]\},\{u[\m{$\nu$}]\}] //\mmaDef{gOrd}//\mmaDef{gSimplify}
\end{mmaCell} 
\begin{mmaCell}{Output}
		\mmaSup{$\eta$}{$\mu$,$\nu$}I+\mmaSup{$\gamma$}{$\mu$,$\nu$}
\end{mmaCell}
\begin{mmaCell}{Output}
		\mmaSub{F}{$\lambda$1,$\lambda$2}\mmaSup{$\gamma$}{$\lambda$1,$\lambda$2}
\end{mmaCell}
\begin{mmaCell}{Output}
		\mmaSub{\mmaSup{F}{$\lambda$1}}{$\lambda$1}I
\end{mmaCell}
Here in the last example the function \mtd{gSimplify} is used to simplify result (see section \ref{sect_Functions_Simplify}), in effect deleting the part where symmetric $F_{\mu \nu}$ is contracted with antisymmetric $\gamma^{\mu \nu}$. Notice also that \mtd{gOrd} replaces the dummy indices $\mu$ and $\nu$ with new dummy indices \mtd{$\lambda$1} and \mtd{$\lambda$2}. This is done so that terms that differ only in the names of dummy indices are put into the same form, and can thus be simplified. As a further example, consider the following.
\begin{mmaCell}{Input}
		\mmaDef{gForm}[F,\{\m{$\mu$},\m{$\nu$}\}]*\mmaDef{y}[\{u[\m{$\mu$}]\},\{u[\m{$\nu$}]\}]+\mmaDef{gForm}[F,\{a,b\}]*\mmaDef{y}[\{u[a]\},\{u[b]\}]//
		  \mmaDef{gOrd}
\end{mmaCell} 
\begin{mmaCell}{Output}
		2\mmaSub{F}{$\lambda$1,$\lambda$2}\mmaSup{$\gamma$}{$\lambda$1,$\lambda$2}
\end{mmaCell}
If there are repeated indices in the gamma matrices, one should be a lower and other an upper index. Then \mtd{gOrd} sums over repeated indices. However, indices which are integers are naturally not summed over.
\begin{mmaCell}{Input}
		\mmaDef{gSetRep}[\{-1,1,1,1,1,1,1\},-1,1];
		\mmaDef{gOrd}[\mmaDef{y}[\{u[\m{$\mu$},\m{$\nu$}]\},\{u[\m{$\sigma$}],d[\m{$\mu$}]\}]]
		\mmaDef{gOrd}[\mmaDef{y}[\{u[\m{$\mu$},1]\},\{1\}]]	
\end{mmaCell} 
\begin{mmaCell}{Output}
		6\mmaSup{$\hat{\eta}$}{$\nu$,$\sigma$}I+5\mmaSup{$\gamma$}{$\mu$,$\nu$}
\end{mmaCell}
\begin{mmaCell}{Output}
		\mmaSub{\mmaSup{$\delta$}{*}}{\{1\},\{$\mu$\}}\mmaSup{$\gamma$}{$\mu$}
\end{mmaCell}
Assumptions are taken into account by \mtd{gOrd} (for more information on using assumptions, see section \ref{sect_Functions_Assumptions}). Here in the first line we make the assumption that $\mu$ and $\nu$ are not equal. Thus, when expressing the product $\gamma_{\mu} \gamma_{\nu}$ in terms of antisymmetrised products, only the term which is proportional to $\gamma_{\mu \nu}$ appears.
\begin{mmaCell}{Input}
		\mmaDef{gAddAssumptions}[\mmaDef{gUnequal}[\m{$\mu$},\m{$\nu$}]];
		\mmaDef{y}[\{d[\m{$\mu$}]\},\{d[\m{$\nu$}]\}]//\mmaDef{gOrd}
\end{mmaCell} 
\begin{mmaCell}{Output}
		\mmaSub{$\gamma$}{$\mu$,$\nu$}
\end{mmaCell}
\vspace{-5pt}
\subsection{Real and Imaginary Parts}
\vspace{-5pt}
Commutative objects, such as tensors, scalars and bilinears can be automatically split into their real and imaginary parts by using the command \mtd{gReIm}.
\begin{funcDefnN}[label=gReIm_infobox]{\mtd{gReIm}}
	\mtd{gReIm[expr]}\\
	Splits the commutative objects appearing in expression \mtd{expr} to their real and imaginary parts.\\[3pt]
	\textbf{Arguments}\\
	\mtd{expr} is the expression to be modified.	
\end{funcDefnN}
For example a tensor $F_{a,b}$ can be split into its real and imaginary parts as follows.
\begin{mmaCell}{Input}
		\mmaDef{gReIm}[\mmaDef{gTensor}[F,\{a,b\}]]				
\end{mmaCell} 
\begin{mmaCell}{Output}
		$\Ii$\mmaDef{gIm}[\mmaSub{$\ddot{F}$}{a,b}]+\mmaDef{gRe}[\mmaSub{$\ddot{F}$}{a,b}]		
\end{mmaCell} 
The real and imaginary parts are denoted by \mtd{gRe} and \mtd{gIm}, respectively.
\begin{objDefnN}[label=gRe_infobox]{\mtd{gRe}}
	\mtd{gRe[expr]}\\
	Denotes the real part of \mtd{expr}.\\[3pt]
	\textbf{Arguments}\\
	\mtd{expr} is the expression whose real part \mtd{gRe[expr]} denotes.	
\end{objDefnN}
\begin{objDefnN}[label=gIm_infobox]{\mtd{gIm}}
	\mtd{gIm[expr]}\\
	Denotes the imaginary part of \mtd{expr}.\\[3pt]
	\textbf{Arguments}\\
	\mtd{expr} is the expression whose imaginary part \mtd{gIm[expr]} denotes.	
\end{objDefnN}
The objects \mtd{gRe} and \mtd{gIm} have some additional functionality compared to the basic \mtd{Re} and \mtd{Im}. For example, they check automatically whether a tensor or a spinor is assumed to be real. Here \mtd{gAddAssumptions} is a function that is used to make assumptions (see section \ref{sect_Functions_Assumptions}), and \mtd{gImaginaries} denotes the set of imaginary scalars, matrices, tensors and spinors.
\begin{mmaCell}{Input}
		\mmaDef{gAddAssumptions}[\{F$\in$Reals, G$\in$gImaginaries\}];
		\mmaDef{gRe}[\mmaDef{gTensor}[F,\{a,b\}]]
		\mmaDef{gRe}[\mmaDef{gTensor}[G,\{a,b\}]]
		\mmaDef{gRe}[\mmaDef{gTensor}[H,\{a,b\}]]	
		\mmaDef{gIm}[\mmaDef{gTensor}[F,\{a,b\}]]
		\mmaDef{gIm}[\mmaDef{gTensor}[G,\{a,b\}]]
		\mmaDef{gIm}[\mmaDef{gTensor}[H,\{a,b\}]]	
\end{mmaCell} 
\begin{mmaCell}{Output}
		\mmaSub{$\ddot{F}$}{a,b}
\end{mmaCell} 
\begin{mmaCell}{Output}
		0
\end{mmaCell} 
\begin{mmaCell}{Output}
		\mmaDef{gRe}[\mmaSub{$\ddot{H}$}{a,b}]	
\end{mmaCell} 
\begin{mmaCell}{Output}
		0	
\end{mmaCell} 
\begin{mmaCell}{Output}
		\mmaSub{$\ddot{G}$}{a,b}	
\end{mmaCell} 
\begin{mmaCell}{Output}
		\mmaDef{gIm}[\mmaSub{$\ddot{H}$}{a,b}]		
\end{mmaCell} 
These also simplify the expressions by, for example, eliminating imaginary units inside \mtd{gRe} and \mtd{gIm}.
\begin{mmaCell}{Input}
		\mmaDef{gRe}[I*a]
\end{mmaCell} 
\begin{mmaCell}{Output}
		-\mmaDef{gIm}[a]	
\end{mmaCell} 
Furthermore, in order to ensure automatic simplification of similar terms whenever possible, the expressions inside \mtd{gRe} and \mtd{gIm} are put into a standard form, by conjugating the expressions, if needed. The standard form is defined so that the object which is first in alphabetical order is not conjugated. 
\begin{mmaCell}{Input}
		\mmaDef{gRe}[Conjugate[a]*b]
		\mmaDef{gRe}[a*Conjugate[b]]	
		\mmaDef{gRe}[Conjugate[a]*b]-\mmaDef{gIm}[I*a*Conjugate[b]]	
\end{mmaCell} 
\begin{mmaCell}{Output}
		\mmaDef{gRe}[a\mmaSup{(b)}{*}]		
\end{mmaCell}
\begin{mmaCell}{Output}
		\mmaDef{gRe}[a\mmaSup{(b)}{*}]			
\end{mmaCell} 
\begin{mmaCell}{Output}
		0			
\end{mmaCell}  
The command \mtd{gReIm} is especially useful if only the real or imaginary part of an object appears. For instance, consider the following.
\begin{mmaCell}{Input}
		\mmaDef{gRe}Im[\mmaDef{gForm}[X,\{u[\m{$\mu$},\m{$\nu$}]\}]+Conjugate[\mmaDef{gForm}[X,\{u[\m{$\mu$},\m{$\nu$}]\}]]]		
\end{mmaCell} 
\begin{mmaCell}{Output}
		2\mmaDef{gRe}[\mmaSup{X}{$\mu$,$\nu$}]		
\end{mmaCell} 
\vspace{-5pt}
\subsection{Conjugations} \label{sect_functions_conjugations}
\vspace{-5pt}
\subsubsection{Complex Conjugation}
\vspace{-5pt}
Complex conjugation of objects, including gamma matrices, forms and other objects is still done by using \mtd{Conjugate}. This has, however, been modified to take into account, for example, whether a tensor is assumed to be real. Also the notation has been changed in order to make long expressions more readable.
\begin{funcDefnN}[label=Conjugate_infobox]{\mtd{Conjugate}}
	\mtd{Conjugate[expr]}\\
	Gives the complex conjugate of \mtd{expr}.\\[3pt]
	\textbf{Arguments}\\
	\mtd{expr} is the expression to be complex conjugated.\\[3pt]
	\textbf{Notation}\\
	\mtd{\mmaSup{expr}{*}}
\end{funcDefnN}
For example, we can assume that the form $F$ and scalar $a$ are real, and form $G$ and scalar $b$ are imaginary. Then the redefined \mtd{Conjugate} acts on these as follows.
\begin{mmaCell}{Input}
		\mmaDef{gUseSpecialRep}[False];
		\mmaDef{gClearAssumptions}[];
		\mmaDef{gAddAssumptions}[\{F$\in$Reals,G$\in$gImaginaries,a$\in$Reals\}];	
		Conjugate[\mmaDef{gForm}[F,\{\m{$\mu$},\m{$\nu$}\}]]
		Conjugate[\mmaDef{gForm}[G,\{\m{$\mu$},\m{$\nu$},\m{$\sigma$}\}]]
		Conjugate[\mmaDef{gForm}[H,\{\m{$\mu$}\}]]
		Conjugate[3*I*a+4]
		Conjugate[\mmaDef{m}[A]]
\end{mmaCell} 
\begin{mmaCell}{Output}
		\mmaSub{F}{$\mu$,$\nu$}		
\end{mmaCell} 
\begin{mmaCell}{Output}
		-\mmaSub{G}{$\mu$,$\nu$,$\sigma$}			
\end{mmaCell} 
\begin{mmaCell}{Output}		
		-\mmaSub{\mmaSup{H}{*}}{$\mu$}			
\end{mmaCell} 
\begin{mmaCell}{Output}	
		4-3$\Ii$a		
\end{mmaCell} 
\begin{mmaCell}{Output}	
		\mmaSup{$\mathcal{A}$}{*}	
\end{mmaCell} 
\mtd{Conjugate} can also be used on expressions containing sums, multiplications etc.
\begin{mmaCell}{Input}
		\mmaDef{gBConvention}["ToLeft"];
		\mmaDef{gTimes}[\mmaDef{gH}[\mmaDef{s}[\m{$\xi$},\{i\}]],\mmaDef{y}[\{u[\m{$\mu$}]\}]]*\mmaDef{gForm}[H,\{\m{$\mu$}\}]+
		  a*\mmaDef{gTimes}[\mmaDef{y}[\{u[\m{$\mu$},\m{$\nu$},\m{$\lambda$}]\}],\mmaDef{s}[\m{$\xi$},\{\}]]//Conjugate
\end{mmaCell} 
\begin{mmaCell}{Output}
		a(\mmaSup{(\mmaSup{$\gamma$}{$\mu$,$\nu$,$\lambda$})}{*}$\star$\mmaSup{$\widetilde{\xi}$}{*})+\mmaSub{\mmaSup{H}{*}}{$\mu$}(\mmaSup{\mmaSub{$\widetilde{\xi}$}{i}}{T}$\star$\mmaSup{(\mmaSup{$\gamma$}{$\mu$})}{*})
\end{mmaCell} 
\vspace{-5pt}
\subsubsection{Transpose}
\vspace{-5pt}
Non-commutative objects can be transposed (in spinor indices) using \mtd{gT}. Obviously this does not affect commutative objects such as tensor components or scalars.
\begin{funcDefnN}[label=gT_infobox]{\mtd{gT}}
	\mtd{gT[expr]}\\
	Gives the transpose of \mtd{expr}.\\[3pt]
	\textbf{Arguments}\\
	\mtd{expr} is the expression to be transposed.\\[3pt]
	\textbf{Notation}\\
	\mtd{\mmaSup{expr}{T}}
\end{funcDefnN}
\begin{mmaCell}{Input}
		\mmaDef{gT}[3*a]
		\mmaDef{gT}[\mmaDef{gForm}[F,\{\m{$\mu$},\m{$\nu$}\}]]
		\mmaDef{gT}[\mmaDef{y}[\{u[\m{$\mu$},\m{$\nu$}]\}]]
\end{mmaCell} 
\begin{mmaCell}{Output}
		3a	
\end{mmaCell} 
\begin{mmaCell}{Output}
		\mmaSub{F}{$\mu$,$\nu$}		
\end{mmaCell} 
\begin{mmaCell}{Output}
		\mmaSup{(\mmaSup{$\gamma$}{$\mu$,$\nu$})}{T}	
\end{mmaCell} 
As with other functions, \mtd{gT} can be used on more complicated expressions as well. As usual, transposition reverses the order of transposed matrices and vectors, which results in a change of sign, if the operation exchanges the order of anticommuting spinors. Note that in the second term here, due to the \mtd{gAConvention} setting \mtd{"ToLeft"}, the $A$-intertwiner is automatically commuted to left.
\begin{mmaCell}{Input}
		\mmaDef{gAConvention}["ToLeft"];
		\mmaDef{gSetRep}[\{-1,1,1,1\},1,1];
		\mmaDef{gUseSpecialRep}["False"];
		\mmaDef{y}[\{u[\m{$\mu$},\m{$\nu$}]\}]+3*\mmaDef{gTimes}[\mmaDef{dc}[\m{$\xi$},\{i\}],\mmaDef{y}[\{u[\m{$\mu$},\m{$\nu$}]\}],\mmaDef{s}[\m{$\xi$},\{j\}]]//\mmaDef{gT}
\end{mmaCell} 
\begin{mmaCell}{Output}
		\mmaSup{(\mmaSup{$\gamma$}{$\mu$,$\nu$})}{T}+3(\mmaSup{\mmaSub{$\widetilde{\xi}$}{j}}{T}$star$\mmaSup{$\mathcal{A}$}{*}$\star$\mmaSup{(\mmaSup{$\gamma$}{$\mu$,$\nu$})}{*}$\star$\mmaSup{\mmaSub{$\widetilde{\xi}$}{i}}{*})
\end{mmaCell} 
\vspace{-5pt}
\subsubsection{Hermitian Conjugation}
\vspace{-5pt}
Hermitian conjugation, which is the combination of complex conjugation and transposition, is implemented by function \mtd{gH}. On commutative objects this acts simply as the ordinary complex conjugation.
\begin{funcDefnN}[label=gH_infobox]{\mtd{gH}}
	\mtd{gH[expr]}\\
	Gives the Hermitian conjugate of \mtd{expr}.\\[3pt]
	\textbf{Arguments}\\
	\mtd{expr} is the expression to be Hermitian conjugated.\\[3pt]
	\textbf{Notation}\\
	\mtd{\mmaSup{expr}{$\dagger$}}
\end{funcDefnN}
\begin{mmaCell}{Input}
		\mmaDef{gOperationOrder}[\{\mmaDef{gH},Inverse,Conjugate,\mmaDef{gT}\}];
		\mmaDef{gH}[\mmaDef{gForm}[H,\{\m{$\mu$}\}]]
		\mmaDef{gH}[\mmaDef{y}[\{u[\m{$\mu$},\m{$\nu$}]\}]]
		\mmaDef{gH}[\mmaDef{dc}[\m{$\xi$},\{i\}]]
		\mmaDef{gH}[\mmaDef{m}[B]]		
\end{mmaCell} 
\begin{mmaCell}{Output}
		\mmaSub{\mmaSup{H}{*}}{$\mu$}	
\end{mmaCell} 
\begin{mmaCell}{Output}	
		\mmaSup{(\mmaSup{$\gamma$}{$\mu$,$\nu$})}{$\dagger$}
\end{mmaCell} 
\begin{mmaCell}{Output}
		-($\mathcal{A}$$\star$\mmaSub{$\widetilde{\xi}$}{i})
\end{mmaCell} 
\begin{mmaCell}{Output}
		\mmaSup{$\mathcal{B}$}{$\dagger$}
\end{mmaCell} 
On expressions containing non-commutative multiplications defined using \mtd{gTimes}, the order of non-commutative objects is reversed. Note that in the case of the second term, the $A$-intertwiner is again commuted automatically to left, and therefore Hermitian conjugation acts effectively just as an exchange of indices $i$ and $j$. This relation can be used to simplify some expressions, as will be seen in section \ref{sect_Functions_Bilinears_Options}.
\begin{mmaCell}{Input}
		\mmaDef{gAConvention}["ToLeft"];
		\mmaDef{gSetRep}[\{-1,1,1,1\},1,1];
		I*\mmaDef{gTimes}[\mmaDef{y}[\{u[\m{$\mu$},\m{$\nu$}]\}],\mmaDef{s}[\m{$\xi$},\{\}]]+
		  3*\mmaDef{gTimes}[\mmaDef{dc}[\m{$\xi$},\{i\}],\mmaDef{y}[\{u[\m{$\mu$},\m{$\nu$}]\}],\mmaDef{s}[\m{$\xi$},\{j\}]]//\mmaDef{gH}
\end{mmaCell} 
\begin{mmaCell}{Output}
		-$\Ii$(\mmaSup{$\widetilde{\xi}$}{$\dagger$}$\star$\mmaSup{(\mmaSup{$\gamma$}{$\mu$,$\nu$})}{$\dagger$})-3(\mmaSub{$\bar{\xi}$}{j}$star$\mmaSup{$\gamma$}{$\mu$,$\nu$}$\star$\mmaSub{$\widetilde{\xi}$}{i})
\end{mmaCell}
\vspace{-5pt}
\subsubsection{Associated Options}
\vspace{-5pt}
As can be seen by looking at the relations \eqref{AT=Ac}-\eqref{app_intertwiners_last} in appendix \ref{app_Rep_Properties}, some conjugates of the intertwiners can be expressed in terms of other conjugates. For example the complex conjugated $A^*$ is related to the transposed $A^T$ by a sign. In this situation it is not clear, which form should be used. The form, which is the most desired one to appear in the output is determined by the option \mtd{gOperationOrder}.
\begin{optDefnN}[label=gOperationOrder_infobox]{\mtd{gOperationOrder}}
	\mtd{gOperationOrder[n,opt]}\\
	Determines, which form of a matrix is used, if, for example, the inverse and Hermitian conjugation of a matrix can be related to each other.\\[3pt]
	\textbf{Arguments}\\
	\mtd{opt} is a list containing \mtd{gH, gT, Conjugate} and \mtd{Inverse} in some order. This determines, which form of a matrix should be preferred in the final output if two or more of these operations on the matrix are equivalent up to a sign (or in case of gamma matrices, up to a sign and conjugation by intertwiners). The default value of \mtd{gOperationOrder} is \mtd{\{Inverse, gH, Conjugate, gT\}}, so that if, for example, it is possible to express the transposed matrix $A^T$ in terms of the complex conjugated $A^*$, the latter is then used in the output.\\[3pt]
	\textbf{Optional Arguments}\\
	\mtd{n} is the number of the subrepresentation for which the setting applies (see section \ref{sect_Subalgebras}). If this value is not specified, then the setting is changed for every subrepresentation.
\end{optDefnN}
\begin{mmaCell}{Input}
		\mmaDef{gUseSpecialRep}[False];
		\mmaDef{gOperationOrder}[\{Inverse,\mmaDef{gH},\mmaDef{Conjugate},\mmaDef{gT}\}];
		\mmaDef{gT}[\mmaDef{m}[A]]
		\mmaDef{Conjugate}[\mmaDef{m}[A]]
		\mmaDef{gH}[\mmaDef{m}[C]]
		\mmaDef{Inverse}[\mmaDef{m}[B]]
		\mmaDef{gOperationOrder}[\{\mmaDef{gT},\mmaDef{Conjugate},\mmaDef{Inverse},\mmaDef{gH}\}];
		\mmaDef{gT}[\mmaDef{m}[A]]
		\mmaDef{Conjugate}[\mmaDef{m}[A]]
		\mmaDef{gH}[\mmaDef{m}[C]]
		\mmaDef{Inverse}[\mmaDef{m}[B]]		
\end{mmaCell}
\begin{mmaCell}{Output}
		-\mmaSup{$\mathcal{A}$}{*}
\end{mmaCell}
\begin{mmaCell}{Output}
		\mmaSup{$\mathcal{A}$}{*}
\end{mmaCell}
\begin{mmaCell}{Output}
		\mmaSup{$\mathcal{C}$}{-1}
\end{mmaCell}
\begin{mmaCell}{Output}
		\mmaSup{$\mathcal{B}$}{-1}
\end{mmaCell}
\begin{mmaCell}{Output}
		\mmaSup{$\mathcal{A}$}{T}
\end{mmaCell}
\begin{mmaCell}{Output}
		-\mmaSup{$\mathcal{A}$}{T}
\end{mmaCell}
\begin{mmaCell}{Output}
		-\mmaSup{$\mathcal{C}$}{*}
\end{mmaCell}
\begin{mmaCell}{Output}
		\mmaSup{$\mathcal{B}$}{*}
\end{mmaCell}
This works also on gamma matrices, provided the \mtd{"Unconjugated"} option is enabled for the intertwiner relating the two operations on matrices.
\begin{mmaCell}{Input}
		\mmaDef{gAConvention}["DoNothing"];
		\mmaDef{gCConvention}["DoNothing"];
		\mmaDef{gOperationOrder}[\{Inverse,\mmaDef{gH},\mmaDef{Conjugate},\mmaDef{gT}\}];
		\mmaDef{gT}[\mmaDef{y}[\{u[a,b,c]\}]]
		\mmaDef{gAConvention}["Unconjugated"];
		\mmaDef{gT}[\mmaDef{y}[\{u[a,b,c]\}]]
\end{mmaCell}
\begin{mmaCell}{Output}
		\mmaSup{(\mmaSup{$\gamma$}{a,b,c})}{T}
\end{mmaCell}
\begin{mmaCell}{Output}
		\mmaSup{$\mathcal{A}$}{*}$\star$\mmaSup{(\mmaSup{$\gamma$}{a,b,c})}{*}$\star$\mmaSup{(\mmaSup{$\mathcal{A}$}{*})}{-1}
\end{mmaCell}
In case \mtd{gOperationOrder} is in conflict with \mtd{gIntertwinerOrder}, the latter overrides the former.
\vspace{-5pt}
\subsection{Bilinears} \label{sect_bilinears}
\vspace{-5pt}
As discussed in section \ref{sect_Review_Spinor_BLs_and_Tensors}, bilinears are forms whose components are given by sandwiching an antisymmetrised product of gamma matrices between two spinors. They are denoted by \mtd{gBL}.
\begin{objDefnN}[label=gBL_infobox]{\mtd{gBL}}
	\mtd{gBL[F,\{i\},\{u[$\mu$],d[$\nu$]\},n]}\\
	Denotes components of a form \mtd{F} defined by some bilinear with Lorentz indices \mtd{$\mu$} and \mtd{$\nu$}, and additional indices \mtd{i}, transforming under the \mtd{n}:th subalgebra.\\[3pt]
	\textbf{Arguments}\\
	\mtd{i} is a list of additional indices (can be empty),\\
	\mtd{$\mu$} and \mtd{$\nu$} are lists of Lorentz indices.\\[3pt]
	\textbf{Optional Arguments}\\
	\mtd{n} is the number of the subrepresentation under which the bilinear transforms (see section \ref{sect_Subalgebras}). If this is left unspecified then the default value 0, corresponding to the full representation, is used. In practice this parameter can be almost always left unspecified, and the program will take care of this automatically. The only exceptions are the situations, in which the spinor appears on the left-hand side of a function assignment, in rules or in assumptions.\\[3pt]
	\textbf{Notation}\\
	\mtd{\mmaSup{\mmaSub{F}{i:$\nu$}}{$\mu$}}
\end{objDefnN}
For example, we can consider the form $F_{12}^{\mu \nu}$, whose components are given by expression $\xi_1^\dagger \gamma^{\mu \nu} \xi_2$. Then, we can denote the bilinear corresponding to this expression by
\begin{mmaCell}{Input}
		\mmaDef{gBL}[F,\{1,2\},\{u[\m{$\mu$},\m{$\nu$}]\}]
\end{mmaCell} 
\begin{mmaCell}{Output}
		\mmaSup{\mmaSub{F}{1,2:}}{$\mu$,$\nu$}
\end{mmaCell}
This is just a more compact notation for the following, completely equivalent expression.
\begin{mmaCell}{Input}
		\mmaDef{gTimes}[\mmaDef{dc}[\m{$\xi$},\{1\}],\mmaDef{y}[\{u[\m{$\mu$},\m{$\nu$}]\}],\mmaDef{s}[\m{$\xi$},\{2\}]]
\end{mmaCell} 
\begin{mmaCell}{Output}
		\mmaSub{$\overline{\xi}$}{1}$\star$\mmaSup{$\gamma$}{$\mu,\nu$}$\star$\mmaSub{$\xi$}{2}
\end{mmaCell}
However, there is some additional functionality for calculations with bilinears in \mtd{gBL} notation (see section \ref{sect_Functions_Bilinears_Options}). To be able to translate between the \mtd{gBL} notation and product notation, we can define the correspondence between the two by using the function \mtd{gSetBilinearNames}.
\begin{funcDefnN}[label=gSetBilinearNames_infobox]{\mtd{gSetBilinearNames}}
	\mtd{gSetBilinearNames[n,bilinearNames,ccBilinearNames,$\gamma$5bilinearNames, cc$\gamma$5BilinearNames,$\xi$,$\eta$]}
	Defines the correspondence between bilinears denoted by product of spinors and gamma matrices, and more condensed notation using \mtd{gBL}.\\[3pt]
	\textbf{Arguments}\\
	\mtd{bilinearNames} is a list of names for bilinears where gamma matrices are sandwiched between a Dirac conjugate spinor and another spinor, i.e. bilinears of the form $\bar{\xi} \gamma^{\mu_1\dotsc\mu_m} \eta$. The first argument in the list denotes the name for the scalar $\bar{\xi} \eta$, the second entry is the name for the one-form $\bar{\xi} \gamma^{\mu} \eta$, and so on, the $m$:th entry corresponding to bilinears of the form $\bar{\xi} \gamma^{(m-1)} \eta$.\\
	\mtd{ccBilinearNames} is a list of names of bilinears which are formed by sandwiching gamma matrices between a charge conjugate spinor and another spinor, i.e. bilinears of the form $\bar{\xi}^c \gamma^{\mu_1\dotsc\mu_m} \eta$. The first argument denotes again the scalar, and the $m$:th entry the bilinear $\bar{\xi}^c \gamma^{(m-1)}\eta$.\\
	\mtd{$\gamma$5BilinearNames} is a list of names of bilinears which are formed by sandwiching an antisymmetrised product of gamma matrices and $\gamma_*$ between a Dirac conjugated spinor and another spinor, i.e. bilinears of the form $\bar{\xi} \gamma_* \gamma^{\mu_1\dotsc\mu_m} \eta$. Note that these make sense only in even dimensions. The $m$:th entry denotes the name for the bilinear $\bar{\xi} \gamma_* \gamma^{(m-1)}\eta$.\\
	\mtd{cc$\gamma$5BilinearNames} is a list of names of bilinears which are formed by sandwiching an antisymmetrised product of gamma matrices and $\gamma_*$ between a charge conjugate spinor and another spinor, i.e. bilinears of the form $\bar{\xi}^c \gamma^* \gamma^{\mu_1\dotsc\mu_m} \eta$. The $m$:th entry denotes the name for the bilinear $\bar{\xi}^c \gamma_* \gamma^{(m-1)}\eta$.\\
	\mtd{$\xi$} is the name of the spinor which is on the left side of the bilinear.\\[3pt]
	\textbf{Optional Arguments}\\
	\mtd{n} is the number of the subalgebra under which the spinors forming the bilinears transform. If this parameter is not specified, then \mtd{0}, corresponding to the full algebra is used as the default value.\\
	\mtd{$\eta$} is the name of the spinor which is on the right side of the bilinear. If this is not specified, $\xi$ is used as the default value.
\end{funcDefnN}
For example, in a 3-dimensional space we could define the following notation for different bilinears
\begin{align}
S_{ij} &= \overline{\xi}_i \xi_j , ~~~ K_{ij}^{\mu} = \overline{\xi}_i \gamma^\mu\xi_, j~~~ U_{ij}^{\mu \nu} = \overline{\xi}_i \gamma^{\mu \nu} \xi_j, ~~~ X_{ij}^{\mu \nu \lambda} = \overline{\xi}_i \gamma^{\mu \nu \lambda} \xi_j,\\
A_{ij} &= \overline{\xi}_i^c \xi_j , ~~~ B_{ij}^{\mu} = \overline{\xi}_i^c \gamma^\mu\xi_, j~~~ V_{ij}^{\mu \nu} = \overline{\xi}_i^c \gamma^{\mu \nu} \xi_j, ~~~ Y_{ij}^{\mu \nu \lambda} = \overline{\xi}_i^c \gamma^{\mu \nu \lambda} \xi_j,
\end{align}
using the following command.
\begin{mmaCell}{Input}
		\mmaDef{gSetRep}[\{1,1,1\},1,1]
		\mmaDef{gSetBilinearNames}[\{S,K,U,X\},\{A,B,V,Y\},\{\},\{\},\m{$\xi$}]
\end{mmaCell} 
Translating between these two equivalent notations can now be done by using two functions, \mtd{gProductToBL} and \mtd{gBLToProduct}.
\begin{funcDefnN}[label=gProductToBL_infobox]{\mtd{gProductToBL}}
	\mtd{gProductToBL[expr]}\\
	Expresses every bilinear in \mtd{expr} in the \mtd{gBL} notation.\\[3pt]
	\textbf{Arguments}\\
	\mtd{expr} is the expression to be modified.
\end{funcDefnN}
\begin{funcDefnN}[label=gBLToProduct_infobox]{\mtd{gBLToProduct}}
	\mtd{gBLToProduct[expr]}\\
	Expresses every bilinear in \mtd{expr} in the product notation.\\[3pt]
	\textbf{Arguments}\\
	\mtd{expr} is the expression to be modified.
\end{funcDefnN}
After having previously defined the correspondence between the two notations for bilinears, we can now easily translate between the two. Note in particular that changing the notation preserves the lowered and raised indices, and takes into account the extra indices of spinors.
\begin{mmaCell}{Input}
		\mmaDef{gProductToBL}[\mmaDef{gTimes}[\mmaDef{gH}[\mmaDef{s}[\m{$\xi$},\{\}]],\mmaDef{s}[\m{$\xi$},\{\}]]]
		\mmaDef{gProductToBL}[\mmaDef{gTimes}[\mmaDef{gH}[\mmaDef{s}[\m{$\xi$},\{1\}]],\mmaDef{s}[\m{$\xi$},\{2\}]]]
		\mmaDef{gProductToBL}[\mmaDef{gTimes}[\mmaDef{gH}[\mmaDef{s}[\m{$\xi$},\{1\}]],\mmaDef{y}[\{u[\m{$\mu$},\m{$\nu$}]\}],\mmaDef{s}[\m{$\xi$},\{2\}]]]
		\mmaDef{gProductToBL}[\mmaDef{gTimes}[\mmaDef{gH}[\mmaDef{s}[\m{$\xi$},\{1\}]],\mmaDef{y}[\{u[\m{$\mu$}],d[\m{$\nu$}]\}],\mmaDef{s}[\m{$\xi$},\{2\}]]]	
\end{mmaCell} 
\begin{mmaCell}{Output}
		S
\end{mmaCell} 
\begin{mmaCell}{Output}
		\mmaSub{S}{1,2}
\end{mmaCell} 
\begin{mmaCell}{Output}
		\mmaSup{\mmaSub{U}{1,2:}}{$\mu$,$\nu$}
\end{mmaCell} 
\begin{mmaCell}{Output}
		\mmaSub{\mmaSup{\mmaSub{U}{1,2:}}{$\mu$}}{$\nu$}
\end{mmaCell} 
This works also the other way round. The \mtd{gBL} notation is changed to product notation with \mtd{gBLToProduct}.
\begin{mmaCell}{Input}
		\mmaDef{gBLToProduct}[\mmaDef{gBL}[S,\{\},\{\}]]
		\mmaDef{gBLToProduct}[\mmaDef{gBL}[S,\{1,2\},\{\}]]
		\mmaDef{gBLToProduct}[\mmaDef{gBL}[U,\{1,2\},\{u[\m{$\mu$},\m{$\nu$}]\}]]
		\mmaDef{gBLToProduct}[\mmaDef{gBL}[U,\{1,2\},\{d[\m{$\mu$},\m{$\nu$}]\}]]	
\end{mmaCell} 
\begin{mmaCell}{Output}
		$\overline{\xi}$$\star$$\widetilde{\xi}$	
\end{mmaCell}
\begin{mmaCell}{Output}
		\mmaSub{$\overline{\xi}$}{1}$\star$\mmaSub{$\widetilde{\xi}$}{2}		
\end{mmaCell}
\begin{mmaCell}{Output}	
		\mmaSub{$\overline{\xi}$}{1}$\star$\mmaSup{$\gamma$}{$\mu$,$\nu$}$\star$\mmaSub{$\widetilde{\xi}$}{2}		
\end{mmaCell}
\begin{mmaCell}{Output}
		\mmaSub{$\overline{\xi}$}{1}$\star$\mmaSub{\mmaSup{$\gamma$}{$\mu$}}{$\nu$}$\star$\mmaSub{$\widetilde{\xi}$}{2}		
\end{mmaCell}
If we want to define a different notation for bilinears that have spinor $\xi$ on the left side, and spinor $\eta$ on the right side, this can be done as follows.
\begin{mmaCell}{Input}
		\mmaDef{gSetBilinearNames}[\{S1,K1,U1,X1\},\{A1,B1,V1,Y1\},\{\},\{\},\m{$\xi$},\m{$\eta$}]
\end{mmaCell} 
Now bilinears of the form $\bar{\xi}\gamma^{(n)}\eta$ and $\bar{\xi}^c\gamma^{(n)}\eta$ can also be expressed in both notations.
\begin{mmaCell}{Input}
		\mmaDef{gProductToBL}[\mmaDef{gTimes}[\mmaDef{dc}[\m{$\xi$},i],\mmaDef{y}[\{u[\m{$\mu$}]\}],\mmaDef{s}[\m{$\eta$},j]]]
		\mmaDef{gBLToProduct}[\mmaDef{gBL}[K1,\{i,j\},\{u[\m{$\mu$}]\}]]	
\end{mmaCell} 
\begin{mmaCell}{Output}
		\mmaSup{\mmaSub{K1}{i,j:}}{$\mu$}
\end{mmaCell} 
\begin{mmaCell}{Output}
		\mmaSub{$\bar{\xi}$}{i}$\star$\mmaSup{$\gamma$}{$\mu$}$\star$\mmaSub{$\eta$}{j}	
\end{mmaCell} 
Note that this does not override the previous definitions for bilinears involving $\xi$ only. We can still use both definitions simultaneously.
\begin{mmaCell}{Input}
		\mmaDef{gBLToProduct}[\mmaDef{gBL}[K,\{i,j\},\{u[\m{$\mu$}]\}]]	
		\mmaDef{gBLToProduct}[\mmaDef{gBL}[K1,\{i,j\},\{u[\m{$\mu$}]\}]]	
\end{mmaCell} 
\begin{mmaCell}{Output}
		\mmaSub{$\bar{\xi}$}{i}$\star$\mmaSup{$\gamma$}{$\mu$}$\star$\mmaSub{$\xi$}{j}	
\end{mmaCell} 
\begin{mmaCell}{Output}	
		\mmaSub{$\bar{\xi}$}{i}$\star$\mmaSup{$\gamma$}{$\mu$}$\star$\mmaSub{$\eta$}{j}	
\end{mmaCell} 
All these definitions can be deleted using \mtd{gClearBilinearNames}.
\begin{funcDefnN}[label=gClearBilinearNames_infobox]{\mtd{gClearBilinearNames}}
	\mtd{gClearBilinearNames[]}\\
	Clears all definitions made by \mtd{gSetBilinearNames}.\\[3pt]
	\textbf{Arguments}\\
	None.
\end{funcDefnN}
In practice this means that after using \mtd{gClearBilinearNames} the functions \mtd{gProductToBL} and \mtd{gBLToProduct} no longer recognise the notation set earlier.
\begin{mmaCell}{Input}
		\mmaDef{gClearBilinearNames}[];
		\mmaDef{gProductToBL}[\mmaDef{gTimes}[\mmaDef{gH}[\mmaDef{s}[\m{$\xi$},\{1\}]],\mmaDef{y}[\{u[\m{$\mu$},\m{$\nu$}]\}],\mmaDef{s}[\m{$\xi$},\{2\}]]]	
		\mmaDef{gBLToProduct}[\mmaDef{gBL}[U,\{1,2\},\{u[\m{$\mu$},\m{$\nu$}]\}]]
\end{mmaCell} 
\begin{mmaCell}{Output}	
		\mmaSub{$\overline{\xi}$}{1}$\star$\mmaSup{$\gamma$}{$\mu$,$\nu$}$\star$\mmaSub{$\widetilde{\xi}$}{2}		
\end{mmaCell}
\begin{mmaCell}{Output}
		\mmaSup{\mmaSub{U}{1,2:}}{$\mu$,$\nu$}
\end{mmaCell} 
\vspace{-5pt}
\subsubsection{Associated options} \label{sect_Functions_Bilinears_Options}
\vspace{-5pt}
Compared to other tensors, in certain cases bilinears have additional properties due to their construction, as mentioned in section \ref{sect_Review_BL_relations}. For instance, the bilinear of the form $ \bar{\xi_1} \gamma^{(n)} \xi_2$ is related to the bilinear $\bar{\xi_2} \gamma^{(n)} \xi_1$ by relation \eqref{Eq_Complex_conjugate_Bilinear}. In order to aid with simplification of expressions, these relations can be used to automatically put bilinears to a standard form. The following options define the behaviour of bilinears in these cases.
\begin{optDefnN}[label=gBLConvention_infobox]{\mtd{gBLConvention}}
	\mtd{gBLConvention[n,opt]}\\
	Determines the automated behaviour of bilinears in cases in which bilinear relations, such as \eqref{Eq_Complex_conjugate_Bilinear}, can be used to express the bilinears in different equivalent ways.\\[3pt]
	\textbf{Arguments}\\
	\mtd{opt} is one of the options listed below.\\[3pt]
	\textbf{Optional Arguments}\\
	\mtd{n} is the number of the subrepresentation for which the setting applies (see section \ref{sect_Subalgebras}). If this value is not specified, then the setting is changed for every subrepresentation.\\[3pt]
	\textbf{Options}\\
	\mtd{"Alphabetical"}: the bilinears of the form $\bar{\xi_i} \gamma^{(n)} \xi_j$, that are expressed in \mtd{gBL} notation are automatically expressed in a form in which the indices \mtd{i} and \mtd{j} are alphabetically ordered, by conjugating the bilinears if needed.\\
	\mtd{"Unconjugated"}: the bilinears of the form $\bar{\xi_i} \gamma^{(n)} \xi_j$, when expressed using \mtd{gBL} notation, are expressed in a form in which they are not complex conjugated, if possible. This is done by reversing the order of spinors, if necessary.\\
	\mtd{"DoNothing"}: bilinears are not expressed in another form. This is the default setting.
\end{optDefnN}
For example, consider the bilinear $\overline{\xi}_2 \gamma^{\mu \nu} \xi_1$. Then, by taking the complex conjugate of this, we arrive at relation $\overline{\xi}_2 \gamma^{\mu \nu} \xi_1=-(\overline{\xi}_1 \gamma^{\mu \nu} \xi_2)^*$. The function \mtd{gBLConvention} determines, which form is used in the output. First we use the option \mtd{"Alphabetical"}, which expresses the bilinears in such a way that the spinor whose additional index is first in alphabetical order, is always on the left side.
\begin{mmaCell}{Input}
		\mmaDef{gSetRep}[\{-1,1,1,1\},1,1];
		\mmaDef{gSetBilinearNames}[\{S,K,U,X,Z\},\{A,B,V,Y,W\},\{S5,K5,X5,Z5\},
		\{A5,B5,V5,Y5,W5\},\m{$\xi$}];
		\mmaDef{gBLConvention}["Alphabetical"];
		\mmaDef{gBL}[U,\{2,1\},\{u[\m{$\mu$},\m{$\nu$}]\}]
		\mmaDef{gBL}[U,\{1,2\},\{u[\m{$\mu$},\m{$\nu$}]\}]
		\mmaDef{gBL}[S,\{2,1\},\{\}]
\end{mmaCell} 
\begin{mmaCell}{Output}
		\mmaSup{(\mmaSup{\mmaSub{U}{1,2:}}{$\mu$,$\nu$})}{*}
\end{mmaCell} 
\begin{mmaCell}{Output}
		\mmaSup{\mmaSub{U}{1,2:}}{$\mu$,$\nu$}	
\end{mmaCell} 
\begin{mmaCell}{Output}
		-\mmaSup{(\mmaSub{S}{1,2})}{*}
\end{mmaCell} 
This affects the bilinears that are of the form $\bar{\xi} \gamma^{(n)} \eta$ or $\bar{\xi} \gamma_* \gamma^{(n)} \eta$, but not the charge conjugate bilinears, i.e. bilinears of the form $\bar{\xi}^c \gamma^{(n)} \eta$ or $\bar{\xi}^c \gamma_* \gamma^{(n)} \eta$, since the relation these have to their complex conjugates depends on the chosen representation.
\begin{mmaCell}{Input}
		\mmaDef{gBL}[U,\{j,i\},\{u[\m{$\mu$}],d[\m{$\nu$}]\}]
		\mmaDef{gBL}[U5,\{j,i\},\{u[\m{$\mu$}],d[\m{$\nu$}]\}]
		\mmaDef{gBL}[V,\{j,i\},\{u[\m{$\mu$}],d[\m{$\nu$}]\}]
		\mmaDef{gBL}[V5,\{j,i\},\{u[\m{$\mu$}],d[\m{$\nu$}]\}]
\end{mmaCell}
\begin{mmaCell}{Output}
		\mmaSup{(\mmaSub{\mmaSup{U}{$\mu$}}{i,j:$\nu$})}{*}	
\end{mmaCell} 
\begin{mmaCell}{Output}
		-\mmaSup{(\mmaSub{\mmaSup{U5}{$\mu$}}{i,j:$\nu$})}{*}	
\end{mmaCell} 
\begin{mmaCell}{Output}
		\mmaSub{\mmaSup{V}{$\mu$}}{j,i:$\nu$}
\end{mmaCell} 
\begin{mmaCell}{Output}
		\mmaSub{\mmaSup{V5}{$\mu$}}{j,i:$\nu$}	
\end{mmaCell} 
The second option, \mtd{"Unconjugated"}, expresses the bilinears in such a form that they are not complex conjugated, if this is possible. This also affects only on the bilinears that do not contain charge conjugate spinors.
\begin{mmaCell}{Input}
		\mmaDef{gBLConvention}["Unconjugated"];
		\mmaDef{gBL}[U,\{j,i\},\{u[\m{$\mu,\nu$}]\}]
		Conjugate[\mmaDef{gBL}[U,\{i,j\},\{u[\m{$\mu,\nu$}]\}]]
		Conjugate[\mmaDef{gBL}[U,\{j,i\},\{u[\m{$\mu,\nu$}]\}]]			
		Conjugate[\mmaDef{gBL}[U5,\{i,j\},\{u[\m{$\mu,\nu$}]\}]]
		Conjugate[\mmaDef{gBL}[V,\{i,j\},\{u[\m{$\mu,\nu$}]\}]]		
\end{mmaCell} 
\begin{mmaCell}{Output}
		\mmaSup{\mmaSub{U}{j,i:}}{$\mu$,$\nu$}	
\end{mmaCell} 
\begin{mmaCell}{Output}
		\mmaSup{\mmaSub{U}{j,i:}}{$\mu$,$\nu$}	
\end{mmaCell} 
\begin{mmaCell}{Output}
		\mmaSup{\mmaSub{U}{i,j:}}{$\mu$,$\nu$}	
\end{mmaCell} 
\begin{mmaCell}{Output}	
		-\mmaSup{\mmaSub{U5}{j,i:}}{$\mu$,$\nu$}
\end{mmaCell} 
\begin{mmaCell}{Output}
		\mmaSup{(\mmaSup{\mmaSub{V}{j,i:}}{$\mu$,$\nu$})}{*}	
\end{mmaCell} 
The option \mtd{"DoNothing"} ensures that the bilinears are not modified from the form in which they are input.
\begin{mmaCell}{Input}
		\mmaDef{gBLConvention}["DoNothing"];
		\mmaDef{gBL}[U,\{2,1\},\{u[\m{$\mu,\nu$}]\}]
		\mmaDef{Conjugate}[\mmaDef{gBL}[U,\{2,1\},\{u[\m{$\mu,\nu$}]\}]]
\end{mmaCell} 
\begin{mmaCell}{Output}
		\mmaSup{\mmaSub{U}{2,1:}}{$\mu$,$\nu$}	
\end{mmaCell} 
\begin{mmaCell}{Output}
		\mmaSup{(\mmaSup{\mmaSub{U}{2,1:}}{$\mu$,$\nu$})}{*}			
\end{mmaCell} 
Note that this works only for the bilinears expressed in the \mtd{gBL} notation. This is so because, for example, when not expressed in the \mtd{gBL} notation, the complex conjugate of a bilinear is automatically expressed in terms of complex conjugates of the spinors and gamma matrices appearing in it. One of the main benefits of using the \mtd{gBL} notation is that this does not happen.
\begin{mmaCell}{Input}
		\mmaDef{gBLConvention}["Alphabetical"];
		\mmaDef{gTimes}[\mmaDef{dc}[\m{$\xi$},\{2\}],\mmaDef{y}[\{u[\m{$\mu$},\m{$\nu$}]\}],\mmaDef{s}[\m{$\xi$},\{1\}]]
		\mmaDef{Conjugate}[\mmaDef{gTimes}[\mmaDef{dc}[\m{$\xi$},\{2\}],\mmaDef{y}[\{u[\m{$\mu$},\m{$\nu$}]\}],\mmaDef{s}[\m{$\xi$},\{1\}]]]		
\end{mmaCell} 
\begin{mmaCell}{Output}
		\mmaSub{$\overline{\xi}$}{2}$\star$\mmaSup{$\gamma$}{$\mu$,$\nu$}$\star$\mmaSub{$\widetilde{\xi}$}{1}	
\end{mmaCell} 
\begin{mmaCell}{Output}
		-(\mmaSup{\mmaSub{$\widetilde{\xi}$}{2}}{T}$\star$$\mathcal{A}$$\star$\mmaSup{(\mmaSup{$\gamma$}{$\mu$,$\nu$})}{*}$\star$\mmaSup{\mmaSub{$\widetilde{\xi}$}{1}}{*})	
\end{mmaCell} 
Using these options is especially useful for simplifying expressions that include bilinears that look at a first glance different, but are actually related by complex conjugation. For example
\begin{mmaCell}{Input}
		\mmaDef{gBL}[K,\{2,1\},\{\m{$\mu$}\}]+Conjugate[\mmaDef{gBL}[K,\{1,2\},\{\m{$\mu$}\}]]
		\mmaDef{gBL}[K,\{2,1\},\{\m{$\mu$}\}]+
		  \mmaDef{gTimes}[\mmaDef{dc}[\m{$\xi$},\{1\}],\mmaDef{y}[\{u[\m{$\mu$}]\}],\mmaDef{s}[\m{$\xi$},\{2\}]]//\mmaDef{gProductToBL}//\mmaDef{gReIm}		
\end{mmaCell} 
\begin{mmaCell}{Output}
		2\mmaSup{(\mmaSup{\mmaSub{K}{1,2:}}{$\mu$})}{*}	
\end{mmaCell} 
\begin{mmaCell}{Output}
		2\mmaDef{gRe}[\mmaSup{\mmaSub{K}{1,2}}{$\mu$}]
\end{mmaCell} 
There are also various other identities involving bilinears, such as \eqref{Eq_Majorana_spinor_exchange} for bilinears formed out of Majorana spinors. Relation \eqref{Eq_Majorana_spinor_exchange} is always used by \mtd{gSimplify} (see section \ref{sect_Functions_Simplify}) to express the bilinear in such a form that the spinors appearing are ordered alphabetically. For this, and other similar relations, there is no need for options similar to \mtd{gBLConvention}, since this does not introduce complex conjugation or other similar complication. Thus we can always use the other relations to order the spinors in alphabetical order.
\vspace{-5pt}
\subsection{Dual Elements in Clifford Algebra}
\vspace{-5pt}
As has been noted before in section \ref{sect_Review_Basis}, every $2^m \times 2^m$ (where $d=2m$ or $d=2m+1$) matrix can be expressed in terms of antisymmetrised products of gamma matrices $\gamma^{(1)},\dotsc,\gamma^{(d)}$. As can be ascertained by a simple counting argument, not every matrix in this set can be independent. Indeed, the linearly independent set is provided by matrices $\gamma^{(1)},\dotsc,\gamma^{(m)}$, and the relation between matrices of rank $r$ and $d-r$ is given by formulae \eqref{Basis_dual_odd} and \eqref{Basis_dual_Even}.\\

Exchanging between an antisymmetrised product and its dual is implemented by function \mtd{gDual}.
\begin{funcDefnN}[label=gDual_infobox]{\mtd{gDual}}
	\mtd{gDual[expr]}\\
	Expresses every rank $r$ antisymmetrised product of gamma matrices in \mtd{expr} in terms of antisymmetrised products of rank $d-r$, using formulae \eqref{Basis_dual_odd} or \eqref{Basis_dual_Even}. \\[3pt]
	\textbf{Arguments}\\
	\mtd{expr} is any expression.
\end{funcDefnN}
For example, to reproduce the formulae \eqref{Basis_dual_Even} and \eqref{Basis_dual_odd} for the signatures $(-+++)$ and $(+++++++)$, we can use the following.
\begin{mmaCell}{Input}
		\mmaDef{gSetRep}[\{-1,1,1,1\},1,1];
		\mmaDef{gDual}[\mmaDef{y}[\{u[a,b]\}]]
		\mmaDef{gSetRep}[\{1,1,1,1,1,1,1\},-1,1];
		\mmaDef{gDual}[\mmaDef{y}[\{u[a,b,c,d]\}]]	
\end{mmaCell} 
\begin{mmaCell}{Output}
		-\mmaFrac{1}{2}$\Ii$\mmaSub{\mmaSup{$\epsilon$}{a,b}}{$\lambda$1,$\lambda$2} (\mmaSub{$\gamma$}{*}$\star$\mmaSup{$\gamma$}{$\lambda$1,$\lambda$2})		
\end{mmaCell}
\begin{mmaCell}{Output}
		\mmaFrac{1}{6}$\Ii$\mmaSub{\mmaSup{$\epsilon$}{a,b,c,d}}{$\lambda$1,$\lambda$2,$\lambda$3} \mmaSup{$\gamma$}{$\lambda$1,$\lambda$2,$\lambda$3}		
\end{mmaCell}
This works not only for basic gamma matrices, but also for the expressions where the gamma matrices are multiplied by other objects. It also works on sums, products and so on.
\begin{mmaCell}{Input}
		\mmaDef{gDual}[\mmaDef{gTimes}[\mmaDef{gH}[\mmaDef{s}[\m{$\xi$},\{\}]],\mmaDef{y}[\{u[a,b,c,d]\}],\mmaDef{s}[\m{$\xi$},\{\}]]]	
		\mmaDef{gDual}[\mmaDef{gForm}[F,\{a,b\}]*\mmaDef{gTimes}[\mmaDef{gH}[\mmaDef{s}[\m{$\xi$}]],\mmaDef{y}[\{u[a,b,c,d]\}],\mmaDef{s}[\m{$\xi$},\{\}]]+
		  3*\mmaDef{gTimes}[\mmaDef{y}[\{u[a,b]\}],\mmaDef{s}[\m{$\xi$},\{\}]]]
		\mmaDef{gDual}[\mmaDef{gTimes}[\mmaDef{gT}[\mmaDef{s}[\m{$\xi$},\{\}]],\mmaDef{y}[\{u[a,b,c]\}],\mmaDef{s}[\m{$\xi$},\{\}]]*\mmaDef{y}[\{d[a,b,c]\}]]
\end{mmaCell} 
\begin{mmaCell}{Output}
		\mmaFrac{1}{6}$\Ii$\mmaSub{\mmaSup{$\epsilon$}{a,b,c,d}}{$\lambda$1,$\lambda$2,$\lambda$3} ($\bar{\xi}$$\star$\mmaSup{$\gamma$}{$\lambda$1,$\lambda$2,$\lambda$3}$\star$$\widetilde{\xi}$)
\end{mmaCell}
\begin{mmaCell}{Output}
		-\mmaFrac{1}{40}$\Ii$\mmaSub{\mmaSup{$\epsilon$}{a,b}}{$\lambda$1,$\lambda$2,$\lambda$3,$\lambda$4,$\lambda$5} (\mmaSup{$\gamma$}{$\lambda$1,$\lambda$2,$\lambda$3,$\lambda$4,$\lambda$5}$\star$$\widetilde{\xi}$)+
		
		  \mmaFrac{1}{6}$\Ii$\mmaSub{F}{$\lambda$1,$\lambda$2}\mmaSub{\mmaSup{$\epsilon$}{c,d,$\lambda$1,$\lambda$2}}{$\lambda$3,$\lambda$4,$\lambda$5} ($\bar{\xi}$$\star$\mmaSup{$\gamma$}{$\lambda$3,$\lambda$4,$\lambda$5}$\star$$\widetilde{\xi}$)
\end{mmaCell}
\begin{mmaCell}{Output}
		-\mmaFrac{1}{576}\mmaSub{$\epsilon$}{$\lambda$1,$\lambda$2,$\lambda$3,$\lambda$4,$\lambda$5,$\lambda$6,$\lambda$7}\mmaSub{\mmaSup{$\epsilon$}{$\lambda$1,$\lambda$2,$\lambda$3}}{$\lambda$10,$\lambda$11,$\lambda$8,$\lambda$9}(\mmaSup{$\widetilde{\xi}$}{T}$\star$\mmaSup{$\gamma$}{$\lambda$10,$\lambda$11,$\lambda$8,$\lambda$9}$\star$$\widetilde{\xi}$)\mmaSup{$\gamma$}{$\lambda$4,$\lambda$5,$\lambda$6,$\lambda$7}
\end{mmaCell}
This works also on bilinears defined using \mtd{gBL} notation, as long as correspondence between the \mtd{gBL} and product notation for bilinears has been defined (see \ref{sect_bilinears}).
\begin{mmaCell}{Input}
		\mmaDef{gSetBilinearNames}[\{S,K,V,X,L,M,N,Y\},\{\},\{\},\{\},\m{$\xi$}];
		\mmaDef{gDual}[\mmaDef{gBL}[V,\{i,j\},\{u[\m{$\mu$},\m{$\nu$}]\}]]
		\mmaDef{gClearBilinearNames}[];
\end{mmaCell} 
\begin{mmaCell}{Output}
		-\mmaFrac{1}{120}$\Ii$\mmaSup{\mmaSub{M}{i,j:}}{$\lambda$1,$\lambda$2,$\lambda$3,$\lambda$4,$\lambda$5}\mmaSub{\mmaSup{$\epsilon$}{$\mu$,$\nu$}}{$\lambda$1,$\lambda$2,$\lambda$3,$\lambda$4,$\lambda$5} 
\end{mmaCell}
\vspace{-5pt}
\subsection{Fierz Rearrangement}
\vspace{-5pt}
In section \ref{sect_Review_Basis}, we used the fact that the matrices $\gamma^{(1)},\dotsc,\gamma^{(m)}$ form a basis of $2^m \times 2^m$ matrices in dimensions $d=2m+1$, to derive the Fierz arrangement formulae \eqref{sect_Review_Basis_Fierz_basic_odd} and \eqref{sect_Review_Basis_Fierz_odd}, and similar relations \eqref{sect_Review_Basis_Fierz_basic_even} and \eqref{sect_Review_Basis_Fierz_even} for even-dimensional representations.

Given a product of two bilinears, or other expression that can be rearranged using \eqref{sect_Review_Basis_Fierz_basic_odd} or \eqref{sect_Review_Basis_Fierz_basic_even}, we can get an alternative form for the expression using the Fierz rearrangement relations. This is done by using \mtd{gFierz}.
\begin{funcDefnN}[label=gFierz_infobox]{\mtd{gFierz}}
	\mtd{gFierz[expr]}\\
	Gives an equivalent form of \mtd{expr}, where every product of two spinor bilienars, or other suitable expression has been rearranged using \eqref{sect_Review_Basis_Fierz_basic_odd}, \eqref{sect_Review_Basis_Fierz_basic_even}, or an analogous relation.\\[3pt]
	\textbf{Arguments}\\
	\mtd{expr} is the expression to be rearranged.
\end{funcDefnN}
To illustrate the basic usage of this function, we can find the Fierz identity involving the contraction of two vector bilinears in a 5-dimensional representation, and a contraction of a 2-form with a 1-form.
\begin{mmaCell}{Input}
		\mmaDef{gSetRep}[\{-1,1,1,1,1\},1,-1]
		\mmaDef{gSpinorType}["Commutative"];
		\mmaDef{gFierz}[\mmaDef{gTimes}[\mmaDef{dc}[\m{$\xi$},\{i\}],y[\{u[\m{$\mu$}]\},s[\m{$\xi$},\{j\}]]]
		\mmaDef{gTimes}[\mmaDef{dc}[\m{$\xi$},\{i\}],y[\{d[\m{$\mu$}]\},s[\m{$\xi$},\{j\}]]]]
		\mmaDef{gFierz}[\mmaDef{gTimes}[\mmaDef{dc}[\m{$\xi$},\{i\}],y[\{u[\m{$\mu$},\m{$\nu$}]\},s[\m{$\xi$},\{j\}]]]
		\mmaDef{gTimes}[\mmaDef{dc}[\m{$\xi$},\{i\}],y[\{d[\m{$\mu$}]\},s[\m{$\xi$},\{j\}]]]]	
\end{mmaCell}
\begin{mmaCell}{Output}
		\mmaFrac{5}{4}\mmaSup{(\mmaSub{$\overline{\xi}$}{i}$\star$\mmaSub{$\widetilde{\xi}$}{j})}{2}-\mmaFrac{3}{4}(\mmaSub{$\overline{\xi}$}{i}$\star$\mmaSub{$\gamma$}{$\lambda$1}$\star$\mmaSub{$\widetilde{\xi}$}{j})(\mmaSub{$\overline{\xi}$}{i}$\star$\mmaSup{$\gamma$}{$\lambda$1}$\star$\mmaSub{$\widetilde{\xi}$}{j})-\mmaFrac{1}{8}(\mmaSub{$\overline{\xi}$}{i}$\star$\mmaSub{$\gamma$}{$\lambda$1,$\lambda$2}$\star$\mmaSub{$\widetilde{\xi}$}{j})(\mmaSub{$\overline{\xi}$}{i}$\star$\mmaSup{$\gamma$}{$\lambda$1,$\lambda$2}$\star$\mmaSub{$\widetilde{\xi}$}{j})
\end{mmaCell}
\begin{mmaCell}{Output}
		(\mmaSub{$\overline{\xi}$}{i}$\star$\mmaSup{$\gamma$}{$\lambda$1}$\star$\mmaSub{$\widetilde{\xi}$}{j})(\mmaSub{$\overline{\xi}$}{i}$\star$\mmaSub{\mmaSup{$\gamma$}{$\nu$}}{$\lambda$1}$\star$\mmaSub{$\widetilde{\xi}$}{j})
\end{mmaCell}
If the short \mtd{gBL} notation for bilinears has been defined using \mtd{gSetBilinearNames} (see section \ref{sect_bilinears}), then this will be automatically used. 
\begin{mmaCell}{Input}
		\mmaDef{gSetBilinearNames}[\{S,K,U,X,Z\},\{\},\{\},\{\},\m{$\xi$}]
		\mmaDef{gFierz}[\mmaDef{gTimes}[\mmaDef{dc}[\m{$\xi$},\{i\}],y[\{u[\m{$\mu$}]\},s[\m{$\xi$},\{j\}]]]
		\mmaDef{gTimes}[\mmaDef{dc}[\m{$\xi$},\{i\}],y[\{d[\m{$\mu$}]\},s[\m{$\xi$},\{j\}]]]]
		\mmaDef{gFierz}[\mmaDef{gTimes}[\mmaDef{dc}[\m{$\xi$},\{i\}],y[\{u[\m{$\mu$},\m{$\nu$}]\},s[\m{$\xi$},\{j\}]]]
		\mmaDef{gTimes}[\mmaDef{dc}[\m{$\xi$},\{i\}],y[\{d[\m{$\mu$}]\},s[\m{$\xi$},\{j\}]]]]	
\end{mmaCell}
\begin{mmaCell}{Output}
		-\mmaFrac{3}{4}\mmaSub{K}{i,j:$\lambda$1}\mmaSup{\mmaSub{K}{i,j:}}{$\lambda$1}+\mmaFrac{5}{4}\mmaSup{\mmaSub{S}{i,j}}{2}-\mmaFrac{1}{8}\mmaSub{U}{i,j:$\lambda$1,$\lambda$2}\mmaSup{\mmaSub{U}{i,j:}}{$\lambda$1,$\lambda$2}
\end{mmaCell}
\begin{mmaCell}{Output}
		\mmaSup{K}{i,j:$\lambda$1}\mmaSub{\mmaSup{\mmaSub{U}{i,j:}}{$\nu$}}{$\lambda$1}
\end{mmaCell}
The \mtd{gBL} notation can be also used in input.
\begin{mmaCell}{Input}
		\mmaDef{gFierz}[\mmaDef{gBL}[K,\{i,j\},\{u[\m{$\mu$}]\}]\mmaDef{gBL}[K,\{i,j\},\{d[\m{$\mu$}]\}]]
\end{mmaCell}
\begin{mmaCell}{Output}
		-\mmaFrac{3}{4}\mmaSub{K}{i,j:$\lambda$1}\mmaSup{\mmaSub{K}{i,j:}}{$\lambda$1}+\mmaFrac{5}{4}\mmaSup{\mmaSub{S}{i,j}}{2}-\mmaFrac{1}{8}\mmaSub{U}{i,j:$\lambda$1,$\lambda$2}\mmaSup{\mmaSub{U}{i,j:}}{$\lambda$1,$\lambda$2}
\end{mmaCell}
As is the case with virtually any function, \mtd{gFierz} can be used on more complicated expressions as well.
\begin{mmaCell}{Input}
		\mmaDef{gFierz}[3*a*\mmaDef{gBL}[S,\{i,j\},\{\}]*\mmaDef{gBL}[K,\{i,j\},\{u[\m{$\mu$}]\}]-
		5*\mmaDef{gTimes}[\mmaDef{dc}[\m{$\xi$},\{i\}],y[\{u[\m{$\sigma$},\m{$\mu$}]\},s[\m{$\xi$},\{j\}]]]
		\mmaDef{gTimes}[\mmaDef{dc}[\m{$\xi$},\{i\}],y[\{d[\m{$\sigma$}]\},s[\m{$\xi$},\{j\}]]]]
\end{mmaCell}
\begin{mmaCell}{Output}
		\mmaFrac{3}{2}a\mmaSup{\mmaSub{K}{i,j:}}{$\mu$}\mmaSub{S}{i,j}+5\mmaSub{K}{i,j:$\lambda$1}\mmaSup{\mmaSub{U}{i,j:}}{$\lambda$1,$\mu$}-\mmaFrac{3}{8}a\mmaSub{U}{i,j:$\lambda$1,$\lambda$2}\mmaSup{\mmaSub{X}{i,j:}}{$\lambda$1,$\lambda$2,$\mu$}
\end{mmaCell}
\mtd{gFierz} does not need to be used on a product of two bilinears, but can also be used on other expressions that can be manipulated using relations \eqref{sect_Review_Basis_Fierz_basic_odd} or \eqref{sect_Review_Basis_Fierz_odd}.
\begin{mmaCell}{Input}
		\mmaDef{gFierz}[\mmaDef{gTimes}[\mmaDef{s}[\m{$\xi$},\{\}],\mmaDef{gT}[\mmaDef{s}[\m{$\eta$},\{1\}]],\mmaDef{y}[\{u[\m{$\mu$}]\}],\mmaDef{s}[\m{$\eta$},\{\}]]]
\end{mmaCell} 
\begin{mmaCell}{Output}
		\mmaFrac{1}{4}(\mmaSup{\mmaSub{$\widetilde{\eta}$}{1}}{T}$\star$\mmaSub{$\widetilde{\xi}$}{i})(\mmaSup{$\gamma$}{$\mu$}$\star$\mmaSub{$\widetilde{\eta}$}{2})-\mmaFrac{1}{4}(\mmaSub{\mmaSup{$\gamma$}{$\mu$}}{$\lambda$1}$\star$\mmaSub{$\widetilde{\eta}$}{2})(\mmaSup{\mmaSub{$\widetilde{\eta}$}{1}}{T}$\star$\mmaSup{$\gamma$}{$\lambda$1}$\star$\mmaSub{$\widetilde{\xi}$}{i})+\mmaFrac{1}{4}(\mmaSup{$\gamma$}{$\lambda$1}$\star$\mmaSub{$\widetilde{\eta}$}{2})(\mmaSup{\mmaSub{$\widetilde{\eta}$}{1}}{T}$\star$\mmaSub{\mmaSup{$\gamma$}{$\mu$}}{$\lambda$1}$\star$\mmaSub{$\widetilde{\xi}$}{i})-
		\mmaFrac{1}{8}(\mmaSub{\mmaSup{$\gamma$}{$\mu$}}{$\lambda$1,$\lambda$2}$\star$\mmaSub{$\widetilde{\eta}$}{2})(\mmaSup{\mmaSub{$\widetilde{\eta}$}{1}}{T}$\star$\mmaSup{$\gamma$}{$\lambda$1,$\lambda$2}$\star$\mmaSub{$\widetilde{\xi}$}{i})+\mmaFrac{1}{4}(\mmaSup{\mmaSub{$\widetilde{\eta}$}{1}}{T}$\star$\mmaSup{$\gamma$}{$\mu$}$\star$\mmaSub{$\widetilde{\xi}$}{i})\mmaSub{$\widetilde{\eta}$}{2}
\end{mmaCell}
\subsection{Simplifying expressions} \label{sect_Functions_Simplify}
Expressions containing gamma matrices and other non-commutative objects, as well as objects with Lorentz indices, can be simplified by using the function \mtd{gSimplify}.
\begin{funcDefnN}[label=gSimplify_infobox]{\mtd{gSimplify}}
	\mtd{gSimplify[expr]}\\
	Simplifies the expression \mtd{expr} that contains non-commutative objects, such as gamma matrices, spinors or intertwiners, or objects with Lorentz indices.\\[3pt]
	\textbf{Arguments}\\
	\mtd{expr} is the expression to be simplified.
\end{funcDefnN}
The function \mtd{gSimplify} has multiple distinct features. First of all, it recognises the situations in which an expression simplifies due to the symmetry properties of the tensors involved. For instance, if an antisymmetric tensor is contracted with a symmetric one, the term containing the contraction is automatically simplified to be zero.
\begin{mmaCell}{Input}
		\mmaDef{gSimplify}[\mmaDef{gForm}[F,\{a,b\}]*\mmaDef{gTensor}[S,\{u[a]\}]*\mmaDef{gTensor}[S,\{u[b]\}]]	
		\mmaDef{gSimplify}[\mmaDef{gForm}[F,\{a,b\}]*\mmaDef{gSymm}[S,\{u[a,b]\}]]
		\mmaDef{gSimplify}[\mmaDef{gSymm}[S,\{u[a,b]\}]*(\mmaDef{gTimes}[\mmaDef{y}[\{u[a,b,c]\}],\mmaDef{s}[\m{$\xi$},\{i\}]]
		  +\mmaDef{s}[\m{$\xi$},\{i\}])]
\end{mmaCell}
\begin{mmaCell}{Output}
		0
\end{mmaCell}
\begin{mmaCell}{Output}
		0
\end{mmaCell}
\begin{mmaCell}{Output}
		\mmaSup{$\hat{S}$}{a,b}\mmaSub{$\widetilde{\xi}$}{i}
\end{mmaCell}
Note that this also works with tensors with user-defined properties. For example, if we define the tensor to be symmetric in its first and third indices, this is taken into account by \mtd{gSimplify}.
\begin{mmaCell}{Input}
		\mmaDef{gTensor}[{T,\{d[\mmaPat{a\_},\mmaPat{b\_},\mmaPat{c\_}]\}}]:=\mmaDef{gTensor}[T,\{\mmaPat{c},\mmaPat{b},\mmaPat{a}\}]/;
		  !\mmaDef{OrderedQ}[\{\mmaPat{a},\mmaPat{c}\}]
		\mmaDef{gSimplify}[\mmaDef{gForm}[F,\{u[\m{$\mu$},\m{$\nu$}]\}]*\mmaDef{gTensor}[T,\{\m{$\mu$},\m{$\alpha$},\m{$\nu$}\}]]
\end{mmaCell}
\begin{mmaCell}{Output}
		0
\end{mmaCell}
It also automatically raises and lowers indices of tensors that are contracted with the metric $\eta^{\mu \nu}$ or its inverse. 
\begin{mmaCell}{Input}
		\mmaDef{gSimplify}[\mmaDef{gSymm}[\m{$\eta$},\{u[b,c]\}](\mmaDef{gForm}[F,\{u[a,b]\}]+\mmaDef{y}[\{d[a,b]\}])]
		\mmaDef{gSimplify}[\mmaDef{gSymm}[\m{$\eta$},\{u[a,\m{$\alpha$}]\}]*\mmaDef{gSymm}[\m{$\eta$},\{u[b,\m{$\beta$}]\}]*\mmaDef{gSymm}[S,\{d[\m{$\alpha$},\m{$\beta$}]\}]]
		\mmaDef{gSimplify}[\mmaDef{gSymm}[\m{$\eta$},\{u[a,b]\}]*\mmaDef{gSymm}[\m{$\eta$},\{u[a,b]\}]]
\end{mmaCell}
\begin{mmaCell}{Output}
		-\mmaSub{\mmaSup{F}{c}}{a}-\mmaSub{\mmaSup{$\gamma$}{c}}{a}
\end{mmaCell}
\begin{mmaCell}{Output}
		\mmaSup{$\hat{S}$}{a,b}
\end{mmaCell}
\begin{mmaCell}{Output}
		\mmaSub{\mmaSup{$\hat{\eta}$}{$\lambda$1}}{$\lambda$1}
\end{mmaCell}
The contracted indices are raised or lowered so that the terms which differ only by the placement of dummy indices are recognised as equal.
\begin{mmaCell}{Input}
		\mmaDef{gSimplify}[\mmaDef{gForm}[F,\{u[a,b]\}]*\mmaDef{y}[\{d[a,b]\}]+\mmaDef{gForm}[F,\{a,b\}]*
		  \mmaDef{y}[\{u[a,b]\}]]
		\mmaDef{gSimplify}[\mmaDef{gTensor}[F,\{u[a,b],d[c],u[f]\}]*
		  \mmaDef{gTimes}[\mmaDef{dc}[\m{$\xi$},\{i\}],\mmaDef{y}[\{d[a,f],u[c]\},\{d[b]\}]]]		
\end{mmaCell}
\begin{mmaCell}{Output}
		2\mmaSub{F}{$\lambda$1,$\lambda$2}\mmaSup{$\gamma$}{$\lambda$1,$\lambda$2}
\end{mmaCell}
\begin{mmaCell}{Output}
		-\mmaSub{$\ddot{F}$}{$\lambda$1,$\lambda$2,$\lambda$3,$\lambda$4}(\mmaSub{$\overline{\xi}$}{i}$\star$(\mmaSup{$\gamma$}{$\lambda$1,$\lambda$3,$\lambda$4}\mmaSup{$\gamma$}{$\lambda$2}))
\end{mmaCell}
As already seen in the last example, the names of dummy indices are updated when using \mtd{gSimplify} so that if two expressions differ only by naming of the dummy indices, they are recognised as equal.
\begin{mmaCell}{Input}
		\mmaDef{gSimplify}[\mmaDef{gForm}[F,\{a,b\}]*\mmaDef{gTensor}[G,\{u[a,b]\}]+
		  \mmaDef{gForm}[F,\{i,j\}]*\mmaDef{gTensor}[G,\{u[i,j]\}]]
		\mmaDef{gSimplify}[\mmaDef{gForm}[F,\{a,b\}]*\mmaDef{y}[\{u[a,b,c]\}]]		
\end{mmaCell}
\begin{mmaCell}{Output}
		2\mmaSup{F}{$\lambda$1,$\lambda$2}\mmaSub{$\ddot{G}$}{$\lambda$1,$\lambda$2}
\end{mmaCell}
\begin{mmaCell}{Output}
		\mmaSub{F}{$\lambda$1,$\lambda$2}\mmaSup{$\gamma$}{c,$\lambda$1,$\lambda$2}
\end{mmaCell}
Note that \mtd{gSimplify} is not simply an improved version of \mtd{Simplify}. On the contrary, on many expressions \mtd{gSimplify} works like \mtd{Expand}, since this tends to result in more readable output, when these expressions contain non-commutative objects. \mtd{Simplify} can be used if further simplification is needed.
\begin{mmaCell}{Input}
		\mmaDef{gSimplify}[\mmaDef{gTensor}[F,\{a,b\}]*(\mmaDef{gSymm}[H,\{u[a,b]\}-\mmaDef{y}[\{u[a,b]\}]])]
		\mmaDef{gSimplify}[\mmaDef{gTensor}[F,\{a,b\}]*(\mmaDef{gSymm}[H,\{u[a,b]\}-\mmaDef{y}[\{u[a,b]\}]])]//
		  \mmaDef{Simplify}
\end{mmaCell}
\begin{mmaCell}{Output}
		\mmaSup{$\hat{H}$}{$\lambda$1,$\lambda$2}\mmaSub{$\ddot{F}$}{$\lambda$1,$\lambda$2}-\mmaSub{$\ddot{F}$}{$\lambda$1,$\lambda$2}\mmaSup{$\gamma$}{$\lambda$1,$\lambda$2}
\end{mmaCell}
\begin{mmaCell}{Output}
		\mmaSub{$\ddot{F}$}{$\lambda$1,$\lambda$2}(\mmaSup{$\hat{H}$}{$\lambda$1,$\lambda$2}-\mmaSup{$\gamma$}{$\lambda$1,$\lambda$2})
\end{mmaCell}
\vspace{-5pt}
\subsection{Using Assumptions} \label{sect_Functions_Assumptions}
\vspace{-5pt}
We have already seen examples where to achieve some functionality it has been necessary to use assumptions. To a large extent this works just as usual in Mathematica: assumptions can be made by including statements in the list named \mtd{\$Assumptions}. For example, assuming that variable \mtu{a} belongs to the set of real numbers can be done by using the following code.
\begin{mmaCell}{Input}
		\$Assumptions=\{a$\in$Reals\};
\end{mmaCell}
There are, however, some additional features in the package that make working with gamma matrices and bilinears easier. In particular, we have introduced a new command called \mtd{gAddAssumptions} that can be used to add new assumptions.
\begin{funcDefnN}[label=gAddAssumptions_infobox]{\mtd{gAddAssumptions}}
	\mtd{gAddAssumptions[assumptions]}\\
	Adds a list of new assumptions, \mtd{assumptions}, to the list of all assumptions, \mtd{\$Assumptions}. This function also checks in many cases whether the new assumptions are consistent with other assumptions made, and derives other assumptions that follow from \mtd{assumptions} and assumptions made earlier.\\[3pt]
	\textbf{Arguments}\\
	\mtd{assumptions} is a list of new assumptions to be added to the list of all assumptions \mtd{\$Assumptions}. If only one assumption is to be added, then the argument can also be the assumption to be added (without the need to include it in a list of one element).
\end{funcDefnN}
The list of all assumptions can be cleared using \mtd{gClearAssumptions}
\begin{funcDefnN}[label=gClearAssumptions_infobox]{\mtd{gClearAssumptions}}
	\mtd{gClearAssumptions[]}\\
	Clears all the assumptions made earlier.\\[3pt]
	\textbf{Arguments}\\
	None.
\end{funcDefnN}
As we have already shown in an previous example, making an assumption that a tensor, whether it be \mtd{gForm}, \mtd{gSymm} or \mtd{gTensor}, is real (i.e all of its components are real) is done by entering the following.
\begin{mmaCell}{Input}
		\mmaDef{gAddAssumptions}[F$\in$Reals]
\end{mmaCell}
\begin{mmaCell}{Output}
		\{a$\in$Reals,F$\in$Reals\}
\end{mmaCell}
where \mtd{F} is the name of the tensor. The output here is just the list containing all the assumptions made, including also all the assumptions made earlier, which in this case is the assumption that $a$ be real (which was manually added in the beginning of this section). In this case this is equivalent to the usual way of making assumptions in Mathematica.

After using either of these ways of inputting an assumption, it is now assumed that \textit{any tensor} (\mtd{gForm}, \mtd{gSymm} or \mtd{gTensor} with any number of indices) that has name \mtd{F} is real.
\begin{mmaCell}{Input}
		\mmaDef{gRe}[\mmaDef{gForm}[F,\{a,b\}]]
		\mmaDef{gRe}[\mmaDef{gSymm}[F,\{a,b,c\}]]	
		\mmaDef{gRe}[\mmaDef{gTensor}[F,\{a\}]]			
\end{mmaCell}
\begin{mmaCell}{Output}
		\mmaSub{F}{a,b}
\end{mmaCell}
\begin{mmaCell}{Output}
		\mmaSub{$\hat{F}$}{a,b,c}
\end{mmaCell}
\begin{mmaCell}{Output}
		\mmaSub{$\ddot{F}$}{a}
\end{mmaCell}
To assume that a tensor is imaginary (i.e. all of its components are purely imaginary), we need to write the following.
\begin{mmaCell}{Input}
		\mmaDef{gClearAssumptions}[];
		\mmaDef{gAddAssumptions}[F$\in$gImaginaries];
\end{mmaCell}
Here the first line clears the assumption made earlier that $F$ is real. Now any tensor with name \mtd{F} is assumed to be imaginary.
\begin{mmaCell}{Input}	
		\mmaDef{gIm}[\mmaDef{gTensor}[F,\{a\}]]			
\end{mmaCell}
\begin{mmaCell}{Output}		
		\mmaSub{$\ddot{F}$}{a}		
\end{mmaCell}
Note that here we introduced \mtu{gImaginaries} to denote the set of imaginary objects.
\begin{objDefnN}[label=gImaginaries_infobox]{\mtd{gImaginaries}}
	\mtd{gImaginaries}\\
	Denotes the set of all purely imaginary scalars, tensors, matrices and spinors. \\[3pt]
	\textbf{Arguments}\\
	None.
\end{objDefnN}
\vspace{-5pt}
\subsubsection{Assumptions on Indices}
\vspace{-5pt}
Another new thing that can be done with the assumptions, is to assume that some indices are equal, or to assume that they cannot be equal. This is done by making use of two new objects, \mtd{gEqual} and \mtd{gUnequal}, which are defined as follows.
\begin{objDefnN}[label=gEqual_infobox]{\mtd{gEqual}}
	\mtd{gEqual[$\mu$1,...,$\mu$n]}\\
	When entered as an assumption, assumes that the indices \mtd{$\mu$1,...,$\mu$n} are all equal. \\[3pt]
	\textbf{Arguments}\\
	\mtd{$\mu$1,...,$\mu$n} are the indices (such as those appearing in \mtd{gForm} or \mtd{y}) that are assumed to be equal.
\end{objDefnN}
Of course in many situations this could be achieved by simply setting the indices equal. However, if we for example have a gamma matrix product of the form $\gamma^{\mu \nu} \gamma_\sigma$, where we want to assume that $\mu = \sigma$, but do not want to sum over $\mu$, we cannot simply set $\mu=\sigma$, since then this would be interpreted as summing over the repeated index $\sigma$. The way to simplify this expression without summing is to use assumptions.
\begin{mmaCell}{Input}
		\mmaDef{gSetRep}[\{1,1,1,1\},1,1]
		\m{$\mu$}=\m{$\sigma$};
		\mmaDef{y}[\{u[\m{$\mu$},\m{$\nu$}]\},\{d[\m{$\sigma$}]\}]//\mmaDef{gOrd}
		Clear[\m{$\mu$}];
		\mmaDef{gAddAssumptions}[\mmaDef{gEqual}[\m{$\mu$},\m{$\sigma$}]];
		\mmaDef{y}[\{u[\m{$\mu$},\m{$\nu$}]\},\{u[\m{$\sigma$}]\}]//\mmaDef{gOrd}	
		\mmaDef{gForm}[F,\{\m{$\mu$},\m{$\sigma$}\}]	
\end{mmaCell}
\begin{mmaCell}{Output}		
		-3\mmaSup{$\gamma$}{$\nu$}
\end{mmaCell}
\begin{mmaCell}{Output}		
		-\mmaSub{\mmaSup{$\delta$}{*}}{\{$\mu$\},\{$\nu$\}}\mmaSub{\mmaSup{$\hat{\eta}$}{$\mu$}}{$\sigma$}\mmaSup{$\gamma$}{$\nu$}
\end{mmaCell}
\begin{mmaCell}{Output}		
		0
\end{mmaCell}
Similarly, assuming that the indices are never equal is done by using \mtd{gUnequal}.
\begin{objDefnN}[label=gUnequal_infobox]{\mtd{gUnequal}}
	\mtd{gUnequal[$\mu$1,..,$\mu$n]}\\
	When entered as an assumption, assumes that the indices \mtd{$\mu$1,...,$\mu$n} are never equal. \\[3pt]
	\textbf{Arguments}\\
	\mtd{$\mu$1,...,$\mu$n} are the indices that are assumed not to be equal.
\end{objDefnN}
\begin{mmaCell}{Input}
		\mmaDef{gClearAssumptions}[];
		\mmaDef{gAddAssumptions}[\mmaDef{gUnequal}[\m{$\mu$},\m{$\sigma$}]];
		\mmaDef{y}[\{u[\m{$\mu$},\m{$\nu$}]\},\{u[\m{$\sigma$}]\}]//\mmaDef{gOrd}	
\end{mmaCell}
\begin{mmaCell}{Output}		
		\mmaSup{$\hat{\eta}$}{$\nu$,$\sigma$}\mmaSup{$\gamma$}{$\nu$} + \mmaSup{$\gamma$}{$\mu$,$\nu$,$\sigma$}
\end{mmaCell}
The main reason to use the function \mtd{gAddAssumptions}, instead of manually updating the list \mtd{\$Assumptions}, is that \mtd{gAddAssumptions} automatically also adds assumptions that follow from the assumptions just added in conjunction with those that have already been made earlier. For example, assuming that $\mu = \nu$ and $\mu = \sigma$ naturally means also that $\nu = \sigma$. \mtd{gAddAssumption} notices this, and adds $\nu = \sigma$ to the list of assumptions even if this is not explicitly added. This works whether the assumptions are made at the same time or if the other is made first, and the second only at a later point. We can easily demonstrate this, recalling that \mtd{gAddAssumptions} prints as its output the list of all assumptions made.
\begin{mmaCell}{Input}
		\mmaDef{gClearAssumptions}[];
		\mmaDef{gAddAssumptions}[\{\mmaDef{gEqual}[\m{$\mu$},\m{$\nu$}],\mmaDef{gEqual}[\m{$\mu$},\m{$\sigma$}]\}]
		\mmaDef{gClearAssumptions}[];
		\mmaDef{gAddAssumptions}[\mmaDef{gEqual}[\m{$\mu$},\m{$\nu$}]]
		\mmaDef{gAddAssumptions}[\mmaDef{gEqual}[\m{$\mu$},\m{$\sigma$}]]		
\end{mmaCell}
\begin{mmaCell}{Output}		
		\{\mmaDef{gEqual}[$\mu$,$\nu$,$\sigma$]\}	
\end{mmaCell}
\begin{mmaCell}{Output}		
		\{\mmaDef{gEqual}[$\mu$,$\nu$]\}		
\end{mmaCell}
\begin{mmaCell}{Output}		
		\{\mmaDef{gEqual}[$\mu$,$\nu$,$\sigma$]\}		
\end{mmaCell}
This also works with more complicated cases. Consider, for instance, setting $\mu \neq \nu$, $\nu = \sigma$ and $\mu = \lambda$. Then from these assumptions also follows that $\sigma \neq \lambda$, $\sigma \neq \mu$, and so on. These all are automatically derived by \mtd{gAddAssumptions}.
\begin{mmaCell}{Input}
		\mmaDef{gClearAssumptions}[];
		\mmaDef{gAddAssumptions}[\{\mmaDef{gUnequal}[\m{$\mu$},\m{$\nu$}],\mmaDef{gEqual}[\m{$\nu$},\m{$\sigma$}],\mmaDef{gEqual}[\m{$\mu$},\m{$\lambda$}]\}]	
\end{mmaCell}
\begin{mmaCell}{Output}		
		\{\mmaDef{gEqual}[$\mu$,$\lambda$],\mmaDef{gEqual}[$\nu$,$\sigma$],
		\mmaDef{gUnequal}[$\lambda$,$\sigma$],\mmaDef{gUnequal}[$\lambda$,$\nu$],\mmaDef{gUnequal}[$\mu$,$\nu$],\mmaDef{gUnequal}[$\mu$,$\sigma$]\}	
\end{mmaCell}
In addition to all this, \mtd{gAddAssumptions} also checks in many cases whether the assumptions entered are consistent with each other, gives a warning if this is not so, and does not add the inconsistent assumptions.
\begin{mmaCell}{Input}
		\mmaDef{gClearAssumptions}[];
		\mmaDef{gAddAssumptions}[\mmaDef{gEqual}[\m{$\mu$},\m{$\nu$}]]
		\mmaDef{gAddAssumptions}[\mmaDef{gUnequal}[\m{$\mu$},\m{$\nu$}]]	
\end{mmaCell}
\begin{mmaCell}{Output}		
		\{\mmaDef{gEqual}[$\mu$,$\nu$]\}
		\mmaOpt{gAddAssumptions}: Assumptions are not consistent with each other.	
\end{mmaCell}
\vspace{-5pt}
\subsubsection{Assumptions on spinors}
\vspace{-5pt}
Assumptions can also be used to define irreducible spinors, i.e. Majorana or Weyl spinors. For this, there are thee new objects, that denote the set of Weyl spinors with positive and negative chiralities, and Majorana spinors.
\begin{objDefnN}[label=gPositiveChiral_infobox]{\mtd{gPositiveChiral}}
	\mtd{gPositiveChiral}\\
	Denotes the set of all Weyl spinors with positive chirality. \\[3pt]
	\textbf{Arguments}\\
	None.
\end{objDefnN}
\begin{objDefnN}[label=gNegativeChiral_infobox]{\mtd{gNegativeChiral}}
	\mtd{gNegativeChiral}\\
	Denotes the set of all Weyl spinors with negative chirality. \\[3pt]
	\textbf{Arguments}\\
	None.
\end{objDefnN}
\begin{objDefnN}[label=gMajorana_infobox]{\mtd{gMajorana}}
	\mtd{gMajorana}\\
	Denotes the set of all Majorana spinors. \\[3pt]
	\textbf{Arguments}\\
	None.
\end{objDefnN}
For example, if a spinor is assumed to be a Weyl spinor, this is automatically taken into account when multiplying the spinor by $\gamma_*$ and simplifying the resulting expression.
\begin{mmaCell}{Input}
		\mmaDef{gSetRep}[\{-1,1,1,1,1,1,1,1,1,1\},1,1];
		\mmaDef{gAddAssumptions}[\{s[\m{$\xi$},\{\}]$\in$gPositiveChiral\}];
		\mmaDef{g}$\gamma$\mmaDef{5Convention}["ToRight"];
		\mmaDef{gTimes}[\mmaDef{m}[\m{$\gamma$5}],\mmaDef{s}[\m{$\xi$},\{\}]]//\mmaDef{gSimplify}
\end{mmaCell}
\begin{mmaCell}{Output}
		$\widetilde{\xi}$	
\end{mmaCell}
Also, after making this assumption the program can use simplifying identities like \eqref{App_bl_relations_vanishing_chiral} that are specific to chiral spinors.
\begin{mmaCell}{Input}
		\mmaDef{gTimes}[\mmaDef{dc}[\m{$\xi$},\{\}],\mmaDef{s}[\m{$\xi$},\{\}]]//\mmaDef{gSimplify}
\end{mmaCell}
\begin{mmaCell}{Output}
		0	
\end{mmaCell}
The assumption that a spinor is Majorana works in a completely similar fashion. For example, for Majorana spinors the product $B^{-1}\xi^*$ is automatically simplified to $\xi$, the left charge conjugate is automatically expressed in terms of the Dirac conjugate, and the identities specific to Majorana spinors, such as \eqref{app_bl_relations_Majorana_exchange} are used.
\begin{mmaCell}{Input}
		\mmaDef{gUseSpecialRep}[False];
		\mmaDef{gAddAssumptions}[\{\mmaDef{s}[\m{$\eta$},\{\}]$\in$gMajorana,\mmaDef{s}[\m{$\xi$},\{\}]$\in$gMajorana\}];
		\mmaDef{gBConvention}["ToRight"];
		\mmaDef{gTimes}[Inverse[\mmaDef{m}[B]],Conjugate[\mmaDef{s}[\m{$\eta$},\{\}]]]//\mmaDef{gSimplify}
		\mmaDef{gBConvention}["ToLeft"];
		\mmaDef{gCConvention}["ToLeft"];			
		\mmaDef{lcc}[\m{$\eta$},\{\}]//\mmaDef{gSimplify}
		\mmaDef{gTimes}[\mmaDef{dc}[\m{$\xi$},\{\}],\mmaDef{y}[\{u[\m{$\mu$}]\}],\mmaDef{s}[\m{$\eta$},\{\}]]//\mmaDef{gSimplify}
\end{mmaCell}
\begin{mmaCell}{Output}
		$\widetilde{\eta}$	
\end{mmaCell}
\begin{mmaCell}{Output}
		$\overline{\eta}$	
\end{mmaCell}
\begin{mmaCell}{Output}
		-($\overline{\eta}$$\star$\mmaSup{$\gamma$}{$\mu$}$\star$$\widetilde{\xi}$)
\end{mmaCell}
Note that the Majorana condition and reality condition are treated as different conditions. Assuming that the spinor is real makes the program treat all components of the spinor as real (which is not, in general compatible with Lorentz transformations). So for example
\begin{mmaCell}{Input}
		\mmaDef{gAddAssumptions}[\{\mmaDef{s}[\m{$\eta$},\{1\}]$\in$gMajorana,\mmaDef{s}[\m{$\eta$},\{2\}]$\in$Reals\}];
		Conjugate[\mmaDef{s}[\m{$\eta$},\{1\}]]
		Conjugate[\mmaDef{s}[\m{$\eta$},\{2\}]]	
		\mmaDef{gBConvention}["ToRight"];
		\mmaDef{gTimes}[Inverse[\mmaDef{m}[B]],Conjugate[\mmaDef{s}[\m{$\eta$},\{1\}]]]//\mmaDef{gSimplify}
		\mmaDef{gTimes}[Inverse[\mmaDef{m}[B]],Conjugate[\mmaDef{s}[\m{$\eta$},\{2\}]]]	
\end{mmaCell}
\begin{mmaCell}{Output}
		\mmaSub{$\widetilde{\eta}$}{1}
\end{mmaCell}
\begin{mmaCell}{Output}
		\mmaSup{\mmaSub{$\widetilde{\eta}$}{2}}{*}		
\end{mmaCell}
\begin{mmaCell}{Output}
		\mmaSub{$\widetilde{\eta}$}{1}
\end{mmaCell}
\begin{mmaCell}{Output}
		\mmaSup{B}{-1}$\star$\mmaSub{$\widetilde{\eta}$}{2}
\end{mmaCell}
Note also that whether these assumptions can be consistently used depends on the dimension and possibly the intertwiners chosen (see table \ref{Table_possible_spinors}). The spinors that can be used for the chosen dimension and representation can be checked by using the function \mtd{gAllowedSpinors}.
\begin{funcDefnN}[label=gAllowedSpinors_infobox]{\mtd{gAllowedSpinors}}
	\mtd{gAllowedSpinors[n]}\\
	Gives a list of irreducible spinors that can be used for \mtd{n}:th subrepresentation.\\[3pt]
	\textbf{Arguments}\\
	\mtd{n} is the number of subrepresentation. If this is left empty, then the function tells allowed spinors for the full representation.
	\tcbline
	\mtd{gAllowedSpinors[signature,$\eta$]}\\
	Gives a list of allowed spinors for a Clifford algebra defined by signature \mtd{signature}, and the intertwiners defined by $\eta$.\\[3pt]
	\textbf{Arguments}\\
	\mtd{signature} is the signature defining the Clifford algebra.\\
	\mtd{$\eta$} us the sign defining the properties of the intertwiners, appearing in equation \eqref{C_properties general}. 
\end{funcDefnN}
For example, if we wish to know which spinors are allowed for the 10-dimensional Clifford algebra with a Lorentzian signature, we could check this as follows.
\begin{mmaCell}{Input}
		\mmaDef{gSetRep}[\{-1,1,1,1,1,1,1,1,1,1\},1,1];
		\mmaDef{gAllowedSpinors}[]
\end{mmaCell}
\begin{mmaCell}{Output}
		\{Majorana,Weyl,Majorana-Weyl\}
\end{mmaCell}
This function can also be used to check which spinors are possible for other representations without changing the representation. This is done by using the second form of the function.
\begin{mmaCell}{Input}
		\mmaDef{gAllowedSpinors}[\{-1,1,1,1\},1]
\end{mmaCell}
\begin{mmaCell}{Output}
		\{Majorana,Weyl\}
\end{mmaCell}
Also the function \mtd{gAddAssumptions} will check whether the assumptions made about the spinors are consistent.
\begin{mmaCell}{Input}
		\mmaDef{gSetRep}[\{-1,1,1,1\},1,1];
		\mmaDef{gAddAssumptions}[\{\mmaDef{s}[$\xi$,\{\}]$\in$gMajorana,\mmaDef{s}[$\xi$,\{\}]$\in$gPositiveChiral\}]
\end{mmaCell}
\begin{mmaCell}{Output}
		\mmaOpt{gAddAssumptions}: Majorana-Weyl spinors are not allowed for this
		representation.
\end{mmaCell}
\newpage
\section{Explicit Representations} \label{sect_Explicit_reps}
\vspace{-5pt}
This far we have only seen the package working with completely abstract expressions for gamma matrices and intertwiners. In many situations, however, it is also useful to be able to consider explicit representations for gamma matrices and intertwiners, as well as for spinors and tensors. For this purpose, the package includes functions for defining explicit expressions for different objects, and for displaying symbolic expressions using these.
\subsection{Defining Explicit Representation}
\vspace{-5pt}
\subsubsection{Gamma Matrices and Intertwiners} \label{sect_Explicit_Reps_Gammas_Intertwiners}
\vspace{-5pt}
Explicit expressions for the gamma matrices can be set by using \mtd{gSetRepMatrices}.
\begin{funcDefnN}[label=gSetRepMatrices_infobox]{\mtd{gSetRepMatrices}}
	\mtd{gSetRepMatrices[gammas,C]}	\\
	Sets an explicit representation for the gamma matrices and the $A$-, $C$- and $B$-intertwiners.\\[3pt]
	\textbf{Arguments}\\
	\mtd{gammas} is a list of explicit expressions for the gamma matrices  with indices up. The first matrix in the list gives the value for $\gamma^0$ or $\gamma^1$, depending on whether the numbering of indices starts from 0 or 1 (see section \ref{gSetRep}) and the following matrices give values to the subsequent gamma matrices.\\
	\mtd{C} is the matrix for $C$-intertwiner.
\end{funcDefnN}
Note that the values for $A$- and $B$-intertwiners are not needed, since these will be automatically determined using the gamma matrices and $C$-intertwiner, as explained in appendix \ref{app_Rep_Properties}.

If we would like to setup a representation for 2-dimensional Lorentzian Clifford algebra with the following definitions:
\begin{equation}
\gamma^0 = i \sigma_1, \qquad \gamma^1 = \sigma_2, \qquad C = \sigma_1,
\end{equation}
this would be done as follows: First we need to use \mtd{gSetRep} to define the abstract Clifford algebra.
\begin{mmaCell}{Input}
		\mmaDef{gSetRep}[\{-1,1\},-1,-1,1];
\end{mmaCell} 
Then we can set up the actual explicit representation for the gamma matrices and intertwiners.
\begin{mmaCell}{Input}
		s1=\{\{0,1\},\{1,0\}\};
		s2=\{\{0,-I\},\{I,0\}\};
		s3=\{\{1,0\},\{0,-1\}\};
		\mmaDef{gSetRepMatrices}[\{I*s1,s2\},s1];
\end{mmaCell} 
\vspace{-5pt}
\subsubsection{Coordinates}
\vspace{-5pt}
To be able to use derivatives, we must also define coordinates. This is done by function \mtd{gSetCoordinates}.
\begin{funcDefnN}[label=gSetCoordinates_infobox]{\mtd{gSetCoordinates}}
	\mtd{gSetCoordinates[{x1,x2,...,xd}]}\\
	Sets explicit expressions for coordinates that are used for computing explicit expressions for derivatives.\\[3pt]
	\textbf{Arguments}\\
	\mtd{x1,...,xd} are coordinates. The first, \mtd{x1}, corresponds to the index 1 (or 0 if numbering of indices is set to begin from 0), the second, \mtd{x2}, corresponds to index 2 (or 1), and so on.
\end{funcDefnN}
To set two coordinates x0 and x1 corresponding to two indices 0 and 1, we can use the following code.
\begin{mmaCell}{Input}
		\mmaDef{gSetCoordinates}[\{x0,x1\}];
\end{mmaCell}
\vspace{-5pt}
\subsubsection{Spinors}
\vspace{-5pt}
Explicit expressions for spinors can be set using function \mtd{gSetSpinor}.
\begin{funcDefnN}[label=gSetSpinor_infobox]{\mtd{gSetSpinor}}
	\mtd{gSetSpinor[s[x,\{i\},n],components]}\\
	Sets explicit expression \mtd{components} for spinor \mtd{\mmaSub{$\widetilde{x}$}{(n)i}}.\\[3pt]
	\textbf{Arguments}\\
	\mtd{s[x,\{i\},n]} is the spinor (with name \mtd{x} and additional indices \mtd{i} transforming under the \mtd{n}:th subrepresentation) for which the explicit expressions \mtd{components} will be set.\\
	\mtd{components} is a list containing the components of the spinor.
\end{funcDefnN}
For instance, setting explicit expression $(a_1,a_2)^{\top}$ for a spinor named $\xi_1$ is done with the following code.
\begin{mmaCell}{Input}
		\mmaDef{gSetSpinor}[\mmaDef{s}[\m{$\xi$},\{1\}],\{\mmaSub{a}{1},\mmaSub{a}{2}\}];
\end{mmaCell} 
All the explicit expressions for spinors can be deleted using \mtd{gClearSpinors}.
\begin{funcDefnN}[label=gClearSpinors_infobox]{\mtd{gClearSpinors}}
	\mtd{gClearSpinors[]}\\
	Clears all explicit expressions for spinors.\\[3pt]
	\textbf{Arguments}\\
	None.
\end{funcDefnN}
\vspace{-5pt}
\subsubsection{Tensors}
\vspace{-5pt}
For tensors, including symmetric \mtd{gSymm} and antisymmetric \mtd{gForm}, explicit components can be set by using \mtd{gSetTensor}.
\begin{funcDefnN}[label=gSetTensor_infobox]{\mtd{gSetTensor}}
	\mtd{gSetTensor[F,components]}\\
	Sets an explicit expression \mtd{components} for tensor \mtd{F}.\\[3pt]
	\textbf{Arguments}\\
	\mtd{F} is the name of the tensor, which can be either \mtd{gTensor}, \mtd{gForm} or \mtd{gSymm},\\
	\mtd{components} is a table containing the components of the tensor with the indices down.
\end{funcDefnN}
For example, in a 2-dimensional space, if we would like to define a two-index antisymmetric tensor $F$ as follows.
\begin{align}
[F_{ab}] &= \begin{pmatrix}
0 & -2 \\
2  & 0
\end{pmatrix}. \label{Explicit_F_defn}
\end{align}
We could do this by using the following code.
\begin{mmaCell}{Input}
		\mmaDef{gSetTensor}[F,\{\{0,-2\},\{2,0\}\}];
\end{mmaCell} 
Similarly, in a 3-dimensional space, an arbitrary 2-index tensor can be defined by
\begin{align}
[G_{ab}] &= \begin{pmatrix}
1 & -2  & 3\\
2 & 0  & 4\\
2 & 1 & 3
\end{pmatrix}. \label{Explicit_G_defn}
\end{align}
The code for making this identification is (we need to temporarily define a 3-dimensional representation).
\begin{mmaCell}{Input}
		\mmaDef{gSetRep}[\{1,1,1\},-1,-1]
		\mmaDef{gSetTensor}[G,\{\{1,-2,3\},\{2,0,4\},\{2,1,3\}\}];
		\mmaDef{gSetRep}[\{-1,1\},-1,-1,1];
		\mmaDef{gSetCoordinates}[\{x0,x1\}]
\end{mmaCell}
Note that setting a new value for tensor components of a tensor will override their old values, but changing the dimension of the representation leaves the old explicit expressions for the tensors. These expressions can be cleared using \mtd{gClearTensors}.
\begin{funcDefnN}[label=gClearTensors_infobox]{\mtd{gClearTensors}}
	\mtd{gClearTensors[]}\\
	Clears all explicit expressions for tensors.\\[3pt]
	\textbf{Arguments}\\
	None.
\end{funcDefnN}
\vspace{-5pt}
\subsection{Using Explicit Representations} \label{sect_Explicit_reps_using}
\vspace{-5pt}
Once the explicit expressions for the gamma matrices, intertwiners, and possibly spinors and tensors have been defined, any expression containing them can be given in the explicit form by using \mtd{gExplicit}.
\begin{funcDefnN}[label=gExplicit_infobox]{\mtd{gExplicit}}
	\mtd{gExplicit[expr]}\\
	Gives an explicit form of a symbolic expression \mtd{expr} using the explicit representations for the gamma matrices, intertwiners, spinors, and tensors set by using \mtd{gSetRepMatrices}, \mtd{gSetSpinor} and \mtd{gSetTensor}.\\[3pt]
	\textbf{Arguments}\\
	\mtd{expr} is the expression to be expressed in explicit form.
\end{funcDefnN}
The simplest use of \mtd{gExplicit} is to give explicit forms for products of gamma matrices, spinors, or forms. For example, accessing the explicit definitions made in the previous section can be done as follows.
\begin{mmaCell}{Input}
		\mmaDef{gExplicit}[\mmaDef{y}[\{u[0,1]\}]]
		\mmaDef{gExplicit}[\mmaDef{s}[\m{$\xi$},\{1\}]]
		\mmaDef{gExplicit}[\mmaDef{gForm}[F,\{0,1\}]]
		\mmaDef{gExplicit}[\mmaDef{dc}[\m{$\xi$},\{1\}]]
		\mmaDef{gExplicit}[\mmaDef{gTimes}[\mmaDef{gH}[\mmaDef{s}[\m{$\xi$},\{1\}]],\mmaDef{m}[A]]]
\end{mmaCell}
\begin{mmaCell}{Output}
		\{\{-1,0\},\{0,1\}\}	
\end{mmaCell}
\begin{mmaCell}{Output}
		\{\mmaSub{a}{1},\mmaSub{a}{2}\}
\end{mmaCell}
\begin{mmaCell}{Output}
		-2
\end{mmaCell}
\begin{mmaCell}{Output}
		\{$\Ii$\mmaSup{(\mmaSub{a}{2})}{*},$\Ii$\mmaSup{(\mmaSub{a}{1})}{*}\}	
\end{mmaCell}
\begin{mmaCell}{Output}
		\{$\Ii$\mmaSup{(\mmaSub{a}{2})}{*},$\Ii$\mmaSup{(\mmaSub{a}{1})}{*}\}	
\end{mmaCell}
As mentioned in the previous section, the explicit components defined for the gamma matrices are those with indices up, and for tensors those with indices down. Therefore we get the appropriate signs when using \mtd{gExplicit} to give explicit forms for objects with lowered or raised indices.
\begin{mmaCell}{Input}
		\mmaDef{gExplicit}[\mmaDef{y}[\{d[0]\}]]
		\mmaDef{gExplicit}[\mmaDef{gForm}[F,\{u[0,1]\}]]
		\mmaDef{gExplicit}[\mmaDef{gForm}[F,\{u[0],d[1]\}]]
\end{mmaCell}
\begin{mmaCell}{Output}
		\{\{0,-$\Ii$\},\{-$\Ii$,0\}\}
\end{mmaCell}
\begin{mmaCell}{Output}
		2
\end{mmaCell}
\begin{mmaCell}{Output}
		2
\end{mmaCell}
If the tensors or spinors have no value set by \mtd{gSetTensor} or \mtd{gSetSpinor}, an alternative notation will be used to represent the components of these objects.
\begin{mmaCell}{Input}
		\mmaDef{gExplicit}[\mmaDef{s}[\m{$\eta$},\{\}]]
		\mmaDef{gExplicit}[\mmaDef{gForm}[H,\{0,1\}]]
\end{mmaCell}
\begin{mmaCell}{Output}
		\{$\eta$[1],$\eta$[2]\}
\end{mmaCell}
\begin{mmaCell}{Output}
		H[0,1]
\end{mmaCell}
The function \mtd{gExplicit} is not, however, limited only to simple expressions like these. Indeed, sums and products, both commutative, and non-commutative are given an explicit expression just as easily.
\begin{mmaCell}{Input}
		\mmaDef{gExplicit}[\mmaDef{gTimes}[\mmaDef{gH}[\mmaDef{s}[\m{$\xi$},\{1\}]],\mmaDef{y}[\{u[0,1]\}],\mmaDef{s}[\m{$\xi$},\{1\}]]]
		\mmaDef{gExplicit}[3*\mmaDef{s}[\m{$\xi$},\{1\}]+2*\mmaDef{gTimes}[\mmaDef{y}[\{u[0]\}],\mmaDef{s}[\m{$\xi$},\{1\}]]]
\end{mmaCell}
\begin{mmaCell}{Output}
		-\mmaSup{(\mmaSub{a}{1})}{*}\mmaSub{a}{1}+\mmaSup{(\mmaSub{a}{2})}{*}\mmaSub{a}{2}
\end{mmaCell}
\begin{mmaCell}{Output}
		\{3\mmaSub{a}{1}+2$\Ii$\mmaSub{a}{2},2$\Ii$\mmaSub{a}{1}+3\mmaSub{a}{2}\}
\end{mmaCell}
Although expressions with symbolic indices cannot be expressed in explicit form, \mtd{gExplicit} can still be used on expressions with dummy indices, and these will be converted into sums, and can thus be evaluated explicitly. 
\begin{mmaCell}{Input}
		\mmaDef{gExplicit}[\mmaDef{gForm}[F,\{a,b\}]*\mmaDef{y}[\{u[a,b]\}]]
		\mmaDef{gSetTensor}[G,\{\{-1,0\},\{0,0\}\}];
		\mmaDef{gExplicit}[\mmaDef{gTensor}[G,\{0,b\}]*\mmaDef{y}[\{u[0,b]\}]]
\end{mmaCell}
\begin{mmaCell}{Output}
		\{\{4,0\},\{0,-4\}\}
\end{mmaCell}
\begin{mmaCell}{Output}
		0
\end{mmaCell}
There are two tensors, the form \mtd{gForm[\mtu{$\epsilon$},\{\mtu{$\mu$1},...,\mtu{$\mu$n}\}]} and symmetric tensor \mtd{gSymm[\mtu{$\eta$},\{\mtu{$\mu$},\mtu{$\nu$}\}]}, that have pre-defined explicit expressions in any representation. Form $\epsilon$ is interpreted as the Levi-Civita symbol, with the convention that $\epsilon_{1,\dotsc,d}=1$ or $\epsilon_{0,..,d-1}=1$ (depending on whether the numbering of the gamma matrices starts from 1 or 0). The symmetric tensor $\eta$ is interpreted as the metric $\eta^{\mu \nu}$ appearing in the definition of the Clifford algebra \eqref{Clifford_algebra_anticommutation}.
\begin{mmaCell}{Input}
		\mmaDef{gExplicit}[\mmaDef{gForm}[\m{$\epsilon$},\{0,1\}]]
		\mmaDef{gExplicit}[\mmaDef{gSymm}[\m{$\eta$},\{0,0\}]]
		\mmaDef{gExplicit}[\mmaDef{gSymm}[\m{$\eta$},\{1,1\}]]
\end{mmaCell}
\begin{mmaCell}{Output}
		-1
\end{mmaCell}
\begin{mmaCell}{Output}
		-1
\end{mmaCell}
\begin{mmaCell}{Output}
		1
\end{mmaCell}
Notice, however, that, for example, the \textit{form} $\eta$ and \textit{symmetric tensor} $\epsilon$ have no predefined meaning.
\begin{mmaCell}{Input}
		\mmaDef{gExplicit}[\mmaDef{gSymm}[\m{$\epsilon$},\{0,1\}]]
		\mmaDef{gExplicit}[\mmaDef{gForm}[\m{$\eta$},\{0,1\}]]
\end{mmaCell}
\begin{mmaCell}{Output}
		$\epsilon$[0,1]
\end{mmaCell}
\begin{mmaCell}{Output}
		$\eta$[0,1]
\end{mmaCell}
Recalling that we have set coordinates previously, we can also make use of derivatives. On expressions containing coordinates, these work as ordinary derivatives.
\begin{mmaCell}{Input}
		\mmaDef{gExplicit}[\mmaDef{gD}[\{0\},x0]]
		\mmaDef{gExplicit}[\mmaDef{gD}[\{0\},x0]]
\end{mmaCell}
\begin{mmaCell}{Output}
		1
\end{mmaCell}
\begin{mmaCell}{Output}
		0
\end{mmaCell}
However, unlike with the symbolic expressions, now any symbol that does not have explicit functional dependence on the coordinates, will be treated as a constant.
\begin{mmaCell}{Input}
		\mmaDef{gExplicit}[\mmaDef{gD}[\{1\},a]]
		\mmaDef{gExplicit}[\mmaDef{gD}[\{1\},a[x1]]]
\end{mmaCell}
\begin{mmaCell}{Output}
		0
\end{mmaCell}
\begin{mmaCell}{Output}
		a'[x1]
\end{mmaCell}
We can also take derivatives of tensors, spinors, and other objects. For example, by contracting indices, we can calculate the divergence of a tensor, or represent a box operator.
\begin{mmaCell}{Input}
		\mmaDef{gSetTensor}[T,\{x0^2*x1,1/x1\}]
		\mmaDef{gSetSpinor}[s[\m{$\xi$},\{i\}],\{x0,x1\}]
		\mmaDef{gD}[\{\m{$\mu$}\},\mmaDef{gTensor}[T,\{u[\m{$\mu$}]\}]]//\mmaDef{gExplicit}
		\mmaDef{gD}[\{u[\m{$\mu$}],d[\m{$\mu$}]\},x1^2]//\mmaDef{gExplicit}
		\mmaDef{gD}[\{\m{$\mu$}\},\mmaDef{gTimes}[\mmaDef{dc}[\m{$\xi$},\{i\}],\mmaDef{y}[\{u[\m{$\mu$}]\}],s[\m{$\xi$},\{i\}]]]//\mmaDef{gExplicit}
\end{mmaCell}
\begin{mmaCell}{Output}
		-\mmaFrac{1}{\mmaSup{x1}{2}}-2x0x1
\end{mmaCell}
\begin{mmaCell}{Output}
		2
\end{mmaCell}
\begin{mmaCell}{Output}
		-x0+x1-\mmaSup{(x0)}{*}+\mmaSup{(x1)}{*}
\end{mmaCell}
\newpage
\section{Subalgebras} \label{sect_Subalgebras}
\vspace{-5pt}
As discussed in section \ref{sect_Review_Subalgebras} Clifford algebras $\text{Cliff}(d)$ can be broken into subalgebras $\text{Cliff}(d_1)\times\text{Cliff}(d_2)\times\dotsc\times\text{Cliff}(d_m)$, with $d_1+\dotsc+d_m=d$. This also induces a decomposition of the associated special orthogonal groups, $\text{SO}(d) \to \text{SO}(d_1) \times \dotsc \times \text{SO}(d_m)$ A concrete realisation of this decomposition in terms of gamma matrices can be found by using formulae in appendix \ref{App_Matrix_Properties_Decomposition}, and iterating these if needed. These decompositions  and many properties of the corresponding decomposition of special orthogonal groups are also implemented in the package, providing useful tools for example for computing expressions related to dimensional reduction.
\vspace{-5pt}
\subsection{Defining Subalgebras}
\vspace{-5pt}
Setting subrepresentations can be done by using the function \mtd{gSetSubRep}.
\begin{funcDefnN}[label=gSetSubRep_infobox]{\mtd{gSetSubRep}}
	\mtd{gSetSubRep[n,dimension,$\epsilon$,$\eta$,$\zeta$]}\\
	Defines the \mtd{n}:th subalgebra of the full Clifford algebra, and the properties of the intertwiners.\\[3pt]
	\textbf{Arguments}\\
	\mtd{n} is the number of the subalgebra,\\
	\mtd{dimension} is the dimension of the subalgebra,\\
	$\epsilon$ is the parameter with values $\pm 1$ appearing in \eqref{C_properties general},\\
	$\eta$ is the parameter with values $\pm 1$ appearing in \eqref{C_properties general}.\\[3pt] 
	\textbf{Optional Arguments}\\
	$\zeta$ is the parameter with values $\pm 1$ that determines which one of the two inequivalent representations in odd dimensions is used. Note that this is relevant only in odd dimensions. If this value is not specified for an representation in an odd dimension, 1 is used as the default value.
\end{funcDefnN}
The function works essentially like \mtd{gSetRep} (\ref{gSetRep_infobox}), except for two differences. The subalgebras must be defined in increasing order (so that first one should define the first subalgebra, then the second, and so on). Secondly, unlike with \mtd{gSetRep}, the numbering of indices cannot be modified, but this is instead determined by the numbering of indices of the full algebra, so that if for example, the full algebra has indices $0,\dotsc,d-1$, then the first subalgebra will have indices $0,\dotsc,d_1-1$, the second will have indices $d_1-1,\dotsc$ and so on.

For example, defining subalgebras $\text{Cliff}(0,3) \times \text{Cliff}(0,3) \times \text{Cliff}(1,3) \subset \text{Cliff}(1,9)$ would be done as follows: First we need to set up the full representation $\text{Cliff}(1,9)$ by using
\begin{mmaCell}{Input}
	\mmaDef{\mmaDef{gSetRep}}[\{-1,1,1,1,1,1,1,1,1,1\},1,1]
\end{mmaCell}
Then, we define the subalgebras, first the $\text{Cliff}(1,3)$ (since this corresponds to the first four terms of the signature of the full algebra), then the first $\text{Cliff}(0,3)$, and after that the second $\text{Cliff}(0,3)$.
\begin{mmaCell}{Input}
		\mmaDef{\mmaDef{gSetSubRep}}[1,4,1,1]
		\mmaDef{\mmaDef{gSetSubRep}}[2,3,1,1]
		\mmaDef{\mmaDef{gSetSubRep}}[3,3,1,1]
\end{mmaCell}
By default, an index can take any value $1,\dotsc,d$, but when considering subalgebras and their representations, it is useful to define variables that can, for example, only take values $1,\dotsc,d_1$ corresponding to the first subalgebra. Indices like this can be defined by using the function \mtd{gSetIndices}.
\begin{funcDefnN}[label=gSetIndices_infobox]{\mtd{gSetIndices}}
	\mtd{gSetIndices[n,indices]}\\
	Defines, which variables represent indices that take only values corresponding to the \mtd{n}:th subalgebra.\\[3pt]
	\textbf{Arguments}\\
	\mtd{n} is the number of the subalgebra.\\
	\mtd{indices} is a list of indices that take only values corresponding to the \mtd{n}:th subalgebra.
\end{funcDefnN}
For example, making the variables $\alpha$ and $\beta$ take values $0,\dotsc,3$ corresponding to $\text{Cliff}(1,3)$, $\mu$ and $\nu$ take values $4,\dotsc,6$ corresponding to the first $\text{Cliff}(0,3)$, and indices $a$ and $b$ take values $7,\dotsc,9$ corresponding to the second $\text{Cliff}(0,3)$, is done by the following code.
\begin{mmaCell}{Input}
		\mmaDef{\mmaDef{gSetIndices}}[1,\{\m{$\alpha$},\m{$\beta$}\}]
		\mmaDef{\mmaDef{gSetIndices}}[2,\{\m{$\mu$},\m{$\nu$}\}]
		\mmaDef{\mmaDef{gSetIndices}}[3,\{a,b\}]
\end{mmaCell}
The indices that do not appear in the any of the lists, can correspond to any direction, as before.
\vspace{-5pt}
\subsection{Decomposing Expressions}
\vspace{-5pt}
Before we begin discussing subrepresentations, we need to introduce notation for tensor products, \mtd{gTensorProduct}.
\begin{objDefnN}[label=gTensorProduct_infobox]{\mtd{gTensorProduct}}
	\mtd{gTensorProduct[x1,...,xm]}\\
	Denotes tensor product of objects \mtd{x1},...,\mtd{xm}.\\[3pt]
	\textbf{Arguments}\\
	\mtd{x1,...,xm} are the objects whose tensor product \mtd{gTensorProduct[x1,...,xm]} denotes.
\end{objDefnN}
Tensor product can be taken of any objects, such as spinors, matrices or scalars and products thereof. The commutative objects are automatically recognised and commuted to the front.
\begin{mmaCell}{Input}
		\mmaDef{gTensorProduct}[\mmaDef{m[A]},\mmaDef{s}[\m{$\xi$},\{\}],3,\mmaDef{m}[\m{$\sigma$3}]]
		\mmaDef{gTensorProduct}[\mmaDef{s}[\m{$\xi$},\{\}],\mmaDef{gTimes}[\mmaDef{y}[\{u[\m{$\mu$},\m{$\nu$}]\}],2*\mmaDef{s}[\m{$\eta$},\{\}]],3,
		\mmaDef{m}[\m{$\sigma$3}]*\mmaDef{gForm}[A,\{\m{$\mu$}\}]]				
\end{mmaCell}
\begin{mmaCell}{Output}
		3(\mmaSub{A}{(1)}$\otimes$\mmaSub{$\widetilde{\xi}$}{(2)}$\otimes$\mmaSub{I}{(3)}$\otimes$\mmaSup{\mmaSub{$\sigma$}{(4)}}{3})
\end{mmaCell}
\begin{mmaCell}{Output}
		6\mmaSub{A}{$\mu$}\mmaSub{$\widetilde{\xi}$}{(1)}$\otimes$(\mmaSup{\mmaSub{$\gamma$}{(2)}}{$\mu$,$\nu$}$\star$\mmaSub{$\widetilde{\eta}$}{(2)})$\otimes$\mmaSub{I}{(3)}$\otimes$\mmaSup{\mmaSub{$\sigma$}{(4)}}{3}
\end{mmaCell}
Also sums and products of tensor products are handled using the usual rules of summing and multiplying tensor products. Recall that the non-commutative product can be expressed as $\star$ also in input, which is what we have done in the second line.
\begin{mmaCell}{Input}
		\mmaDef{gTensorProduct}[\mmaDef{m}[Id],\mmaDef{m}[Id]+\mmaDef{m}[C],\mmaDef{m}[Id],\mmaDef{m}[Id]]
		\mmaDef{gTensorProduct}[\mmaDef{y}[\{u[\m{$\alpha$},\m{$\beta$}]\}],\mmaDef{m}[Id],\mmaDef{m}[Id],\mmaDef{m}[\m{$\sigma$1}]]$\star$
		  \mmaDef{gTensorProduct}[\mmaDef{s}[\m{$\xi$},\{\}],\mmaDef{m}[Id],\mmaDef{m}[Id],\mmaDef{m}[\m{$\sigma$2}]]
\end{mmaCell}
\begin{mmaCell}{Output}
		\mmaSub{I}{(1)}$\otimes$\mmaSub{C}{(2)}$\otimes$\mmaSub{I}{(3)}$\otimes$\mmaSub{I}{(4)}+I
\end{mmaCell}
\begin{mmaCell}{Output}
		$\Ii$(\mmaSup{\mmaSub{$\gamma$}{(1)}}{$\alpha$,$\beta$}$\star$\mmaSub{$\widetilde{\xi}$}{(1)})$\otimes$\mmaSub{I}{(2)}$\otimes$\mmaSub{I}{(3)}$\otimes$\mmaSup{\mmaSub{$\sigma$}{(4)}}{3}
\end{mmaCell}
Any functions can be used on the tensor products.
\begin{mmaCell}{Input}
		\mmaDef{gD}[\{M\},\mmaDef{gTensorProduct}[\mmaDef{s}[\m{$\eta$},\{\},1],\mmaDef{s}[\m{$\xi$},\{\},2],\mmaDef{m}[A,3],\mmaDef{m}[Id,4]]]
		\mmaDef{gSymm}[\m{$\eta$},\{\m{$\alpha$},\m{$\beta$}\}]\mmaDef{gTensorProduct}[\mmaDef{gTimes}[y[\{u[\m{$\alpha$}]\}],\mmaDef{s}[\m{$\xi$},\{\}]],
		\mmaDef{gTimes}[\mmaDef{y}[\{u[\m{$\mu$}]\},\{u[\m{$\nu$}]\}],\mmaDef{s}[\m{$\xi$},\{\}]],3,\mmaDef{m}[\m{$\sigma$3}]*\mmaDef{gForm}[A,\{\m{$\mu$}\}]]//\mmaDef{gOrd}//
		  \mmaDef{gSimplify}
\end{mmaCell}
\begin{mmaCell}{Output}
		(\mmaSub{d}{M}\mmaSub{$\widetilde{\eta}$}{(1)})$\otimes$\mmaSub{$\widetilde{\xi}$}{(2)}$\otimes$\mmaSub{A}{(3)}$\otimes$\mmaSub{I}{(4)}+\mmaSub{$\widetilde{\eta}$}{(1)}$\otimes$(\mmaSub{d}{M}\mmaSub{$\widetilde{\xi}$}{(2)})$\otimes$\mmaSub{A}{(3)}$\otimes$\mmaSub{I}{(4)}
\end{mmaCell}
\begin{mmaCell}{Output}
		3\mmaSub{A}{$\lambda$2\$1}(\mmaSub{$\gamma$}{(1)$\beta$}$\star$\mmaSub{$\widetilde{\xi}$}{(2)})$\otimes$(\mmaSup{\mmaSub{$\gamma$}{(2)}}{$\lambda$2\$1,$\nu$}$\star$\mmaSub{$\widetilde{\xi}$}{(2)})$\otimes$\mmaSub{I}{(3)}$\otimes$\mmaSup{\mmaSub{$\sigma$}{(4)}}{3}+
		
		  3\mmaSup{A}{$\nu$}($\otimes$\mmaSub{$\gamma$}{(1)$\beta$}$\star$\mmaSub{$\widetilde{\xi}$}{(1)})$\otimes$\mmaSub{$\widetilde{\xi}$}{(2)}$\otimes$\mmaSub{I}{(3)}$\otimes$\mmaSup{\mmaSub{$\sigma$}{(4)}}{3}
\end{mmaCell}
Here the index \mtd{$\lambda$2\$1} runs over values corresponding to the 2nd subalgebra i.e. over 4, 5, and 6. This default notation can be customised by modifying option \mtd{gSumIndexIndicator}.
\begin{optDefnN}[label=gSumIndexIndicator_infobox]{\mtd{gSumIndexIndicator}}
	\mtd{gSumIndexIndicator[n,indicator]}\\
	Sets the name for variables indicating sum indices corresponding to the \mtd{n}:th subalgebra.\\[3pt]
	\textbf{Arguments}\\
	\mtd{indicator} is a string with which the name of every sum index corresponding to \mtd{n}:th subalgebra will begin.\\
	\mtd{n} is the number of the subrepresentation for which the setting applies.
\end{optDefnN}
For example, if we would like the indices of the form \mtd{$\gamma$n}, \mtd{$\sigma$n}, and \mtd{cn} denote the sum indices corresponding to the first, second and third subalgebra, respectively, we could do this by using the following.
\begin{mmaCell}{Input}
		\mmaDef{gSumIndexIndicator}[1,"$\gamma$"]		
		\mmaDef{gSumIndexIndicator}[2,"$\sigma$"]	
		\mmaDef{gSumIndexIndicator}[3,"c"]								
\end{mmaCell}
After this, for example the expression $A_{M}A^{M}$ looks much nicer when decomposed in terms of the subrepresentations than it would look had we left the default notation for indices corresponding to the subalgebras.
\begin{mmaCell}{Input}
		\mmaDef{gForm}[A,\{u[M]\}]\mmaDef{gForm}[A,\{M\}]//\mmaDef{gDecomposeToSubReps}
\end{mmaCell}
\begin{mmaCell}{Output}
		\mmaSub{A}{c1}\mmaSup{A}{c1}+\mmaSub{A}{$\gamma$1}\mmaSup{A}{$\gamma$1}+\mmaSub{A}{$\sigma$1}\mmaSup{A}{$\sigma$1}
\end{mmaCell}
Like in the previous example, expressions can be automatically expressed in terms of subrepresentations using \mtd{gDecomposeToSubReps}.
\begin{funcDefnN}[label=gDecomposeToSubReps_infobox]{\mtd{gDecomposeToSubReps}}
	\mtd{gDecomposeToSubReps[expr]}\\
	Expresses the tensors, spinors and matrices in terms of those associated to the subrepresentations whenever possible, if subalgebras have been defined.\\[3pt]
	\textbf{Arguments}\\
	\mtd{expr} is the expression to be modified.
\end{funcDefnN}
In practice this does several different things: First of all, all spinors, gamma matrices and intertwiners are all expressed in terms of tensor products of spinors, gamma matrices and intertwiners associated to the subalgebras, using expressions like those  listed in appendix \ref{App_Matrix_Properties_Decomposition}.
\begin{mmaCell}{Input}
		\mmaDef{m}[A]//\mmaDef{gDecomposeToSubReps}
		\mmaDef{s}[\m{$\xi$},\{i\}]//\mmaDef{gDecomposeToSubReps}	
		\mmaDef{gTimes}[\mmaDef{y}[\{u[\m{$\mu$}]\}],\mmaDef{m}[A],\mmaDef{s}[\m{$\xi$}]]//\mmaDef{gDecomposeToSubReps}	
\end{mmaCell}
\begin{mmaCell}{Output}
		\mmaSub{A}{(1)}$\otimes$\mmaSub{I}{(2)}$\otimes$\mmaSub{I}{(3)}$\otimes$\mmaSub{I}{(4)}		
\end{mmaCell}
\begin{mmaCell}{Output}
		\mmaSub{$\widetilde{\xi}$}{(1)i}$\otimes$\mmaSub{$\widetilde{\xi}$}{(2)i}$\otimes$\mmaSub{$\widetilde{\xi}$}{(3)i}$\otimes$\mmaSub{$\widetilde{\xi}$}{(4)i}					
\end{mmaCell}
\begin{mmaCell}{Output}
		-((\mmaSub{A}{(1)}$\star$\mmaSub{$\gamma$}{*(1)}$\star$\mmaSub{$\widetilde{\xi}$}{(1)})$\otimes$(\mmaSup{\mmaSub{$\gamma$}{(2)}}{$\mu$}$\star$\mmaSub{$\widetilde{\xi}$}{(2)})$\otimes$\mmaSub{$\widetilde{\xi}$}{(3)}$\otimes$(\mmaSup{\mmaSub{$\sigma$}{(4)}}{2}$\star$\mmaSub{$\widetilde{\xi}$}{(4)}))			
\end{mmaCell}

In case of gamma matrices, this naturally requires that all the antisymmetrised indices appearing in the gamma matrices are those associated to some subalgebra (see \ref{gSetIndices_infobox}).
\begin{mmaCell}{Input}
		\mmaDef{y}[\{u[\m{$\alpha$},\m{$\beta$}]\}] //\mmaDef{gDecomposeToSubReps}
		\mmaDef{y}[\{u[\m{$\alpha$},\m{$\beta$}]\},\{u[\m{$\mu$}]\}] //\mmaDef{gDecomposeToSubReps}
		\mmaDef{y}[\{u[\m{$\alpha$},\m{$\beta$}]\},\{u[k]\}] //\mmaDef{gDecomposeToSubReps}
		\mmaDef{y}[\{u[\m{$\alpha$},\m{$\beta$},k]\}] //\mmaDef{gDecomposeToSubReps}					
\end{mmaCell}
\begin{mmaCell}{Output}
		\mmaSup{\mmaSub{$\gamma$}{(1)}}{$\alpha$,$\beta$}$\otimes$\mmaSub{I}{(2)}$\otimes$\mmaSub{I}{(3)}$\otimes$\mmaSub{I}{(4)}					
\end{mmaCell}
\begin{mmaCell}{Output}
		(\mmaSub{$\gamma$}{*(1)}$\star$\mmaSup{\mmaSub{$\gamma$}{(1)}}{$\alpha$,$\beta$})$\otimes$\mmaSup{\mmaSub{$\gamma$}{(2)}}{$\mu$}$\otimes$\mmaSub{I}{(3)}$\otimes$\mmaSup{\mmaSub{$\sigma$}{(4)}}{1}					
\end{mmaCell}
\begin{mmaCell}{Output}
		\mmaSup{\mmaSub{$\gamma$}{(1)}}{$\alpha$,$\beta$}$\otimes$\mmaSub{I}{(2)}$\otimes$\mmaSub{I}{(3)}$\otimes$\mmaSub{I}{(4)}$\star$\mmaSup{$\gamma$}{k}			
\end{mmaCell}
\begin{mmaCell}{Output}
		\mmaSup{$\gamma$}{$\alpha$,$\beta$,k}				
\end{mmaCell}
Here in the last example the gamma matrices were not decomposed into tensor products, since among the antisymmetrised indices there appears an index \mtd{k}, which is not associated to any particular subalgebra.

Even though the decompositions we have seen so far have all used pre-programmed expressions, it is also possible to define custom decompositions for matrices and spinors by using the function \mtd{gSetDecomposition}.
\begin{funcDefnN}[label=gSetDecomposition_infobox]{\mtd{gSetDecomposition}}
	\mtd{gSetDecomposition[obj,decomposition]}\\
	Sets a custom decomposition in terms of subrepresentations for spinors and matrices other than the gamma matrices.\\[3pt]
	\textbf{Arguments}\\
	\mtd{obj} is the object (a matrix or a spinor), for which the custom decomposition is defined.\\
	\mtd{decomposition} is the decomposed expression for the object.
	\tcbline
	\mtd{gSetDecomposition[y,n,decomposition]}\\
	Sets a custom decomposition in terms of subrepresentations for the gamma matrices.\\[3pt]
	\textbf{Arguments}\\
	\mtd{n} is the number of the subalgebra to which the gamma matrix is associated.\\
	\mtd{decomposition} is the decomposed expression for the gamma matrix. Here the free index must be denoted by \mtd{gx\$}, and this index must be up. In addition, the \mtd{n}:th argument in the tensor product \mtd{gTensorProduct} must be an expression involving the gamma matrix if \mtd{n}:th subrepresentation.
\end{funcDefnN}
For example, using the first form of \mtd{gSetDecomposition}, we could define a custom decomposition for a matrix called $X$, or for a spinor $\xi_{i}$.
\begin{mmaCell}{Input}
		\mmaDef{gSetDecomposition}[\mmaDef{m}[X],\mmaDef{gTensorProduct}[\mmaDef{m}[X],2,\mmaDef{m}[A],\mmaDef{m}[Z]]]
		\mmaDef{gSetDecomposition}[\mmaDef{s}[\m{$\xi$},\{i\}],\mmaDef{gTensorProduct}[\mmaDef{s}[\m{$\eta$},\{i\}],\mmaDef{s}[\m{$\lambda$},\{\}],
		  \mmaDef{s}[\m{$\zeta$},\{\}],\mmaDef{s}[\m{$\theta$},\{\}]]]
		\mmaDef{m}[X]//\mmaDef{gDecomposeToSubReps}		
		\mmaDef{s}[\m{$\xi$},\{i\}]//\mmaDef{gDecomposeToSubReps}
		\mmaDef{gAConvention}["ToLeft"];
		\mmaDef{dc}[\m{$\xi$},\{i\}]//\mmaDef{gDecomposeToSubReps}			
\end{mmaCell}
\begin{mmaCell}{Output}
		2\mmaSub{X}{(1)}$\otimes$\mmaSub{I}{(2)}$\otimes$\mmaSub{I}{(3)}$\otimes$\mmaSub{Z}{(4)}
\end{mmaCell}
\begin{mmaCell}{Output}
		\mmaSub{$\widetilde{\eta}$}{(1)i}$\otimes$\mmaSub{$\widetilde{\lambda}$}{(2)}$\otimes$\mmaSub{$\widetilde{\zeta}$}{(3)}$\otimes$\mmaSub{$\widetilde{\theta}$}{(4)}	
\end{mmaCell}
\begin{mmaCell}{Output}
		\mmaSub{$\overline{\eta}$}{(1)i}$\otimes$\mmaSub{$\overline{\lambda}$}{(2)}$\otimes$\mmaSub{$\overline{\zeta}$}{(3)}$\otimes$\mmaSup{\mmaSub{$\theta$}{(4)}}{$\dagger$}					
\end{mmaCell}
Custom decompositions for the gamma matrices can be set by using the second form of \mtd{gSetDecomposition}. For example, if we would like to use a different decomposition in the case $\text{Even} \to \text{Odd} \times \text{Odd}$, in which the gamma matrices are given by
\begin{equation}
\begin{split}
\gamma^{\mu} &= \gamma^{\mu}_{(1)} \otimes I_{(2)} \otimes \sigma_2 \text{ for $\mu=0,\dotsc,n-1$},\\
\gamma^{\alpha} &= I_{(1)} \otimes \gamma^{\alpha}_{(2)} \otimes \sigma_3 \text{ for $\alpha=n,\dotsc,d-1$},
\end{split}
\end{equation}
we could do this by doing the following.
\begin{mmaCell}{Input}
		\mmaDef{gSetRep}[\{-1,1,1,1,1,1,1,1\},-1,-1];
		\mmaDef{gSetSubRep}[1,3,1,1];
		\mmaDef{gSetSubRep}[2,5,1,-1];
		\mmaDef{gSetIndices}[1,\{\m{$\mu$}\}];	
		\mmaDef{gSetIndices}[2,\{\m{$\alpha$}\}];	
		\mmaDef{gSetDecomposition}[\mmaDef{y},1,\mmaDef{gTensorProduct}[\mmaDef{y}[\{u[gx\$]\}],\mmaDef{m}[Id],
		  \mmaDef{m}[\m{$\sigma$2}]]]	
		\mmaDef{gSetDecomposition}[\mmaDef{y},2,\mmaDef{gTensorProduct}[\mmaDef{m}[Id],\mmaDef{y}[\{u[gx\$]\}],
		  \mmaDef{m}[\m{$\sigma$3}]]]
		\mmaDef{gDecomposeToSubReps}[\mmaDef{y}[\{u[\m{$\mu$}]\}]]
		\mmaDef{gDecomposeToSubReps}[\mmaDef{y}[\{u[\m{$\alpha$}]\}]]
		\mmaDef{gDecomposeToSubReps}[\mmaDef{y}[\{d[\m{$\alpha$}]\}]]
		\mmaDef{gDecomposeToSubReps}[\mmaDef{y}[\{u[\m{$\mu$},\m{$\alpha$}]\}]]		
\end{mmaCell}
\begin{mmaCell}{Output}
		\mmaSup{\mmaSub{$\gamma$}{(1)}}{$\mu$}$\otimes$\mmaSub{I}{(2)}$\otimes$\mmaSup{\mmaSub{$\sigma$}{(3)}}{2}
\end{mmaCell}
\begin{mmaCell}{Output}
		\mmaSub{I}{(1)}$\otimes$\mmaSup{\mmaSub{$\gamma$}{(2)}}{$\alpha$}$\otimes$\mmaSup{\mmaSub{$\sigma$}{(3)}}{3}
\end{mmaCell}
\begin{mmaCell}{Output}
		\mmaSub{I}{(1)}$\otimes$\mmaSub{\mmaSub{$\gamma$}{(2)}}{$\alpha$}$\otimes$\mmaSup{\mmaSub{$\sigma$}{(3)}}{3}
\end{mmaCell}
\begin{mmaCell}{Output}
		$\Ii$\mmaSup{\mmaSub{$\gamma$}{(1)}}{$\mu$}$\otimes$\mmaSup{\mmaSub{$\gamma$}{(2)}}{$\alpha$}$\otimes$\mmaSup{\mmaSub{$\sigma$}{(3)}}{1}
\end{mmaCell}
Custom decompositions can be deleted by using \mtd{gClearDecompositions}.
\begin{funcDefnN}[label=gClearDecomposition_infobox]{\mtd{gClearDecompositions}}
	\mtd{gClearDecompositions[]}\\
	Clears custom decompositions that have been defined for spinors and matrices by using \mtd{gSetDecompositions}.\\[3pt]
	\textbf{Arguments}\\
	None.
\end{funcDefnN}
\begin{mmaCell}{Input}
		\mmaDef{gClearDecompositions}[];
		\mmaDef{gDecomposeToSubReps}[\mmaDef{y}[\{u[\m{$\alpha$}]\}]]
		\mmaDef{s}[\m{$\xi$},\{i\}]//\mmaDef{gDecomposeToSubReps}	
\end{mmaCell}
\begin{mmaCell}{Output}
		\mmaSub{I}{(2)}$\otimes$\mmaSup{\mmaSub{$\gamma$}{(2)}}{$\alpha$}$\otimes$\mmaSup{\mmaSub{$\sigma$}{(3)}}{2}
\end{mmaCell}
\begin{mmaCell}{Output}
		\mmaSub{$\widetilde{\xi}$}{(1)i}$\otimes$\mmaSub{$\widetilde{\xi}$}{(2)i}$\otimes$\mmaSub{$\widetilde{\xi}$}{(3)i}	
\end{mmaCell}
Apart from decomposing the matrices and spinors with spinor indices into tensor products corresponding to the subalgebras, the function \mtd{gDecomposeToSubReps} also decomposes tensors into parts that transform differently under the subalgebras. For example, we can decompose the expression $F_{MN}F^{MN}$ in terms of parts that transform differently under $\text{SO}(1,2)\times \text{SO}(7)$, as in formula \eqref{review_F_decomposition}. 
\begin{mmaCell}{Input}
		\mmaDef{gSetRep}[\{-1,1,1,1,1,1,1,1,1,1\},1,1];
		\mmaDef{gSetSubRep}[1,3,1,1];
		\mmaDef{gSetSubRep}[2,7,-1,1];		
		\mmaDef{gForm}[F,\{u[M,N]\}]\mmaDef{gForm}[F,\{M,N\}]//\mmaDef{gDecomposeToSubReps}
\end{mmaCell}
\begin{mmaCell}{Output}
		\mmaSub{F}{$\gamma$1,$\gamma$2}\mmaSup{F}{$\gamma$1,$\gamma$2}+2\mmaSub{F}{$\gamma$1,$\sigma$1}\mmaSup{F}{$\gamma$1,$\sigma$1}+\mmaSub{F}{$\sigma$1,$\sigma$2}\mmaSup{F}{$\sigma$1,$\sigma$2}
\end{mmaCell}
Recall that here the indices \mtd{$\gamma$n} are those transforming under the the first SO subalgebra, $\text{SO}(1,2)$, and the indices \mtd{$\sigma$n} transform under $\text{SO}(7)$.
\vspace{-5pt}
\subsection{Using Assumptions}
\vspace{-5pt}
Coordinate dependence of different objects, including tensors, spinors and scalars, and the decomposition properties of tensors can be set using \mtd{g$\Omega$}.
\begin{objDefnN}[label=gOmega_infobox]{\mtd{g$\Omega$}}
	\mtd{g$\Omega$[a1,...,an][x1,...,xm]}\\
	Used in assumptions to denote the decomposition of an antisymmetric or symmetric tensor under subalgebra $\text{SO}(d_1) \times \dotsc \times \text{SO}(d_m) \subset \text{SO}(d)$, and the coordinate dependence of tensors, spinors, and scalars. Specifically, this denotes the set of tensors that have \mtd{a1} indices corresponding to (transforming under) the first subalgebra, \mtd{a2} indices corresponding to the second subalgebra, and so on. The second list denotes the coordinate dependence of a tensor with these indices, specifically, such a tensor can depend on coordinates \mtd{x1},...,\mtd{xm}. \\[3pt]
	\textbf{Arguments}\\
	\mtd{a1,...,an} are numbers of indices in the tensor that transform under different subalgebras. The first, \mtd{a1} denotes the number of indices corresponding to the first subalgebra, the second the number of indices corresponding to the second subalgebra and so on.\\
	\mtd{x1,...,xm} are the coordinates, on which the components of tensors that have \mtd{a1} indices corresponding to the first subalgebra, \mtd{a2} corresponding to the second subalgebra and so on, can depend on. If the first list is left empty, then this applies for every component of the tensor. If this list is left empty, then the corresponding components of the tensor can depend on any coordinate.
\end{objDefnN}
To get an idea of how this works, let us concentrate first on the case where we make no assumptions on the transformation properties of $F$ under the subalgebras. In this case, we leave the first list empty, then the second list just defines the coordinate dependence, so that \mtd{g$\Omega$[][x1,...,xm]} denotes the set of tensors that depend on coordinates \mtd{x1,...,xm}. For example, assuming that a 2-form with components $F_{M,N}$ depends on the coordinates \mtd{x0,x1,x2,x3}, associated to the first subalgebra $SO(1,3)$ in the decomposition $SO(1,9) \to SO(1,3) \times SO(3) \times SO(3)$, is done by using the following command.
\begin{mmaCell}{Input}
		\mmaDef{\mmaDef{gSetRep}}[\{-1,1,1,1,1,1,1,1,1,1\},1,1];		
		\mmaDef{\mmaDef{gSetSubRep}}[1,4,1,1];
		\mmaDef{\mmaDef{gSetSubRep}}[2,3,1,1];
		\mmaDef{\mmaDef{gSetSubRep}}[3,3,1,1];	
		\mmaDef{gSetCoordinates}[\{x0,x1,x2,x3,x4,x5,x6,x7,x8,x9\}];
		\mmaDef{gAddAssumptions}[F$\in$\mmaDef{g}$\Omega$[][x0,x1,x2,x3]];
\end{mmaCell}
Now we can, for example take derivative $\partial_\mu$ of $F$, and define that $\mu$ takes values 4,5 and 6. Then $\mu$ corresponds to directions \mtd{x4,x5} and \mtd{x6}, which are associated to the second subalgebra $SO(3) \subset SO(1,9)$. Thus we see that this derivative is zero. On the other hand, since F can a priori depend on coordinates corresponding to the first direction, $\partial_\alpha F_{MN}$ is not necessarily zero if $\alpha$ denotes an index that corresponds to one of coordinates \mtd{x1,x2,x3} and \mtd{x4}.
\begin{mmaCell}{Input}
		\mmaDef{gSetIndices}[1,\{\m{$\alpha$}\}];
		\mmaDef{gSetIndices}[2,\{\m{$\mu$}\}];
		\mmaDef{gD}[\{\m{$\mu$}\},\mmaDef{gForm}[F,\{M,N\}]]
		\mmaDef{gD}[\{\m{$\alpha$}\},\mmaDef{gForm}[F,\{M,N\}]]		
\end{mmaCell}
\begin{mmaCell}{Output}
		0
\end{mmaCell}
\begin{mmaCell}{Output}
		\mmaSub{d}{$\alpha$}\mmaSub{F}{M,N}
\end{mmaCell}
This usage of \mtd{g$\Omega$} extends also to objects without Lorentz indices, such as spinors and scalars. For example, we can assume that scalar $\Delta$ and spinor $\xi$ depend only on the coordinates \mtd{x4,x5} and \mtd{x6}.
\begin{mmaCell}{Input}
		\mmaDef{gAddAssumptions}[\{\m{$\Delta$}$\in$\mmaDef{g}$\Omega$[][x4,x5,x6],\mmaDef{s}[\m{$\xi$},\{\}]$\in$\mmaDef{g}$\Omega$[][x4,x5,x6]\}];
		\mmaDef{gD}[\{\m{$\alpha$}\},\m{$\Delta$}]
		\mmaDef{gD}[\{\m{$\mu$}\},\m{$\Delta$}]	
		\mmaDef{gD}[\{\m{$\alpha$}\},\mmaDef{s}[\m{$\xi$},\{\}]]				
\end{mmaCell}
\begin{mmaCell}{Output}
		0
\end{mmaCell}
\begin{mmaCell}{Output}
		\mmaSub{d}{$\mu$}$\Delta$
\end{mmaCell}
\begin{mmaCell}{Output}
		0
\end{mmaCell}
The first part of the object \mtd{g$\Omega$} determines the decomposition of the form under the subalgebras. For example, consider again the subalgebra $\text{Cliff}(1,3) \times \text{Cliff}(0,3) \times \text{Cliff}(0,3) \subset \text{Cliff}(1,9)$. If we wish to assume that the form $F$ transforms non-trivially only under the first induced $\text{SO}$ subalgebra, $\text{SO}(1,3)$, then we will need to assume that the form has two indices corresponding to these directions, and no indices corresponding to other directions. Thus, we can do this by using the following.
\begin{mmaCell}{Input}
		\mmaDef{gClearAssumptions}[];	
		\mmaDef{gAddAssumptions}[F$\in$\mmaDef{g}$\Omega$[2,0,0][]];
\end{mmaCell}
This is reflected in the way in which the contraction $F_{MN}F^{MN}$ decomposes to parts that transform differently under the subalgebras.
\begin{mmaCell}{Input}
		\mmaDef{gForm}[F,\{M,N\}]\mmaDef{gForm}[F,\{u[M,N]\}]//\mmaDef{gDecomposeToSubReps}
\end{mmaCell}
\begin{mmaCell}{Output}
		\mmaSub{F}{$\gamma$1,$\gamma$2}\mmaSup{F}{$\gamma$1,$\gamma$2}
\end{mmaCell}
If we wanted to assume that there are terms that contain only indices corresponding to the first subalgebra, but also terms that contain an index corresponding to the first subrepresentation, and another to the second or third subrepresentation, we could do this by adding additional assumptions as follows.
\begin{mmaCell}{Input}
		\mmaDef{gAddAssumptions}[\{F$\in$\mmaDef{g}$\Omega$[1,1,0][],F$\in$\mmaDef{g}$\Omega$[1,0,1][]\}];
\end{mmaCell}
Now these additional terms will also be included when considering the decomposition of $F_{MN}F^{MN}$.
\begin{mmaCell}{Input}
		\mmaDef{gForm}[F,\{M,N\}]\mmaDef{gForm}[F,\{u[M,N]\}]//\mmaDef{gDecomposeToSubReps}
\end{mmaCell}
\begin{mmaCell}{Output}
		2\mmaSub{F}{c1,$\gamma$1}\mmaSup{F}{c1,$\gamma$1}+\mmaSub{F}{$\gamma$1,$\gamma$2}\mmaSup{F}{$\gamma$1,$\gamma$2}+2\mmaSub{F}{$\gamma$1,$\sigma$1}\mmaSup{F}{$\gamma$1,$\sigma$1}
\end{mmaCell}
These two functionalities can also be combined to make assumptions on the coordinate dependence of components transforming differently under the subalgebra. For example, we can assume that, in the previous example, the components transforming only under $\text{SO}(1,3)$ depend only on coordinates \mtd{x1,x2,x3} and \mtd{x4} corresponding to the $\text{SO}(1,3)$ subalgebra, and that the components transforming under both $\text{SO}(1,3)$ and $\text{SO}(3)$ depend on coordinates \mtd{x4,x5} and \mtd{x6}, by entering the following assumption. Note that we do not make any assumptions on the coordinate dependence of the remaining terms, so this part can depend on any coordinate.
\begin{mmaCell}{Input}
		\mmaDef{gClearAssumptions}[];
		\mmaDef{gAddAssumptions}[\{F$\in$\mmaDef{g}$\Omega$[2,0,0][x0,x1,x2,x3],F$\in$\mmaDef{g}$\Omega$[1,1,0][x4,x5,x6],
		  F$\in$\mmaDef{g}$\Omega$[1,0,1][]\}];
\end{mmaCell}
Now we see that, for example, the part transforming only under $\text{SO}(1,3)$ indeed depends only on the coordinates corresponding to that subalgebra, and the other terms have similar properties.
\begin{mmaCell}{Input}
		\mmaDef{gSetIndices}[3,\{a\}]
		\mmaDef{gD}[\{\m{$\alpha$}\},\mmaDef{gForm}[F,\{\m{$\gamma$1},\m{$\gamma$2}\}]]			
		\mmaDef{gD}[\{\m{$\mu$}\},\mmaDef{gForm}[F,\{\m{$\gamma$1},\m{$\gamma$2}\}]]	
		\mmaDef{gD}[\{\m{a}\},\mmaDef{gForm}[F,\{\m{$\gamma$1},\m{$\gamma$2}\}]]
		\mmaDef{gD}[\{\m{$\alpha$}\},\mmaDef{gForm}[F,\{\m{$\gamma$1},\m{$\sigma$1}\}]]	
		\mmaDef{gD}[\{\m{$\mu$}\},\mmaDef{gForm}[F,\{\m{$\gamma$1},\m{$\sigma$1}\}]]	
		\mmaDef{gD}[\{\m{a}\},\mmaDef{gForm}[F,\{\m{$\gamma$1},\m{$\sigma$1}\}]]	
		\mmaDef{gD}[\{1\},\mmaDef{gForm}[F,\{\m{$\gamma$1},\m{$\sigma$1}\}]]	
\end{mmaCell}
\begin{mmaCell}{Output}
		\mmaSub{d}{$\alpha$}\mmaSub{F}{$\gamma$1,$\gamma$2}
\end{mmaCell}
\begin{mmaCell}{Output}
		0
\end{mmaCell}
\begin{mmaCell}{Output}
		0
\end{mmaCell}
\begin{mmaCell}{Output}
		0
\end{mmaCell}
\begin{mmaCell}{Output}
		\mmaSub{d}{$\mu$}\mmaSub{F}{$\gamma$1,$\sigma$1}
\end{mmaCell}
\begin{mmaCell}{Output}
		0
\end{mmaCell}
\begin{mmaCell}{Output}
		0
\end{mmaCell}
\subsection{Explicit Representations}
Explicit representations for the gamma matrices generating the subalgebras can be set by using the function \mtd{gSetSubRepMatrices} that is very similar to \mtd{gSetRepMatrices}.
\begin{funcDefnN}[label=gSetSubRepMatrices_infobox]{\mtd{gSetSubRepMatrices}}
	\mtd{gSetSubRepMatrices[n,gammas,C]}	\\
	Sets an explicit representation for the gamma matrices and the $A$-, $C$- and $B$-intertwiners forming a representation of the \mtd{n}:th subalgebra.\\[3pt]
	\textbf{Arguments}\\
	\mtd{n} is the number of the subrepresentation.\\
	\mtd{gammas} is a list of explicit expressions for gamma matrices  with indices up. The first matrix in the list gives the value for $\gamma^i$ where $i$ is the first index corresponding to the \mtd{n}:th subalgebra, and the following matrices give values to the subsequent gamma matrices.\\
	\mtd{C} is the matrix for $C$-intertwiner.
\end{funcDefnN}
The function \mtd{gSetSubRepMatrices} works in just the same way as \mtd{gSetRepMatrices}. As an example, we consider a Clifford algebra $\text{Cliff}(1,4)$ and its subalgebra $\text{Cliff}(1,2)\times \text{Cliff}(2)$. We first define explicit three- and two-dimensional gamma matrices, and use them as representations of $\text{Cliff}(1,2)$ and $\text{Cliff}(2)$, building a 5-dimensional representation of $\text{Cliff}(1,4)$ out of these. Note that here \mtd{s1}, \mtd{s2} and \mtd{s3} denote sigma matrices as we defined them earlier in section \ref{sect_Explicit_Reps_Gammas_Intertwiners}.
\begin{mmaCell}{Input}
		\mmaDef{gSetRep}[\{-1,1,1,1,1\},1,-1]
		\mmaDef{gSetSubRep}[1,3,1,1]
		\mmaDef{gSetSubRep}[2,2,1,1]
		
		\mmaDef{gSetSubRepMatrices}[1,\{I*\mmaDef{s1},\mmaDef{s2},\mmaDef{s3}\},\mmaDef{s2}];	
		\mmaDef{gSetSubRepMatrices}[2,\{\mmaDef{s1},\mmaDef{s2}\},\mmaDef{s2}];
\end{mmaCell}
We can now, as an example, check that the 5-dimensional representation provided by these matrices is indeed a representation of the Clifford algebra. We do this by first writing the anticommutator $\gamma^{i} \gamma^j + \gamma^j \gamma^i$, then express this in terms of subrepresentations, set this equal to $2 \eta^{ij}$, and take the explicit form of this equality. Then we take a table of all the results and delete duplicates.
\begin{mmaCell}{Input}
		Table[\mmaDef{gExplicit}[\mmaDef{gDecomposeToSubReps}[y[\{u[\mmaFnc{i}]\},\{u[\mmaFnc{j}]\}]+
		 y[\{u[\mmaFnc{j}]\},\{u[\mmaFnc{i}]\}]]==\mmaDef{gSymm}[\m{$\eta$},\{u[\mmaFnc{i},\mmaFnc{j}]\}]\mmaDef{m}[Id]],\{\mmaFnc{i},0,4\},\{\mmaFnc{j},0,4\}]//
		   Flatten//\mmaDef{DeleteDuplicates}
\end{mmaCell}
\begin{mmaCell}{Output}
		\{True,\{\{0,0,0,0\},\{0,0,0,0\},\{0,0,0,0\},\{0,0,0,0\}\}==0\}
\end{mmaCell}
This shows that these matrices indeed form a representation of the five-dimensional Clifford algebra. Note that besides \mtd{True}, there is also an equation in the resulting list. This is because Mathematica differentiates between zero scalar and a zero matrix. For our purposes, however, they are equal.
\vspace{-5pt}
\subsection{Options and Subrepresentations} \label{sect_Subalgebras_Options}
\vspace{-5pt}
For the most part, using assumptions with subrepresentations is no different from using them with no subrepresentations. However, there are a few things that should be taken into consideration when dealing with subrepresentations. First of all, it is possible to define different options for different subrepresentations. For example, if we have two subalgebras, we could make the $A$-intertwiners commute automatically to right for the first subrepresentation, and commute automatically to left for the second subrepresentation.
\begin{mmaCell}{Input}
		\mmaDef{gSetRep}[\{-1,1,1,1,-1,1,1,1\},-1,-1]
		\mmaDef{gSetSubRep}[1,4,1,1]
		\mmaDef{gSetSubRep}[2,4,1,1]
		\mmaDef{gAConvention}[1,"ToRight"];
		\mmaDef{gAConvention}[2,"ToLeft"];
		\mmaDef{gTensorProduct}[\mmaDef{gTimes}[\mmaDef{y}[\{u[\m{$\mu$},\m{$\nu$}]\},1],\mmaDef{m}[A,1],\mmaDef{y}[\{d[\m{$\mu$},\m{$\nu$}]\},1]],
		  \mmaDef{gTimes}[\mmaDef{y}[\{u[\m{$\mu$},\m{$\nu$}]\},2],\mmaDef{m}[A,2],\mmaDef{y}[\{d[\m{$\mu$},\m{$\nu$}]\},2]]]
\end{mmaCell}
\begin{mmaCell}{Output}
		(\mmaSup{\mmaSub{$\gamma$}{(1)}}{$\mu$,$\nu$}$\star$\mmaSup{(\mmaSub{$\gamma$}{(1)$\mu$,$\nu$})}{$\dagger$}$\star$\mmaSub{A}{(1)})$\otimes$(\mmaSub{A}{(2)}$\star$\mmaSup{(\mmaSup{\mmaSub{$\gamma$}{(2)}}{$\mu$,$\nu$})}{$\dagger$}$\star$\mmaSub{$\gamma$}{(2)$\mu$,$\nu$})
\end{mmaCell}
Changing the value of \mtd{gAConvention} for all subrepresentations can be done by leaving the subrepresentation number out of \mtd{gAConvention}. For example, we could make the $A$-intertwiner not to move automatically for both subrepresentations by using the following.
\begin{mmaCell}{Input}
		\mmaDef{gAConvention}["DoNothing"];
		\mmaDef{gTensorProduct}[\mmaDef{gTimes}[\mmaDef{y}[\{u[\m{$\mu$},\m{$\nu$}]\},1],m[A,1],\mmaDef{y}[\{d[\m{$\mu$},\m{$\nu$}]\},1]],
		\mmaDef{gTimes}[\mmaDef{y}[\{u[\m{$\mu$},\m{$\nu$}]\},2],m[A,2],\mmaDef{y}[\{d[\m{$\mu$},\m{$\nu$}]\},2]]]
\end{mmaCell}
\begin{mmaCell}{Output}
		(\mmaSup{\mmaSub{$\gamma$}{(1)}}{$\mu$,$\nu$}$\star$\mmaSub{A}{(1)}$\star$\mmaSub{$\gamma$}{(1)$\mu$,$\nu$})$\otimes$(\mmaSup{\mmaSub{$\gamma$}{(2)}}{$\mu$,$\nu$}$\star$\mmaSub{A}{(2)}$\star$\mmaSub{$\gamma$}{(2)$\mu$,$\nu$})
\end{mmaCell}
\newpage
\section{Other Topics}
\vspace{-5pt}
\subsection{Naming Conventions}
\vspace{-5pt}
Almost all new functions introduced in the package begin with prefix \mtd{g}. This is to avoid any clashes with the names of built-in or user-defined functions. Therefore new functions can still be defined without problems provided that their names do not begin with prefix \mtd{g}. The only exceptions to this naming rule are a few common objects whose names have been kept short in order to keep giving input as easy as possible. Specifically, the functions that do not begin with \mtd{g} are \mtd{y}, \mtd{s}, \mtd{m}, \mtd{dc}, \mtd{lcc} and \mtd{rcc}.

The variables and functions that are used internally begin with the prefix \mtd{\$g}. Users do not usually need to modify or access the internal variables directly, but their values should instead be set through such functions as \mtd{gSetRepMatrices}, and accessed trough other functions, such as \mtd{gExplicit}. The variables that users should modify directly begin, like functions, with prefix \mtd{g}.

Apart from functions and variables, there are other reserved expressions that have a pre-defined significance, and should not therefore be used. For example, the symbols \mtu{u} and \mtu{d} are reserved for denoting upper and lower indices. Variables of the form \mtu{$\lambda$} and \mtu{$\lambda$x} (where \mtu{x} is any other symbol) are reserved for sum dummy indices that are automatically generated by functions such as \mtd{gOrd} or \mtd{gSimplify}, unless this has been modified using \mtd{gSumIndexIndicator}. In order to avoid irregularities when simplifying expressions, users should avoid using indices beginning with \mtu{$\lambda$}, if they are not dummy indices.

There are also a few other expressions with pre-defined special meaning. For example \mtd{gSymm[\mtu{$\eta$},\{\mtu{$\mu$},\mtu{$\nu$}\}]} denotes the metric $\eta_{\mu \nu}$, \mtd{gForm[\mtu{$\epsilon$},\{\mtu{$\mu$1},...,\mtu{$\mu$n}\}]} denotes the Levi-Civita symbol, and \mtd{m[\mtu{A}]}, \mtd{m[\mtu{B}]}, and \mtd{m[C]} denote the intertwiners. Therefore, to ensure that these work properly, \mtu{$\eta$}, \mtu{$\epsilon$} \mtu{A}, \mtu{B}, and \mtd{C} should not be used as variables. For the full list of reserved expressions like this, see appendix \ref{App_Reserved_Expressions}.
\vspace{-5pt}
\subsection{Limitations and Known Issues}
\vspace{-5pt}
Here we list some of the limitations of the program, and some known issues with implementations. We hope to be able to fix many of the issues listed here in future.
\begin{itemize}
	\item While most of the output notation can be also used as input, sometimes copying output and using it as a new input causes errors.
	
	\item There are some checks on inputs but many functions do not give any warnings, when given faulty input, and therefore may give unexpected output if the input is even slightly wrong.
	
	\item There is no symbolic notation for all combinations of up- and down indices for tensors without special symmetry properties. Similarly, there is no symbolic notation for products of more than five antisymmetrised products of gamma matrices.
	
	\item Unlike \mtd{gForm}, \mtd{gSymm} and \mtd{gTensor}, \mtd{y} and \mtd{gBL} do not have default positioning of indices, so every index appearing in these objects must be inside of either \mtd{u} or \mtd{d}.
\end{itemize}
\vspace{-5pt}
\subsection{Bug Reports and Feature Requests}
\vspace{-5pt}
Bug reports or requests for new features can be either sent using the bug report feature on GitHub, or alternatively sent directly to the author via email to  \href{mailto:pyry.r.kuusela@gmail.com}{\mtd{pyry.r.kuusela@gmail.com}}. When reporting bugs, please include a brief description of the problem, and if possible instructions how it can be reproduced, or a short Mathematica notebook illustrating the problem. This ensures that the bug can be fixed as soon as possible.

\newpage
\section{Examples} \label{sect_Examples}
\vspace{-5pt}
In this section, we give a few more involved examples of computations that can be performed with the package. This will both serve as a non-trivial demonstration of the properties of objects and functions presented in the previous sections, and also shows how these objects can be used in an optimal fashion when performing calculations that appear often in practical applications of gamma matrices and Clifford algebras.
While we will present the examples here for completeness, although it may be more convenient to use the accompanying Mathematica file \mtd{GammaMaP\_examples.nb} to inspect the code used in these examples.
\vspace{-5pt}
\subsection{Torsion Conditions for Supersymmetric SUGRA Compactification}
\vspace{-5pt}
In this example we reproduce the torsion conditions for AdS$_3\times \mathcal{M}_7$ compactifications of IIB supergravity, preserving at least (0,2), following \cite{Couzens:2017nnr}. In particular, we will reproduce the equations (2.14)-(2.21) of that paper. The basic idea is to start from 10-dimensional supersymmetry conditions, the use the decomposition $\text{SO}(1,9) \to \text{SO}(1,2) \times \text{SO}(7)$ to reduce the equations to 7 dimensions. We can use the 7-dimensional spinors parametrising the supersymmetry transformations to form bilinears. The supersymmetry conditions can then be expressed equivalently as conditions on the bilinears obtained in this fashion. It can then be shown that these in turn impose conditions on so-called torsion tensor. In the process we will demonstrate how to use some features related to dimensional reduction, and how to use derivatives, bilinears and relations involving spinors.

We start with the 10-dimensional IIB supergravity, whose NS-NS sector of consists of a real scalar $\phi$, called dilaton, the metric $g$, and a real two-form $B^{(2)}$. The RR sector contains a real scalar and a two-form, $C^{(0)}$ and $C^{(2)}$, and a real four-form $C^{(4)}$ with a self-dual field strength $F^{(5)}$. For the purposes of this example, we will assume that only the scalars, the metric and the four-form are non-zero. 

The scalars can be combined into a single field, $\tau$, called axio-dilaton
\begin{align}
\tau = \tau_1 + i \tau_2 \equiv C^{(0)} + i e^{- \phi},
\end{align}
which can then be used to define a one-form field $P$, and a derivative that is covariant with respect to both Lorentz and $U(1)$ transformations.
\begin{equation}
\begin{split}
P &=\frac{i}{2 \tau_{2}} d \tau,\\
\mathcal{D}_\mu &= \nabla_\mu + i q \frac{1}{2 \tau_{2}} d \tau_{1}.
\end{split}
\end{equation}
We are looking for supersymmetric solutions, so we must require that the supersymmetry variations of the fermionic fields vanish. The supersymmetry transformations of fermionic fields are, in our case, given by
\begin{align}
\delta\psi_{M}&=\mathcal{D}_{M}\epsilon+\frac{i}{192}\Gamma^{P_{1}\dotsc P_{4}}F_{MP_{1}\dotsc P_{4}}^{(5)}\epsilon=0~,\label{susy1}\\[3pt]
\delta \lambda&= i \Gamma^{M}P_{M}\epsilon^{c}=0\label{susy2}.
\end{align}
Here $\epsilon$ is the type IIB supersymmetry parameter, i.e. a Weyl spinor satisfying $\Gamma_{11}\epsilon = - \epsilon$.

For the purposes of compactification, we take the metric to be
\begin{align}
\text{d}s^2 = e^{2 \Delta} (\text{ds}^2(\text{AdS}_3) + \text{ds}^2(\mathcal{M}_7)),
\end{align}
where $\mathcal{M}_7$ is some internal 7-dimensional manifold. As discussed in \cite{Couzens:2017way}, the appropriate spinor decomposition in this case is given by
\begin{align}
\epsilon = \eta \otimes e^{\Delta/2} \xi \otimes \theta,
\end{align}
where $\theta$ satisfies $\sigma_3 \theta = \theta$\footnote{Note that we use different decomposition for gamma matrices, which is the origin of a different sign here.} and is purely imaginary, and $\eta$ satisfies the Killing spinor equation on $\text{AdS}_3$.
\begin{align}
\nabla_a \eta = \frac{\alpha n}{2} \gamma_{(1)a} \eta, \label{Killing_spinor_eq_AdS3}
\end{align}
where $\alpha= \pm 1$, and $n$ is a real parameter. 
\vspace{-5pt}
\subsubsection{Reducing the 10-dimensional Supersymmetry Conditions to 7-dimensions}
\vspace{-5pt}
We begin by defining the full Clifford algebra to be $\text{Cliff}(1,9)$, and the two subalgebras to be $\text{Cliff}(1,2)$ and $\text{Cliff}(7)$. The coordinates on the underlying manifold are defined to be \mtd{x0},$\dotsc$,\mtd{x9}. In addition, we wish to keep in mind the explicit representations for subalgebras defined in \cite{Couzens:2017way}, so we need to use $\zeta=-1$ for the first subrepresentation.
\begin{mmaCell}{Input}
		\mmaDef{gSetRep}[\{-1,1,1,1,1,1,1,1,1,1\},1,1]
		\mmaDef{gSetSubRep}[1,3,1,1,-1]
		\mmaDef{gSetSubRep}[1,7,-1,1]
		\mmaDef{gSetCoordinates}[\{x0,x1,x2,x3,x4,x5,x6,x7,x8,x9\}]		
\end{mmaCell}
After this, we define the indices \mtd{$\alpha$} and \mtd{$\beta$} to correspond to the first subalgebra, and  the indices \mtd{$\mu$}, \mtd{$\nu$}, and \mtd{$\rho$} to the second. In addition, we define the sum indices corresponding to the first representation to be of the form \mtd{$\delta$n}, and those corresponding to the second representation to be of the form \mtd{$\sigma$n}. 
\begin{mmaCell}{Input}
		\mmaDef{gSetIndices}[1,\{\m{$\alpha$},\m{$\beta$}\}]
		\mmaDef{gSetIndices}[2,\{\m{$\mu$},\m{$\nu$},\m{$\rho$}\}]	
		\mmaDef{gSumIndexIndicator}[1,"$\delta$"]
		\mmaDef{gSumIndexIndicator}[2,"$\sigma$"]	
\end{mmaCell}
At this point, it is convenient to define the decomposition of the supersymmetry parameter $\epsilon$. As discussed above, we take the decomposition to be
\begin{align}
\epsilon = \eta \otimes e^{\Delta/2} \xi \otimes \theta.
\end{align}
\begin{mmaCell}{Input}
		\mmaDef{gClearDecompositions}[];
		\mmaDef{gSetDecomposition}[\mmaDef{s}[\m{$\epsilon$},\{\}],\mmaDef{gTensorProduct}[\mmaDef{s}[\m{$\eta$},\{\}],
		  Exp[\m{$\Delta$}/2]*\mmaDef{s}[\m{$\xi$},\{\}],\mmaDef{s}[\m{$\theta$},\{\}]]]
\end{mmaCell}
Imposing that $\sigma_3 \theta = \theta$ can be done, for instance, with the following function definition. Alternative ways are, among others, using rules like we will do for components of $F^{(5)}$ below.
\begin{mmaCell}{Input}
		\mmaDef{gTimes}[\mmaPat{a\_\_\_},\mmaDef{m}[\m{$\sigma$1},3],\mmaDef{s}[\m{$\theta$},\{\},3],\mmaPat{b\_\_\_}]:=\mmaDef{gTimes}[\mmaPat{a},\mmaDef{s}[\m{$\theta$},\{\},3],\mmaPat{b}]
\end{mmaCell}
Then we move on to making assumptions about the fluxes, spinors, and scalars appearing in the computation. The warp factor $\Delta$ is a real scalar that can only depend on the coordinates on the internal space $\mathcal{M}_7$. The five-form flux $F_{(5)}$ has two parts, as discussed above. One has indices only along $\mathcal{M}_7$, and the other has three indices along $\text{AdS}_3$, and two indices along $\mathcal{M}_7$. These both parts can only depend on coordinates on $\mathcal{M}_7$. The flux $P$ has only indices along $\mathcal{M}_7$, and depends only on coordinates on $\mathcal{M}_7$. The spinor $\eta$ appearing in the decomposition of $\epsilon$ depends only on coordinates along $\text{AdS}_3$, $\xi$ depends only on coordinates along $\mathcal{M}_7$, and $\theta$ is a purely imaginary constant. 
\begin{mmaCell}{Input}
		\mmaDef{gClearAssumptions}[];
		\mmaDef{gAddAssumptions}[\{\m{$\Delta$}$\in$Reals,\m{$\Delta$}$\in$\mmaDef{g}$\Omega$[][x3,x4,x5,x6,x7,x8,x9],F5$\in$Reals,
		  F5$\in$\mmaDef{g}$\Omega$[0,5][x3,x4,x5,x6,x7,x8,x9],
		  F5$\in$\mmaDef{g}$\Omega$[3,2][x3,x4,x5,x6,x7,x8,x9],
		  P$\in$\mmaDef{g}$\Omega$[0,1][x3,x4,x5,x6,x7,x8,x9],\mmaDef{s}[\m{$\eta$},\{\},1]$\in$\mmaDef{g}$\Omega$[][x0,x1,x2],
		  \mmaDef{s}[\m{$\xi$},\{\},2]$\in$\mmaDef{g}$\Omega$[][x3,x4,x5,x6,x7,x8,x9],\mmaDef{s}[\m{$\theta$},\{\},3]$\in$Constants,
		  \mmaDef{s}[\m{$\theta$},\{\},3]$\in$gImaginaries\}];
\end{mmaCell}
We need also to make $\eta$ satisfy the Killing spinor equation \eqref{Killing_spinor_eq_AdS3}. In what follows, we will denote the covariant derivative on $\eta$ simply by \mtd{gD}, so that we can use the following line to make $\eta$ automatically satisfy the Killing spinor equation.
\begin{mmaCell}{Input}
		\mmaDef{gClearDerivative}[];
		\mmaDef{gD}[\{u[],d[\mmaPat{a\_}]\},\mmaDef{s}[\m{$\eta$},\{\},1]]:=n*\m{$\alpha_1$}/2*\mmaDef{gTimes}[\mmaDef{y}[\{d[\mmaPat{a}]\},1],\mmaDef{s}[\m{$\eta$},\{\},1]]
\end{mmaCell}
After this, we are almost ready to begin the proper computation. First, we just make sure that both \mtd{gBConvention} and \mtd{gCConvention} are set to \mtd{"ToRight"}, as that will be most convenient for us in what follows.
\begin{mmaCell}{Input}
		\mmaDef{gBConvention}["ToRight"];
		\mmaDef{gCConvention}["ToRight"];
\end{mmaCell}
Even though we will not need the reduced dilatino equation for deriving the torsion conditions, we will still derive it as a simple example. Since there are no free indices, there is only one case we need to consider, so deriving the reduced conditions can be done almost entirely automatically. We just input the dilatino variation, and use \mtd{gDecomposeToSubReps}.
\begin{mmaCell}{Input}
		I*Exp[\m{$\Delta$}/2]*\mmaDef{gForm}[P,\{P1\}]*\mmaDef{gTimes}[\mmaDef{y}[\{u[P1]\}],\mmaDef{rcc}[\m{$\epsilon$},\{\}]]//
		  \mmaDef{gDecomposeToSubReps}
\end{mmaCell}
\begin{mmaCell}{Output}
		\mmaSup{e}{$\Delta$/2}\mmaSub{P}{$\sigma$1}\mmaSup{\mmaSub{$\widetilde{\eta}$}{(1)}}{c}$\otimes$(\mmaSup{\mmaSub{$\gamma$}{(2)}}{$\sigma$1}$\star$\mmaSup{\mmaSub{$\widetilde{\xi}$}{(2)}}{c})$\otimes$\mmaSub{$\widetilde{\theta}$}{(3)}
\end{mmaCell}		
The condition that this is zero is clearly equivalent to
\begin{align}
P_{\mu}\gamma^\mu \xi^c = 0,
\end{align}
which is precisely the reduced dilatino supersymmetry condition.

Then we move on to the reduced gravitino variations \eqref{susy1}. Let us first consider the case in which the index $M$ takes values corresponding to the first subalgebra. We can denote these values by $\alpha$ due to the definitions we have made previously. For clarity, let us consider the expression for the gravitino variation term-by-term, and put everything together only in the end. 

One of the few things in this calculation that we cannot do automatically using this package is to derive the expression for the covariant derivative $D_\alpha$ in the local frame on $\text{AdS}_3$. The difficulty here is the warp factor $e^{2\Delta}$, which corresponds to a Weyl transformation of the metric. Including this warp factor can, however, be easily done by hand, and after a short calculation, we find that in terms of the local frame
\begin{align}
D_{\hat{\alpha}} \epsilon = (\nabla_{\alpha} + \frac{1}{2} \partial_\mu \Delta \Gamma_\alpha^{\phantom{\alpha}\mu}) \epsilon,
\end{align}
where $\hat{\alpha}$ is a curved index, and $\alpha$ is the corresponding flat (local frame) index. Thus the first term in the gravitino variation can be decomposed as follows.
\begin{mmaCell}{Input}
		\mmaDef{gD}[\{\m{$\alpha$}\},\mmaDef{s}[\m{$\epsilon$},\{\}]]+1/2*\mmaDef{gD}[\{\m{$\mu$}\},$\Delta$]\mmaDef{gTimes}[\mmaDef{y}[\{d[\m{$\alpha$}],u[\m{$\mu$}]\}],\mmaDef{s}[\m{$\epsilon$},\{\}]]//
		  \mmaDef{gDecomposeToSubReps}
\end{mmaCell}
\begin{mmaCell}{Output}
		-\mmaFrac{1}{2}$\Ii$\mmaSup{e}{$\Delta$/2}(\mmaSub{d}{$\sigma$1}$\Delta$)(\mmaSub{$\gamma$}{(1)$\alpha$}$\star$\mmaSub{$\widetilde{\eta}$}{(1)})$\otimes$(\mmaSup{\mmaSub{$\gamma$}{(2)}}{$\sigma$1}\mmaSub{$\widetilde{\xi}$}{(2)})$\otimes$\mmaSub{$\widetilde{\theta}$}{(3)}+\mmaFrac{1}{2}\mmaSup{e}{$\Delta$/2}n$\alpha$(\mmaSub{$\gamma$}{(1)$\alpha$}$\star$\mmaSub{$\widetilde{\eta}$}{(1)})$\otimes$\mmaSub{$\widetilde{\xi}$}{(2)}$\otimes$\mmaSub{$\widetilde{\theta}$}{(3)}
\end{mmaCell}
We can then move on to the second term. We first decompose the 10-dimensional term. Here we have also multiplied $F_{(5)}$ by $e^{4\Delta}$, which takes into account the Weyl transformation.
\begin{mmaCell}{Input}
		I/192*\mmaDef{gForm}[F5,\{\m{$\alpha$},P1,P2,P3,P4\}]*\mmaDef{gTimes}[\mmaDef{y}[\{u[P1,P2,P3,P4]\}],
		  \mmaDef{s}[\m{$\epsilon$},\{\}]]//\mmaDef{gDecomposeToSubReps}
\end{mmaCell}
\begin{mmaCell}{Output}
		\mmaFrac{1}{32}$\Ii$\mmaSup{e}{9/2$\Delta$}\mmaSub{F5}{$\alpha$,$\delta$1,$\delta$2,$\sigma$1,$\sigma$2}(\mmaSup{\mmaSub{$\gamma$}{(1)}}{$\delta$1,$\delta$2}$\star$\mmaSub{$\widetilde{\eta}$}{(1)})$\otimes$(\mmaSup{\mmaSub{$\gamma$}{(2)}}{$\sigma$1,$\sigma$2}$\star$\mmaSub{$\widetilde{\xi}$}{(2)})$\otimes$\mmaSub{$\widetilde{\theta}$}{(3)}
\end{mmaCell}
At this point we must recall that $F_{(5)}$ is given by $(1+*)\text{dVol}(\text{AdS}_3)\wedge F$, which means, in terms of components, that $F_{(5)\alpha,\delta_1,\delta_2,\sigma_1,\sigma_2}=\epsilon_{\alpha \delta_1 \delta_2} F_{\sigma_1 \sigma_2}$.
We therefore substitute this form to the expression above.
\begin{mmaCell}{Input}
		\%/.\{\mmaDef{gForm}[F5,\{u[],d[\m{$\alpha$},\m{$\delta$1},\m{$\delta$2},\m{$\sigma$1},\m{$\sigma$2}]\}]->
		  \mmaDef{gForm}[\m{$\epsilon$},\{\m{$\alpha$},\m{$\delta$1},\m{$\delta$2}\}]*\mmaDef{gForm}[F,\{\m{$\sigma$1},\m{$\sigma$2}\}]\}
\end{mmaCell}
\begin{mmaCell}{Output}
		\mmaFrac{1}{32}$\Ii$\mmaSup{e}{9/2$\Delta$}\mmaSub{F}{$\sigma$1,$\sigma$2}\mmaSub{$\epsilon$}{$\alpha$,$\delta$1,$\delta$2}(\mmaSup{\mmaSub{$\gamma$}{(1)}}{$\delta$1,$\delta$2}$\star$\mmaSub{$\widetilde{\eta}$}{(1)})$\otimes$(\mmaSup{\mmaSub{$\gamma$}{(2)}}{$\sigma$1,$\sigma$2}$\star$\mmaSub{$\widetilde{\xi}$}{(2)})$\otimes$\mmaSub{$\widetilde{\theta}$}{(3)}
\end{mmaCell}
Here we recognise the familiar $\epsilon_{\alpha \delta_1 \delta_2} \gamma^{\delta_1 \delta_2}$ that looks like the dual element of $\gamma_\alpha$, possibly up to some prefactor. We can easily verify this, and find the prefactor by using \mtd{gDual}.
\begin{mmaCell}{Input}
		\mmaDef{gDual}[\mmaDef{y}[\{d[\m{$\alpha$}]\},1]]
\end{mmaCell}
\begin{mmaCell}{Output}
		-\mmaFrac{1}{2}\mmaSub{$\epsilon$}{$\alpha$,$\delta$1,$\delta$2}\mmaSup{\mmaSub{$\gamma$}{(1)}}{$\delta$1,$\delta$2}
\end{mmaCell}
We can easily solve this for $\epsilon_{\alpha \delta_1 \delta_2} \gamma^{\delta_1 \delta_2}$, and then we substitute the resulting expression to the previous form for the reduced gravitino variation by using the following rule.
\begin{mmaCell}{Input}
		\%\%/.\{\mmaDef{gForm}[\m{$\epsilon$},\{\m{$\alpha$},\m{$\delta$1},\m{$\delta$2}\}]->1,\mmaDef{y}[\{u[\m{$\delta$1},\m{$\delta$2}],d[]\}]->2*\mmaDef{y}[\{d[\m{$\alpha$}]\},1]\}
\end{mmaCell}
\begin{mmaCell}{Output}
		-\mmaFrac{1}{16}$\Ii$\mmaSup{e}{9/2$\Delta$}\mmaSub{F}{$\sigma$1,$\sigma$2}(\mmaSub{$\gamma$}{(1)$\alpha$}$\star$\mmaSub{$\widetilde{\eta}$}{(1)})$\otimes$(\mmaSup{\mmaSub{$\gamma$}{(2)}}{$\sigma$1,$\sigma$2}$\star$\mmaSub{$\widetilde{\xi}$}{(2)})$\otimes$\mmaSub{$\widetilde{\theta}$}{(3)}
\end{mmaCell}
We are now ready to put everything we have done so far together. By noting that every term is of the form $\gamma_{(1)\alpha} \eta \otimes (\dotsc) \xi_{(2)} \otimes \theta_{(3)}$, we easily see that the gravitino supersymmetry condition reduces in this case to
\begin{align}
\left(\frac{1}{2} \slashed{\partial} \Delta - i \frac{m \alpha}{2} + \frac{e^{4 \Delta}}{8} \slashed{F} \right) \xi = 0, \label{gravitino_variation_alpha}
\end{align}
where we have denoted $1/2 \gamma^{\mu \nu} F_{\mu \nu} \equiv \slashed{F}$.

Then, we can move on to the case, where the index $M$ in the gravitino variation \eqref{susy1} takes values $\mu$, corresponding to $\mathcal{M}_7$. The computation itself proceeds along similar lines as the previous one. We start again with the covariant derivative $\mathcal{D}_\mu$ in the local frame, which is given by
\begin{align}
\mathcal{D}_{\hat{\mu}} \epsilon =(\mathcal{D}_\mu + \frac{1}{2} \partial_\nu \Delta \Gamma_\mu^{\phantom{\mu} \nu})\epsilon,
\end{align}
where $\hat{\mu}$ is again a curved index, and $\mu$ is the flat one. Therefore the first term can be calculated by using the following.
\begin{mmaCell}{Input}
		\mmaDef{gD}[\{\m{$\mu$}\},\mmaDef{s}[\m{$\epsilon$},\{\}]]+1/2*\mmaDef{gD}[\{\m{$\nu$}\},$\Delta$]\mmaDef{gTimes}[\mmaDef{y}[\{d[\m{$\mu$}],u[\m{$\nu$}]\}],\mmaDef{s}[\m{$\epsilon$},\{\}]]//
		  \mmaDef{gDecomposeToSubReps}
\end{mmaCell}
\begin{mmaCell}{Output}
		\mmaSup{e}{$\Delta$/2}\mmaSub{$\widetilde{\eta}$}{(1)}$\otimes$(\mmaSub{d}{$\mu$}\mmaSub{$\widetilde{\xi}$}{(2)})$\otimes$\mmaSub{$\widetilde{\theta}$}{(3)}+\mmaFrac{1}{2}\mmaSup{e}{$\Delta$/2}(\mmaSub{d}{$\mu$}$\Delta$)\mmaSub{$\widetilde{\eta}$}{(1)}$\otimes$\mmaSub{$\widetilde{\xi}$}{(2)}$\otimes$\mmaSub{$\widetilde{\theta}$}{(3)}-
		  \mmaFrac{1}{2}\mmaSup{e}{$\Delta$/2}(\mmaSub{d}{$\sigma$1}$\Delta$)\mmaSub{$\widetilde{\eta}$}{(1)}$\otimes$(\mmaSub{\mmaSup{\mmaSub{$\gamma$}{(2)}}{$\sigma$1}}{$\mu$}$\star$\mmaSub{$\widetilde{\xi}$}{(2)})$\otimes$\mmaSub{$\widetilde{\theta}$}{(3)}
\end{mmaCell}
Then, we can again write out the second term, and decompose it.
\begin{mmaCell}{Input}
		I/192*\mmaDef{gForm}[F5,\{\m{$\mu$},P1,P2,P3,P4\}]*\mmaDef{gTimes}[\mmaDef{y}[\{u[P1,P2,P3,P4]\}],
		\mmaDef{s}[\m{$\epsilon$},\{\}]]//\mmaDef{gDecomposeToSubReps}
\end{mmaCell}
\begin{mmaCell}{Output}
		\mmaFrac{1}{48}\mmaSup{e}{9/2$\Delta$}\mmaSub{F5}{$\delta$1,$\delta$2,$\delta$3,$\mu$,$\sigma$1}(\mmaSup{\mmaSub{$\gamma$}{(1)}}{$\delta$1,$\delta$2,$\delta$3}$\star$\mmaSub{$\widetilde{\eta}$}{(1)})$\otimes$(\mmaSup{\mmaSub{$\gamma$}{(2)}}{$\sigma$1}$\star$\mmaSub{$\widetilde{\xi}$}{(2)})$\otimes$\mmaSub{$\widetilde{\theta}$}{(3)}+
		  \mmaFrac{1}{192}$\Ii$\mmaSup{e}{9/2$\Delta$}\mmaSub{F5}{$\mu$,$\sigma$1,$\sigma$2,$\sigma$3,$\sigma$4}\mmaSub{$\widetilde{\eta}$}{(1)}$\otimes$(\mmaSup{\mmaSub{$\gamma$}{(2)}}{$\sigma$1,$\sigma$2,$\sigma$3,$\sigma$4}$\star$\mmaSub{$\widetilde{\xi}$}{(2)})$\otimes$\mmaSub{$\widetilde{\theta}$}{(3)}
\end{mmaCell}
We need to once again use the facts we know about the form of $F^{(5)}$ to be able to simplify this expression. This time both terms in $(1+*)\text{dVol}(\text{AdS}_3)\wedge F^{(2)}$ contribute. In the above result, $F^{(5)}_{\delta_1, \delta_2, \delta_3 \mu \sigma_1}$ appearing in the first term can again be written as
\begin{align}
F^{(5)}_{\delta_1, \delta_2, \delta_3 \mu \sigma_1} = \epsilon_{\delta_1, \delta_2, \delta_3 } F^{(2)}_{\mu \sigma_1}.
\end{align}
The second term $F^{(5)}_{\mu \sigma_1 \sigma_2 \sigma_3 \sigma_4}$ comes from $*\text{dVol}(\text{AdS}_3)\wedge F^{(2)} = *_7 F^{(2)}$, and can thus be written as
\begin{align}
F^{(5)}_{\mu \sigma_1 \sigma_2 \sigma_3 \sigma_4} = \frac{1}{2} \epsilon_{\sigma_1 \sigma_2 \sigma_3 \sigma_4 \mu}^{\phantom{\sigma_1 \sigma_2 \sigma_3 \sigma_4 \mu} \nu \rho} F^{(2)}_{\nu \rho}.
\end{align}
We can substitute these to the result of the last computation by using suitable rules.
\begin{mmaCell}{Input}
		\% /.\{\mmaDef{gForm}[F5,\{\m{$\mu$},\m{$\sigma$1},\m{$\sigma$2},\m{$\sigma$3},\m{$\sigma$4}\}]->
		  1/2*\mmaDef{gForm}[\m{$\epsilon$},\{d[\m{$\sigma$1},\m{$\sigma$2},\m{$\sigma$3},\m{$\sigma$4},\m{$\mu$}],u[\m{$\nu$},\m{$\rho$}]\}]*gForm[F,\{\m{$\nu$},\m{$\rho$}\}],
		  \mmaDef{gForm}[F5,\{\m{$\delta$1},\m{$\delta$2},\m{$\delta$3},\m{$\mu$},\m{$\sigma$1}\}]->
		  \mmaDef{gForm}[\m{$\epsilon$},\{\m{$\delta$1},\m{$\delta$2},\m{$\delta$3}\}]*\mmaDef{gForm}[F,\{\m{$\mu$},\m{$\sigma$1}\}]\}
\end{mmaCell}
\begin{mmaCell}{Output}
		\mmaFrac{1}{48}\mmaSup{e}{9/2$\Delta$}\mmaSub{F}{$\mu$,$\sigma$1}\mmaSub{$\epsilon$}{$\delta$1,$\delta$2,$\delta$3}(\mmaSup{\mmaSub{$\gamma$}{(1)}}{$\delta$1,$\delta$2,$\delta$3}$\star$\mmaSub{$\widetilde{\eta}$}{(1)})$\otimes$(\mmaSup{\mmaSub{$\gamma$}{(2)}}{$\sigma$1}$\star$\mmaSub{$\widetilde{\xi}$}{(2)})$\otimes$\mmaSub{$\widetilde{\theta}$}{(3)}+
		\mmaFrac{1}{384}$\Ii$\mmaSup{e}{9/2$\Delta$}\mmaSub{F}{$\nu$,$\rho$}\mmaSub{\mmaSup{$\epsilon$}{$\nu$,$\rho$}}{$\mu$,$\sigma$1,$\sigma$2,$\sigma3$,$\sigma$4}\mmaSub{$\widetilde{\eta}$}{(1)}$\otimes$(\mmaSup{\mmaSub{$\gamma$}{(2)}}{$\sigma$1,$\sigma$2,$\sigma$3,$\sigma$4}$\star$\mmaSub{$\widetilde{\xi}$}{(2)})$\otimes$\mmaSub{$\widetilde{\theta}$}{(3)}
\end{mmaCell}
Here we can again recognise that the contractions of the Levi-Civita symbol and gamma matrices look like those in formula \eqref{Basis_dual_odd}. In fact, we have that
\begin{equation}
\begin{split}
\epsilon_{\delta_1 \delta_2 \delta_3} \gamma^{\delta_1 \delta_2 \delta_3} &= 6 \, I,\\
\epsilon_{\sigma_1 \sigma_2 \sigma_3 \sigma_4 \mu}^{\phantom{\sigma_1 \sigma_2 \sigma_3 \sigma_4 \mu} \nu \rho} \gamma^{\sigma_1 \sigma_2 \sigma_3 \sigma_4} &= 24i \gamma_{\mu}^{\phantom{\mu} \nu \rho}.
\end{split}
\end{equation}
This is again easy to check by using \mtd{gDual}.
\begin{mmaCell}{Input}
		\mmaDef{gDual}[y[\{d[\m{$\mu$}],u[\m{$\nu$,$\rho$}]\},2]]
\end{mmaCell}
\begin{mmaCell}{Output}
		-\mmaFrac{1}{24}$\Ii$\mmaSub{\mmaSup{$\epsilon$}{$\nu$,$\rho$}}{$\sigma$1,$\sigma$2,$\sigma$3,$\sigma$4,$\mu$}\mmaSup{\mmaSub{$\gamma$}{(2)}}{$\sigma$1,$\sigma$2,$\sigma$3,$\sigma$4}
\end{mmaCell}
We can make these substitutions, once again, by using rules.
\begin{mmaCell}{Input}
		\%\% /.\{\mmaDef{gForm}[\m{$\epsilon$},\{u[\m{$\mu$},\m{$\nu$},\m{$\rho$}],d[\m{$\sigma$1},\m{$\sigma$2},\m{$\sigma$3},\m{$\sigma$4}],\}]->1,
		\mmaDef{gForm}[\m{$\epsilon$},\{\m{$\delta$1},\m{$\delta$2},\m{$\delta$3}\}]->1,
		\mmaDef{y}[\{u[\m{$\delta$1},\m{$\delta$2},\m{$\delta$3}]\},1]->6*\mmaDef{m}[Id,1],
		\mmaDef{y}[\{u[\m{$\sigma$1},\m{$\sigma$2},\m{$\sigma$3},\m{$\sigma$4}]\},2]->24*I*\mmaDef{y}[\{d[\m{$\mu$}],u[\m{$\nu$},\m{$\rho$}]\}]\}
\end{mmaCell}
\begin{mmaCell}{Output}
		\mmaFrac{1}{8}\mmaSup{e}{9/2$\Delta$}\mmaSub{F}{$\mu$,$\sigma$1}\mmaSub{$\widetilde{\eta}$}{(1)}$\otimes$(\mmaSup{\mmaSub{$\gamma$}{(2)}}{$\sigma$1}$\star$\mmaSub{$\widetilde{\xi}$}{(2)})$\otimes$\mmaSub{$\widetilde{\theta}$}{(3)}-\mmaFrac{1}{16}\mmaSup{e}{9/2$\Delta$}\mmaSub{F}{$\nu$,$\rho$}\mmaSub{$\widetilde{\eta}$}{(1)}$\otimes$(\mmaSub{\mmaSup{\mmaSub{$\gamma$}{(2)}}{$\nu$,$\rho$}}{$\mu$}$\star$\mmaSub{$\widetilde{\xi}$}{(2)})$\otimes$\mmaSub{$\widetilde{\theta}$}{(3)}
\end{mmaCell}
We can now put everything together. By looking at this result and the one derived before, we see that now all the terms appearing in the supersymmetry variation of the gravitino are of the form $\eta \otimes (\dotsc) \xi \otimes \theta$. Collecting these terms, we find the reduced supersymmetry condition.
\begin{align}
\left(\mathcal{D}_\mu + \frac{1}{2} \partial_\mu \Delta - \frac{1}{2} \partial_\nu \Delta \gamma^{\nu}_{\phantom{\nu} \mu} - \frac{e^{-4 \Delta}}{16} F_{\nu \rho} \gamma_{\mu}^{\phantom{\mu} \nu \rho} + \frac{e^{-4 \Delta}}{8} F_{\mu \nu} \gamma^{\nu}\right)\xi = 0. \label{gravitino_variation_mu_unsimplified}
\end{align}
This is not, however, the simplest form that we are after. We can still multiply the  condition \eqref{gravitino_variation_alpha} by $\gamma_\mu$, and add this to the equation above, which leads to a simpler form. To do this, it is useful to switch to using the 7-dimensional representation, since we do not need to deal with the spinors $\eta$ and $\theta$ any more.
\begin{mmaCell}{Input}
		\mmaDef{gSetRep}[\{1,1,1,1,1,1,1\},-1,1]
\end{mmaCell}
Then we store the relations \eqref{gravitino_variation_alpha} and \eqref{gravitino_variation_mu_unsimplified} that we have derived above to variables \mtd{rel1} and \mtd{rel2}. To make manipulation of these expressions easier, we do not include the left-hand side $=0$ in these relations, and just remember that these are supposed to be equal to zero.
\begin{mmaCell}{Input}
		rel1=1/2*\mmaDef{gD}[\{\m{$\rho$}\},\m{$\Delta$}]*\mmaDef{gTimes}[\mmaDef{y}[\{u[\m{$\rho$}]\}],\mmaDef{s}[\m{$\xi$},\{\}]]-I*\m{$\alpha$}*n/2+
		  Exp[-4*\m{$\Delta$}]/16*\mmaDef{gForm}[F,\{\m{$\nu$},\m{$\rho$}\}]*\mmaDef{gTimes}[\mmaDef{y}[\{u[\m{$\nu$},\m{$\rho$}]\}],\mmaDef{s}[\m{$\xi$},\{\}]];
		rel2=\mmaDef{gD}[\{\m{$\mu$}\},\mmaDef{s}[\m{$\xi$},\{\}]]+1/2*\mmaDef{gD}[\{\m{$\mu$}\},\m{$\Delta$}]*\mmaDef{s}[\m{$\xi$},\{\}]-
		  1/2*\mmaDef{gD}[\{\m{$\nu$}\},\m{$\Delta$}]*\mmaDef{gTimes}[\mmaDef{y}[\{u[\m{$\rho$}],d[\m{$\mu$}]\}],\mmaDef{s}[\m{$\xi$},\{\}]]-
		  Exp[-4*\m{$\Delta$}]/16*\mmaDef{gForm}[F,\{\m{$\nu$},\m{$\rho$}\}]*\mmaDef{gTimes}[\mmaDef{y}[\{d[\m{$\mu$}],u[\m{$\nu$},\m{$\rho$}]\}],\mmaDef{s}[\m{$\xi$},\{\}]]+
		  Exp[-4*\m{$\Delta$}]/8*\mmaDef{gForm}[F,\{\m{$\mu$},\m{$\nu$}\}]*\mmaDef{gTimes}[\mmaDef{y}[\{u[\m{$\nu$}]\}],\mmaDef{s}[\m{$\xi$},\{\}]];
\end{mmaCell}
To derive a more useful form of the second relation, we can multiply the first relation by $\gamma_\mu$, and then subtract this from the second relation. 
\begin{mmaCell}{Input}
		\mmaDef{rel2}-\mmaDef{gTimes}[\mmaDef{y}[\{d[\m{$\mu$}]\}],\mmaDef{rel1}]//\mmaDef{gOrd}//\mmaDef{gSimplify}
\end{mmaCell}
\begin{mmaCell}{Output}
		\mmaSub{d}{$\mu$}$\widetilde{\xi}$-\mmaFrac{1}{8}\mmaSup{e}{-4$\Delta$}\mmaSub{F}{$\lambda$1,$\lambda$2}(\mmaSub{\mmaSup{\mmaSub{$\gamma$}{(2)}}{$\lambda$1,$\lambda$2}}{$\mu$}$\star$\mmaSub{$\widetilde{\xi}$}{(2)})+\mmaFrac{1}{2}$\Ii$n$\alpha$\mmaSub{$\gamma$}{$\mu$}
\end{mmaCell}
This is the simple form that we were after.
\begin{align}
\left(\mathcal{D}_\mu + \frac{i n \alpha}{2} \gamma_{\mu}- \frac{e^{-4 \Delta}}{8} F_{\nu \rho} \gamma_{\mu}^{\phantom{\mu} \nu \rho} \right)\xi = 0. \label{gravitino_variation_mu}
\end{align}
\vspace{-5pt}
\subsubsection{Deriving the 7-dimensional Torsion Conditions}
\vspace{-5pt}
Since we are now making computations with spinors that used to form a part of a decomposition of the original spinor $\epsilon$, we need to switch to using commutative spinors.
\begin{mmaCell}{Input}
		\mmaDef{gSetRep}[\{1,1,1,1,1,1,1\},-1,1]
		\mmaDef{gSpinorType}["Commutative"]
\end{mmaCell}
Then we need to make a few assumptions about the scalars and forms appearing in the following computations. We know that $\Delta$, $m$, and $\alpha$ are all real scalars, from which obviously follows that the exterior derivative form $\text{d} \Delta$ is also real. $F$ is a real two-form.
\begin{mmaCell}{Input}
		\mmaDef{gClearAssumptions}[];
		\mmaDef{gAddAssumptions}[\{\m{$\Delta$}$\in$Reals, m$\in$Reals, \m{$\alpha$}$\in$Reals, d\m{$\Delta$}$\in$Reals, F$\in$Reals\}];
\end{mmaCell}
We are now ready to input the relations \eqref{gravitino_variation_alpha} and \eqref{gravitino_variation_mu}. We will leave out =0 from both relations for computational simplicity, and store the relations to variables \mtd{reld1} and \mtd{relD1}, respectively.
\begin{mmaCell}{Input}
		reld1=1/2*\mmaDef{gD}[\{\m{$\mu$}\},\m{$\Delta$}]*\mmaDef{gTimes}[\mmaDef{y}[\{u[\m{$\mu$}]\}],\mmaDef{s}[\m{$\xi$},\{\}]]-
		  1/2*I*\m{$\alpha$}*n*s[\m{$\xi$},\{\}]+
		  Exp[-4\m{$\Delta$}]/16*\mmaDef{gForm}[F,\{\m{$\mu$},\m{$\nu$}\}]*\mmaDef{gTimes}[\mmaDef{y}[\{u[\m{$\mu$},\m{$\nu$}]\},\mmaDef{s}[\m{$\xi$},\{\}]]]
		  
		relD1=\mmaDef{gD}[\{\m{$\mu$}\},\mmaDef{s}[\m{$\xi$},\{\}]]+1/2*I*\m{$\alpha$}*n*\mmaDef{gTimes}[y[\{d[\m{$\mu$}]\}],\mmaDef{s}[\m{$\xi$},\{\}]]-
		  Exp[-4\m{$\Delta$}]/8*\mmaDef{gForm}[F,\{\m{$\nu$},\m{$\sigma$}\}]*\mmaDef{gTimes}[y[\{d[\m{$\mu$}],u[\m{$\nu$},\m{$\sigma$}]\}],\mmaDef{s}[\m{$\xi$},\{\}]]
\end{mmaCell}
\begin{mmaCell}{Output}
		\mmaFrac{1}{2}(\mmaSub{d}{$\mu$}$\Delta$)(\mmaSup{$\gamma$}{$\mu$}$\star$$\widetilde{\xi}$)+\mmaFrac{1}{16}\mmaSup{e}{-4$\Delta$}\mmaSub{F}{$\mu$,$\nu$}(\mmaSup{$\gamma$}{$\mu$,$\nu$}$\star$$\widetilde{\xi}$)-\mmaFrac{1}{2}$\Ii$n$\widetilde{\xi}$$\alpha$
\end{mmaCell}
\begin{mmaCell}{Output}
		\mmaSub{d}{$\mu$}$\widetilde{\xi}$+\mmaFrac{1}{2}$\Ii$n(\mmaSub{$\gamma$}{$\mu$}$\star$$\widetilde{\xi}$)$\alpha$-\mmaFrac{1}{8}\mmaSup{e}{-4$\Delta$}\mmaSub{F}{$\nu$,$\sigma$}(\mmaSub{\mmaSup{$\gamma$}{$\nu$,$\sigma$}}{$\mu$}$\star$$\widetilde{\xi}$)
\end{mmaCell}
These relations imply a few additional relations. We can derive these by taking Hermitian conjugates of the relations above, by taking complex conjugates of them, and multiplying the relations by $B$-intertwiner. We store these relations to new variables. Note that in order to get the nicest possible forms in each case, we have to modify the \mtd{gBConvention} to commute the $B$-matrices to right or left, depending on which results in the most useful form.
\begin{mmaCell}{Input}
		reld2=\mmaDef{gH}[\mmaDef{reld1}]
		relD2=\mmaDef{gH}[\mmaDef{relD1}]
		\mmaDef{gBConvention}["ToRight"];
		reld3=\mmaDef{gTimes}[\mmaDef{m}[\mmaDef{B}],\mmaDef{Conjugate}[\mmaDef{reld1}]]
		relD3=\mmaDef{gTimes}[\mmaDef{m}[\mmaDef{B}],\mmaDef{Conjugate}[\mmaDef{relD1}]]
		\mmaDef{gBConvention}["ToLeft"];
		reld4=\mmaDef{gH}[reld3]
		relD4=\mmaDef{gH}[relD3]
\end{mmaCell}
\begin{mmaCell}{Output}
		\mmaFrac{1}{2}$\Ii$n$\bar{\xi}$$\alpha$+\mmaFrac{1}{2}(\mmaSub{d}{$\mu$}$\Delta$)($\bar{\xi}$$\star$\mmaSup{$\gamma$}{$\mu$})-\mmaFrac{1}{16}\mmaSup{e}{-4$\Delta$}\mmaSub{F}{$\mu$,$\nu$}($\bar{\xi}$$\star$\mmaSup{$\gamma$}{$\mu$,$\nu$})
\end{mmaCell}
\begin{mmaCell}{Output}
		\mmaSub{d}{$\mu$}$\bar{\xi}$-\mmaFrac{1}{2}$\Ii$n($\bar{\xi}$$\star$\mmaSub{$\gamma$}{$\mu$})$\alpha$+\mmaFrac{1}{8}\mmaSup{e}{-4$\Delta$}\mmaSub{F}{$\nu$,$\sigma$}($\bar{\xi}$$\star$\mmaSub{\mmaSup{$\gamma$}{$\nu$,$\sigma$}}{$\mu$})
\end{mmaCell}
\begin{mmaCell}{Output}
		-\mmaFrac{1}{2}(\mmaSub{d}{$\mu$}$\Delta$)(\mmaSup{$\gamma$}{$\mu$}$\star$\mmaSup{$\widetilde{\xi}$}{c})+\mmaFrac{1}{16}\mmaSup{e}{-4$\Delta$}\mmaSub{F}{$\mu$,$\nu$}(\mmaSup{$\gamma$}{$\mu$,$\nu$}$\star$\mmaSup{$\widetilde{\xi}$}{c})+\mmaFrac{1}{2}$\Ii$n$\alpha$\mmaSup{$\widetilde{\xi}$}{c}
\end{mmaCell}
\begin{mmaCell}{Output}
		B$\star$\mmaSup{(\mmaSub{d}{$\mu$}$\widetilde{\xi}$)}{*}+\mmaFrac{1}{2}$\Ii$n$\alpha$(\mmaSub{$\gamma$}{$\mu$}$\star$\mmaSup{$\widetilde{\xi}$}{c})+\mmaFrac{1}{8}\mmaSup{e}{-4$\Delta$}\mmaSub{F}{$\nu$,$\sigma$}(\mmaSub{\mmaSup{$\gamma$}{$\nu$,$\sigma$}}{$\mu$}$\star$\mmaSup{$\widetilde{\xi}$}{c})
\end{mmaCell}
\begin{mmaCell}{Output}
		-\mmaFrac{1}{2}(\mmaSub{d}{$\mu$}$\Delta$)(\mmaSup{$\bar{\xi}$}{c}$\star$\mmaSup{$\gamma$}{$\mu$})-\mmaFrac{1}{16}\mmaSup{e}{-4$\Delta$}\mmaSub{F}{$\mu$,$\nu$}(\mmaSup{$\bar{\xi}$}{c}$\star$\mmaSup{$\gamma$}{$\mu$,$\nu$})-\mmaFrac{1}{2}$\Ii$n$\alpha$\mmaSup{$\bar{\xi}$}{c}
\end{mmaCell}
\begin{mmaCell}{Output}
		(\mmaSub{d}{$\mu$}\mmaSup{$\xi$}{T})$\star$B-\mmaFrac{1}{2}$\Ii$n$\alpha$(\mmaSup{\mmaSub{$\bar{\xi}$}{i}}{c}$\star$\mmaSub{$\gamma$}{$\mu$})-\mmaFrac{1}{8}\mmaSup{e}{-4$\Delta$}\mmaSub{F}{$\nu$,$\sigma$}(\mmaSup{$\bar{\xi}$}{c}$\star$\mmaSub{\mmaSup{$\gamma$}{$\nu$,$\sigma$}}{$\mu$})
\end{mmaCell}
Having exhausted all additional relations, we can now read the derivatives of spinors $\xi_i$, $\bar{\xi}_i$, and $\bar{\xi}_i^c$ from the relations containing the covariant derivative $\mathcal{D}$. We store the derivatives of these spinors into variables \mtd{rightDerivative}, \mtd{leftDerivative}, and \mtd{leftDerivativecc}, respectively.
\begin{mmaCell}{Input}
		\mmaDef{gD}[\{u[],d[\mmaPat{a\_}]\},\mmaDef{s}[\m{$\xi$},\{\},0]]:=-\mmaFrac{1}{2}$\Ii$n\m{$\alpha$}(\mmaSub{\m{$\gamma$}}{\mmaPat{a}}$\star$\m{$\widetilde{\xi}$})+\mmaFrac{1}{8}\mmaSup{\mmaDef{e}}{-4\m{$\Delta$}}\mmaSub{F}{\m{$\nu$},\m{$\sigma$}}(\mmaSub{\mmaSup{\m{$\gamma$}}{\m{$\nu$},\m{$\sigma$}}}{\mmaPat{a}}$\star$\m{$\widetilde{\xi}$});
		\mmaDef{gD}[\{u[],d[\mmaPat{a\_}]\},\mmaDef{dc}[\m{$\xi$},\{\},0]]:=\mmaFrac{1}{2}$\Ii$n\m{$\alpha$}(\m{$\bar{\xi}$}$\star$\mmaSub{\m{$\gamma$}}{\mmaPat{a}})-\mmaFrac{1}{8}\mmaSup{\mmaDef{e}}{-4\m{$\Delta$}}\mmaSub{F}{\m{$\nu$},\m{$\sigma$}}(\m{$\bar{\xi}$}$\star$\mmaSub{\mmaSup{\m{$\gamma$}}{\m{$\nu$},\m{$\sigma$}}}{\mmaPat{a}});
		\mmaDef{gD}[\{u[],d[\mmaPat{a\_}]\},\mmaDef{lcc}[\m{$\xi$},\{\},0]]:=\mmaFrac{1}{2}$\Ii$n\m{$\alpha$}(\mmaSup{\m{$\bar{\xi}$}}{c}$\star$\mmaSub{\m{$\gamma$}}{\mmaPat{a}})+\mmaFrac{1}{8}\mmaSup{\mmaDef{e}}{-4\m{$\Delta$}}\mmaSub{F}{\m{$\nu$},\m{$\sigma$}}(\mmaSup{\m{$\bar{\xi}$}}{c}$\star$\mmaSub{\mmaSup{\m{$\gamma$}}{\m{$\nu$},\m{$\sigma$}}}{\mmaPat{a}})
\end{mmaCell}
As a last step before proceeding to a proper computation, we set names for different bilinears. As mentioned in section \ref{sect_bilinears}, the bilinears of type $\bar{\xi}_i \gamma^{\mu_1\dotsc\mu_n} \xi_j$ are in the first list, beginning from scalar, and ending in 7-form. Similarly, the bilinears of the form $\bar{\xi}_i^c \gamma^{\mu_1\dotsc\mu_n} \xi_j$ are in the second list, again beginning from the scalar and ending in the 7-form.
\begin{mmaCell}{Input}
		\mmaDef{gSetBilinearNames}[\{S,K,U,X,Z,C,T,O\},\{A,B,V,Y,Q,J,H,L\},\{\},\{\},\m{$\xi$}]
\end{mmaCell}
\vspace{-5pt}
\subsubsection{Scalar Bilinear S}
\vspace{-5pt}
We can begin the proper computations with the computation of the exterior derivative of the scalar bilinear. It can be straightforwardly shown that when doing the computation, we can just replace the ordinary derivatives d with the covariant derivatives $\mathcal{D}$, and therefore we can use for the derivatives the expressions we have derived above. The exterior derivative of the scalar $S_{i,i}=\bar{\xi}_i \xi_i$ is given simply by $\text{d}S= \text{d}(\bar{\xi}_i \xi_i) = \mathcal{D}_\mu(\bar{\xi}_i \xi_i) e^\mu$, which we can easily calculate. We store the result in variable \mtd{dS}.
\begin{mmaCell}{Input}
		dS=\mmaDef{gD}[\{\m{$\mu$}\},\mmaDef{gTimes}[\mmaDef{dc}[\m{$\xi$},\{\}],\mmaDef{s}[\m{$\xi$},\{\}]]]*\mmaDef{gForm}[e,\{u[\m{$\mu$}]\}]
\end{mmaCell}
\begin{mmaCell}{Output}
		0
\end{mmaCell}
This is already enough to prove the first result, (2.15), of \cite{Couzens:2017nnr}.
\vspace{-5pt}
\subsubsection{One-form K}
\vspace{-5pt}
The calculation of the exterior derivative of the one-form starts similarly to the scalar case. Now we have just to remember to include the gamma matrix appearing in the definition of $K$. Recalling that we work in the orthonormal frame, we do not need to take derivative of the gamma matrix, as it is constant, and therefore the exterior derivative of $K$ is simply given by $\text{d}K= \mathcal{D}_\mu(\bar{\xi}_i \gamma_a \xi_i) e^{\mu a}$
\begin{mmaCell}{Input}
		dK=\mmaDef{gD}[\{\m{$\mu$}\},\mmaDef{gTimes}[\mmaDef{dc}[\m{$\xi$},\{\}],\mmaDef{y}[\{d[a]\}],\mmaDef{s}[\m{$\xi$},\{\}]]]*\mmaDef{gForm}[e,\{u[\m{$\mu$},a]\}]
\end{mmaCell}
\begin{mmaCell}{Output}
		-\mmaSup{e}{a,$\mu$}$\big($-\mmaFrac{1}{2}$\Ii$n$\alpha$($\bar{\xi}$$\star$\mmaSub{$\gamma$}{a}\mmaSub{$\gamma$}{$\mu$}$\star$$\widetilde{\xi}$)+\mmaFrac{1}{8}\mmaSup{\mmaDef{e}}{-4$\Delta$}\mmaSub{F}{$\nu$,$\sigma$}($\bar{\xi}$$\star$(\mmaSub{$\gamma$}{a}\mmaSub{\mmaSup{$\gamma$}{$\nu$,$\sigma$}}{$\mu$})$\star$$\widetilde{\xi}$)+
		  \mmaFrac{1}{2}$\Ii$n$\alpha$($\bar{\xi}$$\star$\mmaSub{$\gamma$}{$\mu$}\mmaSub{$\gamma$}{a}$\star$$\widetilde{\xi}$)-\mmaFrac{1}{8}\mmaSup{e}{-4$\Delta$}\mmaSub{F}{$\nu$,$\sigma$}($\bar{\xi}$$\star$\mmaSub{\mmaSup{$\gamma$}{$\nu$,$\sigma$}}{$\mu$}\mmaSub{$\gamma$}{a}$\star$$\widetilde{\xi}$)$\big)$
\end{mmaCell}
Now we can use \mtd{gOrd} to express the gamma matrix products in the standard form, and then \mtd{gProductToBL} to get the expression to a nicer form. It is also useful to use \mtd{gSimplify} to simplify the expression a bit further. We store the result back in the variable \mtd{dK}.
\begin{mmaCell}{Input}
		\mmaDef{dK}=\mmaDef{dK}//\mmaDef{gOrd}//\mmaDef{gSimplify}
\end{mmaCell}
\begin{mmaCell}{Output}
		$\Ii$n$\alpha$\mmaSup{\mmaSub{U}{:}}{$\lambda$1,$\lambda$2}\mmaSub{e}{$\lambda$1,$\lambda$2}-\mmaFrac{1}{4}\mmaSup{\mmaDef{e}}{-4$\Delta$}\mmaSup{\mmaSub{Z}{:}}{$\lambda$1,$\lambda$2,$\lambda$3,$\lambda$4}\mmaSub{e}{$\lambda$1,$\lambda$2}\mmaSub{F}{$\lambda$3,$\lambda$4}
\end{mmaCell}
This time, we are not yet ready, since we have to include the effect of $e^{-4\Delta}$ in the calculation. As can be easily seen, the effect of including this factor is adding an extra term of $4 \text{d}\Delta \wedge K$. To calculate this term, we use the relations \mtd{reld1}-\mtd{reld4} that we have derived earlier. We multiply the expressions \mtd{reld1} and \mtd{reld2} by a suitable combination of gamma matrices and spinors that we have chosen so that we will, in the end, get an expression for $4 \text{d}\Delta \wedge K$. At first we get two different linear combinations that are both equal to zero.
\begin{mmaCell}{Input}
		reldrK=\mmaDef{gTimes}[\mmaDef{dc}[\m{$\xi$},\{i\}],\mmaDef{y}[\{u[a,b]\}],\mmaDef{reld1}]
		reldlK=\mmaDef{gTimes}[\mmaDef{reld2},\mmaDef{y}[\{u[a,b]\}],\mmaDef{s}[\m{$\xi$},\{i\}]]
\end{mmaCell}
\begin{mmaCell}{Output}
		-\mmaFrac{1}{2}$\Ii$n$\alpha$($\bar{\xi}$$\star$\mmaSup{$\gamma$}{a,b}$\star$$\widetilde{\xi}$)+\mmaFrac{1}{2}(\mmaSub{d}{$\mu$}$\Delta$)($\bar{\xi}$$\star$(\mmaSup{$\gamma$}{a,b}\mmaSup{$\gamma$}{$\mu$})$\star$$\widetilde{\xi}$)+\mmaFrac{1}{16}\mmaSup{e}{-4$\Delta$}\mmaSub{F}{$\mu$,$\nu$}($\bar{\xi}$$\star$(\mmaSup{$\gamma$}{a,b}\mmaSup{$\gamma$}{$\mu$,$\nu$})$\star$$\widetilde{\xi}$)
\end{mmaCell}
\begin{mmaCell}{Output}
		\mmaFrac{1}{2}$\Ii$n$\alpha$($\bar{\xi}$$\star$\mmaSup{$\gamma$}{a,b}$\star$$\widetilde{\xi}$)+\mmaFrac{1}{2}(\mmaSub{d}{$\mu$}$\Delta$)($\bar{\xi}$$\star$(\mmaSup{$\gamma$}{a,b}\mmaSup{$\gamma$}{$\mu$})$\star$$\widetilde{\xi}$)-\mmaFrac{1}{16}\mmaSup{e}{-4$\Delta$}\mmaSub{F}{$\mu$,$\nu$}($\bar{\xi}$$\star$(\mmaSup{$\gamma$}{$\mu$,$\nu$}\mmaSup{$\gamma$}{a,b})$\star$$\widetilde{\xi}$)
\end{mmaCell}
We can then take the difference of the two expressions, contract the result with the vielbein $e_{a,b}=e_a \wedge e_b$ that we can represent here by \mtd{gForm} (since it is antisymmetric in indices $a$ and $b$), and then use \mtd{gOrd} to simplify the gamma matrix products. Finally, \mtd{gProductToBL} expresses the bilinears in the \mtd{gBL} notation. To get the simplest possible form, we also use \mtd{gSimplify}.
\begin{mmaCell}{Input}
		(\mmaDef{reldrK}-\mmaDef{reldlK})*\mmaDef{gForm}[e,\{d[a,b]\}]//\mmaDef{gOrd}//\mmaDef{gProductToBL}//
		  \mmaDef{gSimplify}
\end{mmaCell}
\begin{mmaCell}{Output}
		-$\Ii$n$\alpha$\mmaSup{\mmaSub{U}{:}}{$\lambda$1,$\lambda$2}\mmaSub{e}{$\lambda$1,$\lambda$2}-2\mmaSup{\mmaSub{K}{:}}{$\lambda$2}(\mmaSub{d}{$\lambda$1}$\Delta$)\mmaSub{\mmaSup{e}{$\lambda$1}}{$\lambda$2}+\mmaFrac{1}{8}\mmaSup{e}{-4$\Delta$}\mmaSup{\mmaSub{Z}{:}}{$\lambda$1,$\lambda$2,$\lambda$3,$\lambda$4}\mmaSub{e}{$\lambda$1,$\lambda$2}\mmaSub{F}{$\lambda$3,$\lambda$4}-
		  \mmaFrac{1}{4}\mmaSup{e}{-4$\Delta$}S\mmaSub{e}{$\lambda$1,$\lambda$2}\mmaSup{F}{$\lambda$1,$\lambda$2}
\end{mmaCell}
Here, we recognise the first term as $-2\text{d}\Delta \wedge K$. Recalling that the whole expression is equal to zero, we can solve for $\text{d}\Delta \wedge K$, and store the result in a variable called \mtd{d$\Delta$K}.
\begin{mmaCell}{Input}
		d\m{$\Delta$K}=1/2*(\%+2*\mmaDef{gBL}[K,\{\},\{u[\m{$\lambda$2}]\}]*\mmaDef{gD}[\{d[\m{$\lambda$1}]\},\m{$\Delta$}]*
		  \mmaDef{gForm}[e,\{u[\m{$\lambda$1}],d[\m{$\lambda$2}]\}])//\mmaDef{Expand}
\end{mmaCell}
\begin{mmaCell}{Output}
		-\mmaFrac{1}{2}$\Ii$n$\alpha$\mmaSup{\mmaSub{U}{:}}{$\lambda$1,$\lambda$2}\mmaSub{e}{$\lambda$1,$\lambda$2}+\mmaFrac{1}{16}\mmaSup{e}{-4$\Delta$}\mmaSup{\mmaSub{Z}{:}}{$\lambda$1,$\lambda$2,$\lambda$3,$\lambda$4}\mmaSub{e}{$\lambda$1,$\lambda$2}\mmaSub{F}{$\lambda$3,$\lambda$4}-\mmaFrac{1}{8}\mmaSup{e}{-4$\Delta$}\mmaSub{S}{}\mmaSub{e}{$\lambda$1,$\lambda$2}\mmaSup{F}{$\lambda$1,$\lambda$2}
\end{mmaCell}
Now we are ready to compute the quantity $e^{-4\Delta}\text{d}(e^{4\Delta}K)$ that we are interested in.
\begin{mmaCell}{Input}
		\mmaDef{dK}+4\mmaDef{d}$\Delta$\mmaDef{K}
\end{mmaCell}
\begin{mmaCell}{Output}
		-$\Ii$n$\alpha$\mmaSup{\mmaSub{U}{i,i:}}{$\lambda$1,$\lambda$2}\mmaSub{e}{$\lambda$1,$\lambda$2}-\mmaFrac{1}{2}\mmaSup{e}{-4$\Delta$}\mmaSub{S}{i,i:}\mmaSub{e}{$\lambda$1,$\lambda$2}\mmaSup{F}{$\lambda$1,$\lambda$2}
\end{mmaCell}
This is exactly the result (2.16) once we recall that in \cite{Couzens:2017nnr} $S=1$ and $\alpha=1$.
\vspace{-5pt}
\subsubsection{Higher Forms}
\vspace{-5pt}
Computing the supersymmetry conditions for the higher forms proceeds in exactly analogous way to the two computations presented above. For this reason we do not include them here. However, they can be found on the file \mtd{GammaMaP\_examples.nb} that comes with the package.
\vspace{-5pt}
\subsection{Supersymmetric Yang-Mills Theory}
\vspace{-5pt}
In this section, we prove a couple of well-known properties of the supersymmetric Yang-Mills theories. Specifically, we first show that the 10-dimensional Yang-Mills action is invariant under supersymmetry transformations, and then we consider supersymmetry transformations of dimensionally reduced $\mathcal{N}=1$ $D=4$ SYM, and show that the supersymmetry transformations form the correct supersymmetry algebra. In the process, we will see examples of how to handle derivatives, Fierz identities, and different types of spinors.
\vspace{-5pt}
\subsubsection{Invariance of 10D SYM-action Under SUSY Transformations}
\vspace{-5pt}
The Lagrangian of supersymmetric Yang-Mills theory in 10 dimensions is given by \cite{Brink:1976bc}.
\begin{align}
 \mathcal{L} = \frac{1}{2 g^2} \text{Tr}(F_{\mu \nu}F^{\mu \nu}) - \frac{i}{2} \text{Tr}(\overline{\psi} \slashed{D} \psi),
\end{align}
where the field strength $F_{\mu \nu}$ is given by the usual expression $F_{\mu \nu} = \partial_\mu A_\nu - \partial_\nu A_\mu - i [A_\mu, A_\nu]$, and both the gauge field $A_\mu$, and the spinor $\psi$ transform under the adjoint representation of the gauge group. The supersymmetry transformations are given by.
\begin{equation}
\begin{split}
\delta_\epsilon A_\mu = i \overline{\epsilon} \gamma_\mu \psi,\\[3pt]
\delta_\epsilon \psi = \frac{1}{2} F_{\mu \nu} \gamma^{\mu \nu} \epsilon.
\end{split}
\end{equation}
There is a slight complication arising from the fact that for a generic gauge group the fields $A_\mu$ and $\psi$ do not necessarily commute. In Mathematica, there is limited support for non-commutative expressions, so it is perhaps easiest to consider the Abelian case first, and only later amend it to include the terms which arise from taking non-commutativity into account. In principle some techniques, like non-commutative multiplication \mtd{NonCommutativeMultiply} could be used to partially circumvent the problems, but here we just calculate the non-commutative parts by hand, since these computations will be relatively easy, and deal with the rest of the computations in Mathematica. 

We begin our computations, as usual, by defining the relevant representation, and the properties of the spinors and tensors. We work in the 10-dimensional Lorentzian signature. The spinor $\psi$ is Majorana-Weyl (we can take its chirality to be positive), and the field strength and the gauge field, $F_{\mu \nu}$ and $A_\mu$, are both real. Also, the spinor $\epsilon$ parametrising the supersymmetry transformations is Majorana-Weyl and constant.
\begin{mmaCell}{Input}
		\mmaDef{gSetRep}[\{-1,1,1,1,1,1,1,1,1,1\},1,1];
		\mmaDef{gAddAssumptions}[\{\mmaDef{s}[\m{$\psi$},\{\}]$\in$gPositiveChiral,\mmaDef{s}[\m{$\psi$},\{\}]$\in$gMajorana,
		  \mmaDef{s}[\m{$\epsilon$},\{\}]$\in$gPositiveChiral,\mmaDef{s}[\m{$\epsilon$},\{\}]$\in$gMajorana,\mmaDef{s}[\m{$\epsilon$},\{\}]$\in$Constants,
		  F$\in$Reals,A$\in$Reals\}];
\end{mmaCell}
After this, we can define a derivative $\delta_\epsilon$ representing the supersymmetry transformations. We cheat a little and treat $\epsilon$ as if it were a Lorentz index, since this does not matter here.
\begin{mmaCell}{Input}
		\mmaDef{gD}[\m{$\delta$},\{u[],d[\m{$\epsilon$}]\},\mmaDef{gForm}[A,\{u[],d[\mmaPat{a\_}]\}]]:=
		  I*\mmaDef{gTimes}[\mmaDef{s}[\m{$\epsilon$},\{\}],y[\{d[\mmaPat{a}]\}],\mmaDef{s}[\m{$\psi$},\{\}]]
		\mmaDef{gD}[\m{$\delta$},\{u[],d[\m{$\epsilon$}]\},\mmaDef{gForm}[A,\{u[\mmaPat{a\_}],d[]\}]]:=
		  I*\mmaDef{gTimes}[\mmaDef{s}[\m{$\epsilon$},\{\}],y[\{u[\mmaPat{a}]\}],\mmaDef{s}[\m{$\psi$},\{\}]]
				
		\mmaDef{gD}[\m{$\delta$},\{u[],d[\m{$\epsilon$}]\},\mmaDef{s}[\m{$\psi$},\{\},0]]:=
		  I*\mmaDef{gForm}[F,\{a,b\}]\mmaDef{gTimes}[y[\{u[a,b]\}],\mmaDef{s}[\m{$\epsilon$},\{\}]]	
\end{mmaCell}
It is convenient to use both $F_{\mu \nu}$ and its expression in terms of $A_\mu$. Therefore we define a set of rules expressing $F_{\mu \nu}$ in terms of $A_\mu$. This has to be done for all the possible position of indices.
\begin{mmaCell}{Input}
		frule=\{\mmaDef{gForm}[F,\{u[],d[\mmaPat{a\_},\mmaPat{b\_}]\}]:>
		\mmaDef{gD}[\{\mmaPat{a}\},\mmaDef{gForm}[A,\{\mmaPat{b}\}]]-\mmaDef{gD}[\{\mmaPat{b}\},\mmaDef{gForm}[A,\{\mmaPat{a}\}]],
		\mmaDef{gForm}[F,\{u[\mmaPat{a\_},\mmaPat{b\_}],d[]\}]:>
		\mmaDef{gD}[\{u[\mmaPat{a}]\},\mmaDef{gForm}[A,\{u[\mmaPat{b}]\}]]-\mmaDef{gD}[\{u[\mmaPat{b}]\},\mmaDef{gForm}[A,\{u[\mmaPat{a}]\}]],
		\mmaDef{gForm}[F,\{u[\mmaPat{a\_}],d[\mmaPat{b\_}]\}]:>
		\mmaDef{gD}[\{u[\mmaPat{a}]\},\mmaDef{gForm}[A,\{\mmaPat{b}\}]]-\mmaDef{gD}[\{\mmaPat{b}\},\mmaDef{gForm}[A,\{u[\mmaPat{a}]\}]]\};
\end{mmaCell}
We can now begin computing the supersymmetry transformations. Let us first look at the variation of the Dirac term $\frac{i}{2}\overline{\psi} \slashed{D} \psi$.
\begin{mmaCell}{Input}
		\mmaDef{gD}[\m{$\delta$},\{\m{$\epsilon$}\},I/2*\mmaDef{gTimes}[\mmaDef{dc}[\m{$\psi$},\{\}],\mmaDef{y}[\{u[\m{$\mu$}]\}],\mmaDef{gD}[\{\m{$\mu$}\},\mmaDef{s}[\m{$\psi$}],\{\}]]]//\mmaDef{Expand}
\end{mmaCell}
\begin{mmaCell}{Output}
		-\mmaFrac{1}{4}$\Ii$\mmaSub{F}{a,b}($\overline{\epsilon}$$\star$(\mmaSup{$\gamma$}{a,b}\mmaSup{$\gamma$}{$\mu$})$\star$(\mmaSub{d}{$\mu$}$\widetilde{\psi}$))+\mmaFrac{1}{4}$\Ii$(\mmaSub{d}{$\mu$}\mmaSub{F}{a,b})($\overline{\psi}$$\star$(\mmaSup{$\gamma$}{$\mu$}\mmaSup{$\gamma$}{a,b})$\star$$\widetilde{\epsilon}$)
\end{mmaCell}
Here we can immediately see that it might be useful to add an total derivative $\frac{i}{4} \partial_\mu(F_{a,b} \overline{\epsilon}\gamma^{a,b}\gamma^\mu \psi)$ to the expression. After doing this, we simplify the expression.
\begin{mmaCell}{Input}
		\%+\mmaDef{gD}[\{\m{$\mu$}\},\mmaDef{gTimes}[\mmaDef{dc}[\m{$\epsilon$},\{\}],\mmaDef{y}[\{u[a,b]\},\{u[\m{$\mu$}]\}],\mmaDef{s}[\m{$\psi$},\{\}]]]//\mmaDef{gOrd}
		  //\mmaDef{gSimplify}
\end{mmaCell}
\begin{mmaCell}{Output}
		$\Ii$\mmaSub{F}{$\lambda$1,$\lambda$2}($\overline{\epsilon}$$\star$\mmaSup{$\gamma$}{$\lambda$2}$\star$(\mmaSup{d}{$\lambda$1}$\widetilde{\psi}$))-\mmaFrac{1}{2}$\Ii$(\mmaSub{F}{$\lambda$1,$\lambda$2})($\overline{\epsilon}$$\star$(\mmaSup{$\gamma$}{$\lambda$1,$\lambda$2,$\lambda$3})$\star$(\mmaSub{d}{$\lambda$3}$\widetilde{\psi}$))
\end{mmaCell}
Now we add the variation of the Yang-Mills term. For this, we need to use the rule that expresses $F_{\mu \nu}$ in terms of $A_\mu$, before we take the variation, since the variation is defined for $A_\mu$ and not $F_{\mu \nu}$. We also express $F_{\mu \nu}$ in terms of $A_\mu$ in the first term to be able to compare terms in the two expressions.
\begin{mmaCell}{Input}
		(\%/.\mmaDef{frule})-\mmaDef{gD}[\m{$\delta$},\{d[\m{$\epsilon$}]\},1/4*\mmaDef{gForm}[F,\{$\mu$,$\nu$\}]
		  *\mmaDef{gForm}[F,\{u[$\mu$,$\nu$]\}]/.\mmaDef{frule}]//\mmaDef{gSimplify}
\end{mmaCell}
\begin{mmaCell}{Output}
		$\Ii$(\mmaSub{d}{$\lambda$1}\mmaSub{A}{$\lambda$})($\overline{\epsilon}$$\star$(\mmaSup{$\gamma$}{$\lambda$1,$\lambda$2,$\lambda$3})$\star$(\mmaSub{d}{$\lambda$2}$\widetilde{\psi}$))
\end{mmaCell}
We note that this term vanishes after partial integration. Thus we have shown that the supersymmetry variation of the Lagrangian vanishes up to a total derivative.

We then move on to discuss the non-Abelian case. After a few lines of computation that goes through essentially like the one above, the main difference being that derivatives are replaced with covariant derivatives, we will find out that there is an additional term that is not a total derivative. 
\begin{align}
\delta_\epsilon \mathcal{L} = \partial_\mu(\dotsc) + \frac{i}{2} \text{Tr}(\overline{\psi}\gamma^\mu[\overline{\epsilon}\gamma_\mu \psi,\psi])
\end{align}
By using some group theory, we can express the trace over the gauge group indices as
\begin{align}
\frac{i}{2} \text{Tr}(\overline{\psi}\gamma^\mu[\overline{\epsilon}\gamma_\mu \psi,\psi]) = \frac{1}{2} f_{abc}\overline{\psi}_a \gamma_\mu \psi_b \overline{\epsilon} \gamma^\mu \psi_c. \label{non-abelian_term}
\end{align}
The supersymmetry variation of the action vanishes if we can show that this term must be zero. To do so, we use the Fierz rearrangement. Let us consider rearranging the two bilinears appearing in the expression. For this, we need to remember to assume that all spinors are Majorana-Weyl, and to set \mtd{g$\gamma$5Convention} to \mtd{"ToRight"} to make sure that the expressions are simplified using the Weyl property of the spinors.
\begin{mmaCell}{Input}
		\mmaDef{gAddAssumptions}[\{\mmaDef{s}[\m{$\psi$},\{a\}]$\in$gPositiveChiral,
		  \mmaDef{s}[\m{$\psi$},\{b\}]$\in$gPositiveChiral,\mmaDef{s}[\m{$\psi$},\{c\}]$\in$gPositiveChiral,
		  \mmaDef{s}[\m{$\psi$},\{a\}]$\in$gMajorana,\mmaDef{s}[\m{$\psi$},\{b\}]$\in$gMajorana,,\mmaDef{s}[\m{$\psi$},\{c\}]$\in$gMajorana\}];
	
		\mmaDef{gFierzRearrange}[\mmaDef{gTimes}[\mmaDef{dc}[\m{$\psi$},\{a\}],y[\{d[\m{$\mu$}]\}],\mmaDef{s}[$\psi$,\{b\}]]*
		\mmaDef{gTimes}[\mmaDef{dc}[\m{$\epsilon$},\{\}],y[\{u[\m{$\mu$}]\}],\mmaDef{s}[$\psi$,\{c\}]]]//\mmaDef{gOrd}//\mmaDef{gSimplify}	
\end{mmaCell}
\begin{mmaCell}{Output}
		\mmaFrac{1}{2}($\bar{\epsilon}$$\star$\mmaSub{$\gamma$}{$\lambda$1}$\star$\mmaSub{$\widetilde{\psi}$}{b})(\mmaSub{$\bar{\psi}$}{a}$\star$\mmaSup{$\gamma$}{$\lambda$1}$\star$\mmaSub{$\widetilde{\psi}$}{c})-\mmaFrac{1}{24}($\bar{\epsilon}$$\star$\mmaSub{$\gamma$}{$\lambda$1,$\lambda$2,$\lambda$3}$\star$\mmaSub{$\widetilde{\psi}$}{b})(\mmaSub{$\bar{\psi}$}{a}$\star$\mmaSup{$\gamma$}{$\lambda$1,$\lambda$2,$\lambda$3}$\star$\mmaSub{$\widetilde{\psi}$}{c})
\end{mmaCell}
Since $\psi_a$ and $\psi_c$ are Majorana spinor, we have the relation 
\begin{align}
\overline{\psi}_a \gamma^{(3)} \widetilde{\psi}_c = \overline{\psi}_c \gamma^{(3)} \widetilde{\psi}_a,
\end{align}
as we can easily verify.
\begin{mmaCell}{Input}
		(\mmaSub{\m{$\overline{\psi}$}}{a}\m{$\star$}\mmaSup{\m{$\gamma$}}{\m{$\mu$},\m{$\nu$},\m{$\sigma$}}\m{$\star$}\mmaSub{\m{$\widetilde{\psi}$}}{c})==(\mmaSub{\m{$\overline{\psi}$}}{c}\m{$\star$}\mmaSup{\m{$\gamma$}}{\m{$\mu$},\m{$\nu$},\m{$\sigma$}}\m{$\star$}\mmaSub{\m{$\widetilde{\psi}$}}{a})//\mmaDef{gSimplify}
\end{mmaCell}
\begin{mmaCell}{Output}
		True
\end{mmaCell}
Using this, and the previous Fierz identity, we are easily able to derive the following identity.
\begin{align}
\frac{1}{2} f_{abc}\overline{\psi}_a \gamma_\mu \psi_b \overline{\epsilon} \gamma^\mu \psi_c = \frac{1}{4} f_{abc} \overline{\psi}_a \gamma_\mu \psi_c \overline{\epsilon} \gamma^\mu \psi_b  = -\frac{1}{4} f_{abc}  \overline{\psi}_a \gamma_\mu \psi_b \overline{\epsilon} \gamma^\mu \psi_c,
\end{align}
from which it follows that the term \eqref{non-abelian_term} vanishes. We have therefore shown that the action is supersymmetric also in the non-Abelian case.
\vspace{-5pt}
\subsubsection{Supersymmetry Transformations of D=4 $\mathcal{N}=1$ SYM} \label{sect_Examples_D=4_SYM}
\vspace{-5pt}
In this section, we will consider the four-dimensional supersymmetric Yang-Mills theory given by the Lagrangian
\begin{align}
\mathcal{L} = - \frac{1}{4} F_{\mu \nu} F^{\mu \nu} - \frac{1}{2} \overline{\psi} \slashed{D} \psi + \frac{1}{2} H^2,
\end{align}
where we have kept the auxiliary field $H$, as we will need this for the supersymmetry algebra. The supersymmetry transformations on the fields are as follows \cite{Freedman:2012zz}.
\begin{equation}
\begin{split}
\delta_\epsilon A_\mu &= - \frac{1}{2} \overline{\epsilon} \gamma_\mu \psi,\\[3pt]
\delta_\epsilon \lambda &= \frac{1}{4} \gamma^{\mu \nu} F_{\mu \nu} \epsilon + \frac{i}{2} \gamma_* H \epsilon,\\[3pt]
\delta_\epsilon H &= \frac{i}{2} \overline{\epsilon} \gamma_* \slashed{D} \psi.
\end{split}
\end{equation}
We will now endeavour to find the algebra satisfied by these transformations. This can be done by computing the commutators of the transformations $\delta_{\epsilon_1}$ and $\delta_{\epsilon_2}$ on the fields $A_\mu$, $\psi$ and $H$. We will this time confine ourselves to the Abelian case, since the generalisation to the non-Abelian case is again straightforward.

Let us first define the Clifford algebra that we are using. We work in four dimensions with Lorentzian signature, so we can take $\epsilon=\eta=1$
\begin{mmaCell}{Input}
		\mmaDef{gSetRep}[\{-1,1,1,1\},1,1];
\end{mmaCell}
We then proceed, as in the previous example, to define derivatives corresponding to the supersymmetry transformations.
\begin{mmaCell}{Input}
		\mmaDef{gClearDerivatives}[]
		\mmaDef{gD}[\m{$\delta$},\{u[],d[\mmaPat{e\_}]\},\mmaDef{gForm}[A,\{u[],d[\mmaPat{a\_}]\}]]:=
		   -1/2*\mmaDef{gTimes}[\mmaDef{s}[\mmaPat{e},\{\}],y[\{d[\mmaPat{a}]\}],\mmaDef{s}[\m{$\psi$},\{\}]]
		\mmaDef{gD}[\m{$\delta$},\{u[],d[\m{$\epsilon$}]\},\mmaDef{gForm}[A,\{u[\mmaPat{a\_}],d[]\}]]:=
		   -1/2*\mmaDef{gTimes}[\mmaDef{s}[\m{$\epsilon$},\{\}],y[\{u[\mmaPat{a}]\}],\mmaDef{s}[\m{$\psi$},\{\}]]
				
		\mmaDef{gD}[\m{$\delta$},\{u[],d[\m{$\epsilon$}]\},\mmaDef{s}[\m{$\psi$},\{\},0]]:=
		  I/2*\mmaDef{gForm}[F,\{a,b\}]\mmaDef{gTimes}[y[\{u[a,b]\}],\mmaDef{s}[\m{$\epsilon$},\{\}]]+
		  1/2*I*D*\mmaDef{gTimes}[m[\m{$\gamma$}5],s[\m{$\epsilon$},\{\}]]
		  
		\mmaDef{gD}[\m{$\delta$},\{u[],d[\mmaPat{e\_}]\},H]:=
		  1/2*I*\mmaDef{gTimes}[\mmaDef{dc}[\mmaPat{e},\{\}],\mmaDef{m}[\m{$\gamma$5}],\mmaDef{y}[\{u[c]\}],\mmaDef{gD}[\{c\},\mmaDef{s}[\m{$\psi$},\{\}]]]
		
		\mmaDef{gD}[\m{$\delta$},\{u[],d[\mmaPat{e\_}]\},\mmaDef{s}[\m{$\epsilon$1},\{\},0]]:=0
		\mmaDef{gD}[\m{$\delta$},\{u[],d[\mmaPat{e\_}]\},\mmaDef{s}[\m{$\epsilon$2},\{\},0]]:=0
\end{mmaCell}
Then we will make the required assumptions on different spinors. We are assuming that both $\psi$, and the spinors parametrising supersymmetry transformations, $\epsilon_1$ and $\epsilon_2$, are Majorana. In addition, we are still assuming that $F_{\mu \nu}$ and $A_\mu$ are real.
\begin{mmaCell}{Input}
		\mmaDef{gClearAssumptions}[];
		\mmaDef{gAddAssumptions}[\{A$\in$Reals,F$\in$Reals,\mmaDef{s}[\m{$\psi$},\{\}]$\in$gMajorana,
		\mmaDef{s}[\m{$\epsilon$1},\{\}]$\in$gMajorana,\mmaDef{s}[\m{$\epsilon$2},\{\}]$\in$gMajorana\}]
\end{mmaCell}
After this, the computations can be done in few lines. We begin with computing $[\delta_{\epsilon_2},\delta_{\epsilon_1}]A_\mu$. We can do this simply by taking the variation $\delta_{\epsilon_1} A_\mu$, and then taking the variation $\delta_{\epsilon_2} \delta_{\epsilon_1} A_\mu$ of it. After this, we exchange the order of variations, and subtract the two terms. Finally, \mtd{gOrd} can be used to simplify the resulting expressions that contains products of gamma matrices. 
\begin{mmaCell}{Input}
		\mmaDef{gD}[\m{$\delta$},\{\m{$\epsilon$2}\},\mmaDef{gD}[\m{$\delta$},\{\m{$\epsilon$1}\},\mmaDef{gForm}[A,\{\m{$\mu$}\}]]]
		  -\mmaDef{gD}[\m{$\delta$},\{\m{$\epsilon$1}\},\mmaDef{gD}[\m{$\delta$},\{\m{$\epsilon$2}\},\mmaDef{gForm}[A,\{\m{$\mu$}\}]]]//\mmaDef{gOrd}//\mmaDef{gSimplify}
\end{mmaCell}
\begin{mmaCell}{Output}
		\mmaFrac{1}{2}\mmaSub{F}{$\lambda$1,$\mu$}($\overline{\epsilon}$1$\star$\mmaSup{$\gamma$}{$\lambda$1}$\star$$\widetilde{\epsilon}$2)
\end{mmaCell}
This is already enough to establish the first commutation relation.
\begin{align}
[\delta_{\epsilon_2},\delta_{\epsilon_1}]A_\mu = \frac{1}{2} \overline{\epsilon_1} \gamma^\lambda \epsilon_2 F_{\lambda \mu}.
\end{align}
Next, we consider the supersymmetry transformations on the spinor $\psi$. We begin similarly to the previous computation. There is a small difference, though. Since the variation of $\psi$ contains $F_{\mu \nu}$, we must express this in terms of $A_\mu$ so that we get the correct second variation. We do this by using the rule \mtd{frule}, which we have defined previously to express $F_{\mu \nu}$ in terms of $A_\mu$. 
\begin{mmaCell}{Input}
		\mmaDef{gD}[\m{$\delta$},\{\m{$\epsilon$2}\},\mmaDef{gD}[\m{$\delta$},\{\m{$\epsilon$1}\},\mmaDef{s}[\m{$\psi$},\{\}]]/.\mmaDef{frules}]
		  -\mmaDef{gD}[\m{$\delta$},\{\m{$\epsilon$1}\},\mmaDef{gD}[\m{$\delta$},\{\m{$\epsilon$2}\},\mmaDef{s}[\m{$\psi$},\{\}]]/.\mmaDef{frules}]//\mmaDef{gOrd}//\mmaDef{gSimplify}
\end{mmaCell}
\begin{mmaCell}{Output}
		\mmaFrac{1}{4}(\mmaSup{$\gamma$}{$\lambda$1,$\lambda$2}$\star$$\widetilde{\epsilon}2$)($\overline{\epsilon}1$$\star$\mmaSub{$\gamma$}{$\lambda$2}$\star$(\mmaSub{d}{$\lambda$1}$\widetilde{\psi}$))-\mmaFrac{1}{4}(\mmaSup{$\gamma$}{$\lambda$1,$\lambda$2}$\star$$\widetilde{\epsilon}$1)($\overline{\epsilon}$2$\star$\mmaSub{$\gamma$}{$\lambda$2}$\star$(\mmaSub{d}{$\lambda$1}$\widetilde{\psi}$))+
		  \mmaFrac{1}{4}(\mmaSub{$\gamma$}{*}$\star$$\widetilde{\epsilon}$2)($\overline{\epsilon}$1$\star$\mmaSub{$\gamma$}{*}$\star$\mmaSup{$\gamma$}{$\lambda$1}$\star$(\mmaSub{d}{$\lambda$1}$\widetilde{\psi}$))-\mmaFrac{1}{4}(\mmaSub{$\gamma$}{*}$\star$$\widetilde{\epsilon}$1)($\overline{\epsilon}$2$\star$\mmaSub{$\gamma$}{*}$\star$\mmaSup{$\gamma$}{$\lambda$1}$\star$(\mmaSub{d}{$\lambda$1}$\widetilde{\psi}$))
\end{mmaCell}
This is not, however, quite the form that we are looking for. We would like to have something of the form $\overline{\epsilon}_2 \gamma^\mu \epsilon_1 (\dotsc)$, so it looks like using the Fierz rearrangement formula could be useful to get this into that form.
\begin{mmaCell}{Input}
		\% //\mmaDef{gFierz}
\end{mmaCell}
\begin{mmaCell}{Output}
		\mmaFrac{1}{2}(\mmaSub{d}{$\lambda$1}$\widetilde{\psi}$)($\overline{\epsilon}$1$\star$\mmaSup{$\gamma$}{$\lambda$1}$\star$$\widetilde{\epsilon}$2)
\end{mmaCell}
This is the second commutation relation.
\begin{align}
[\delta_{\epsilon_2},\delta_{\epsilon_1}]\psi = \frac{1}{2} \overline{\epsilon_1} \gamma^\lambda \epsilon_2 \partial_\lambda \psi.
\end{align}
The last remaining relation is that involving the auxiliary field $H$.
\begin{mmaCell}{Input}
		\mmaDef{gD}[\m{$\delta$},\{\m{$\epsilon$2}\},\mmaDef{gD}[\m{$\delta$},\{\m{$\epsilon$1}\},H]]
		-\mmaDef{gD}[\m{$\delta$},\{\m{$\epsilon$1}\},\mmaDef{gD}[\m{$\delta$},\{\m{$\epsilon$2}\},H]]//\mmaDef{gOrd}//\mmaDef{gSimplify}
\end{mmaCell}
\begin{mmaCell}{Output}
		\mmaFrac{1}{2}(\mmaSub{d}{$\lambda$1}H)($\overline{\epsilon}$1$\star$\mmaSup{$\gamma$}{$\lambda$1}$\star$$\widetilde{\epsilon}$2)+\mmaFrac{I}{4}(\mmaSub{d}{$\lambda$1}\mmaSub{F}{$\lambda$2,$\lambda$3})($\overline{\epsilon}$1$\star$\mmaSub{$\gamma$}{*}$\star$\mmaSup{$\gamma$}{$\lambda$1,$\lambda$2,$\lambda$3}$\star$$\widetilde{\epsilon}$2)
\end{mmaCell}
The last term here vanishes by the Bianchi identity, as can be easily verified by expressing $F_{\mu \nu}$ in terms of $A_\mu$. 
\begin{mmaCell}{Input}
 	\%/.\mmaDef{frule}//\mmaDef{gSimplify}
\end{mmaCell}
\begin{mmaCell}{Output}
		\mmaFrac{1}{2}(\mmaSub{d}{$\lambda$1}H)($\overline{\epsilon}$1$\star$\mmaSup{$\gamma$}{$\lambda$1}$\star$$\widetilde{\epsilon}$2)
\end{mmaCell}
This establishes the final commutation relation.
\begin{align}
[\delta_{\epsilon_2},\delta_{\epsilon_1}]H = \frac{1}{2} \overline{\epsilon_1} \gamma^\lambda \epsilon_2 \partial_\lambda H.
\end{align}

\vspace{-5pt}
\section*{Acknowledgements}
\vspace{-5pt}
We wish to thank Max H\"ubner and Marieke van Beest for useful discussions, Philip Candelas and Xenia de la Ossa for comments and suggestions, and Mikko Lahdensivu for comments on the manuscript. We also thank Emil Aaltosen s\"a\"ati\"o for providing computer resources used in completing this work. The author is supported by Osk. Huttusen s\"a\"ati\"o.
\newpage
\appendix

\numberwithin{equation}{subsection}

\setcounter{equation}{0}

\section{List of Objects} \label{app_Objects}
\tcbset{colframe=blue!75!black,colback=blue!5!white,colupper=black,fonttitle=\bfseries,nobeforeafter,center title}
\tcbox[left=0mm,right=0mm,top=0mm,bottom=0mm,boxsep=0mm,toptitle=0.5mm,bottomtitle=0.5mm,title=Objects]{
	\begin{tabular}{|l|c|}
	\hline
	Object & Description \\ \hline
	\mtd{gBL[F,\{i\},\{u[$\mu$],d[$\nu$]\},n]}& \ref{gBL_infobox}  \\ \hline		
	\mtd{gCase[\{X1\},...,\{Xn\}]} & \ref{gCase_infobox}  \\ \hline
	\mtd{gD[\{u[$\mu$],d[$\nu$]\},expr]}& \ref{gD_infobox}  \\ \hline	
	\mtd{dc[x,\{i\},n]} & \ref{dc_infobox}   \\\hline	
	\mtd{gEqual[$\mu$1,...,$\mu$n]}& \ref{gEqual_infobox}  \\ \hline	
	\mtd{gForm[F,{u[$\mu$],d[$\nu$]}]}& \ref{gForm_infobox}  \\ \hline
	\mtd{gIm[expr]}& \ref{gIm_infobox}  \\ \hline
	\mtd{lcc[x,\{i\},n]}&  \ref{lcc_infobox} \\ \hline	
	\mtd{m[X,n]}& \ref{m_infobox}  \\ \hline
	\mtd{gMajorana}& \ref{gMajorana_infobox}  \\ \hline
	\mtd{gNegativeChiral}& \ref{gNegativeChiral_infobox}  \\ \hline
	\mtd{gPositiveChiral}& \ref{gPositiveChiral_infobox}  \\ \hline
	\mtd{rcc[x,\{i\},n]}&  \ref{rcc_infobox} \\ \hline	
	\mtd{gRe[expr]}& \ref{gRe_infobox}\\ \hline	
	\mtd{s[x,\{i\},n]} & \ref{s_infobox}  \\ \hline	
	\mtd{gSymm[F,{u[$\mu$],d[$\nu$]}]}& \ref{gSymm_infobox}  \\ \hline
	\mtd{gTensor[F,{u[$\mu$],d[$\nu$]}]}& \ref{gTensor_infobox}  \\ \hline	
	\mtd{gTensorProduct[x1,...,xm]}& \ref{gTensorProduct_infobox}  \\ \hline	
	\mtd{gUnequal[$\mu$1,...,$\mu$n]}& \ref{gUnequal_infobox}  \\ \hline	
	\mtd{y[\{a1\},...,\{an\},n]} & \ref{y_infobox} \\ \hline
	\mtd{g$\Omega$[a1,...,an][x1,...,xm]} & \ref{gOmega_infobox} \\ \hline										
\end{tabular}
}\hfill

\section{List of Functions} \label{app_Functions}

\tcbset{colframe=red!75!black,colback=red!5!white,colupper=black,fonttitle=\bfseries,nobeforeafter,center title}

\tcbox[left=0mm,right=0mm,top=0mm,bottom=0mm,boxsep=0mm,toptitle=0.5mm,bottomtitle=0.5mm,title=Functions]{
	\begin{tabular}{|l|c|}
	\hline
	Function & Description \\ \hline
	\mtd{gAddAssumptions[expr]}& \ref{gAddAssumptions_infobox} \\ \hline
	\mtd{gAllowedSpinors[n]}& \ref{gAllowedSpinors_infobox} \\ \hline
	\mtd{gAllowedSpinors[signature,$\epsilon$,$\eta$]}& \ref{gAllowedSpinors_infobox} \\ \hline		
	\mtd{gBLToProduct[expr]}& \ref{gBLToProduct_infobox} \\ \hline			
	\mtd{gClearAssumptions[]}& \ref{gClearAssumptions_infobox} \\ \hline	
	\mtd{gClearBilinearNames[]}& \ref{gClearBilinearNames_infobox} \\ \hline
	\mtd{gClearDecomposition[]} & \ref{gClearDecomposition_infobox} \\ \hline			
	\mtd{gClearDerivatives[]} & \ref{gClearDerivatives_infobox} \\ \hline	
	\mtd{gClearSpinors[]}& \ref{gClearSpinors_infobox} \\ \hline		
	\mtd{gClearTensorProperties[]}& \ref{gClearTensorProperties_infobox} \\ \hline	
	\mtd{gClearTensors[]}& \ref{gClearTensors_infobox} \\		 \hline	
	\mtd{Conjugate[expr]}& \ref{Conjugate_infobox}  \\ \hline	
	\mtd{gDecomposeToSubReps[expr]}& \ref{gDecomposeToSubReps_infobox}  \\ \hline
	\mtd{gDual[expr]}& \ref{gDual_infobox}  \\ \hline	
	\mtd{gExplicit[expr]}& \ref{gExplicit_infobox} \\ \hline	
	\mtd{gFierz[expr]}& \ref{gFierz_infobox}  \\ \hline			
	\mtd{gH[expr]}& \ref{gH_infobox}  \\ \hline
	\mtd{gOrd[expr]}& \ref{gOrd_infobox}  \\ \hline	
	\mtd{gProductToBL[expr]}& \ref{gProductToBL_infobox}  \\ \hline	
	\mtd{gReIm[expr]}& \ref{gReIm_infobox}    \\ \hline
	\mtd{gSetBilinearNames[n,bilinearNames,ccBilinearNames,$\gamma$5bilinearNames,}& \ref{gSetBilinearNames_infobox} \\
	\mtd{cc$\gamma$5BilinearNames,$\xi$,$\eta$]}&  \\ \hline
	\mtd{gSetCoordinates[{x1,...,xd}]}& \ref{gSetCoordinates_infobox} \\ \hline	
	\mtd{gSetDecomposition[obj,decomposition]}& \ref{gSetDecomposition_infobox} \\ \hline	
	\mtd{gSetDecomposition[y,n,decomposition]}& \ref{gSetDecomposition_infobox} \\ \hline				
	\mtd{gSetIndices[n,indices]} & \ref{gSetIndices_infobox} \\ \hline
	\mtd{gSetRep[signature,$\epsilon$,$\eta$,$\zeta$,startFrom0]} & \ref{gSetRep_infobox} \\ \hline
	\mtd{gSetRepMatrices[gammas,C]} & \ref{gSetRepMatrices_infobox} \\ \hline		
	\mtd{gSetSpinor[s[x,\{i\},n],components]}& \ref{gSetSpinor_infobox} \\ \hline	
	\mtd{gSetSubRep[n,dimension]} & \ref{gSetSubRep_infobox} \\ \hline	
	\mtd{gSetSubRepMatrices[n,gammas,C]} & \ref{gSetSubRepMatrices_infobox} \\ \hline		
	\mtd{gSetTensor[F,components]}& \ref{gSetTensor_infobox} \\ \hline	
	\mtd{gSimplify[expr]}& \ref{gSimplify_infobox} \\ \hline	
	\mtd{gT[expr]}& \ref{gT_infobox} \\ \hline	
	\mtd{gTimes[x1,x2,...,xn]}& \ref{gTimes_infobox} \\ \hline			
				
\end{tabular}
}\hfill

\section{List of Options} \label{app_Options}

\tcbset{colframe=green!75!black,colback=green!5!white,colupper=black,fonttitle=\bfseries,nobeforeafter,center title}

\tcbox[left=0mm,right=0mm,top=0mm,bottom=0mm,boxsep=0mm,toptitle=0.5mm,bottomtitle=0.5mm,title=Options]{
	\begin{tabular}{|l|c|}
	\hline
	Object & Description \\ \hline
	\mtd{gAConvention}& \ref{gAConvention_infobox}  \\ \hline	
	\mtd{gBConvention} & \ref{gBConvention_infobox}   \\ \hline
	\mtd{gBLConvention}& \ref{gBLConvention_infobox}  \\ \hline			
	\mtd{gCConvention} & \ref{gCConvention_infobox}  \\ \hline	
	\mtd{gIntertwinerOrder}&  \ref{gIntertwinerOrder_infobox} \\ \hline
	\mtd{gOperationOrder}& \ref{gOperationOrder_infobox}  \\ \hline
	\mtd{gSpinorType} & \ref{gSpinorType_infobox} \\ \hline		
	\mtd{gSumIndexIndicator} & \ref{gSumIndexIndicator_infobox} \\ \hline			
	\mtd{gUseSpecialRep} & \ref{gUseSpecialRep_infobox} \\ \hline
	\mtd{g$\gamma$5Convention}&  \ref{ggamma5Convention_infobox} \\ \hline		
\end{tabular}
}\hfill

\section{List of Other Reserved Expressions} \label{App_Reserved_Expressions}
\tcbset{colframe=black!75!white,colback=black!5!white,colupper=black,fonttitle=\bfseries,nobeforeafter,center title}

\tcbox[left=0mm,right=0mm,top=0mm,bottom=0mm,boxsep=0mm,toptitle=0.5mm,bottomtitle=0.5mm,title=Other reserved expressions]{
	\begin{tabular}{|l|c|}
		\hline
		Expression & Description \\ \hline
		\mtu{u} & Denoted upper indices. \\ \hline
		\mtu{d} & Denotes lower indices. \\ \hline				
		\mtu{$\lambda$x} & Denotes dummy indices. \\ \hline
		\mtu{$\gamma$5} & \mtd{m[\mtu{$\gamma$5}]} denotes the highest rank gamma matrix in even dimensions, $\gamma_*$. \\ \hline
		\mtu{$\sigma$1} & \mtd{m[\mtu{$\sigma$1}]} denotes the sigma matrix $\sigma_1$. \\ \hline
		\mtu{$\sigma$2} & \mtd{m[\mtu{$\sigma$2}]} denotes the sigma matrix $\sigma_2$. \\ \hline
		\mtu{$\sigma$3} & \mtd{m[\mtu{$\sigma$3}]} denotes the sigma matrix $\sigma_3$. \\ \hline		
		\mtu{Id} & \mtd{m[\mtu{Id}]} denotes the identity matrix. \\ \hline	
		\mtu{A} & \mtd{m[\mtu{A}]} denotes the $A$-intertwiner. \\ \hline	
		\mtu{B} & \mtd{m[\mtu{B}]} denotes the $B$-intertwiner. \\ \hline	
		\mtd{C} & \mtd{m[\mtd{C}]} denotes the $C$-intertwiner. \\ \hline														
		\mtu{$\eta$}& \mtd{gTensor[\mtu{$\eta$},\{\mtu{$\mu$},\mtu{$\nu$}\}]} denotes the metric $\eta_{\mu \nu}$ appearing in \eqref{Clifford_algebra_anticommutation}.  \\ \hline
		\mtu{$\epsilon$} & \mtd{gForm[\mtu{$\epsilon$},\{\mtu{$\mu$1},...,\mtu{$\mu$d}\}]} denotes the Levi-Civita symbol $\epsilon_{\mu_1,\dotsc,\mu_d}$.  \\ \hline				
	\end{tabular}
}\hfill

\newpage

\section{Properties of Special Matrices} \label{app_Rep_Properties}
\vspace{-5pt}
In this appendix we summarise the most relevant properties of sigma and gamma matrices and $A$- $B$- and $C$-intertwiners, and show how they can be decomposed in terms of subrepresentations in different cases.
\vspace{-5pt}
\subsection{Properties of Special Matrices}
\vspace{-5pt}
\subsubsection{Sigma Matrices}
\vspace{-5pt}
Sigma matrices are given by
\begin{align}
\sigma_1 = \begin{pmatrix}
0 & 1 \\
1 & 0 
\end{pmatrix},~~~
\sigma_2 = \begin{pmatrix}
0 & -i \\
i & 0 
\end{pmatrix},~~~
\sigma_3 = \begin{pmatrix}
1 & 0 \\
0 & -1 
\end{pmatrix}.
\end{align}
and satisfy the following commutation and anticommutation relations
\begin{equation}
\begin{split}
[\sigma_i,\sigma_j] &= 2i \epsilon_{ijk} \sigma_k,\\[3pt]
\{\sigma_i, \sigma_j\} &= 2 \delta_{ij}. \label{sigma_commutation_anticommutation}
\end{split}
\end{equation}
All of the sigma matrices are Hermitian, $\sigma_1$ and $\sigma_3$ are real and symmetric, while $\sigma_2$ is imaginary and antisymmetric.
\begin{equation}
\begin{split}
\sigma_i^\dag &= \sigma_i,\\
\sigma_1^T &= \sigma_1^* = \sigma_1,\\
\sigma_3^T &= \sigma_3^* = \sigma_3 \\
\sigma_2^T &= \sigma_2^* = -\sigma_2.
\end{split}
\end{equation}
\vspace{-5pt}
\subsubsection{Gamma Matrices}
\vspace{-5pt}
We assume that for any representation the gamma matrices have been chosen so that they have the following properties. Here $\alpha$ is a timelike index and $a$ a spacelike one.
\begin{equation}
\begin{split}
\gamma_\alpha^\dagger &= - \gamma_\alpha,\\
\gamma_a^\dagger &= \gamma_a.
\end{split}
\end{equation}
A particular representation that we call the special representation, can be built out of tensor products of sigma matrices. For the Euclidean signature $(+\dotsc+)$ it is given by
\begin{equation}
\begin{split}
\gamma^1 &= \sigma_1 \otimes I \otimes I \otimes \dotsc,\\
\gamma^2 &= \sigma_2 \otimes I \otimes I \otimes \dotsc,\\
\gamma^3 &= \sigma_3 \otimes \sigma_1 \otimes I \otimes \dotsc,\\
\gamma^4 &= \sigma_3 \otimes \sigma_2 \otimes I \otimes \dotsc,\\
\gamma^5 &= \sigma_3 \otimes \sigma_3 \otimes \sigma_1 \otimes \dotsc,\\
&\dotsc\\
\gamma^d &= \sigma_3 \otimes \sigma_3 \otimes \dotsc \otimes \sigma_2 \text{ in even dimensions,}\\
\gamma^d &= \sigma_3 \otimes \sigma_3 \otimes \dotsc \otimes \sigma_3 \text{ in odd dimensions.} \label{App_special_rep_gamma_matrices}
\end{split}
\end{equation}
In case with timelike directions, the matrices corresponding to the timelike directions can be multiplied by $i$ to get a representation with the correct signature.

In even dimensions there is only one representation up to unitary transformations, so that any representation is given by matrices of the form
\begin{align}
\gamma'_\mu = U \gamma_\mu U^{-1}
\end{align}
In odd dimensions there are two inequivalent representations that are related by exchanging the signs of the last gamma matrix, $\gamma^d \to -\gamma^d$.
\vspace{-5pt}
\subsubsection{A-intertwiner}
\vspace{-5pt}
In a generic representation the $A$-intertwiner is given in terms of gamma matrices by taking a product of all gamma matrices with timelike indices $A = \gamma^1 \gamma^2 \dotsc \gamma^t$. It satisfies the following relations.
\begin{equation}
\begin{split}
\gamma_n^\dagger &= (-1)^t A \gamma_n A^{-1},\\
A^{-1} &= A^\dagger = (-1)^{t(t+1)/2} A,\\
A^T &= (-1)^{t(t+1)/2} A^*. \label{AT=Ac}
\end{split}
\end{equation}
In addition to these, the special representation with the gamma matrices given by \eqref{App_special_rep_gamma_matrices} has the following properties.
\begin{equation}
\begin{split}
A^*&= (-1)^{\left\lfloor{t/2}\right\rfloor+t} A,\\
A^T &= (-1)^{\left\lfloor{t/2}\right\rfloor + t(t-1)/2 } A. \label{special_A_relations}
\end{split}
\end{equation}
\vspace{-5pt}
\subsubsection{C-intertwiner}
\vspace{-5pt}
In a generic representation the $C$-intertwiner has the following properties.
\begin{equation}
\begin{split}
C^T &= -\epsilon \, C, \label{C_properties_first}\\
\gamma^T_n &= - \eta \, C \gamma_n C^{-1},\\
C^{-1} &= C^\dagger =-\epsilon \, C^*.\\
\end{split}
\end{equation}
In addition for the special representation, for which the gamma matrices are given by \eqref{App_special_rep_gamma_matrices}, the $C$-intertwiners are given by.
\begin{equation}
\begin{split}
C &= \sigma_2 \otimes \sigma_1 \otimes \sigma_2 \otimes \dotsc \text{ for $\eta=1$},\\
C &= \sigma_1 \otimes \sigma_2 \otimes \sigma_1 \otimes \dotsc \text{ for $\eta=-1$}. \label{App_C_special_rep}
\end{split}
\end{equation}
And these satisfy the relations.
\begin{equation}
\begin{split}
C&= C^\dagger = C^{-1},\\
C^* &= C^T = - \epsilon \, C. \label{C_properties_last}
\end{split}
\end{equation}
\vspace{-5pt}
\subsubsection{B-intertwiner}
\vspace{-5pt}
In a generic representation the $B$-intertwiner is defined by the relation $B^T = C A^{-1}$, from which follows that $B = (-1)^{t(t+1)/2+1} \epsilon A^T C$.
$B$ satisfies the following relations.
\begin{equation}
\begin{split}
\gamma_n^* &= -(-1)^t\eta \, B \gamma_n B^{-1},\\
B^\dagger &= B^{-1},\\
B^* &= - \epsilon \eta^t (-1)^{t(t+1)/2} B^\dagger,\\
B &=  - \epsilon \eta^t (-1)^{t(t+1)/2} B^T.
\end{split}
\end{equation}
For the special representation $B$ can be expressed as $B=(-1)^{t^2+1+\left\lfloor{t/2}\right\rfloor} \epsilon A C$. 
\begin{equation}
\begin{split}
B^{-1} &= \eta^t (-1)^{\left\lfloor{t/2}\right\rfloor + t(3t+1)/2} B,\\
B^T &= -(-\eta)^t(-1)^{3t(t-1)/2} \epsilon \, B,\\
B^* &= -(-1)^{(\left\lfloor{t/2}\right\rfloor +t)}\epsilon \, B,\\
B^\dagger &= (-\eta)^t(-1)^{(\left\lfloor{t/2}\right\rfloor +3t(t-1)/2+t)}B.
\end{split}
\end{equation}
\vspace{-5pt}
\subsubsection{Highest Rank Gamma Matrix $\gamma_*$}
\vspace{-5pt}
In even dimensions $\gamma_*$ is defined by $\gamma_* = (-i)^{d/2+t} \gamma^0\dotsc\gamma^{d-1}$. Therefore, in any representation, it has the following properties:
\begin{equation}
\begin{split}
\gamma_*\gamma_n \gamma_*^{-1} &= - \gamma_n,\\
\gamma_*^\dagger &= \gamma_*^{-1} =\gamma_*,\\
\gamma_*^T &= \gamma_*^*.
\end{split}
\end{equation}
In addition, for the special representation, we have
\begin{align}
\gamma_*^T = \gamma_*^* = \gamma_*.
\end{align}
\vspace{-5pt}
\subsection{Commutation Relations of Intertwiners}
\vspace{-5pt}
For a generic representation the following commutation relations between different intertwiners hold.
\begin{equation}
\begin{split}
CA &= (-\eta)^t (-1)^{t(t-1)/2} A^T C,\\
BA &= \eta^t (-1)^{t+t^2} A^* B,\\
A\gamma_* &= (-1)^{t} \, \gamma_* A,\\
C\gamma_* &= (-\eta)^d (-1)^{d(d-1)/2} \, \gamma_*^T C,\\
B\gamma_* &= (-1)^{d/2+t} \, \gamma_*^* B.
\end{split}
\end{equation}
For the special representation, we have the following relations.
\begin{equation}
\begin{split}
AC &= (-\eta)^t (-1)^{\left\lfloor{t/2}\right\rfloor + t(t-1)} CA,\\
AB &= (-\eta)^t(-1)^{\left\lfloor{t/2}\right\rfloor+ t(t-1)} BA,\\
A\gamma_* &= (-1)^{t} \gamma_* A,\\
CB &= (-\eta)^t (-1)^{\left\lfloor{t/2}\right\rfloor + t(t-1)} BC,\\
C\gamma_* &= (-\eta)^d (-1)^{d(d-1)/2} \gamma_* C,\\
B\gamma_* &= (-1)^{d/2+t} \gamma_* B. \label{app_intertwiners_last}
\end{split}
\end{equation}
\vspace{-5pt}
\subsection{Bilinear Relations} \label{app_bilinear_relations}
\vspace{-5pt}
Here we summarise the relations that hold for commutative spinor bilinears in any representation. The relations for the anticommutative spinor can be obtained by simply changing the sign on the right-hand-side of the relations.\\
 
For any spinors $\xi_1$ and $\xi_2$ we have the following relations.
\begin{equation}
\begin{split}
(\overline{\xi}_1 \gamma^{(n)} \xi_2)^* &= (-1)^{nt}(-1)^{n(n-1)/2}(-1)^{t(t+1)/2} \, \overline{\xi}_2 \, \gamma^{(n)} \, \xi_1,\\[3pt]
(\overline{\xi}_1 \gamma_* \gamma^{(n)} \xi_2)^* &= (-1)^{t(t+1)/2}(-1)^{n(n+1)/2}(-1)^{t(d-1)}(-1)^{tn} \, \overline{\xi}_2 \, \gamma_* \gamma^{(n)} \, \xi_1
\end{split}
\end{equation}
From this follows that 
\begin{equation}
\begin{split}
&\overline{\xi} \gamma^{(n)} \xi \text{ is }
\begin{cases}
\text{ real if } (-1)^{nt}(-1)^{1/2n(n+1)}(-1)^{t(t+1)/2}=1,\\[3pt]
\text{ imaginary if } (-1)^{nt}(-1)^{1/2n(n+1)}(-1)^{t(t+1)/2}=-1.
\end{cases}\\[5pt]
&\overline{\xi}  \gamma_* \gamma^{(n)} \xi \text{ is }
\begin{cases}
\text{ real if } (-1)^{t(t+1)/2}(-1)^{n(n+1)/2}(-1)^{t(d-1)}(-1)^{tn}=1,\\[3pt]
\text{ imaginary if } (-1)^{t(t+1)/2}(-1)^{n(n+1)/2}(-1)^{t(d-1)}(-1)^{tn}=-1.
\end{cases}
\end{split}
\end{equation}
We also have the following relations for charge conjugate bilinears.
\begin{equation}
\begin{split}
\overline{\xi}_1^c \gamma^{(n)} \xi_2 &= -\epsilon(-\eta)^n(-1)^{n(n-1)/2} \, \overline{\xi}_2^c \gamma^{(n)} \xi_1,\\[3pt]
\overline{\xi}_1^c \gamma_* \gamma^{(n)} \xi_2 &= -\epsilon \eta^{n+d}(-1)^{d(d+1)/2}(-1)^{n(n-1)/2} \, \overline{\xi}_2^c \gamma_* \gamma^{(n)} \xi_1
\end{split}
\end{equation}
In particular these allow us to deduce when certain charge conjugate bilinears vanish automatically.
\begin{equation}
\begin{split}
\overline{\xi}^c \gamma^{(n)} \xi &= 0 \qquad \text{ if } \epsilon(-\eta)^n(-1)^{n(n-1)/2} = 1,\\[3pt]
\overline{\xi}^c \gamma_* \gamma^{(n)} \xi &= 0 \qquad \text{ if } \epsilon \eta^{n+d}(-1)^{d(d+1)/2}(-1)^{n(n-1)/2} = 1,
\end{split}
\end{equation}
If $\xi_1$ and $\xi_2$ are Weyl spinors with chiralities $(\pm)_1$ and $(\pm)_2$, so that $\gamma_* \xi_1 = (\pm)_1 \xi_1, \gamma_* \xi_2 = (\pm)_2 \xi_2$, we get the following relations.
\begin{equation}
\begin{split}
\overline{\xi}_2 \gamma^{(n)} \xi_1 &= (-1)^{n+t} (\pm)_1 (\pm)_2 \overline{\xi}_2 \gamma^{(n)} \xi_1,\\[3pt]
\overline{\xi}_2 \gamma^{(n)} \xi_1 &= 0 \text{ if }
\begin{cases}
\xi_1 \text{ and } \xi_2 \text{ have the same chirality, and } (-1)^{n+t}=-1,\\[3pt]
\xi_1 \text{ and } \xi_2 \text{ have the oposite chiralities, and } (-1)^{n+t}=1.
\end{cases} \label{App_bl_relations_vanishing_chiral}
\end{split}
\end{equation}
If $\xi_1$ and $\xi_2$ are Majorana spinors, then.
\begin{equation}
\begin{split}
\overline{\xi}_1 \gamma^{(n)} \xi_2 &= -\epsilon (-\eta)^n (-1)^{n(n-1)/2} \, \overline{\xi_2} \gamma^{(n)} \xi_1 \label{app_bl_relations_Majorana_exchange}.\\[3pt]
\overline{\xi}_1 \gamma_* \gamma^{(n)} \xi_2 &= -\epsilon (-\eta)^{n+d} (-1)^{n(n+1)/2} \, (-1)^{d(d-1)/2} \overline{\xi}_2 \gamma_* \gamma^{(n)} \xi_1.
\end{split}
\end{equation}
Bilinears with charge conjugate spinors on both sides can be related to the "normal" bilinears trough the following relations.
\begin{equation}
\begin{split}
\overline{\xi}_1^c \, \gamma^{(n)} \, \xi_2^c &= \eta^{n+t}(-1)^{3t(t+1)/2}(-1)^{n(n+1)/2} \, \overline{\xi}_2 \, \gamma^{(n)} \, \xi_1,\\[3pt]
\overline{\xi}_1^c \, \gamma_* \, \gamma^{(n)} \xi_2^c &= \eta^{n+t+d}(-1)^{t(t-1)/2}(-1)^{d(d-1)/2}(-1)^{n(n+1)/2} \, \overline{\xi}_2 \, \gamma_* \gamma^{(n)} \, \xi_1.
\end{split}
\end{equation}
\vspace{-5pt}
\subsection{Decomposition to Subrepresentations} \label{App_Matrix_Properties_Decomposition}
\vspace{-5pt}
We have three qualitatively different cases when decomposing representation of Clifford algebras under $\text{Cliff}(d) \to \text{Cliff}(n) \times \text{Cliff}(d-n)$, depending whether we start with odd or even dimensional Clifford algebra, and whether the subalgebras are odd or even.
\vspace{-5pt}
\subsubsection{Odd $\to$ Even $\times$ Odd}
\vspace{-5pt}
In the case $d$ is odd, we have only one case to consider, since one of the subalgebras must have an even dimension and the other odd. Let us choose the first subalgebra to have an even dimension. The gamma matrices can be decomposed as:
\begin{equation}
\begin{split}
\gamma^\mu &= \gamma_{(1)}^\mu \otimes I_{(2)} \text{ for $\mu = 0,\dotsc,n-1$},\\[3pt]
\gamma^\alpha &= \gamma_{*(1)} \otimes \gamma_{(2)}^\alpha \text{ for $\alpha = n,\dotsc,d$}.
\end{split}
\end{equation}
Then the $A$-intertwiner can be written as
\begin{align}
A = \gamma^{\mu_0}\dotsc\gamma^{\mu_{t_1-1}}\gamma^{\alpha_0}\dotsc\gamma^{\alpha_{t_2-1}} = A_{(1)} \gamma_{*(1)}^{t_2} \otimes A_{(2)},
\end{align}
where $\mu_0,\dotsc,\mu_{t_1-1}$ and $\alpha_0,\dotsc,\alpha_{t_2-1}$ denote the timelike directions corresponding to the first and second subalgebras, respectively.

The C intertwiner can be written as
\begin{align}
C = C_{(1)} \otimes C_{(2)},
\end{align}
if the signs $\eta_1$ and $\eta_2$ are given in terms of $\eta$ of the full representation as
\begin{equation}
\begin{split}
\eta_1 &= \eta,\\
\eta_2 &= (-1)^{n(n-1)/2} \eta.
\end{split}
\end{equation}
The signs $\epsilon_1$ and $\epsilon_2$ are then determined by the values of $\eta_1$ and $\eta_2$ and the dimensions of corresponding representations (see table \ref{Table_epsilon_eta}).

Finally, the $B$-intertwiner is given, as before, by expression $B=(A^{-1}C)^T$, which in this case gives
\begin{align}
B = B_{(1)} (\gamma_{*(1)}^{t_2})^T \otimes B_{(2)}.
\end{align}
\vspace{-5pt}
\subsubsection{Even $\to$ Odd $\times$ Odd}
\vspace{-5pt}
In the second case the full algebra is odd-dimensional, and we decompose it into two even-dimensional subalgebras. In this case we can decompose the gamma matrices as
\begin{equation}
\begin{split}
\gamma^{\mu} &= \gamma^{\mu}_{(1)} \otimes I_{(2)} \otimes \sigma_1 \text{ for $\mu=0,\dotsc,n-1$},\\[3pt]
\gamma^{\alpha} &= I_{(1)} \otimes \gamma^{\alpha}_{(2)} \otimes \sigma_2 \text{ for $\alpha=n,\dotsc,d-1$}. 
\end{split} \label{gamma_matrix_decomposition_Even->OddxOdd}
\end{equation}
The $A$-matrix can then be expressed as
\begin{align}
A &= A_{(1)} \otimes A_{(2)} \otimes \sigma_1^{t_1} \sigma_2^{t_2},
\end{align}
where $t_1$ and $t_2$ again denote the number of timelike directions corresponding to the first and second subalgebras, respectively. 

In this case, the dimensions $n$ and $d-n$ completely determine the allowed $\eta_1$, $\eta_2$, $\epsilon_1$, and $\epsilon_2$. The decomposition of $C$ matrix is slightly different in each of these cases. For the representations of gamma matrices we chose above, we have that
\begin{equation}
\begin{tabular}{ l c l }
	$C = C_{(1)} \otimes C_{(2)} \otimes I;$ & \qquad & $\eta_1 = \eta_2 = \eta$, \\[3pt]
	$C = C_{(2)} \otimes C_{(2)} \otimes \sigma_2;$ & \qquad & $\eta_1 = -\eta, \eta_2 = \eta,$ \\[3pt]
	$C = C_{(2)} \otimes C_{(2)} \otimes \sigma_1;$ & \qquad & $\eta_1 = \eta, \eta_2 = -\eta,$ \\[3pt]
	$C = C_{(2)} \otimes C_{(2)} \otimes \sigma_3;$ & \qquad & $\eta_1 = \eta_2 = -\eta.$\\
\end{tabular}
\end{equation}
The $B$-intertwiner can be calculated using the expressions for the $A$- and $C$-intertwiners. We have again four different cases.

\begin{equation}
\begin{tabular}{ l c l }
$B = (-1)^{t_2} B_{(1)} \otimes B_{(2)} \otimes \sigma_1^{t_1} \sigma_2^{t_2};$ & \qquad & $\eta_1 = \eta_2 = \eta$, \\[3pt]
$B = (-1)^{t_2} B_{(1)} \otimes B_{(2)} \otimes \sigma_1^{t_1} \sigma_2^{t_2+1};$ & \qquad & $\eta_1 = -\eta, \eta_2 = \eta,$ \\[3pt]
$B = (-1)^{t_2} B_{(1)} \otimes B_{(2)} \otimes \sigma_1^{t_1+1} \sigma_2^{t_2};$ & \qquad & $\eta_1 = \eta, \eta_2 = -\eta,$ \\[3pt]
$B = (-1)^{t_2} B_{(1)} \otimes B_{(2)} \otimes \sigma_3 \sigma_1^{t_1} \sigma_2^{t_2};$ & \qquad & $\eta_1 = \eta_2 = -\eta.$\\
\end{tabular}
\end{equation}
In principle, we would not need to include the constant prefactors here, but we include them to be consistent in our conventions.

Finally, $\gamma_*$ decomposes as
\begin{equation}
\gamma_* = \zeta_1 \zeta_2 \, I_{(1)} \otimes I_{(2)} \otimes \sigma_3.
\end{equation}
\vspace{-5pt}
\subsubsection{Even $\to$ Even $\times$ Even}
\vspace{-5pt}
The last case is the one where an even-dimensional algebra is decomposed into two even-dimensional subalgebras. In this case, we can represent the gamma matrices as.
\begin{equation}
\begin{split}
\gamma^\mu &= \gamma_{(1)}^\mu \otimes I_{(2)} \text{ for $\mu=0,\dotsc,n-1$,}\\[3pt]
\gamma^\alpha &= \gamma_{*(1)} \otimes \gamma^\alpha \text{ for $\alpha=n,\dotsc,d-1$,}
\end{split}
\end{equation}
The $A$-intertwiner is then given by
\begin{align}
A = A_{(1)} \gamma_{*(1)}^{t_2} \otimes A_{(2)}.
\end{align}
If the signs $\eta_1$ and $\eta_2$ are determined by $\eta$ as
\begin{equation}
\begin{split}
\eta_1 &= \eta,\\
\eta_2 &= (-1)^{1/2n(n-1)} \eta,
\end{split}
\end{equation}
then the remaining signs $\epsilon_1$ and $\epsilon_2$ are determined by these (see table \ref{Table_epsilon_eta}), and the $C$-intertwiner is simply
\begin{align}
C = C_{(1)} \otimes C_{(2)}.
\end{align}
The $B$-intertwiner is
\begin{align}
B = B_{(1)} (\gamma_{*(1)}^{t_2})^T \otimes B_2,
\end{align}
where $t_2$ denotes the number of timelike directions corresponding to the second subalgebra.

The highest rank Clifford algebra element $\gamma_*$ decomposes as
\begin{equation}
\gamma_* = \gamma_{*(1)} \otimes \gamma_{*(2)}.
\end{equation}
\bibliographystyle{JHEP}
\bibliography{Gamma_package}
\end{document}